# Interface-Induced Phenomena in Magnetism

Frances Hellman*
*Department of Physics, University of California, Berkeley, Berkeley, California 94720, USA*
*Materials Sciences Division, Lawrence Berkeley National Laboratory, Berkeley, California 94720, USA*

Axel Hoffmann
*Materials Science Division, Argonne National Laboratory, Argonne, Illinois 60439, USA*

Yaroslav Tserkovnyak
*Department of Physics and Astronomy, University of California, Los Angeles, California 90095, USA*

Geoffrey S. D. Beach
*Department of Materials Science and Engineering, Massachusetts Institute of Technology, Cambridge, Massachusetts 02139, USA*

Eric E. Fullerton
*Center for Memory and Recording Research, University of California, San Diego, 9500 Gilman Drive, La Jolla, California 92093-0401, USA*

Chris Leighton
*Department of Chemical Engineering and Materials Science, University of Minnesota, Minneapolis, Minnesota 55455, USA*

Allan H. MacDonald
*Department of Physics, University of Texas at Austin, Austin, Texas 78712-0264, USA*

Daniel C. Ralph
*Physics Department, Cornell University, Ithaca, New York 14853, USA*
*Kavli Institute at Cornell, Cornell University, Ithaca, New York 14853, USA*

Dario A. Arena
*Department of Physics, University of South Florida, Tampa, Florida 33620-7100, USA*

Hermann A. Dürr
*Stanford Institute for Materials and Energy Sciences, SLAC National Accelerator Laboratory, 2575 Sand Hill Road, Menlo Park, California 94025, USA*

Peter Fischer
*Materials Sciences Division, Lawrence Berkeley National Laboratory, Berkeley, California 94720, USA*
*Physics Department, University of California, 1156 High Street, Santa Cruz, California 94056, USA*

* corresponding author fhellman@berkeley.edu

Julie Grollier
*Unité Mixte de Physique CNRS/Thales and Université Paris Sud 11, 1 Avenue Fresnel, 91767 Palaiseau, France*

Joseph P. Heremans
*Department of Mechanical and Aerospace Engineering, The Ohio State University, Columbus, Ohio 43210, USA*
*Department of Materials Science and Engineering, The Ohio State University, Columbus, Ohio 43210, USA*
*Department of Physics, The Ohio State University, Columbus, Ohio 43210, USA*

Tomas Jungwirth
*Institute of Physics, Academy of Sciences of the Czech Republic, Cukrovarnicka 10, 162 53 Praha 6, Czech Republic*
*School of Physics and Astronomy, University of Nottingham, Nottingham NG7 2RD, United Kingdom*
Alexey V. Kimel
*Radboud University, Institute for Molecules and Materials, Nijmegen 6525 AJ, The Netherlands*

Bert Koopmans
*Department of Applied Physics, Center for NanoMaterials, COBRA Research Institute, Eindhoven University of Technology, P.O. Box 513, 5600 MB Eindhoven, The Netherlands*

Ilya N. Krivorotov
*Department of Physics and Astronomy, University of California, Irvine, California 92697, USA*

Steven J. May
*Department of Materials Science & Engineering, Drexel University, Philadelphia, Pennsylvania 19104, USA*

Amanda K. Petford-Long
*Materials Science Division, Argonne National Laboratory, 9700 South Cass Avenue, Argonne, Illinois 60439, USA*
*Department of Materials Science and Engineering, Northwestern University, 2220 Campus Drive, Evanston, Illinois 60208, USA*

James M. Rondinelli
*Department of Materials Science and Engineering, Northwestern University, Evanston, Illinois 60208, USA*

Nitin Samarth
*Department of Physics, The Pennsylvania State University, University Park, Pennsylvania 16802, USA*

Ivan K. Schuller
*Department of Physics and Center for Advanced Nanoscience, University of California, San*

*Diego, La Jolla, California 92093, USA*
*Materials Science and Engineering Program, University of California, San Diego, La Jolla, California 92093, USA*

Andrei N. Slavin
*Department of Physics, Oakland University, Rochester, Michigan 48309, USA*

Mark D. Stiles
*Center for Nanoscale Science and Technology, National Institute of Standards and Technology, Gaithersburg, Maryland 20899-6202, USA*

Oleg Tchernyshyov
*Department of Physics and Astronomy, The Johns Hopkins University, Baltimore, Maryland 21218, USA*

André Thiaville
*Laboratoire de Physique des Solides, UMR CNRS 8502, Université Paris-Sud, 91405 Orsay, France*

Barry L. Zink
*Department of Physics and Astronomy, University of Denver, Denver, CO 80208, USA*

## Abstract

This article reviews static and dynamic interfacial effects in magnetism, focusing on interfacially-driven magnetic effects and phenomena associated with spin-orbit coupling and intrinsic symmetry breaking at interfaces. It provides a historical background and literature survey, but focuses on recent progress, identifying the most exciting new scientific results and pointing to promising future research directions. It starts with an introduction and overview of how basic magnetic properties are affected by interfaces, then turns to a discussion of charge and spin transport through and near interfaces and how these can be used to control the properties of the magnetic layer. Important concepts include spin accumulation, spin currents, spin transfer torque, and spin pumping. An overview is provided to the current state of knowledge and existing review literature on interfacial effects such as exchange bias, exchange spring magnets, spin Hall effect, oxide heterostructures, and topological insulators. The article highlights recent discoveries of interface-induced magnetism and non-collinear spin textures, non-linear dynamics including spin torque transfer and magnetization reversal induced by interfaces, and interfacial effects in ultrafast magnetization processes.

## Table of Contents

## I. INTRODUCTION

Magnetic materials provide an intellectually rich arena for fundamental scientific discovery and for the invention of faster, smaller, more energy-efficient technologies. The effects of *spin-orbit coupling* and *symmetry breaking* at the interface between a magnet and non-magnet are of particular interest and importance. The discovery, three decades ago, of giant magnetoresistance [Baibich *et al.*, 1988; Binasch *et al.*, 1989] highlighted the intimate relationship of charge transport and magnetic structure, including the importance of interfaces and layered structures, and brought about the now flourishing field of spintronics [Zutic *et al.*, 2004]. Recent focus turned to the significant role played by spin-orbit coupling at interfaces and how this affects the interplay between charge, spin, orbital, and lattice degrees of freedom; quantum confinement; interface and surface states; energies of competing ground states, including exotic spin states; and the effects of strong electron correlations, disorder, and frustration.

Coupling between distinct order parameters across interfaces yields important science

(including proximity effects, exchange bias, and exchange spring-induced hard magnets) that has been studied for decades. However, magnetism is based on strong short-range correlations between electronic spin and orbital degrees of freedom, and these are inherently altered at interfaces, particularly in the presence of strong spin-orbit coupling and during transient responses to stimuli. Interfaces therefore not only modify bulk magnetic properties, but are also capable of creating magnetism from non-magnetic layers, altering the nature of a magnetic state, or impacting its dynamic evolution following an electrical, optical, thermal, or magnetic pulse.

From a technological perspective, the static and dynamic magnetic properties of condensed matter are at the heart of spin-based information and sensing technologies, and are essential for both information storage and power generation and conditioning. Magnetic systems offer various types of manipulatable states for information storage and computation, and their inherent nonvolatility makes them central to energy-related research and associated technologies. Interfaces play a fundamental role in these technologies; recent discoveries suggest they can create new types of magnetic states and new means for their rapid manipulation [Hoffmann and Bader, 2015]. An improved basic understanding of the interplay between the (exchange) spin-spin and (relativistic) spin-orbit interactions at interfaces should result in increased energy efficiency, improved functionality, reduced size, and increased speed of a range of magnetically-based devices.

This article reviews what is currently known about static and dynamic interfacial effects in magnetism, focusing particularly on the recent explosion in interfacially-driven magnetic phenomena associated with spin-orbit coupling and intrinsic symmetry breaking at interfaces, and identifying the most exciting new scientific results and anticipating areas for future research, as summarized in **Fig. 1**. We highlight recent discoveries of emergent magnetic properties at interfaces and in heterostructures, interface-induced non-collinear spin textures, out-of-equilibrium spin and charge response engendered by interfaces, and interfacial effects in ultrafast magnetization dynamics, while providing references to existing review literature on the more established interfacial phenomena such as exchange bias, exchange spring magnets, the spin Hall effect, and topological insulators. The intent is to provide pedagogical material and background literature so that graduate students or researchers interested in entering this field can use this article as a reference, and both experts and non-experts can gain an understanding of the current state and future potential of this exciting field.

## II. EMERGENT MAGNETISM AT INTERFACES

This section provides an overview of *magnetic phenomena at interfaces*, focusing on static, equilibrium effects that are influenced or even created by the interface, leading to the term "emergent". Section II.A presents a brief review of bulk magnetism to provide necessary background, followed in Sec. II.B by discussion of physical phenomena that arise at magnetic interfaces, particularly due to their intrinsic symmetry breaking. Section II.C summarizes the current status and recent developments in magnetic heterostructures followed by Sec. II.D which focuses on magnetic oxide heterostructures, using perovskite oxides as an illustrative example. Following this discussion of *intrinsic* interfacial magnetic phenomena, Sec. II.E then discusses *extrinsic* effects in magnetic films and heterostructures, such as defects, interdiffusion, and roughness, including those produced by intrinsic lattice mismatch, and briefly covers the structural, chemical, and magnetic characterization methods that are crucial to contemporary research in this field. Finally, Sec. II.F provides comments on open frontiers and opportunities in thin film, interfacial, and heterostructure magnetism.

### *A. Overview of bulk magnetism including finite thickness and surface effects*

#### 1. Magnetic moments, exchange and dipolar interactions

In bulk magnets, the dominant magnetic energies derive from the exchange interaction, the interaction between orbital wavefunctions and the local electric fields from neighboring ions (referred to as crystal fields), spin-orbit coupling, and the magnetic dipolar interaction [Fulde, 1995; Wahle *et al*., 1998; Coey, 2010]. The relative values of these energies, as well as the electron kinetic energy (related to band width), fundamentally determine both static and dynamic properties of the magnetic state. The exchange interaction, whether intra- or inter-atomic, is the quantum-induced manifestation of the charge-charge interaction between electrons. It stabilizes the magnetic moments in both isolated atoms (*via* Hund's rules) and solid materials. In solid materials, *s*- and *p*- shell electron wavefunctions are typically hybridized into bands or covalent or ionic bonds, and do not contribute to magnetization, but the less-spatially extended *d*- and *f*-shell wavefunctions retain a more localized nature, with a degree of hybridization that depends on the details of the chemical bonding. Depending on this degree of hybridization (and in turn the band width), either a local moment model or a band model of magnetism is more appropriate.

The *local moment* description is exemplified by rare earth metals and insulators, and by some transition metal systems, particularly insulators but also metals with limited overlap of *d*-wavefunctions and resultingly narrow *d*-electron bands; examples of the latter are discussed in section II.D on oxides. For *isolated atoms*, with one or more outer shell electrons, the *intra*-atomic exchange interaction leads to the Hund's rule splitting of electron energies, with resulting spin, orbital, and total angular momentum **S**, **L**, and **J** = **L**+**S** respectively (capital letters refer to the combined angular momenta of the *one or more* outer shell electrons of each atom or ion). Spin-orbit coupling (discussed below) results in a ground state in which spin and orbital moments are either parallel or antiparallel. This is captured in Hund's third rule which states that for orbitals that are less than half full (*e.g.,* fewer than 7 electrons in the *f*-shell), **S** and **L** are antiparallel resulting in total angular momentum $J=|L-S|$, while for orbitals that are more than half full, they are parallel and $J=L+S$, and for a half-filled orbital, $L=0$ hence $J=S$. The magnetic moment per atom is $gJ\mu_B$ with $\mu_B$ the Bohr magneton and $g$ the Landé $g$ factor: $g=\frac{3}{2}+\frac{S(S+1)-L(L+1)}{2J(J+1)}$.

In a local moment picture of a solid, the isolated atom wavefunctions are no longer exact solutions due to the non-spherically-symmetric electric fields of neighboring ions (called "crystal fields", although also important in amorphous materials). The relatively weak effects on *f*-shell electrons leaves the isolated atom wavefunctions (with $S, L, J, M_J$ quantum numbers) as a good approximation for rare earth elements, but lifts the degeneracy of the *J*-manifold of orbitals (even in zero magnetic field) to produce singly- or doubly-degenerate low-lying energy states, depending on even versus odd numbers of electrons. For even numbers, there are a singlet and a series of doublet states ($M_J = 0, \pm 1, \pm 2, \ldots \pm J$), whose energy depends on the crystal field interaction, including its symmetry, while for odd numbers ($M_J = \pm 1/2, \pm 3/2, \ldots \pm J$), there are only doublets (known as Kramer's doublets). The doublet states are then split by magnetic field or by interatomic exchange interactions. Depending on the symmetry of the crystal structure, the lowest energy state may have magnetic moment per atom of $gJ\mu_B$, as for isolated atoms, or it may have small or even no moment but a large magnetic susceptibility [Coey, 2010, page 124-].

By contrast, the relatively strong crystal field effects (stronger than spin-orbit coupling) on *d*-orbitals cause mixing of the five *d*-orbitals of the original radially-symmetric atom, causing *J* to no longer be a "good" quantum number. In this limit, the orbital angular momentum is largely quenched, which can be understood classically as due to precession of the orbital momentum direction in the non-uniform electric field, or quantum mechanically as a mixing of wavefunctions with different orbital angular momentum directions. The new *d*-orbitals depend on the symmetries of the structure and are given names, *e.g.,* $t_{2g}$ ($d_{xy}$, $d_{xz}$, $d_{yz}$) and $e_g$ ($d_{x^2-y^2}$, $d_{3r^2-z^2}$) for octahedrally-coordinated atomic sites in materials with cubic symmetry; hybridization of these orbitals between neighboring atoms result in either metallic or insulating bands depending on their occupation and the magnitude of electron-electron interactions. The moment per atom is $\approx 2S\mu_B$, where the value of $S$ is determined from the total number of electrons and the order in which these *d*-orbitals are occupied, which in turn depends on the crystal field splitting relative to the Hund's rule exchange splitting (see low spin-high spin transitions as an example).

In a local moment model, the exchange interaction between electrons on different atoms leads to magnetic order; without this *inter*-atomic exchange, the material would be paramagnetic, with a Curie law susceptibility. The interatomic exchange interaction couples the total spin **S** (or when appropriate, angular momentum **J**) of two nearby atoms (labeled i and j) *via* an interaction which, despite its underlying complexity involving overlap of wavefunctions and Coulomb interactions, can be shown to have a relatively simple form as the leading term: $-J_{ij}\mathbf{S}_i \cdot \mathbf{S}_j$, where $J_{ij}$ is termed the exchange integral, or exchange constant. If $J_{ij}$ is positive, $\mathbf{S}_i$ and $\mathbf{S}_j$ couple ferromagnetically, whereas if it is negative, antiferromagnetic coupling results. In rare earth metals, the interatomic exchange is dominated by an indirect exchange, mediated by *sp*-band conduction electrons, known as the RKKY interaction [Ruderman and Kittel, 1954; Kasuya, 1956; Yosida, 1957]. In compounds such as oxides, there are other indirect exchange mechanisms, such as superexchange and double exchange (see Sec. II.D below).

The *band description* is exemplified by metallic transition metal ferromagnets and antiferromagnets such as Fe, Co, and Mn, where the *d*-electrons responsible for magnetism are themselves strongly hybridized. To a first approximation, these electrons can be thought of as weakly interacting and independent. In this limit, polarizing the electrons to form a net magnetic moment requires removing a minority electron and adding a majority electron to a state with higher single particle energy. The ferromagnetic ground state arises when the energy is reduced more by the exchange interaction (a Coulomb interaction at its core) between its polarized electrons than the increased kinetic energy associated with occupying states with higher single particle energies. In most metals the exchange energy reduction is less than the increased kinetic energy, and there is no spontaneous magnetization, but in certain transition metals and their alloys, the net energy is reduced, as captured first by [Stoner, 1938] and [Hubbard, 1963] and more completely by mean-field calculations [Moruzzi *et al*., 1978] based on the local spin density approximation [Kohn and Sham, 1965; von Barth and Hedin, 1972; Gunnarsson and Lundqvist, 1976; Jones and Gunnarsson, 1989]. The energy achieves a minimum value at the saturation magnetization, $M_s$, where the moment per atom calculated from $M_s$ is typically not simply related to the number of $d$-electrons of the atom, and is a non-integer number of Bohr magnetons. The interatomic exchange in such metallic systems is described as *direct exchange*, associated with direct overlap of neighboring atoms' $d$-orbitals. While these systems are frequently modeled using near neighbor pair-wise exchange interactions (between non-integer local moments) as done for local moment systems, this description of the exchange interaction is only approximate. In some materials, particularly the metallic oxides highlighted later in this section, more sophisticated treatments of electron-electron interactions are necessary. Band *anti*ferromagnetism (as in Cr, Mn, FeMn, *etc.*) is best described in reciprocal space, using the framework of a spin density wave model, although it is common to project this onto a local moment-like model with alternating up and down spins on each site.

In a weakly interacting, independent electron model of a band magnet, there would only be spin moments. However, spin-orbit coupling typically gives these materials a small orbital moment, which proves to be extremely important to their properties. The contribution of the orbital moment to the total moment in band ferromagnets is complicated [Kittel, 1949; Van Vleck, 1950]. It is frequently described in terms of $g$-factors, but there are two definitions of "$g$-factors", arising from two ways to measure these. Measurements of the ratio of total magnetic moment to total angular momentum determines the magnetomechanical ratio $g'$ which is closely related to the Landé $g$ factors of free ions with a partially filled shell mentioned above. Measurements of the precession frequency determine the spectroscopic splitting factor $g$, which is the ratio of the total magnetic moment to the *spin* angular momentum. These two $g$-factors are related; the latter proves most relevant to magnetization dynamic effects. In the *absence of band structure effects*, $(1/g) + (1/g') = 1$. For $L$=0, spin-only atoms, $g = g' = 2$, but when $L \neq 0$, $g' < 2$ and $g > 2$ (for transition metals, $g$ is typically 2.1 to 2.4) [Min and Jang, 1991; Morrish, 2001]. Notably, in the Landé $g$ factors of free ions, there is no orientation dependence, but in solid materials, both $g$ and $g'$ become dependent on direction due to crystal field effects.

The effect of band structure on the spectroscopic $g$-factor can be dramatic. In transition metal ferromagnets, these effects are typically small because the energy associated with the magnetic field is small compared to exchange splitting. Exceptions occur near avoided band crossings where there can be strong mixing of spin states across these gaps. In ferromagnetic metals, such regions account for a small fraction of the electronic states and their net contribution is still small. However, in semiconductors, these effects can be quite large. In such systems there is typically no spin splitting, the total number of electrons is small and they can all be at the conduction band minimum. In narrow gap semiconductors, there can be strong mixing between the $s$-like states at the conduction band minimum and the $p$-like states at the valence band maximum (in topological insulators, this coupling is strong enough to cause band inversion). In this case, spin-correlated circulating orbital currents, with large orbital moments, are induced by the spin-orbit interaction, such that the spectroscopic $g$-factor can be as large as 50, can change sign, and is tunable by band gap engineering and/or electrostatic tuning, leading to spintronics-relevant experiments in non-magnetic systems [Salis *et al.*, 2001; Krishtopenko *et al.,* 2011; Weisbuch and Hermann, 1977; van Bree *et al.*, 2014].

The positive interatomic exchange interaction in ferromagnets tends to lock neighboring spins in similar directions, which makes a long wavelength description of ferromagnets often appropriate. In both band and local moment models, fluctuations in the magnitude of the magnetization are energetically costly, so that $M_s$ is typically treated as a (temperature-dependent) constant, with only its direction allowed to vary. At finite temperatures, the magnetization direction fluctuates. In a long wavelength description, the long wave length fluctuations are treated explicitly as fluctuations in the direction of **M** and short wavelength fluctuations treated implicitly by reducing $M_s$, making $M_s$ a temperature dependent constant.

These spatially varying fluctuations propagate as spin waves, with corresponding quasiparticles called magnons. Magnons, similar to phonons, obey boson statistics. With increasing temperature, the spatial variations in **M** become larger, indicating an increased population of magnons with higher average energy. When the energy in these fluctuations becomes comparable to the exchange energy, the Gibbs free energy of the paramagnetic state is equal to that of the magnetic state, and a thermodynamic phase transition between the two occurs at the Curie temperature $T_C$. For antiferromagnets, with negative $J_{ij}$, similar effects cause a Neél transition at $T_N$, although competing antiferromagnetic interactions make the connection between $J_{ij}$ and $T_N$ more complex than is typically the case for ferromagnets. In the absence of strong coupling to the atomic structure, these transitions are second order, with associated fluctuations of $\mathbf{M}$ (or the staggered magnetization vector **N** of an antiferromagnet) near $T_C$ ($T_N$), but there are many instances of first order transitions, typically with discontinuities in lattice constants and in $M_s$ (or $N$). In Sec. VI, we will discuss experiments where strong laser pulses excite materials sufficiently that $M_s$ is significantly changed from its equilibrium value.

Because the exchange interaction is fundamentally linked to the overlap of electron wave functions and electron-electron Coulomb interactions, it is intrinsically short-ranged. The *magnetic dipolar interaction* between spins is much weaker than the exchange interaction at short range, but because it falls off only as $1/r^3$, it is longer ranged, and is quite important in ferromagnets. It is this interaction that leads materials to have zero net moment due to formation of magnetic domain structures, despite the cost of added exchange energy due to the non-collinear spins at domain boundaries. While the consequences of this dipolar interaction can be complicated and strongly dependent on sample geometry [Hubert and Schäfer, 1998] and the magnetocrystalline anisotropy to be discussed below, there are some simple trends. One such trend is that the dipolar energy is lowest when $\mathbf{M}$ lies along the longest dimension of the sample, such as in the plane of a thin film or along the length of an ellipsoid or needle, as this minimizes the magnetization perpendicular to the sample boundaries and the resulting magnetic field outside the sample. The variation of dipolar energy with the orientation of $\mathbf{M}$ is often referred to as *shape anisotropy* (*e.g.*, $\mu_0 M_s^2/2$ for thin films), and is quantified by demagnetizing factors [O'Handley, 2000]. Care must be taken in considering this to be in the same form as magnetocrystalline anisotropy, as it only takes this simple form for uniform $\mathbf{M}$. For example, in thin films, dipolar energy is often minimized by a non-uniform magnetization such as magnetic domains or a wandering magnetization in the plane of the film [Hubert and Schäfer, 1998].

### 2. Spin-orbit coupling

Spin-orbit coupling is a relativistic effect that occurs because at large electron orbital velocity, the electric field due to the positive nucleus is transformed into a magnetic field that couples to the electron spin. This contribution to the Hamiltonian can be approximated in terms of the coupling between the spin and orbital motion of the electrons,

$$\mathcal{H}_{so} \approx \frac{g_e}{2}\frac{e}{m^2c^2}\frac{1}{r}\frac{\partial V(r)}{\partial r}\ \mathbf{s}\cdot\mathbf{l} \qquad \textbf{[2.1]}$$

where $e$ and $m$ are the electronic charge and mass, $c$ is the speed of light, $V(r)$ the Coulomb potential of the core, $g_e$ the electron g-factor ≈ 2.002, and **l** and **s** the orbital and spin angular momenta of an outer electron of a given atom. The strength of spin-orbit coupling for individual electron orbitals depends on the matrix elements of this potential with the radial wave functions of the orbital $\varphi_{nl}(r)$ and can be estimated by

$$\lambda_{nl} \approx \frac{g_e}{2}\frac{l}{2}\frac{\hbar^2 e}{m^2c^2}\int_0^\infty dr\, \varphi_{nl}^2(r)\frac{1}{r}\frac{\partial V(r)}{\partial r} \qquad \textbf{[2.2]}$$

where $\lambda_{nl}$ is the spin-orbit coupling strength for individual orbitals with quantum numbers *n* and *l*. The spin-orbit coupling thus depends on $\frac{\partial V(r)}{\partial r}$ which increases as the nuclear charge increases, but there is a competing effect associated with specifics of the orbitals. For example, based simply on nuclear charge scaling, one would expect 4*d* transition metals to have much stronger spin-orbit coupling than 3*d*; simple arguments based on the hydrogen atom wavefunctions lead to a $Z^4$ dependence. However, 4*d* orbitals have a node in the radial wave function, and so have lower amplitude deep in the core, where the spin-orbit coupling potential is highest. As a result, 4*d* elements in the left part of the periodic table have lower spin-orbit coupling than 3*d* elements in the right part (see **Figure 2**), even though in each column of the periodic table the 4*d* element has stronger spin-orbit coupling than the corresponding 3*d* element. The connection between $\lambda_{nl}$ and the effects of spin-orbit coupling in solids is complicated and generally requires band structure calculations. Nonetheless the overall statement holds that the difference in the effects of spin-orbit coupling energy between *3d, 4d* and *5d* elements is far smaller than the $Z^4$ dependence naively expected [Tanaka *et al.*, 2008]. Spin-orbit matrix elements are generally dominated by the Coulomb potential of the ionic core even in solids with strong interfacial electric potentials or when electric fields are applied, but interfacial or applied electric fields modify the wavefunctions which then modify spin-orbit effects [Shanavas *et al.*, 2014].

For transition metals with unfilled *d*-orbitals, the electric fields from neighboring ions (crystal fields) are strong enough to largely (although typically not entirely) quench the orbital angular momentum. The orbital moments that remain are generally small, and spin-orbit coupling energies are thus smaller than in isolated atoms. The net moments in many transition metal ferromagnets have contributions from orbital moments ≈ 5 % to 10 % of the spin moments [Meyer and Asch, 1961], and typical spectroscopic *g*-factors ≈ 2.1. By contrast, in rare earth (unfilled *f*-band) materials, the spin-orbit coupling energy is typically larger than the crystal field energy; they often have large moments with large orbital contributions, and with *g* given by the Landé *g* factor (neglecting complexities including directional dependence due to crystal field interactions [*e.g.,* Jensen and MacKinstosh, 1991; Marfunin, 1979 p. 92]; and lower $T_c$ due to the weaker indirect exchange interactions of the more shielded *f*-electrons compared to the direct exchange of *d*-electrons.

An important consequence of spin-orbit coupling is that it connects the magnetization direction to the crystal lattice. Crystal fields together with spin-orbit coupling, causes preferred directions of $\mathbf{M}$, and the resulting variation of energy with the (local or global) orientation of $\mathbf{M}$ is referred to as magnetocrystalline anisotropy [Néel, 1954; Daalderop *et al*., 1990; Johnson *et al*., 1996]. This anisotropy is typically weak in materials with cubic symmetry due to the large number of symmetry axes, but can be quite large in materials with large spin-orbit coupling and uniaxial symmetry. In this case, the energy takes the form $-K_i(\mathbf{S}_i \cdot \mathbf{u})^2$ where $\mathbf{u}$ is the unique axis.

In a local moment description and in the absence of magnetic anisotropy, spins, whether ferromagnetically or antiferromagnetically coupled, can be oriented in any direction and are referred to as Heisenberg spins. In the presence of anisotropy, if spins are largely restricted to a plane or along an axis, depending on the sign of $K_i$, the spins are referred to as *x-y* or Ising spins, respectively. Cubic or other symmetries require similar suitable restrictions in simulations.

Importantly, particularly for phenomena discussed in Sec. IV, in materials that lack inversion symmetry (whether due to underlying crystal structure in bulk materials or a bilayer of dissimilar materials), spin-orbit coupling can combine with the exchange interaction to generate an *antisymmetric exchange interaction* that favors a chiral arrangement of the magnetization. This interaction has the form of a Dzyaloshinskii-Moriya (DM) interaction [Dzyaloshinskii, 1957; Moriya, 1960a; Moriya, 1960b], written as $\mathbf{D}_{ij} \cdot (\mathbf{S}_i \times \mathbf{S}_j)$ where $\mathbf{D}_{ij}(= -\mathbf{D}_{ji})$ depends on details of electron wavefunctions and the symmetry of the crystal structure [Moriya, 1960a; Bogdanov and Hubert, 1994]. The inversion symmetry breaking of a bilayer gives rise to one particular form, shown in **Figure 3**, which illustrates how broken inversion symmetry of a bilayer, where (at least) one layer has strong spin-orbit coupling, gives rise to a DM interaction with vector $\mathbf{D}_{ij}$ lying in the plane of the interface.

Inversion symmetry breaking, whether due to underlying crystal structure or interfaces, together with spin-orbit coupling also causes terms in the Hamiltonian of the charge carriers that are anti-symmetric in carrier momentum. This effect was first described by [Dresselhaus, 1955; Rashba and Sheka, 1959; Rashba, 1960] for systems with broken bulk inversion symmetry and [Ohkawa and Uemura, 1974; Vas'ko, 1979; Bychkov and Rashba, 1984] for systems with interfacial or surface inversion symmetry breaking, and leads to various significant spin-dependent transport phenomena. This inversion-symmetry breaking effect is commonly referred to as "Rashba splitting" and/or the "Rashba effect"; for a historical review of this topic, see [Bihlmayer at al., 2015].

### 3. Thin film and surface effects

While this section, II.A, focuses on bulk magnetic properties, there are a number of effects related to finite thickness that we also briefly cover here. The effects of the surface on magnetic systems, including the critical behavior near the transition temperature, have been investigated extensively over the past decades [Mills, 1971; Binder and Hohenberg, 1974; Binder 1983; Diehl, 1986; Binder and Landau, 1990; Dosch, 1992; Binder and Liujten, 2001]. It is common to see a change in $T_C$ or $T_N$ as film thickness is reduced [*e.g.,* Farle *et al*., 1993; Huang *et al*., 1994; Abarra *et al*., 1996; Ambrose and Chien, 1996; Vaz *et al*., 2008; Charilaou and Hellman, 2013, 2014]. This can be due to structural changes with thickness, or interfacial effects such as interdiffusion, roughness, and strain, which will be discussed below, but there are also intrinsic finite size effects. Such intrinsic effects can be due to the fewer neighbors for surface and near-surface atoms, changes in crystal fields [Pothuizenet al., 1995], spin–orbit coupling [Marynowski *et al*., 1999], or correlation effects (due to reduced coordination), which modify the underlying electronic band structure [Essenberger *et al.,* 2011] and can cause either reduced *or* enhanced $J_{ij}$ [Tucker, 2000; Pfeimling and Selke, 1998; Pfeimling, 2004]. As examples, EuTe(111) films exhibit strongly reduced magnetization near the film surface [Schierle *et al*., 2008], the magnetization of NiO(111) and NiO(100) films is stronger at the surfaces, such that surface order persists even above $T_N$ of bulk NiO [Marynowski *et al.*, 1999; Barbier *et al.,* 2004], while $KMnF_3$ (110) [Sinkovic *et al.,* 1985] and MnO(001) [Hermsmeier *et al.,* 1989; Hermsmeier *et al.,* 1990] surfaces exhibit ordering at temperatures that are twice as high as bulk $T_N$. For antiferromagnets, the consequences of surface exchange modifications can be particularly dramatic, since the net spontaneous moment of the material is dominated by surfaces and defects [*e.g.,* Charilaou and Hellman, 2014; Charilaou and Hellman, 2014, 2015a, 2015b].

Quite generally, the magnetic properties of a material ($M_s$, $T_C$, *etc.*) are modified whenever its dimensions become comparable to the relevant magnetic correlation length. As a final comment, since isolated atoms are often magnetic where the associated solid is not, reduced coordination at steps, kinks on steps, and isolated adatoms at surfaces or interfaces between thin films are likely to have significantly different moments (including increased orbital moments) and magnetic anisotropy than interior atoms (Freeman and Wu, 1992; Ney *et al.*, 2001; Charilaou *et al.*, 2016).

## 4. Micromagnetic modeling, magnetization dynamics, and Landau-Lifshitz-Gilbert phenomenology

In a long-wavelength approach known as micromagnetics [Brown, 1963; Hubert and Schäfer, 1998; Fidler and Schrefl, 2000], used to describe the statics and dynamics of magnetic structures at length scales large compared to the lattice spacing, the magnetic moment density is replaced by a continuous vector field, similar to the coarse graining approach of Maxwell's equations in matter. At temperatures sufficiently below the critical (ordering) temperature, the modulus of this vector field is fixed at $M_s$, leaving the local magnetization direction $\mathbf{m}(\mathbf{r},t)$, a unit vector, as the only degree of freedom. The energy density is then constructed from the microscopic terms through a long wavelength Taylor expansion. For example, the interatomic exchange energy density can be written as $A_{ex}\partial_i\,\mathbf{m}(\mathbf{r})\cdot\partial_i\,\mathbf{m}(\mathbf{r})$ where repeated indices are summed over and $A_{ex}$ (proportional to $J_{ij}$) is referred to as the exchange constant (or stiffness). Uniaxial anisotropy energy $-K_i(\mathbf{S}_i\cdot\mathbf{u})^2$ is represented as $-K_u[\mathbf{m}(\mathbf{r})\cdot\mathbf{u}]^2$; related forms exist for cubic and other symmetries. Note that large uniaxial anisotropy occurs *also* in important non-crystalline (amorphous) materials, where single ion anisotropy $K_i$ is not expected to lead to net $K_u$ due to the lack of global symmetries expected in amorphous materials; subtle structural anisotropy is however produced in thin films by the inherent asymmetry of vapor deposition growth which leads to large uniaxial perpendicular $K_u$ [*e.g.,* Gambino *et al.*, 1974; Hellman and Gyorgy, 1992]. The representation of the Dzyaloshinskii-Moriya interaction $\mathbf{D}_{ij}\cdot\left(\mathbf{S}_i\times\mathbf{S}_j\right)$ in the micromagnetic formalism relevant to interfacial phenomena will be discussed in Sec. II.B and further developed in IV.B.

Frequently, magnetic phenomena are investigated through their dynamic behavior. The magnetization is perturbed or driven and its response measured. With few exceptions, the samples are large enough that a micromagnetic description is more appropriate than an atomistic one. The micromagnetic description is based on the Landau-Lifshitz-Gilbert (LLG) equation

$$\mathrm{d}\mathbf{M}/\mathrm{dt} = -\gamma_0\mathbf{M}\times\mathbf{H}_{\text{eff}} + \frac{\alpha}{M_s}\mathbf{M}\times\mathrm{d}\mathbf{M}/\mathrm{dt} = -\gamma_0'\mathbf{M}\times\mathbf{H}_{\text{eff}} - \frac{\alpha\gamma_0'}{M_s}\mathbf{M}\times(\mathbf{M}\times\mathbf{H}_{\text{eff}}) \qquad \textbf{[2.3]}$$

where $\gamma_0 = \mu_0\gamma$, $\gamma$ is the gyromagnetic ratio (=$g\mu_B/\hbar$), $\alpha$ is the Gilbert damping parameter, and the *effective* field $\mathbf{H}_{\text{eff}} = -\frac{1}{\mu_0}\nabla_{\mathbf{M}}\mathcal{E}(\mathbf{M})$ has contributions from all terms that contribute to the energy density of the magnetization, $\mathcal{E}(\mathbf{M})$ (including interaction with an applied field, interatomic exchange, magnetostatic dipole-dipole interaction, magnetocrystalline anisotropy, Dzyaloshinskii-Moriya interaction). The first equation of Eq. 2.3 is written with the damping in the Gilbert form [Gilbert, 2004]; the second is in the original Landau-Lifshitz form [Landau and Lifshitz, 1935]. These two forms are mathematically equivalent with the substitution $\gamma_0' = \gamma_0/(1+\alpha^2)$. While nomenclature is somewhat inconsistent in the literature, here we refer to the first term as a field-like torque (meaning that it includes both the actual applied field and the other terms in the *effective field* described just above) and the second as a damping torque. The two terms in the Landau-Lifshitz form have the same form as spin-transfer and spin-orbit torques introduced in Sec. III, leading to those terms being referred to as field-like torques and damping-like torques respectively. While the two forms of the equation of motion for d**M**/dt (Eq. 2.3) are equivalent and remain equivalent when other torques are introduced, care is required because the additional terms will have slightly different forms in either approach. These equations are frequently written using only the orientation of magnetization, with unit vector **m** = **M**/$M_s$ replacing **M** and all factors of $M_s$ dropped.

In general, magnetization **M** (or **m**) varies spatially, and Eq. 2.3 is solved numerically for specific parameter values. These solutions are facilitated by the growing availability of easily accessible codes [Donahue and Porter, 1999, Fischbacher *et al*., 2007, Vansteenkiste *et al*., 2014]. They require a complete description of the sample, including its geometry and appropriate micromagnetic parameters, and a description of the external driving torques such as those from electrical currents. The introduction of the main micromagnetic parameters from extended Heisenberg models, as sketched above, leads to a first approach for determining the parameters based on electronic structure calculations whose results are mapped to a Heisenberg model [e.g. Heinze *et al*., 2011, Dupé *et al*., 2014]. Recent progress has been particularly dramatic for estimates of the Gilbert damping constant [Starikov *et al*., 2010, Mankovsky *et al*., 2013, Liu *et al*., 2015]. A shortcoming of this approach is that materials are never perfect, and the differences from those considered in the electronic structure calculations can be important. The second, phenomenological, approach fits the micromagnetic parameters to experiments, a recent example being [Romming *et al*., 2015]. Over the years, the community has developed a series of experimental techniques to "measure" the micromagnetic parameters, with an accepted reliability for each parameter. We refer the reader to reviews for parameters like interfacial anisotropy [Johnson *et al*., 1996] and interlayer exchange coupling [Stiles, 1999, Stiles, 2004]. The community has not yet settled on approaches to measure parameters that have more recently become of interest like interfacial micromagnetic energies and interfacial torques. Section III illustrates this debate for the various torques associated with electrical current and their thickness dependencies. For the interfacial DM interaction, the situation is similar. Estimates are presently obtained through domain wall dynamics experiments, by direct observation of domain wall structures, or by a spin-wave technique. All techniques give similar results, but there are cases of disagreement [Soucaille *et al*., 2016].

As these numerical simulations can be rather intensive, several approaches are taken to simplify calculations. One approach is the macrospin approximation, in which the magnetization is assumed to be spatially uniform. In this case the LLG equation simplifies to two coupled differential equations and considerable progress can be made analytically. An alternate approach for situations in which spatial variation is important, *e.g.*, in domain walls, is to make an ansatz for the form of the spatial variation [Schryer and Walker, 1974] and then find the equations of motion for the few degrees of freedom that characterize the ansatz. This approach enables the connection between topology and magnetization dynamics *via* the so-called Thiele equation, discussed in Section IV.C.

### *B. Interfacial magnetic phenomena*

The focus of this article is magnetic behavior that originates at or near interfaces, particularly those between thin films. Many such phenomena arise from the fact that interfaces and surfaces break translational and inversion symmetry, which produces a variety of effects. We focus first on ideal, perfect interfaces, and then turn in II.E to disorder effects *e.g.,* interdiffusion, roughness, strain, defects, *etc.* The discontinuity in atomic structure and consequently in electronic band structure leads to an enormous number of effects at interfaces, particularly if all combinations of metals, semiconductors, and insulators are included. As discussed in Sec. II.A, the surface of a magnet can have significantly different moments than the bulk. Interfaces can create new magnetic states, including creating magnetic moments from non-magnetic materials as local coordination and/or band structure are modified. Dramatic (and in some cases controversial) examples of this are found in oxide heterostructures, as discussed in Sec. II.D.

We first focus on combinations of most topical relevance: metallic and insulating ferromagnets interfaced with non-magnetic metals, particularly those with large spin-orbit coupling such as Pt, Ir, and Ta. An example phenomenon is proximity-induced magnetism at the interface between a non-magnet and a ferromagnet, in which ferromagnetic order in one layer induces a moment in the neighboring non-magnetic layer, which decays away from the interface. In non-magnetic materials that are already close to a ferromagnetic instability, like Pt or Pd, this induced moment can be substantial, although its magnitude is presently somewhat controversial [Wilhelm *et al*., 2000; Wende, 2004; Huang *et al*., 2012; Geprägs *et al*., 2012; Zhang *et al.,* 2015a; Kuschel *et al.*, 2015; Klewe *et al.*, 2016]. This effect is related to the "giant moments" known to exist in Pt and Pd with very dilute Co or Fe substitution [Crangle and Scott 1965]. Interdiffusion between the layers, as discussed below, modifies the idealized proximity effect discussion. The magnitude of induced moment, particularly its orbital component, complicates the interpretation of some experiments discussed in Secs. III.B and III.C, relating to spin caloritronic and spintronic phenomena.

### 1. Interface-induced magnetic anisotropy

An important interfacial magnetic phenomenon occurs because the breaking of inversion symmetry at interfaces ("up" is different than "down" at any bilayer) creates a unique axis perpendicular to the interface, which can in turn induce a uniaxial anisotropy associated with the surface, $-K_\mathrm{s}(\mathbf{S}_i \cdot \mathbf{n})^2$, where $\mathbf{n}$ is the interface normal. For $K_\mathrm{s}$>0, this energy produces perpendicular magnetic anisotropy, favoring a magnetization perpendicular to the interface and counteracting the dipolar-induced shape anisotropy of planar films. This perpendicular anisotropy has been observed in bilayers and superlattices of Co/Pd [Carcia *et al*., 1985], Co/Pt [Carcia, 1988; Engel *et al*., 1991], Co/Ir [den Broeder *et al*., 1991] and Co/Au [Chappert *et al*., 1986], the fundamental origin being anisotropy of the interfacial orbital angular momentum induced by the lowered symmetry [Gay and Richter, 1986; Daalderop *et al*., 1994; Weller *et al*., 1995; Stöhr, 1999]. The presence of non-magnetic heavy metals at such interfaces modifies the interfacial orbital angular momentum of the transition metal, and enhances its spin-orbit interaction, thereby increasing perpendicular magnetic anisotropy. Perpendicular anisotropy is also found in some superlattices involving only transition metals, such as Co/Ni [Daalderop *et al*., 1992], and at interfaces between transition metal ferromagnets and nonmagnetic oxides such as MgO and $AlO_x$, attributed to the nature of the bonding between the metal and oxygen ions at the interface [Monso *et al*., 2002; Rodmacq *et al*., 2003; Rodmacq *et al*., 2009; Yang *et al*., 2011]. Perpendicular magnetic anisotropy has been employed in magnetic recording media for several years, and has seen a recent resurgence in interest due to its application in a variety of spintronic heterostructures exploiting effects such as spin torque (see Sec. III).

It was recently discovered that interfacial perpendicular magnetic anisotropy can be modulated by applying an electric field normal to the interface. Carrier accumulation created at a magnetic metal/non-metal interface by an applied electric field modifies the chemical potential, leading to an unequal change in the occupied density of states for spin up and spin down bands, thus altering the interfacial magnetization [Rondinelli *et al*., 2008]. This change in occupancy includes local changes in the occupancy of *d* orbitals with different symmetry, in turn changing the magnetocrystalline anisotropy, as has been demonstrated in Fe-Pd and Fe-Pt alloy films "gated" with an electrolyte [Weisheit *et al*., 2007], or Fe/MgO heterostructures [Maruyama *et al*., 2009; Niranjan *et al*., 2010]. This method of tuning anisotropy *via* an applied voltage is promising for energy-efficient switching of magnetization in magnetic memories, as has been demonstrated in tunnel junctions based on CoFeB/MgO/CoFeB [Wang *et al*., 2011; Shiota *et al*., 2012]. First principles calculations capture the near-linear change in magnetocrystalline anisotropy with interfacial charge density [Duan *et al*., 2008; Tsujikawa and Oda, 2009], highlighting the key role played by the occupation of selected 3*d* minority spin orbitals [Nakamura *et al*., 2009].

Applied biases not only lead to modifications of electronic properties, but can also drive ionic motion, providing an additional route to the reversible control of interfacial magnetism. An illustrative example from this nascent field of "magnetoionics" comes from the $Co/GdO_x$ interface [Bauer *et al*., 2012; Bi *et al*., 2014; Bauer *et al*., 2015]. The application of an electric field drives reversible oxidation and reduction of a thin Co layer (< 1 nm) through migration of $O^{2-}$ ions within the $GdO_x$ layer. Oxidation of the Co suppresses magnetization, while reduction of $CoO_x$ back to Co, induced by changing the electric field direction, restores the Co moment, and can drive the anisotropy from perpendicular to in-plane. These voltage-driven ionically-controlled effects are nonvolatile and can be spatially patterned *via* local laser-induced heating [Bauer *et al*., 2015].

### 2. Interface-induced Dzyaloshinskii-Moriya interaction

The breaking of inversion symmetry at interfaces, as mentioned above, can also give rise to an interfacial DM interaction [Fert and Levy, 1980; Fert, 1990; Crépieux and Lacroix, 1998], which plays an important role in much of the physics discussed in this review (*e.g.,* non-collinear spin textures in Sec. IV). Interfaces between magnetic materials and materials with large spin-orbit interactions offer promise for giant interfacial DM interactions. The interfacial DM interaction energy, proportional to $(\mathbf{n}\times\mathbf{r}_{ij})\cdot(\mathbf{S}_i\times\mathbf{S}_j)$ in a local moment model (shown in **Fig. 3**) favors canting of spins towards each other around the direction perpendicular to the separation between them ($\mathbf{r}_{ij}$) and the interface normal ($\mathbf{n}$), thereby promoting non-collinear spin textures. DM interactions also occur naturally in band models with broken inversion symmetry and spin-orbit coupling, or out of the spin-polarized Rashba model. In the micromagnetic formalism, the DM interaction has the form $D_m\left[(\hat{\mathbf{z}}\times\hat{\mathbf{x}})\cdot(\mathbf{m}\times\partial_x\mathbf{m})+(\hat{\mathbf{z}}\times\hat{\mathbf{y}})\cdot(\mathbf{m}\times\partial_y\mathbf{m})\right]$ for an interface parallel to $\hat{\mathbf{z}}$. This interaction favors magnetic textures in which the magnetization spirals around an axis perpendicular to the direction of variation and the interface normal.

DM interactions have been computed from a variety of approaches; *e.g.,* Moriya [Moriya, 1960a; Moriya, 1960b] computed the DM interaction from the effect of spin-orbit coupling at magnetic atoms involved in a superexchange interaction. Several groups [Fert and Levy, 1980; Imamura *et al*., 2004; Mross and Johannesson, 2009; Tserkovnyak *et al.*, 2015] have computed the DM interaction analogously to the RKKY exchange interaction in a variety of models. *Ab initio* calculations of DM interactions at interfaces have also been carried out for finite [Bornemann *et al*., 2012] and infinite systems [Freimuth *et al*., 2014a]. These calculation schemes either introduce spin-orbit coupling as a perturbation, or include it *a priori*. The first approach computes the energies of static spin waves with wave vectors close to zero, *i.e.,* for deliberately long periods, giving access to exchange or DM interaction coefficients that treat the same length scales as in micromagnetic structures [Heide *et al*., 2009]. The second approach computes the energies of short period spin waves [Yang *et al*., 2012; Yang *et al*., 2015]. Such calculations give a deeper understanding of the nature of the interfacial DM interaction and its dependence on alloying or oxidation.

It is important to note that, like the exchange interaction, the DM interaction is only measured indirectly *via* its effect on statics and dynamics of spin structure. If the DM interaction is strong enough relative to other interactions, it can lead to chiral ground state structures for the magnetization, including spiral states and skyrmion lattice states. Experimental evidence of significant interfacial DM interactions has been deduced from such ground state magnetic configurations in bilayer and trilayer systems such as Mn/W(110), Fe/Ir(111), Pd/Fe/Ir(111) [Bode *et al.*, 2007; von Bergmann *et al.*, 2014] and in pairs of individual Fe adatoms on Pt(111) surface *via* low-temperature inelastic scanning tunnelling spectroscopy [Khajetoorians *et al.,* 2016]. Even when not strong enough to influence the ground state magnetization configuration, the DM interaction can be strong enough to affect the spin wave dispersion [Udvardi and Szunyogh, 2009; Costa *et al.*, 2010] seen as an asymmetry in the energies of forward and backward moving spin waves, which can be measured *via* spin-polarized electron energy loss spectroscopy [Zakeri *et al.*, 2010] or Brillouin Light Scattering [Cho *et al.,* 2015; Di *et al.,* 2015; Nembach *et al.,* 2015].

### *C. Magnetic heterostructures: bilayers, trilayers and multilayers*

An important phenomenon occurring at the interface between ferromagnets and antiferromagnets is exchange bias which modifies the response of the ferromagnet to an applied magnetic field [Meiklejohn and Bean, 1956; Berkowitz and Takano, 1999; Nogues and Schuller, 1999]. Typically, for small applied field, the spin order in the antiferromagnet does not significantly change, so the exchange coupling shifts the hysteresis loop of the ferromagnet in one field direction, resulting in *unidirectional* anisotropy. This behavior is useful for pinning the magnetization of magnetic layers, such as in “spin valve” heterostructures [Dieny *et al.*, 1991], in which one ferromagnetic layer has its magnetization pinned by exchange bias and the other layer is free to rotate at low fields. This effect is currently used in read heads in hard disk drives, memory elements in magnetic random access memory (MRAM) [Tehrani *et al.*, 1999; Zhu and Park, 2006; Ando *et al.*, 2014], and other applications. The behavior and consequences of exchange bias, and the wealth of related phenomena induced at antiferromagnet/ferromagnet interfaces are found to be remarkably rich and complicated [Berkowitz and Takano, 1999; Nogues and Schuller, 1999; Nogues *et al.,* 2005], and are strongly influenced by many forms of structural disorder, as touched upon in Sec. II.E. Notably, the exchange bias in some ferromagnet/antiferromagnet bilayers, such as $Fe_3O_4$/CoO and Co/$IrMn_3$(111) [Ijiri *et al.*, 2007; Yanes *et al.*, 2013] has been explained by an interfacial DM interaction.

Moving beyond a single interface, there are a variety of important effects associated with *multiple* interfaces in magnetic thin films (such as in trilayers or multilayers of ferromagnetic layers separated by non-magnetic metallic layers). The most obvious of such effects is giant magnetoresistance (GMR), discovered by [Baibich *et al.*, 1988; Binasch *et al.*, 1989] in Fe/Cr trilayers and multilayers, for which Fert and Grünberg received the 2007 Nobel Prize, and subsequently extended to a number of magnetic/non-magnet heterostructures. GMR is based on the transport of spin-polarized electrons through a non-magnetic layer from one interface between non-magnet and ferromagnet to another such interface [Camley and Barnas, 1989; Valet and Fert, 1993; Dieny, 1994]. This transport depends on the relative orientations of **M** in the ferromagnetic layers, and will be discussed in more detail in Sec. III.

While the existence of GMR does not *depend* on exchange coupling between the ferromagnetic layers, its study nevertheless led to the observation in transition metal systems [Parkin *et al*., 1990] of oscillatory interlayer exchange coupling as a function of thickness of the non-magnetic layer [Slonczewski, 1995; Stiles, 1999]. Earlier studies had seen related effects in rare earth systems [Majkrzak *et al.*, 1986, Salamon *et al*., 1986]. The theoretical description of the interlayer exchange coupling [Edwards *et al*., 1991; Bruno and Chappert, 1991] relates this to quantum well states that have been explicitly measured in the non-magnetic layers by photoemission [Ortega and Himpsel, 1992; Garrison *et al*., 1993; Carbone *et al*., 1993], and more generally to the band structure of the two ferromagnetic layers (which may be different) and the non-magnetic layer. Calculations and measurements of interlayer exchange coupling have now matured enough to enable quantitative comparison between experiment [Unguris *et al*., 1997] and calculation [Opitz *et al*., 2001] for the Fe/Au/Fe system, one of the very few systems with close enough lattice matching to allow quantitative comparison. Measurement of the interlayer exchange coupling depends on measurement of the required applied switching field of the layered structure. These indirect exchange interactions and others, such as dipolar coupling, can combine with disorder, discussed in Sec. II.E, to give a wide variety of coupling behavior [Néel, 1962a; Néel,1962b; Slonczewski,1991; Demokritov *et al*.,1994; Demokritov, 1998; Stiles, 2004].

As a final comment, some true *superlattice* effects have been observed in magnetic heterostructures, meaning phenomena that occur only in a structurally coherent multilayer and vanish as the number of bilayers is lowered to unity. Examples include superlattice Bragg peaks in X-ray diffraction [Schuller, 1980], collective behavior of magnons in ferromagnet/nonmagnet superlattices [Grimsditch *et al*., 1983], the opening of superlattice gaps in electronic band structure [Miller *et al*., 1992], and oscillatory transport behavior with Ni and/or Co layer thicknesses in Co/Ni superlattices [Gallego *et al*., 1995]. The latter constitutes direct experimental observation of superlattice effects in transport, confirming a scattering process that exists only in superlattices [Kobayashi *et al*., 1994; Kim *et al*., 1996].

### *D. Complex oxide films, interfaces, and heterostructures*

The majority of the preceding discussion focused on films, interfaces, and heterostructures based on metals and metallic alloys, with simple insulating oxides playing a limited role (as in $AlO_x$- and MgO-based tunnel junctions, or antiferromagnetic oxides in exchange bias structures). The wealth of discoveries in these metallic systems stimulated investigation of magnetic heterostructures based on other constituents, including magnetic semiconductors and oxides. The latter includes binary oxides (see for example the large bodies of work on $Fe_3O_4$, $CrO_2$ [Fong *et al*., 2013], V-O insulator-to-metal transition systems [de la Venta *et al*., 2014], artificially-structured magnetic semiconductors using CoO/ZnO multilayers [Lee *et al.*, 2013a], or EuO [Coey *et al*., 1999]), but a wealth of remarkable behavior is found in complex oxides, concurrent with a general rise in interest in heterostructures and interfaces of these materials [Ramesh and Schlom*,* 2008; Chakalian *et al.,* 2012; Hwang *et al.,* 2012; Bhattacharya and May, 2014; Stemmer and Allen, 2014, Sulpizio *et al.,* 2014]. We provide a brief review, focusing on the perovskite structure magnetic oxides that serve as model systems.

#### 1. Bulk perovskite oxides

Complex transition metal oxides including perovskites are based on multiple cations bonded with oxide ions, and form in remarkably diverse and adaptable structures [Muller and Roy, 1974] [when the ion referred to is, for example, $O^{2-}$, the correct terminology is "oxide ion" (it is a dianion) rather than "oxygen ion"]. Their chemical flexibility allows for a variety of highly tunable electronic and magnetic behaviors [Dagotto, 2005, Khomskii, 2014], including high-temperature superconductivity, colossal magnetoresistance, and multiple coexisting ferroic orders, such as ferroelectricity and ferromagnetism [Spaldin *et al*., 2010]. The presence of oxygen ligands around different cations is key to some of the remarkable behavior found. The oxide anion provides coexisting Coulomb repulsion and intra- and inter-atomic exchange interactions within mixed ionic-covalent materials. The orbital bandwidths are sensitive to both local and long-range structure, owing to the high electronic polarizability of the transition metal-oxygen bonds, which interact with electron correlations from the localized and interacting transition metal *d*-electrons. The balance between these interactions is determined by the metal-oxygen bond network, which is often described in terms of polyhedra. The relevant polyhedron in $ABO_3$ perovskite oxides, where the $A$ cations are typically larger radii alkali, alkali-metal, or lanthanide elements and the $B$ cations are commonly transition metals, is a $BO_6$ octahedron. The perovskite structure is generated by corner-connecting adjacent $BO_6$ octahedra in three-dimensions giving rise to –*B*–O–*B*– chains along all Cartesian directions with the *A* cation occupying the interstices (**Fig. 4**). Few perovskites have ideal 180° *B*–O–*B* bond angles (also referred to as $BO_6$ octahedral rotation angles). Although a seemingly subtle crystallographic effect, rotations away from 180° are important in determining the electronic (metal *vs*. insulator) and magnetic (ferro- or antiferromagnetic) state, as the electronic bandwidth is tied to the angular orbital overlap between the oxygen *p* and transition metal *d* states [Harrison, 1989].

The electronegativity of oxygen leads to (nearly) complete ionization of the *A* and *B* cation valence electrons, and thus, unlike in itinerant metals, a rudimentary understanding of the magnetic state of a transition metal oxide may be gleaned from the number of electrons (spins) and symmetries of the occupied *d*-orbitals of the *B* cation. Also different from metals, the magnetic exchange interactions between spins occur through *d*-orbitals centered at neighboring *B* atoms bridged by an oxide ion, referred to as 'superexchange'. This interaction has been treated extensively in [Goodenough, 1955; Kanamori, 1959; Anderson, 1950]. The guidelines therein can be used to predict the ferromagnetic and antiferromagnetic states for transition metal oxides, and have been dubbed the Goodenough-Kanamori-Anderson rules. These superexchange rules involve kinetic exchange between two spins by virtual transfer of electrons between orbitals, mediated by the *pd*-hybridization. The favored magnetic state therefore depends on the angular overlap between the oxygen *p* and transition metal *d* orbitals (or octahedral rotation angle), orbital orthogonality, and orbital occupancy (filling).

Another type of exchange, termed "double exchange", involves transfer of electrons between *B* cations of different nominal valence, favoring ferromagnetic coupling [Zener, 1951]. This is encountered in perovskite oxide ferromagnetic metals such as $La_{0.67}Sr_{0.33}MnO_3$. The antisymmetric Dzyaloshinskii-Moriya interaction [Dzyaloshinskii, 1957; Moriya, 1960a; Moriya, 1960b] described earlier (Secs. II.A and II.B above) is also operative in transition metal oxides. As in metals, it arises from relativistic spin-orbit interactions; however, in perovskites and related crystal structures with octahedral $BO_6$ units, the atomistic origin is due to asymmetric displacements of the oxygen ligands centered between nearest neighbor *B*-site cations, *i.e.*, the octahedral rotations. The result is that weak ferromagnetism may be induced in antiferromagnetic oxides through a small spin canting, even if global inversion symmetry in the structure is present [Bousquet and Spaldin, 2011]. The DM interaction can also lead to exchange bias at oxide interfaces [Dong *et al*., 2009]. Additionally, this antisymmetric exchange interaction may provide the inverse effect, whereby cycloidal magnetic states in antiferromagnetic compounds can induce polar ionic displacements (broken inversion symmetry) and ferroelectricity [Cheong and Mostovoy, 2007; Malashevich and Vanderbilt, 2008], of interest in the field of multiferroics. Approaches also involve coupling of magnetoelectric or multiferroic antiferromagnets to ferromagnets *via* exchange bias [He *et al*., 2010b].

## 2. Oxide interfaces: electronic and orbital effects

Interfaces between perovskites with dissimilar *B*-site cations create *B*-O-*B*' bonds, across which the exchange interaction can deviate from that found within the adjoined materials. There are cases where *B*-O-*B*' coupling is ferromagnetic even when both *B*-O-*B* and *B*'-O-*B*' interactions are antiferromagnetic. By growing along different crystallographic orientations, the number of *B*-O-*B*' interactions can be tuned, an effect that has been exploited in $(ABO_3)_1/(AB'O_3)_1$ superlattices to engineer long-range ferromagnetism or antiferromagnetism [Ueda *et al.*, 2001]. While simple electron counting serves as a starting point for understanding exchange interactions across a *B*-O-*B*' bond, interface-induced changes to electronic structure, orbital occupancy and atomic structure, as depicted in **Figure 4**, often play critical roles in controlling interfacial magnetism, and can lead to behavior that is not a simple interpolation of the properties of the two adjoined materials.

Given the importance of the electron count and orbital polarization of the metal cations in determining the dominant exchange interactions across *B*-O-*B* bonds in perovskites [Goodenough, 1955; Tokura and Nagaosa, 2000], complex oxide interfaces provide useful illustrations of the consequences of electronic and orbital perturbations on interfacial magnetism [Bhattacharya and May, 2014]. At oxide interfaces, charge transfer can be driven by a difference in chemical potential or by screening of local dipoles. Charge transfer can alter the *B*-site valence states near the interface, enabling magnetic order that is distinct from either constituent, for instance leading to ferromagnetism confined to the interface between two insulating antiferromagnets [Salvador *et al*., 1999; Lin *et al*., 2006; Santos *et al*., 2011] or ferromagnetism from a paramagnetic metal and antiferromagnetic insulator [Takahashi *et al*., 2001; Nanda *et al*., 2007; Freeland *et al*., 2010; He *et al.*, 2012]. The spatial extent of the interfacial magnetism closely matches that of the charge transfer length scale, which is generally quite short, around 0.4 nm to 2 nm [Santos *et al*., 2011; Grutter *et al*., 2013; Hoffman *et al*., 2013], owing to both the large dielectric constant and significant carrier concentration (charge density) that complex oxides often support [Ahn *et al*., 2006].

An interface can alter not only local electronic density but also the orbital occupancy of valence electrons, leading to an orbital reconstruction, as observed at the interface between $La_{2/3}Ca_{1/3}MnO_3$ and $YBa_2Cu_3O_7$. While holes in the non-interfacial cuprate layers are constrained to the $d_{x^2-y^2}$ orbital, at the interface, holes are redistributed between both $d_{3z^2-r^2}$ and $d_{x^2-y^2}$ orbitals allowing for covalent bonding between the Mn $d_{3z^2-r^2}$ and Cu $d_{3z^2-r^2}$ orbitals hybridized with the apical oxygen [Chakhalian *et al*., 2007]. At manganite/cuprate interfaces, the interfacial Mn and Cu cations couple antiferromagnetically [Chakhalian *et al*., 2006; Visani *et al*., 2011], with the Cu moment originating from the same electrons that participate in the orbital reconstruction [Uribe-Laverde *et al*., 2014].

The ability to accumulate and deplete charge at an interface through electrostatic gating provides a *dynamic* means to modify interfacial magnetism. This occurs via changes to exchange energies, arising from modifications to the electron density of states near the Fermi level [Ohno *et al*., 2000]. In simple oxides and low carrier density magnetic semiconductors, conventional gating is effective in modifying carrier density and inducing ferromagnetism [Ohno *et al*., 2000; Lee *et al.,* 2009]. The large carrier concentrations present in perovskite oxides however necessitate the use of non-conventional dielectrics such as electrolytes (*e.g.*, ionic liquids and ion gels), or ferroelectrics, in order to accumulate a carrier density sufficient to alter phase stability. This typically limits the spatial extent of electrostatic modification to less than 1 nm to 2 nm [Ahn *et al*., 2006]. Ionic liquid electrolytes have enabled electric field control of magnetization and $T_C$ in $(Ti,Co)O_2$ and $La_{0.8}Ca_{0.2}MnO_3$, evidenced by magnetotransport measurements [Dhoot *et al*., 2009; Yamada *et al*., 2011]. Similarly, ferroelectric-based modulation of $T_C$ *via* electrostatic manipulation of Mn valence has been demonstrated at $La_{1-x}Sr_xMnO_3/PbZr_{1-x}Ti_xO_3$ interfaces [Vaz *et al*., 2010] and linked to doping (screening) homogeneity and local structural distortions [Spurgeon *et al*., 2014]. When using ionic liquids and ion gels, care must be taken to distinguish between electrostatic and electrochemical processes within the oxide layer [Jeong *et al*., 2013]; recent progress has been made with this issue in specific oxides (Walter *et al*., 2016), but more work is required.

### 3. Oxide interfaces: structural effects

Epitaxial strain and interfacial structural coupling, illustrated in **Figure 4**, provide a means to control both octahedral distortions and rotations [Rondinelli *et al*., 2012], which play a key role in magnetic transition temperatures [Radaelli *et al*., 1997; Chmaissem *et al*., 2001] and can drive transitions from ferromagnetic to antiferromagnetic states [Subramanian *et al*., 1999]. Strain-induced changes in octahedral distortions and rotations persist over the thickness in which the in-plane lattice parameters of the film remain equal to that of the substrate, a length scale determined by the competition of elastic and defect formation energies [Matthews and Blakeslee, 1974]. Therefore, strain enables tuning of magnetic properties coherently throughout films that can be tens of nanometers thick. For example, strained films of $La_{2/3}Sr_{1/3}MnO_3$ grown on a variety of substrates display a decrease in $T_C$ under either tensile or compressive strain [Adamo *et al*., 2009], consistent with theoretical work on strain-induced changes to the hopping of spin-polarized carriers [Millis *et al*., 1998]. For films grown on piezoelectrics, this strain can be dynamically tuned *via* voltage; electric field control of strain was used to alter transition temperatures in $La_{0.7}Sr_{0.3}MnO_3$ films on $Pb(Mg_{1/3}Nb_{2/3})_{0.72}Ti_{0.28}O_3$ substrates [Thiele *et al*., 2007] and FeRh films on $BaTiO_3$ [Cherifi *et al*., 2015].

Interfacial structural coupling, in which the $BO_6$ rotations and distortions are modified at a heterojunction due to the geometric constraints of polyhedral connectivity (**Figure 4**), is particularly important for *interface-induced* magnetism in oxides. Recent structural studies have established a length scale of approximately 2 to 6 unit cells for this octahedral coupling region [Borisevich *et al*., 2010; Rondinelli and Spaldin 2010; He *et al*., 2010a; Zhang *et al*., 2013; Aso *et al*., 2014; Fister *et al*., 2014]. The short length scale for interfacial octahedral coupling can be exploited to induce new magnetic phenomena, distinct from strain-driven effects, not present in compositionally equivalent bulk counterparts. For example, the distance between interfaces in a series of isovalent $(La_{0.7}Sr_{0.3}MnO_3)_n/(Eu_{0.7}Sr_{0.3}MnO_3)_n$ superlattices was tuned to be either greater than or less than the octahedral coupling length scale by changing $n$, leading to modulated or uniform octahedral and magnetic behavior [Moon *et al*., 2014b].

Although bond covalency can drive orbital reconstructions at interfaces, strain can also be used to induce an orbital polarization throughout a film *via* metal-oxygen bond distortions. Tensile strain, for example, may lead to preferential $d_{x^2-y^2}$ occupancy, whereas compressive strain favors occupation of the $d_{3z^2-r^2}$ orbitals owing to the modified crystal field. In half-doped manganite films, strain-induced changes in orbital occupancy have been used to drive the magnetic state from A-type to C-type antiferromagnetism [Konishi *et al*., 1999]. The combination of strain and dimensional confinement also leads to orbital polarization in $LaNiO_3$-$LaAlO_3$-based superlattices (both non-magnetic in bulk), stabilizing non-collinear antiferromagnetism in samples with 2 unit cells of $LaNiO_3$ within each superlattice period [Frano *et al*., 2013], while in superlattices with 3 or more unit cells of $LaNiO_3$, magnetic order is not observed.

The systems described above provide examples of how electronic, orbital and structural modifications at interfaces alter magnetism. At many oxide interfaces, more than one of these play a role, making it challenging to disentangle the roles of electronic or atomic structure. Tuning multiple interfacial properties provides a powerful route to access novel states, as illustrated by work on vanadate superlattices, in which ferromagnetism is predicted to emerge by simultaneously engineering octahedral rotations and bandfilling in $LaVO_3$/$SrVO_3$ superlattices [Dang and Millis, 2013]. Oxide analogues of ferromagnet/heavy metal heterostructures that incorporate strongly spin-orbit coupled 5*d* iridium-based oxides, such as ferromagnetic $La_{0.67}Sr_{0.33}MnO_3$/paramagnetic metal $SrIrO_3$ [Yia *et al.,* 2016], may enable novel interfacial magnetic properties arising from the combination of narrow electron bands, strong correlation effects, and significant spin-orbit coupling.

### *E. Extrinsic effects and characterization*

Sections II.A–D discussed how surfaces and interfaces in magnetic systems induce an extraordinary range of phenomena attributed to finite thickness, symmetry breaking, atomic and electronic reconstructions, and interaction and proximity effects. These effects are *intrinsic* in that they would occur even at ideal interfaces between perfect, defect-free materials; they do not rely on imperfections or defects for their existence. Ideal interfaces between defect-free materials do not generally exist, however. This limit may be approached, *e.g.*, when kinetic limitations dominate thermodynamics, but in most cases non-idealities are inherent. In some cases, the distinction is somewhat artificial, *e.g.,* lattice mismatch and consequent strain in a heterostructure produce defects that are intrinsic to the heterostructure. In this section some of the forms of disorder that occur in films, at interfaces, and in heterostructures are discussed, along with their impact on interface magnetism. We note that disorder, compositional and structural, also occurs *within* individual layers, where *e.g.*, distinguishing between amorphous and nanocrystalline is challenging but generally important for quantitative understanding. It is emphasized that (*a*) non-idealities must often be accounted for to reconcile theory and experiment, (*b*) a detailed appreciation of the physics and chemistry of interfaces is often required to understand interfacial magnetic behavior, and (*c*) defects and disorder are not universally deleterious, but can induce, and even control, novel phenomena. The influence of disorder on interfacial magnetic effects is summarized first, followed by a discussion of intrinsic *vs*. extrinsic contributions to emergent interfacial magnetic phenomena. The focus is placed on illustrative contemporary examples, with reference to earlier literature for more complete discussions of established effects. The section ends with a discussion of key characterization tools that have enabled advances.

### 1. Non-idealities and their influence on interfacial magnetic phenomena

We consider the influence of the various forms of disorder relevant to magnetic interfaces in a hierarchical fashion, starting from a hypothetical ideal interface between two defect-free, epitaxial, single-component materials with zero lattice mismatch, then gradually increasing realism and complexity. Simply permitting that the interface is not exactly parallel to the growth plane (due to vicinality) introduces non-idealities, including terraces and step-edges. These have significant impact, inducing magnetic frustration at ferromagnet/non-magnet and ferromagnet/antiferromagnet interfaces [Escorcia-Aparicio *et al.,* 1999, Himpsel *et al.,* 1998, Vaz *et al.,* 2008], uncompensated spins at ferromagnet/antiferromagnet interfaces [Berkowitz and Takanao, 1999; Nogues and Schuller, 1999; Nogues *et al.,* 2005; Charilaou *et al.*, 2014; Charilaou, Bordel and Hellman, 2014], and step-edge-related magnetic and transport anisotropies in metals [Himpsel *et al.,* 1998; Vaz *et al.,* 2008] and oxides [Wang *et al.,* 2003; Mathews *et al.,* 2005]. Accounting for further non-ideality by introducing roughness, intermixing, interdiffusion, and interfacial reactions has further profound consequences, including differences between chemical and magnetic roughness [Cable *et al.,* 1986; MacKay *et al.,* 1996; Fitzsimmons *et al.,* 2004], and decoupling of chemical and magnetic interface locations [Lund *et al.,* 2004], which play a vital role in many of the effects discussed in Secs. II.A–II.D. Interdiffusion at the well-studied Fe/Cr interface, for example, causes the superlattice to possess significantly different electronic properties than an averaging of the two materials [Revaz *et al.*, 2002]. In heterostructured materials used for spintronic devices, even *sharp* interfaces have tails to their depth profiles on the order of parts per million that modify properties due to Kondo physics [Lee *et al.,* 2007; O'Brien *et al.,* 2014].

Preserving lattice match, purity, and epitaxy, but allowing for intrinsic crystal defects introduces vacancies and interstitials, and potentially stacking faults, twin boundaries, dislocations, *etc.* These different types of defects significantly impact magnetism in films and heterostructures, including magnetocrystalline anisotropies [Weller *et al.,* 2000; Moser *et al.,* 2002], and the wealth of interfacial magnetic phenomena affected by mosaicity. In the non-epitaxial case, grain boundaries are introduced, adding a further layer of complexity due to grain sizes and distributions, grain orientation distributions (*i.e.*, in- and out-of-plane texture), and grain boundary structures. Notably, many applications of interfacial magnetic phenomena (*e.g.*, data storage) operate in this part of "disorder space". Columnar grains are often exploited; their sizes (and thus thermal stabilities), boundary widths (and thus inter-grain couplings), and epitaxial relationships across interfaces enable an impressive degree of control over magnetic properties [Weller *et al.,* 2000; Moser *et al.,* 2002; Piramanayagam, 2007].

Expanding the discussion to include multi-component systems (alloys and compounds) introduces anti-site defects, species-specific intermixing and alloying, and the fascinating phenomena that derive from the interplay between interfaces, charge states, stoichiometry, and defects. The oxides highlighted in Sec. II.D provide a prominent example, where interface polar discontinuities, charge transfer, electronic and orbital reconstructions, formation of a two-dimensional electron gas, and carrier densities are profoundly affected by interface and bulk non-idealities including intermixing, cation defects, oxygen vacancies, *etc.* [Ramesh *et al.,* 2008; Chakalian *et al.,* 2012; Hwang *et al.,* 2012; Bhattacharya *et al,.* 2014; Stemmer *et al.,* 2014; Sulpizio *et al.,* 2014]. At such interfaces, mismatch in lattice parameters (and often also symmetry) is typically present, and can be central to the functionality. Even in simple materials, the issue of how heteroepitaxial strain is relieved (often *via* misfit dislocations [van der Merwe, 1991]) is important, but in complex materials such as perovskites the lattice mismatch accommodation and relaxation mechanisms are rich [Pennycook *et al.,* 2013], as touched upon in the preceding section. They include octahedral distortions and non-equilibrium tilt patterns [Pennycook *et al.,* 2013; Bhattacharya *et al.,* 2014; Moon *et al.,* 2014a], and long-range defect [*e.g.*, oxygen vacancy, **Fig**. **5(a**)] ordering [Torija *et al.,* 2011; Gazquez *et al.,* 2013], providing both challenges and opportunities [Biskup *et al.,* 2014].

Moving beyond bilayers, to multilayers and superlattices, introduces high densities of stacked interfaces. The non-idealities in such structures can develop with thickness in non-trivial ways, leading, for example, to the concept of correlated *vs*. uncorrelated interface disorder [Payne *et al.,* 1993; Fitzsimmons *et al.,* 2004]. Furthermore, under thermodynamic conditions, favorable parameters for growth of material B on material A generally precludes favorable conditions for the growth of A on B, providing a general argument for interface asymmetry. Striking illustrations of interface asymmetry can be found in nominally symmetric complex oxide superlattices [Fitting Kourkoutis *et al*., 2007; May *et al.,* 2008], as shown in **Fig. 5(b)**.

### 2. Intrinsic vs. extrinsic interfacial magnetic phenomena

It is clear that numerous types of defects can, and often do, form in magnetic films and heterostructures and that they can significantly perturb intrinsic interfacial magnetic effects. The interlayer exchange coupling in ferromagnet/non-magnet metallic systems discussed in Sec. II.B provides a classic example of how defects perturb the underlying effect [Himpsel *et al.,* 1998; Schuller *et al.,* 1999; Vaz *et al.,* 2008]. There are also cases, however, where defects play a much greater role, *creating* the interfacial effect, or at least substantially modulating its magnitude. Antiferromagnet/ferromagnet interfaces are illustrative in that regard, where consensus is that the generation of uncompensated spins in the antiferromagnet by point defects, grain boundaries, roughness, surface steps, *etc.*, plays a defining role in exchange bias [Berkowitz and Takano, 1999; Nogues and Schuller, 1999; Nogues *et al.,* 2005]. Despite this extrinsic character, the phenomenon can be controlled to such a level that it forms an essential part of commercial spintronic devices [Zutic *et al.,* 2004; Tsymbal *et al.,* 2011].

There are many examples where defects are even more significant, being the *origin* of the magnetic order itself. These situations push experimental methodologies used to understand the magnetism in these systems to their limits, particularly as the magnetism is often weak and derived from small volumes [Garcia *et al.,* 2009]. Measurement artifacts that mimic magnetic signatures become problematic, as does environmental contamination. The latter is non-trivial, due to the ubiquity of magnetic contaminants in source materials, deposition equipment, substrates, *etc.* [Garcia *et al.,* 2009]. These issues were highlighted in the contentious effort to confirm or eliminate ferromagnetism in dilute magnetic oxides [Coey *et al.,* 2008; Dietl, 2010], and in studies of $d^0$ magnetism in materials such as $SrTiO_3$ [Coey *et al.,* 2016a)] and $CeO_2$ [Coey *et al.,* 2016b], where giant orbital paramagnetism has been hypothesized.

Recent work on indications of magnetism in the nominally $d^0$ perovskite $SrTiO_3$, and at the related $SrTiO_3/LaAlO_3$ interface, provides a high profile example where intrinsic and extrinsic sources are yet to be understood. While evidence for some form of magnetic order in these systems is significant, coming from magnetotransport [Brinkman *et al.,* 2007], magnetization [Ariando *et al.,* 2011], torque magnetometry [Li *et al.,* 2011], X-ray magnetic circular dichroism [Lee *et al.,* 2013b], superconducting quantum interference device (SQUID) microscopy [Bert *et al.,* 2011], *etc.*, other measurements reveal null results [Fitzsimmons *et al.,* 2011]. Sample-to-sample and study-to-study variations exist, along with indications of extrinsic contributions, evidence of some role for oxygen vacancies and perhaps complexes [Rice *et al.,* 2014]), as well as prominent magnetic inhomogeneity [Ariando *et al.,* 2011; Bert *et al.,* 2011; Brinkman *et al.,* 2007, Fitzsimmons *et al.,* 2011; Li *et al.,* 2011; Lee *et al.,* 2013b]. Additional work will be required to understand the origin(s) of this behavior, both from experimental and theoretical perspectives. It is not yet clear whether a single mechanism, intrinsic *or* extrinsic, can explain the variety of observations of local moments [Lee *et al.,* 2011], long-range order, and magnetic inhomogeneity, although recent efforts to map a phase diagram in the moment density-electron density plane provide a promising direction [Sulpizio *et al.,* 2014].

## 3. Characterization tools

While a plethora of structural, chemical, and magnetic characterization tools have been applied to understand interfacial (and bulk) magnetism and disorder, certain techniques have yielded particularly significant advances. These include both real and reciprocal space methods, applied *in* and *ex situ*. In real space, scanning probe microscopies and cross-sectional transmission electron microscopy (TEM) are particularly important. The aberration-corrected, scanning version of the latter (STEM) enables atomic column resolution imaging across interfaces, particularly in complex oxide heterostructures [Torija *et al.,* 2011; Hwang *et al.,* 2012; Gazquez *et al.,* 2013; Pennycook *et al.,* 2013; Biskup *et al.,* 2014]. Chemical, charge state, and even spin-state information can be obtained, utilizing methods such as electron energy loss spectroscopy (EELS) [Varela *et al.,* 2012]. In reciprocal space, a suite of high-resolution X-ray scattering and/or absorption tools is now routinely applied to interface magnetism, in both lab and synchrotron environments. Reciprocal space maps [Gazquez *et al.,* 2013], synchrotron diffraction [Payne *et al.,* 1993; MacKay *et al.,* 1996], and X-ray absorption spectroscopy (XAS) reveal strain-state, structural details, and chemistry, while film and interface specific probes such as grazing-incidence X-ray reflectivity (GIXR) and more advanced methods such as resonant X-ray reflectivity [Macke *et al*., 2014] and coherent Bragg rod analysis [Willmott *et al.,* 2007] enable element-specific depth profiling and interface structure determination. Ultra-high vacuum mainstays such as reflection high-energy electron diffraction (RHEED) also remain important, for *in situ* growth monitoring and control [Himpsel *et al.,* 1998; Ariando *et al.,* 2011].

Modern chemical characterization in magnetic heterostructures employs a variety of spectroscopies. Surface methods such as X-ray photoelectron spectroscopy (XPS) and Auger electron spectroscopy (AES) continue to be important, along with ion beam analysis [*e.g.*, Rutherford back-scattering (RBS)]. Analytical cross-sectional TEM plays a major role, enabling chemical and structural interrogation at the atomic column level [Varela *et al.,* 2012].

Macroscopic tools such as SQUID and magneto-optical Kerr effect (MOKE) magnetometry are often paired with direct, interface-sensitive methods. Real space examples include scanning probe methods such as magnetic force microscopy (MFM) and spin-polarized scanning tunneling microscopy (SP-STM)], scanning electron microscopy with polarization analysis (SEMPA), and MOKE and SQUID microscopies [Dahlberg *et al.,* 1999; Hopster *et al.,* 2005]. These techniques have enabled substantial advances in the understanding of domain structures and interlayer magnetic coupling. Lorentz TEM techniques [Chapman *et al.,* 1999; Petford-Long *et al.,* 2012] and cross-sectional STEM provide spin-state [Gazquez *et al.,* 2011; Varela *et al.,* 2012] or magnetization [Schattschneider *et al.,* 2006] sensitivity. The use of synchrotron methods, often utilizing resonant enhancement, is now widespread in the study of interface magnetism, *via* X-ray magnetic circular dichroism (XMCD) and various forms of coherent or incoherent magnetic scattering [Kortright *et al.,* 1999; Srajer *et al.,* 2006; Kuch *et al.,* 2015]. Polarized neutron reflectometry (PNR) remains a primary method for magnetic depth profiling across interfaces [Fitzsimmons *et al.,* 2004; Lund *et al.,* 2004; Fitzsimmons *et al.,* 2011; Moon *et al.,* 2014b, O'Brien *et al.,* 2014]. Neutron diffraction can measure thin film magnetic order parameters [Fitzsimmons *et al.,* 2004]. Intensity limitations remain a factor, particularly for inelastic scattering; future source and instrument development could overcome these. For further discussion of probes of *dynamic* magnetic effects see Secs. IV and VI.

### *F. Open questions and new directions*

The discussion in Secs. II.A-E leads to a number of exciting new directions in emergent magnetism at interfaces. Here we highlight three particular groups of these.

(*i*) *Interfacial modification of magnetic properties*: this refers to the opportunities associated with exploiting interfaces to modify, control, enhance, or even create and annihilate magnetic order and properties. One example is deliberate introduction of high spin-orbit coupled impurity atoms at interfaces. Another promising example that has already emerged (Secs. II.B and D) is electrostatic or electrochemical control of magnetism *via* large surface or interface electric fields [Ahn *et al*., 2006]. Novel dielectrics, ferroelectrics, and electric double layers in electrolytes are have enabled reversible control of magnetic order, ordering temperature, anisotropy, and coercivity [Ahn *et al*., 2006; Dhoot *et al*., 2009; Vaz *et al*., 2010; Yamada *et al*., 2011]. Substantial further progress is anticipated in the understanding of electrostatic *vs*. electrochemical operation (Walter *et al.*, 2016), relevant length scales, associated disorder, *etc.* Using such methods to manipulate and harness electronic correlations is a related frontier with potential. Stabilization of crystallographic, magnetic, or electronic phases with designed function *via* non-equilibrium epitaxy [Gorbenko *et al*., 2002], interfacial structure manipulation, or static or dynamic strain (Sec. II.D above) is an important and multi-faceted example. First principles electronic structure calculations are essential to guiding and understanding experiments.

(*ii*) *Novel routes to control or enhance interfacial magnetic function:* this area deals not with interfacial *modification* of magnetic properties within a film, but rather control over a truly interfacial effect. External electrical manipulation of interface-induced perpendicular anisotropy in systems such as Co/Pt is an example [Weisheit *et al.*, 2007; Maruyana *et al*., 2009; Niranjan *et al*., 2010; Wang *et al*., 2011; Shiota *et al*., 2012]. Ways to design and enhance specific interactions at magnetic interfaces are of high interest. One could imagine, for instance, controlled enhancement of DM interactions (see Secs. II.A–D above) in order to *design* desirable spin textures, and their associated excitations, at or near interfaces (see Sec. IV), theory and computation playing a vital role. In lower carrier density systems such as oxides, “band-engineered” interfaces are also attractive, but present challenges (see Sec. II.D above) in comparison to conventional semiconductors and metals due to a confluence of the important length scales. A highly attractive, but obviously challenging approach is to develop methods by which the materials required for specific terms in Hamiltonians, desired for a certain problem or application, can be directly predicted. This “Hamiltonian in - material out” approach has been discussed in connection with theory- and/or data-driven efforts such as the material genome and designer materials, and is highly relevant to systems where interfaces are essential to the functionality.

*(iii) Understanding and predicting emergent magnetic behavior due to interfaces or confinement*: effects at interfaces can be classified as emergent in the sense that complex unanticipated phenomena emerge from apparently simple materials and interactions, providing challenges of predictability and design of functionality. To what extent can emergent magnetic phenomena specifically induced by interfaces and/or confinement be predicted? How useful are symmetry- and topology-based phenomenologies *vs*. microscopic theories, including *ab initio*? The possible new directions of research particularly concern magnetism or spin polarized currents created by nominally non-magnetic components (*e.g.,* topological insulators discussed in section III), or predictions of alternate ground states in correlated systems under dimensional confinement. These problems require fresh experimental and theoretical approaches.

## III. SPIN TRANSPORT AT AND THROUGH INTERFACES

In this section we focus on the topic of spin transport near and across interfaces, including spin pumping resulting from microwave fields and thermally-induced spin transport, and how these are affected by, and affect, magnetism in the adjacent layers. Despite many years of study and a strong link between scientific discoveries and technological applications, fundamental questions remain unanswered, and surprising new phenomena continue to challenge current understanding. Well-known effects such as the Nobel Prize-winning GMR and the related tunneling magnetoresistance (TMR) are still subjects of active research. Recent active investigations include spin accumulation at the surface of non-magnetic metals with large spin orbit coupling driven by a charge current parallel to the interface (termed the spin Hall effect, SHE), and the associated spin injection into a neighboring material and consequent transfer of angular momentum into a magnetic layer (known as spin-orbit torque), which enables manipulation of its magnetization. In this section, we limit our attention to *linear* response of charge and spin currents to electrical biases, magnetic fields, and thermal gradients; non-linear responses will be taken up in Sec. V. The surface and interface magnetic and structural phenomena discussed in Sec. II play a vital role in understanding these linear transport phenomena; many effects discussed there strengthen processes that violate spin-conservation and hence modify spin transport and spin torque results.

Section III.A begins by discussing the interplay between magnetization and spin transport at interfaces in phenomena such as GMR using the two-channel (↑-spin and ↓-spin carriers) model of spin transport, spin currents and spin torques, spin transport across antiferromagnetic interfaces, and spin-dependent tunneling. This is followed by a discussion of thermally-generated spin transport in Sec. III.B. Current-induced torques due to spin-orbit interactions at interfaces with strongly spin-orbit coupled materials are discussed in Sec. III.C. The important case of spin torque due to currents carried by the strongly spin-orbit coupled states on the surface of topological insulators is singled out in Sec. III.D. We end in Sec. III.E with an outlook on some open questions.

### *A. Magnetization-dependent transport*

#### 1. Magnetoresistance in heterostructures: two-channel transport and beyond

The discovery of giant magnetoresistance (GMR) with current in the plane of the heterostructures [Baibich *et al*., 1988; Binasch *et al*., 1989] and later with current perpendicular to the plane [Pratt *et al.*, 1991; Bass and Pratt, 1999] prompted an intense focus on the coupling between charge transport and magnetization. In GMR heterostructures, the electrical resistance in a ferromagnet/non-magnet/ferromagnet trilayer (or multilayer) is lower when the magnetization of the two ferromagnets is parallel than when it is antiparallel. This discovery led quickly to applications in magnetic sensors and hard disk drive read heads [Daughton, 1999; Prinz, 1999; Parkin *et al*., 2003], many based on spin valve heterostructures [Dieny *et al*., 1991], which rely on exchange bias to pin the direction of one ferromagnetic layer to a neighboring antiferromagnet while the other ferromagnetic layer is able to switch direction in low applied magnetic field. GMR is based on the transport of spin-polarized electrons through the non-magnetic layer from one non-magnet/ferromagnet interface to another, with in-plane [Camley and Barnas, 1989] or out-of-plane current [Valet and Fert, 1993; Dieny, 1994]. In ferromagnets, the conductivity is different for majority and minority spin channels. The GMR effect acts when electrons sample both ferromagnetic layers before their spin relaxes. Under this circumstance, both spin channels have the same average scattering rate when the magnetizations are anti-parallel but one spin channel sees lower average resistance when the magnetizations are parallel, leading to a difference in resistance for parallel and antiparallel magnetizations, and thus negative GMR. Note that if one ferromagnet has its Fermi level lying in the majority spin channel and the other in the minority spin channel, the identical effect leads to lower resistance in the *anti*-parallel state, and hence an *increased* resistance with applied field, sometimes called inverse GMR [Renard *et al.,* 1996].

A simpler phenomenon directly related to GMR is tunneling magnetoresistance (TMR) [Miyazaki and Tezuka, 1995; Moodera *et al*., 1995], in which the non-magnetic metallic layer in a GMR trilayer is replaced by a thin insulating tunnel barrier such as amorphous $AlO_x$, leading to large MR effects *via* magnetization-dependent tunneling rates. Theoretical predictions [Butler *et al.,* 2001; Mathon and Umerski, 2001] led to the realization [Parkin *et al*., 2004; Yuasa *et al*., 2004] of extremely large TMR values in textured or epitaxial systems such as Fe/MgO/Fe. Such tunnel junctions now play a crucial role in technologies such as read heads in hard disk drives [Zhu and Park, 2006] and MRAM [Tehrani *et al*., 1999; Ando *et al*., 2014], and are important as *detectors* of the magnetic configuration of heterostructures described in other sections of this paper.

Many spin-dependent transport effects can be understood using Mott's two-spin-channel picture of transport in ferromagnetic conductors [Mott, 1936]. This formulation builds on a mean-field description of the many-electron state, in which magnetic order gives rise to a spin- and position-dependent exchange coupling that lowers the potential energy seen by electrons with spin aligned with the collective spin-density (majority spins) relative to that of electrons with spin aligned opposite to the collective spin-density (minority spins). For most magnetic conductors, the mean-field theory description, typically positioned in a density-functional-theory context, is adequate. Magneto-transport phenomena can be qualitatively understood in terms of non-relativistic electronic states with a global spin-quantization axis and differences between tunneling characteristics or bulk transport properties of ↑-spin and ↓-spin electrons. For example, in TMR, the tunnel barrier seen by both ↑-spin and ↓-spin electrons depends on whether magnetization directions on opposite sides of the insulating barrier are parallel or anti-parallel [Julliere, 1975; Miyazaki and Tezuka, 1995; Moodera *et al*., 1995; Moodera and Mathon, 1999].

Central to effects such as GMR and TMR, charge currents flowing in metallic ferromagnets are inevitably (at least partially) spin polarized. Half-metallic materials [de Groot *et al*., 1983; Katsnelson *et al*., 2008] are an extreme case. In these materials, the band structure is such that the Fermi level lies in a gap for one spin orientation, so that only the other spin participates in transport. Examples include $CrO_2$ [Schwarz, 1986; Ji *et al*., 2001] and some Heusler alloys, a large class of materials with structures based on the face-centered-cubic crystal structure [Galanakis and Mavropoulos, 2007]. The impact of these materials on spintronics is in principle large, but in practice has been limited because half-metallic properties are often reduced at surfaces and interfaces and by thermal fluctuations of magnetization direction, which mixes spin directions. However, this remains an active area.

Other more subtle effects in spintronics depend explicitly on either spin-orbit coupling or on spatially non-uniform non-collinear magnetization textures. For example, when spin-orbit coupling is neglected, the resistance of a bulk material with spatially uniform ferromagnetism is independent of magnetization direction; when the magnetization direction changes, the majority and minority spin conductivities and the total conductivity are thus unchanged even though the direction of spin polarization is changed. When spin-orbit coupling (which contributes a non-uniform and non-local effective magnetic field to the Hamiltonian; see Eq. 2.1) is present, exchange and spin-orbit coupling combine to make electronic structure dependent on magnetization direction. When the magnetization direction is fixed, lattice symmetries are reduced and the conductivity tensor has new anisotropies, leading to an effect known as anisotropic magnetoresistance in ferromagnets [Thomson, 1856; McGuire and Potter, 1975], which was used in technology [Daughton, 1992] prior to GMR and TMR. Another related effect in ferromagnets, the anomalous Hall effect [Hall, 1881; Nagaosa *et al*., 2010] requires only spin-orbit coupling and the broken time reversal symmetry found in ferromagnets, not spatial anisotropy.

## 2. Other magnetotransport effects: tunneling and gating

In GMR and TMR, the resistance of a trilayer or multilayer stack (current in plane or perpendicular) depends on the relative magnetization directions of the magnetic layers separated by non-magnetic metallic or insulating layer(s). However, the resistance can depend on the absolute magnetization (or spin-sublattice magnetization) direction of a *single* ferromagnetic (or antiferromagnetic) layer in tunnel structures [Brey *et al.*, 2004; Gould *et al*., 2004; Giraud *et al*., 2005; Gao *et al*., 2007; Moser *et al*., 2007; Park *et al*., 2008; Park *et al*., 2011; Wang *et al.*, 2012b]. This effect, referred to as tunneling anisotropic magnetoresistance, relies on spin-orbit coupling, which is typically enhanced by interfaces.

Furthermore, when a magnetic element, whether ferromagnetic or antiferromagnetic, is used as the gate electrode in a transistor, the chemical potential it induces in the gated layer depends on magnetization orientation. This dependence makes the gating action depend on magnetization direction and can give rise to a large magnetoresistance signal [Wunderlich *et al*., 2006; Ciccarelli *et al*., 2012], even though the magnetic element does not lie in the transport channel.

### 3. Spin currents, non-local spin valves, spin pumping

A key concept in understanding interfacial magnetic and transport properties is the notion of a spin current. Spin currents carry spin from one place to another and apply torques to magnetic materials *via* transfer of spin angular momentum. This can occur in the presence *or* absence of a charge current. Mathematically, spin-currents are tensors $j_i^{\alpha}$, specified by both spin-component ($\alpha$) and current direction (i) labels. The vector $\mathbf{j}^{\alpha}$ is strictly speaking well-defined only when the $\alpha$ component of total spin is a good quantum number, and care is therefore required in applying the spin-current concept to materials with strong spin-orbit coupling. When spin-orbit coupling is negligible, one contribution to the magnetization's time derivative in a material is the net spin current flowing across its boundaries; spin-currents thus contribute to the spin torques that act on the magnetization, to be discussed in Sec. III.C. For weak spin-orbit coupling, the spin-current torque adds to other sources of torque (*e.g.,* applied magnetic field). This statement is valid even though the notions of spin angular momentum transfer and spin current are imprecise because spin-orbit coupling is always present, making it difficult to rigorously define the spin current across an interface, and partially renormalizing the spin current torque [Nuñez and MacDonald, 2006].

Spin current due to polarization of charge current in ferromagnets can be used to inject spin from a ferromagnet into a non-magnetic metal [Johnson and Silsbee, 1985; Chappert, 2008; Ralph and Stiles, 2008] or into a non-magnetic semiconductor [Fiederling *et al*., 1999; Ohno *et al*., 1999]. An important structure in the study of spin currents, which illustrates some key concepts, is the lateral *non-local* spin valve [Johnson and Silsbee, 1985; Jedema *et al*., 2001; Jedema *et al*., 2002], in its modern incarnation essentially a non-magnetic nanowire connecting two ferromagnetic nanowires. Injection of a (partially) spin polarized charge current from one ferromagnet to the non-magnet results in spin injection, creating a non-equilibrium spin population. This population diffuses in the non-magnetic wire, generating a diffusive *pure spin current, with no associated charge current*, which enters the second ferromagnetic nanowire if it is sufficiently close. Exponential damping of the spin population with distance is characterized by the spin diffusion length (the mean distance diffusively travelled between spin-flipping events), determined *via* non-local magnetoresistance measurements in the second ferromagnetic nanowire as a function of distance between the two. Devices with this geometry have facilitated the study of spin injection, relaxation, and detection in metals [Johnson and Silsbee, 1985; Jedema *et al*., 2001; Jedema *et al*., 2002; Ji *et al*., 2007; Fukuma *et al*., 2011; O'Brien *et al.,* 2014], semiconductors [Lou *et al*., 2007], and two-dimensional materials like graphene where spin-orbit coupling can be modified via proximity effects and/or functionalization of the surface [Tombros *et al*., 2007; Pesin and MacDonald, 2012; Han *et al*., 2014]. For spin injection, conductivity mismatch is an issue for both semiconductors and metals [Schmidt *et al*., 2000; Takahashi and Maekawa, 2003], often resolved by insertion of a (spin-conserving) tunnel barrier at the ferromagnet/non-magnet interface. While unresolved issues remain, the understanding of spin relaxation in non-magnets has benefitted greatly from such structures [Bass and Pratt, 2007]. In metals, the Elliott-Yafet spin relaxation mechanism dominates [Elliott, 1954; Yafet, 1963; Beneu and Monod, 1978], whereas in semiconductors such as GaAs, which lack a center of inversion symmetry, the D'yakonov-Perel' mechanism dominates [D'yakonov and Perel', 1971]; see e.g. [Zutic *et al*., 2004; Boross *et al*., 2013]. The spin diffusion length in semiconductors is often much larger than in metals, defects playing an important role in both. In Al, for example, reported values vary from >100 μm in annealed bulk samples [Johnson and Silsbee, 1985; Bass and Pratt, 2007], to ≈100 nm to 1000 nm in polycrystalline films [Bass and Pratt, 2007; O'Brien *et al.,* 2014]. In high *Z* metals, where spin-orbit coupling is strong, Elliott-Yafet relaxation becomes very efficient; metals such as Ta or Pt thus have very short spin diffusion lengths [Vélez *et al*., 2016; Wu *et al*., 2016a].

There are other means to create spin flow not accompanied by charge flow. For example, spin-dependent scattering processes [Smit, 1958; Berger, 1970; Hirsch, 1999] in metals with spin-orbit coupling (either intrinsically or from impurities) can lead to spin currents that flow perpendicular to the direction of charge flow and produce spin accumulations at the sample boundaries. This is one important mechanism for the property known as the spin Hall effect [D'yakonov and Perel', 1971; Hirsch, 1999; Kato *et al*., 2004; Wunderlich *et al*., 2005; Valenzuela and Tinkham, 2006; Hoffmann, 2013; Sinova *et al*., 2015]. Importantly, this spin accumulation can result in a spin current into a neighboring layer, with no accompanying charge current, as shown schematically in **Figure 6**.

The spin Hall effect can result from impurity scattering in the non-magnetic metal, but is sometimes dominated by a Berry-phase related response of occupied quasiparticle wave functions to the lateral electric field producing the charge transport. This intrinsic response [Karplus and Luttinger, 1954; Jungwirth *et al*., 2002; Murakami *et al*., 2003; Sinova *et al*., 2004; Guo *et al*, 2008; Tanaka *et al*., 2008] depends on electronic structure only, not on impurity scattering. The reciprocal effect, in which a pure spin current leads to charge accumulation and development of an electromotive force in a direction transverse to the spin current, is known as the inverse spin Hall effect [Murakami *et al*., 2003; Saitoh *et al*., 2006]. We note that while the spin Hall effect *relies* on spin-orbit coupling, calculation of its value requires detailed band structure calculations, in both the intrinsic and extrinsic (impurity or disorder) limit. There is not a simple proportionality between the atomic spin-orbit coupling parameter $\lambda_{nl}$ previously discussed and the spin Hall effect; as an example, the spin Hall effect in transition metal elements reverses sign with increasing band filling across the periodic table [Tanaka *et al*., 2008].

Although the direct and inverse spin Hall effects were originally discovered experimentally in semiconductors [Chazalviel and Solomon, 1972; Chazalviel, 1975; Bakun *et al*., 1984; Kato *et al*., 2004; Wunderlich *et al*., 2005], the effect is maximized for nonmagnetic heavy-metals such as Pt, Ta, or W with large spin-orbit coupling. We discuss in greater detail in III.C how spin-orbit torques are generated when these materials are layered with ferromagnets.

Spin pumping is another mechanism that can drive a pure spin current. When a ferromagnetic layer is driven into precession, usually *via* microwave excitation, a spin current is generated and flows across a metallic ferromagnet/nonmagnet interface [Tserkovnyak *et al*., 2002; Tserkovnyak *et al*., 2005; Azevedo *et al*., 2005; Saitoh *et al*., 2006; Mosendz *et al*., 2010a]. As illustrated in **Figure 7**, the precessing magnetization in the ferromagnet acts like a peristaltic pump, causing AC and DC spin currents to cross the interface. Instantaneously, the AC spin current is directed along the normal to the ferromagnet/nonmagnet interface with a spin orientation perpendicular to the instantaneous magnetization and its time derivative. The DC spin current also flows perpendicular to the interface and shares the average spin-orientation of the precessing magnet. The spin-pumping spin-current $j_s$ can be expressed [Tserkovnyak *et al.*, 2002; Tserkovnyak *et al*., 2005; Mosendz *et al*., 2010a] in terms of the spin-mixing conductance $g_{\uparrow\downarrow}$:

$$j_s = \frac{\hbar\omega \sin^2(\theta)}{4\pi} Re\left[g_{\uparrow,\downarrow}\right] , \qquad \textbf{[3.1]}$$

where $\omega$ is the precession frequency, and $\theta$ is the precession cone-angle. When a spin-current is driven into a non-magnetic layer with large spin-orbit coupling, the spin current can be converted into a detectable charge current density, $j$, *via* the inverse spin Hall effect. In this case, the direction of $j$ is parallel to the interface between the ferromagnetic and the nonmagnetic conductor [Ando *et al.*, 2011] and perpendicular to the direction of magnetization.

The spin mixing conductance [Brataas *et al.*, 2000, Brataas *et al.*, 2001] characterizes the relaxation of spins transverse to the magnetization at the interface and is an important concept both for spin pumping and spin transfer torques, discussed in Sec. III.C. It describes the transfer of angular momentum between the spin current in a non-magnetic layer and the magnetization in a neighboring ferromagnetic layer. The real part describes a direct transfer of angular momentum and the imaginary part, typically much smaller than the real part, describes rotation of the spin current when scattering from the interface. First principles calculations agree quantitatively with measured values [Xia *et al.*, 2002].

### 4. Spintronics *via* heterostructures with antiferromagnets

Spintronics has recently been enriched by use of antiferromagnets to complement or replace ferromagnets [Shindou and Nagaosa, 2001; Nuñez *at al*., 2006; Haney and MacDonald, 2008; Xu *et al*., 2008; Gomonay and Loktev, 2010; Shick *et al*., 2010; Hals *et al*., 2011; Cheng *et al*., 2014; Železný *et al*., 2014; Chen *et al*., 2014b; Zhang *et al*., 2014; Zhang *et al*., 2015d; Seki *et al*., 2015; Wu *et al*., 2016b]. In many cases, all that is required is an interfacial magnetic moment: the surface uncompensated moments of an antiferromagnet for example are able to polarize the carriers of a semiconductor [Lee *et al*., 2013a]. Antiferromagnets can be used as spin detectors, magnetoresistors and memories and can be employed for highly efficient electrical manipulation of a ferromagnet [Zhang *et al*., 2015d; Tshitoyan *et al*., 2015].

Several recent experiments have focused on transmission and detection of spin-currents in antiferromagnets. Enhanced spin pumping efficiency has been reported using an antiferromagnet near $T_N$ in place of the non-magnet in a bilayer [Frangou *et al*., 2016]. In ferromagnet/antiferromagnet/nonmagnet trilayers, a spin-current was pumped from the ferromagnet and detected by the inverse spin Hall effect in the nonmagnetic layer. Fluctuations of the antiferromagnetic order provide an efficient pathway for spin current transmission [Saglam *et al*., 2016]. Robust spin-transport through the antiferromagnet (insulating NiO) was ascribed to antiferromagnetic moment fluctuations [Wang *et al*., 2014; Hahn *et al*., 2014; Lin *et al.,* 2016; Zink *et al.,* 2016], but uncompensated spins in the antiferromagnet due to defects, grain boundaries, and interfacial roughness may also play a role. Efficient spin transmission through an antiferromagnet (NiO) was inferred from an inverse experiment on a ferromagnet/antiferromagnet/nonmagnet structure [Moriyama *et al*., 2015] in which spin-current was generated by the spin Hall effect in the nonmagnetic layer and absorbed *via* spin-transfer torque in the ferromagnet. Measurements in ferromagnet/antiferromagnet bilayers have demonstrated that a metallic antiferromagnet (*e.g*., PdMn, IrMn, or PtMn) can itself act as an efficient inverse spin Hall effect detector of spin-current injected from the ferromagnet *via* its spin Hall effect. The spin Hall effect in metallic antiferromagnets can be comparable to that in heavy nonmagnetic metals [Mendes *et al*., 2014, Zhang *et al*., 2014, Zhang *et al*., 2015d].

The next step beyond studies of transmission and detection of spin-currents is spin manipulation by antiferromagnets. Using exchange bias, the antiferromagnetic Néel-order spin-axis direction can be controlled by a magnetic field *via* an exchange-coupled ferromagnet [Park *et al*., 2011; Wu *et al*., 2011; Fina *et al*., 2014] or by techniques analogous to heat-assisted magnetic recording [Marti *et al*., 2014]. However, this control mechanism is relatively slow and inefficient [Dieny *et al*., 2010; Prejbeanu *et al*., 2013]. On the other hand, it has been proposed [Nuñez *et al*., 2006; Gomonay and Loktev, 2010; Železný *et al*., 2014] that current-induced magnetic torques of the form $d\mathbf{M}/dt \sim \mathbf{M} \times (\mathbf{M}\times\hat{\mathbf{p}})$ where $\hat{\mathbf{p}}$ is the direction of the electrically-injected carrier polarization, allow for a large angle reorientation of the antiferromagnetic moments. The key to strong coupling to the Néel order is the property that the effective field ($\mathbf{M}_{\mathbf{A,B}} \times \hat{\mathbf{p}}$) should alternate in sign between the A and B spin sublattices of a collinear antiferromagnet. The alternating (staggered) effective field also drives an efficient field-like torque in an antiferromagnet of the form $d\mathbf{M}_{\mathbf{A,B}}/dt \sim \mathbf{M}_{\mathbf{A,B}} \times \hat{\mathbf{p}}_{\mathbf{A,B}}$, where $\hat{\mathbf{p}}_{\mathbf{A,B}}$ is the non-equilibrium staggered spin polarization generated by electrical current. Néel order switching by this latter torque has been recently demonstrated experimentally for the conducting antiferromagnet CuMnAs [Wadley *et al*., 2016].

### *B. Thermal generation of spin current*

The coupling of heat and spin currents is responsible for a broad range of transport phenomena [Bauer *et al.*, 2012; Boona *et al.*, 2014], including spin Seebeck and Peltier effects,

spin-*dependent* Seebeck and Peltier effects, thermally-driven domain-wall motion and reciprocal heat pumping. These add to a range of more established effects including magnon drag thermopower and the anomalous and planar Nernst effects (which convert a heat current into a charge current or voltage in the presence of magnetic fields, similarly to analogous Hall effects). We first briefly overview terminology and some early experiments, then turn to current research, which intersects with interfacial magnetism.

In the well-known Seebeck effect, shown in **Fig. 8a**, a temperature gradient induces electron flow and generates a charge current or voltage (thermo-electric voltage), depending on whether the circuit is closed or open. Conversely, the Peltier effect (**Fig. 8b**) converts a charge current into a heat current that can be used for heating or cooling. In the analogous "spin Seebeck effect", a thermal gradient generates either a spin current or a spin accumulation, shown in **Fig. 8c**, with a reciprocal "spin Peltier effect" shown in **Fig. 8d**. In spin Seebeck experiments to date, the presence of the thermally-generated spin current is detected from voltage measurements on a strip of spin-orbit-coupled non-magnetic material which converts spin current to a transverse electric field $E_{ISHE}$, and a corresponding voltage difference $V_{ISHE}$, *via* the inverse spin Hall effect. When the relationship between $V_{ISHE}$ and temperature difference $\Delta T$ is linear, it is possible to define a spin-Seebeck coefficient $S_{SSE} = V_{ISHE}/\Delta T$.

In the traditional "longitudinal" geometry shown in **Fig. 8c,** the thermal gradient induces a spin gradient parallel to the thermal gradient, which is measured *via* a Pt contact at the end. In a ferromagnetic metal, this could be envisioned as due to different chemical potential for up and down spins, which would lead to a (spin up minus spin down) gradient. This conduction electron effect has two challenges however: it is indistinguishable from conventional transverse Nernst signals, and is severely limited by the spin diffusion length, leading to its only being relevant on short length scales. Magnons also respond to temperature gradients, and can also result in a spin current and spin Seebeck effect. In this longitudinal geometry, the spin Seebeck effect is believed to be reasonably well understood for ferromagnetic insulators such as yttrium iron garnet $Y_3Fe_5O_{12}$ (YIG), with experimental reports from various groups [Uchida *et al.*, 2010a; Qu *et al.*, 2013; Kikkawa *et al.*, 2013; Wu *et al.*, 2014, Wu *et al.*, 2015]. The physical picture of the effect is that the spin current or spin accumulation results from the temperature gradient in a magnetic material driving a thermally-induced magnon flow, which injects spin across the interface *via* incoherent spin pumping. We therefore refer to these as "magnon systems" and give a more detailed overview in the following section.

The spin Seebeck effect was also invoked to explain experiments in a *transverse* geometry, where thermal gradients were intended to be applied in the plane of a thin ferromagnetic film (as in the standard Seebeck experiment of **Fig. 8a**), with Pt contacts on the top surface instead of at the end to measure thermally induced spin accumulation [Uchida *et al.*, 2008; Jaworski *et al.*, 2010; Uchida *et al.*, 2010b; Jaworski *et al.*, 2011]. An explanation was offered for the generation of a spin potential in this "transverse" geometry based on the difference of the effective Seebeck coefficients for up and down spins. However, questions arose regarding the long length scale probed in the experiment, which exceeded the spin diffusion length in the ferromagnetic metal by many orders of magnitude, and recent reports offer different explanations for the signals observed [Huang *et al.,* 2011; Avery *et al.*, 2012; Schmid *et al.,* 2013; Meier *et al.,* 2013b; Jin *et al.,* 2014; Soldatov *et al.,* 2014; Meier *et al.,* 2015]. Many of these focus on the fact that an unintended out-of-plane thermal gradient leads to contamination of the signal by the anomalous Nernst effect in a ferromagnetic metal [Huang *et al.*, 2011], since this thermal analog of the anomalous Hall effect produces a transverse electric field in the presence of mutually perpendicular thermal gradient and magnetization. A measurement in the transverse geometry performed in a suspended sample where the thermal gradient was unambiguously confined to the plane of the metallic ferromagnetic film showed no signature of transverse thermal spin currents [Avery *et al.*, 2012; Schmid, *et al.*, 2013].

Much of the controversy surrounding the earliest transverse experiments focused on the long length scales probed, as the thermal gradients were applied over several millimeters, which is difficult to reconcile with spin polarization of charge current. However, this physical picture of Seebeck coefficients as a source of spin potential is perfectly reasonable if a thermal gradient exists in the ferromagnet *within approximately one spin diffusion length of the interface to the spin detection layer*, as will be discussed just below in section B.2 (Electron systems). In this case the thermal gradient can be used to drive a spin accumulation across this ferromagnetic metal/non-magnetic metal interface. This effect was originally observed in a nanoscale metallic non-local spin valve and was called the *spin-dependent Seebeck effect* [Slachter *et al.*, 2010] to keep it distinct from the long length-scale spin Seebeck effect seen in insulators due to magnons. The corresponding reciprocal effect is the spin-dependent Peltier effect, also experimentally observed [Flipse *et al.*, 2012]. The role of collective excitations, such as phonons and magnons, is a matter of current study, but since the spin-dependent Seebeck effect clearly involves charge transport, we refer to these systems as "electron systems".

Recent work has focused on phenomena where collective effects and electron transport interact. Examples of these "mixed carrier effects" include thermally-induced magnon drag of electrons [Behnia, 2015; Costache *et al.*, 2012; Lucassen *et al.*, 2011; Avery *et al.*, 2011; Watzman *et al.,* 2016]. The relationship between spin-transfer torques in metals and magnon drag was discussed theoretically by [Lucassen *et al.*, 2011], who proposed that such well-known quantities as the thermopower of iron and cobalt, which are ascribed to magnon drag, [Blatt *et al.*, 1967; Watzman *et al.,* 2016] are instead a manifestation of thermal spin-transfer torque, where a heat current in a ferromagnet becomes spin-polarized and exerts a torque on a neighboring ferromagnet. Notably, magnon-drag thermopower in ferromagnetic metals far exceeds the classical diffusive thermopower.

### 1. Magnon systems

In insulators, the longitudinal spin-Seebeck effect is due to magnons. An example [Uchida *et al.*, 2010a] is observed in the YIG/Pt structure illustrated in **Fig. 9(a),** for which a semi-quantitative theory in terms of the thermally-induced magnon flow was developed [Hoffman *et al.*, 2013]. The longitudinal spin Seebeck effect in the YIG/Pt system is commonly attributed to magnon transport impinging on the interface [Xiao *et al.*, 2010; Agrawal *et al.*, 2014; Rezende *et al.*, 2014; Kehlberger *et al.*, 2015]. The magnon contribution to the thermal conductivity $\kappa_M$ in the insulating YIG [Boona and Heremans, 2014] gives a magnon heat current $j_{QM} = -\kappa_M \nabla T$ in the presence of a temperature gradient and thus a magnon particle flux $j_M$. Treating the magnons in the dilute gas approximation and supposing each carries roughly the energy of $k_B T$, one can write $j_M = j_{QM}/k_B T = -\kappa_M \nabla T/k_B T$. At the interface with Pt, this thermally driven magnon flux is converted into an itinerant electron spin current with efficiency parametrized by the spin-mixing conductance $g_{\uparrow,\downarrow}$. More specifically, the spin Seebeck effect at interfaces is governed by the balance between the spin polarization of the electrons in the Pt and the thermal spin pumping in the YIG [Xiao *et al.,* 2010; Uchida *et al.*, 2012]. In a heavy metal such as Pt, the itinerant electron spin current gives rise to an electric field due to the inverse spin Hall effect $E_{ISHE,}$ with a direction given by $\mathbf{j_{QM}} \times \mathbf{M}$, and thus to a measured voltage $V_{ISHE} = |E_{ISHE}|L$, where $L$ is the length of the Pt strip. $V_{ISHE}$ as a function of the applied external field $H$ is shown in **Fig. 9(b)**. The magnon longitudinal spin Seebeck effect has also been observed in (Mn,Zn)$Fe_2O_4$ [Uchida *et al.*, 2010c], $NiFe_2O_4$ [Meier *et al.*, 2013a], $Fe_3O_4$ [Ramos *et al.*, 2013; Wu *et al.*, 2015a], $BaFe_{12}O_{19}$ [Li *et al.*, 2014b], $CoFe_2O_4$ [Niizeki *et al.,* 2015] and in various garnet ferrites [Uchida *et al.,* 2013]. More recently it has also been shown that the longitudinal spin Seebeck effect can be observed in paramagnetic [Wu *et al.*, 2015b] and antiferromagnetic insulators in which some spin polarization is produced by a magnetic field [Seki *et al.*, 2015; Wu *et al.*, 2016b; Lin *et al.*, 2016] or by a ferromagnetic substrate [Prakash *et al.,* 2016].

The Onsager reciprocal of the spin-Seebeck effect, the spin-Peltier effect, has also been observed [Flipse *et al.*, 2014] in the Pt/YIG system. Here, a voltage drives a current in the Pt, and the spin Hall effect produces spin accumulation that launches a magnon spin current in the YIG. The magnon flux couples to the phonons and generates a measurable heat current and temperature gradient. In this experiment, the authors observe a temperature change across the YIG as a current is passed through the Pt; Joule heating effects in the Pt strip are avoided by an appropriate experimental setup.

### 2. Electron systems

In metals, spin-polarized electrons create what is called the spin-dependent Seebeck effect [Slachter *et al.*, 2010; Hu *et al.*, 2014; Yamasaki *et al.,* 2015; Pfeiffer *et al.*, 2015; Choi *et al.*, 2015] and its Onsager-reciprocal, the spin-dependent Peltier effect [Flipse *et al.*, 2012], reviewed by [Boona *et al.*, 2014], using a two-channel transport model. Here, each channel contains charge carriers with a distinct spin polarization, and interactions between channels are ignored. Each channel has partial electrical conductivity $G_\uparrow$ and $G_\downarrow$ and carries its own entropy, giving rise to partial Seebeck coefficients $\alpha_\uparrow$ and $\alpha_\downarrow$. In the presence of a temperature gradient, effective fields $\alpha_\uparrow\nabla T$ and $\alpha_\downarrow\nabla T$ arise in the two channels, which thus generate charge currents. The net charge current is $G_\uparrow\ \alpha_\uparrow\nabla T\ +G_\downarrow\ \alpha_\downarrow\nabla T$, while their difference (divided by the elemental electron charge and multiplied by $\hbar/2$) gives the net spin current, which would generally not vanish even when the net charge current is zero. The difference $\alpha_\uparrow\nabla T$ - $\alpha_\downarrow\nabla T$ therefore drives a net spin accumulation at the ends of the sample, which decays on the length scale of the spin diffusion length. The spin-dependent Seebeck effect coefficient $\alpha_S$ represents the ratio of the spin accumulation density to the temperature gradient. Experimentally the spin-dependent Seebeck effect is measured as shown in **Fig. 10a-c**: a heat current $j_Q$ is driven between a ferromagnet and a non-magnet. As discussed above, this results in a spin current $j_S$ parallel to $j_Q$, which is detected non-locally as a voltage between the normal metal and a second ferromagnet. The spin-dependent Peltier effect (the Onsager-reciprocal to the spin-dependent Seebeck effect) has also been demonstrated experimentally as shown in **Fig. 10d-f** [Flipse *et al*., 2012]. The relative importance of the magnonic spin Seebeck effect to the electron spin-dependent Seebeck effect in these metallic systems is difficult to determine and yet to be established.

### *C. Spin transfer torques and interfacial spin-orbit torques*

Spin currents, created *in a ferromagnet* by spin polarization of a charge current, have long been used to create magnetic torques that can move textures in the magnetic order, for example magnetic domain walls or vortices. These torques will be relevant to section IV and will be discussed there. A spin current that *crosses an interface between a magnetic and nonmagnetic material* can also apply a torque on the magnetization *via* transfer of angular momentum. This is true regardless of the physical mechanism that generates the spin current, and whether the spin current originates in the nonmagnetic material or the magnetic layer. Much of the excitement in the field of interfacial magnetism arises from the recently-developed understanding that interfaces can enable particularly efficient generation of spin currents and hence strong spin-transfer torques which can be used to manipulate the magnetization. We note that electron spins and moments are, by convention, oriented in opposite directions; care is therefore required when combining equations of motion for conduction electron spins and magnetization.

The first-generation of spin-transfer torque heterostructures utilized spin-polarized charge currents [Slonczewski, 1996; Berger, 1996; Tsoi *et al*., 1998; Sun, 1999; Myers *et al*., 1999; Katine *et al*., 2000; Tsoi *et al*., 2000; Kiselev *et al*., 2003; Ralph and Stiles, 2008; Brataas *et al*., 2012]. These structures typically consist of ferromagnet/non-magnetic metal/ferromagnet trilayers or ferromagnet/nonmagnetic insulator/ferromagnet tunnel junctions with spin-polarized charge current perpendicular to the layers. The angular-momentum flow associated with the spin-polarized current from the first layer (commonly referred to as spin filtering) can be absorbed by the second ferromagnetic layer, thereby applying a *spin transfer torque* which can excite and even reverse its magnetization. When optimized, with nanosize contacts, these structures have magnetization switching current densities on the order of $10^{10}$ A/m$^2$ giving switching currents of less than 50 μA, significantly more efficient than magnetic switching driven by a magnetic field due to a nearby current flow (often referred to as an "Oersted field") [Jan *et al.,* 2012; Gajek *et al.,* 2012; Sato *et al.,* 2014]. Spin torque from a spin-polarized current can also be used to generate large-angle steady-state magnetic precession to create magnetic nano-oscillators, potentially useful for frequency-tunable microwave sources and detectors [Kiselev *et al*., 2003; Silva and Rippard, 2008], to be discussed in Sec. V.

The spin-transfer torque(s) $\mathbf{T_{stt}}$ on a ferromagnet with magnetization pointing in the **m** (= $\mathbf{M}/M_s$) direction due to a neighboring ferromagnetic layer with magnetization $M_{fixed}$ and direction $\mathbf{m}_{\mathrm{fixed}}$ can be included in the LLG equation, Eq. 2.3, by adding terms of the form

$$\mathbf{T_{stt}} = \frac{1}{t}\frac{g\mu_B}{e} j_z \varepsilon \mathbf{m}\times(\mathbf{m}\times\mathbf{m}_{\mathrm{fixed}}) + \frac{1}{t}\frac{g\mu_B}{e} j_z \varepsilon' \mathbf{m}\times\mathbf{m}_{\mathrm{fixed}} \qquad \textbf{[3.2]}$$

where $t$ is the thickness of the film; $g$ is the electron g-factor; $j_z$ is the charge current density flowing normal to the interface along the $z$-direction from the fixed layer to the free layer; $\varepsilon$ and $\varepsilon$' are efficiency factors for the two terms, typically less than one, that are functions of the relative orientation of the magnetizations $(\mathbf{m}\cdot\mathbf{m}_{\mathrm{fixed}})$, the geometry, and material parameters including the degree of spin polarization *P* of charge carriers in the layers and layer thicknesses [Slonczewski, 2002; Xiao et al., 2004]. Note that in this expression, the spin current density tensor $j_z^\alpha = P j_z(\mathbf{m}_{fixed})^\alpha$ has been absorbed into various parts of the expression: the total current density flow through the interface, $j_z$, appears explicitly, the polarization, *P*, has been absorbed into the efficiency factors $\varepsilon$ and $\varepsilon'$, and the orientation of the spins (superscript α) is along $\mathbf{m}_{\mathrm{fixed}}$. **Figure 11a** shows the vector representation of Eq. 3.2, and how this acts to switch the magnetization **M** in a ferromagnetic heterostructure, to be discussed further in Sec. V. The first term, which is typically much larger than the second, has the same vector form as the Landau-Lifshitz damping term in Eq. 2.3 and is frequently referred to as the Slonczewski, damping-like, or anti-damping-like (depending on its sign) torque (in **Fig. 11**, we call this the anti-damping torque, to clearly distinguish this from the conventional Landau-Lifshitz damping torque). The second term in Eq. 3.2 has the same form as a precessional torque around a magnetic field (the first term in either version of Eq. 2.3), and is frequently referred to as (effective) field-like torque. Both of these terms act only at the interface between ferromagnetic and non-magnetic layers (unlike the terms in Eq. 2.3); the factor of $1/t$ is included to spread this torque out over the thickness of the film, which assumes that the magnetization is held uniform in the direction of the thickness by the exchange interaction. More care is required when treating these torques in thicker films.

If the spin current from the fixed layer dephases completely when reflecting from or flowing through the free magnetic layer, then only the anti-damping component of torque is expected to be significant, but if the dephasing is not complete, then the effective field component can also be strong [Ralph and Stiles, 2008]. This behavior is described by the real and imaginary parts of the mixing conductance, to which $\varepsilon$ and $\varepsilon'$ are proportional, respectively. The effective-field torque has been found to be negligible for spin-polarized charge currents in all-metallic systems due to scattering-induced dephasing, but it can be significant in magnetic tunnel junctions due to the large contributions from a narrow range of wave vectors of the tunneling electrons. Both torques can also be strongly bias-voltage dependent in magnetic tunnel junctions [Kubota *et al*., 2007; Sankey *et al*., 2008; Chanthbouala *et al*., 2011], reaching up to several millitesla for current densities of $10^{11}$ A/m$^2$.

Spin torque from a spin-polarized current faces an important practical upper limit in efficiency since the strength of the torque is limited by the amount of angular momentum carried by an individual charge carrier, reflected in Eq. 3.2 by efficiency factors being typically less than one. Spin torque from a spin-polarized current also provides poor *energy* efficiency because of the associated charge current; in a typical magnetic tunnel junction used for memory devices, each electron dissipates on average several tenths of an electron volt, while the energy required to excite a magnon and contribute to magnetic dynamics is only about 10 $\mu$eV. It is therefore likely that a different, more efficient mechanism for controlling magnetic devices will be required to make, for example, very-high-density embedded magnetic memory devices.

Current-induced torques with efficiencies greater than one (*i.e*., that exceed $\hbar/2$ angular momentum transfer per unit of charge flow) can be generated by using spin-orbit interactions to transduce a flowing charge current into a pure spin current in the direction *perpendicular* to the charge current. An example of this was shown schematically in **Fig**. **6**. A spin torque arising from spin-orbit interactions has been demonstrated in GaMnAs [Chernyshov *et al*., 2009] and other ferromagnets [*e.g*., Jamali *et al*., 2013; Kurebayashi *et al*., 2014; Ciccarelli *et al*., 2016]. To date, the strongest *spin-orbit-generated torques*, and the ones most readily incorporated into practical device designs, have been measured in bilayers of one material with strong spin-orbit interactions that generates a pure spin current, and a second ferromagnetic layer whose magnetization direction can be controlled by this spin current [Ando *et al*., 2008, Miron *et al*., 2010, Pi *et al*., 2010; Miron *et al*., 2011a; Liu *et al*., 2012a; Liu *et al*., 2012b]. There are two classes of mechanisms that can lead to generation of spin torques from spin-orbit interactions in such samples: three-dimensional (3D) bulk mechanisms such as the spin Hall effect already discussed, and 2D (interfacial) mechanisms that produce an inverse spin galvanic effect, commonly known as the Edelstein (sometimes Rashba-Edelstein) effect.

We first examine spin-torques generated by the spin Hall effect in a heavy metal, called spin-orbit torques. Consider a heavy metal/ferromagnet bilayer of the type depicted in **Fig. 12**. If a charge current is applied within the sample plane, the spin Hall effect in the heavy metal generates a pure spin current flowing vertically toward the ferromagnet. If this spin current is absorbed by the ferromagnet, it creates a spin-transfer torque [Ando *et al*., 2008; Liu *et al*., 2011; Liu *et al*., 2012a; Liu *et al*., 2012b; Kim *et al*., 2013; Garello *et al*., 2013; Fan *et al*., 2014b]. The strength of the spin Hall effect is usually parameterized by a factor called the spin Hall ratio, $\theta_{\mathrm{SH}}$, the ratio of the vertically-flowing spin current density to the applied in-plane charge current density in dimensionless units. This ratio is also commonly called the spin Hall angle, although technically it is the tangent of an angle and not an angle *per se*. The sign convention (for a discussion see e.g. [Schreier et al., 2015]) is that the spin Hall angle is positive for Pt, and negative for Ta. The resulting *spin-orbit* torques to be included into LLG (Eq. 2.3) have a similar form to the *spin-transfer* torques in Eq. 3.2:

$$\mathbf{T}_{\mathrm{sot}} = \frac{1}{t}\frac{g\mu_B}{e}\theta_{\mathrm{SH}}\eta j \mathbf{m}\times(\mathbf{m}\times\hat{\mathbf{p}}) + \frac{1}{t}\frac{g\mu_B}{e}\theta_{\mathrm{SH}}\eta' j \mathbf{m}\times\hat{\mathbf{p}} \qquad \textbf{[3.3]}$$

where $j$ is the in-plane charge current density which flows in direction $\hat{\mathbf{j}}$, $\hat{\mathbf{z}}$ is the direction perpendicular to the interface (from heavy metal to ferromagnet, as shown in **Fig. 6**), and $\hat{\mathbf{p}} = (\hat{\mathbf{z}}\times\hat{\mathbf{j}})$ is the direction of the spin polarization of the spin current which flows along $\hat{\mathbf{z}}$. **Figure 6** shows these directions for a ferromagnet with perpendicular magnetic anisotropy, as well as the resulting torques and time evolution of **M** (further discussed in Section V), which also depends strongly on damping parameter $\alpha$. Similar to Eq. 3.2, the spin current density tensor component $j_z^p$ has been absorbed into $\theta_{\mathrm{SH}}$, $\eta$, and $\eta'$. Spin currents and torques are not typically accessible experimentally, but consequences of them, like spin accumulation and magnetization dynamics, are. Measurements of magnetization dynamics and the dependence on *j, t,* geometry and material parameters, are used to infer contributions to Eq. 3.3 that are compared to theoretical calculations of $\theta_{\mathrm{SH}}$. The efficiency factors $\eta$ and $\eta'$ are typically less than one and are determined by the details of the spin transport in the heavy metal and mixing conductance of the interface, giving $\eta' \ll \eta$ for torques due to the spin Hall effect. As in Eq. 3.2 for spin-transfer torque, Eq. 3.3 for spin-orbit torque contains both damping-like and field-like terms, shown in **Fig. 11**. One important difference between spin-transfer and spin-orbit torque is that the orientation of the damping-like spin-orbit torque is fixed by sample geometry rather than the orientation of a fixed-layer magnetization. Another difference is that spin-orbit torque is determined by an in-plane rather than out-of-plane current, which is critical to the switching efficiency of devices based on each.

The overall efficiency of the spin torque generated by the spin Hall effect is the total rate at which angular momentum is absorbed by the magnetic layer per unit charge current, so the torque efficiency for a spin-Hall heterostructure is

$$\frac{T_{sot}A}{ja} = \eta\theta_{SH}\frac{A}{a}\frac{g\mu_B}{e} \qquad \textbf{[3.4]}$$

where *A* is the large area looking down on the top of the heterostructure through which the spin current flows and *a* is the small in-plane cross-sectional area through which charge current *j* flows (**Fig. 12**). The ratio *A*/*a* can be 30 or more even for a very small structure, allowing the torque resulting from the spin Hall effect to be more than an order of magnitude more efficient than one quantized spin per unit charge, far above conventional spin-transfer torque. Microscopically, the simplest description is that each electron in the in-plane charge current is used many times to transfer torque to the ferromagnet [Pai *et al.*, 2012]. Electrons become polarized by spin-orbit coupling when they deflect toward the ferromagnet. At the interface, they transfer angular momentum to the magnetization through the same mechanisms as for spin transfer torques [Ralph and Stiles, 2008]; here however, since there is no net electron flow into the ferromagnet, they ultimately diffuse back into the heavy metal, and the process repeats (**Fig. 12b**). The relevant length scale is the spin diffusion length in the heavy metal, which can be as short as 1 nm to 2 nm. Therefore, many cycles of spin transfer occur even in a compact magnetic structure. This means that large spin Hall angles occur in materials with large resistivity $\rho$, typically Pt alloys, raising the issue of the power required to achieve large torques, e.g. for switching (discussed in Section V), which in turn depends strongly on device geometry. For *spin Hall* related switching using an optimized geometry, power efficiency is proportional to $\eta\theta_{SH}/(\rho\, t\, j_c)$, and for *spin torque* related switching, power efficiency is proportional to $\epsilon/(RA\, j_c)$, where $RA$ is the resistance area product. The largest spin Hall torque efficiencies measured to date from a heavy metal at room temperature is $\eta\theta_{SH} = 0.5$ [Demasius *et al.*, 2016], comparable to the highest efficiencies for spin torque switching; critical current densities are comparable for the two geometries; and while resistivities are high, devices are sufficiently thin that $RA \gg \rho t$. Thus, for optimized devices, spin Hall switching can be significantly more power efficient than spin torque switching.

Spin-orbit interactions also can generate a current-induced torque *via* 2D interfacial effects [Sklenar *et al*., 2016; Soumyanarayanan *et al*, 2016], either at the interface with a heavy metal or an adjacent two-dimensional electron gas (2DEG) [Edelstein, 1990]. When an ordinary 2DEG is not in a mirror plane, either because of a perpendicular electric field or asymmetry in bonding such as in a bilayer heterostructure, its Hamiltonian contains a momentum-dependent spin-orbit field of the "Rashba" form $\mathbf{B}_{\mathrm{Rashba}} \propto \alpha_R(\mathbf{k}\times\mathbf{z})$, where **k** is the electron momentum, **z** the sample normal direction, and $\alpha_R$ a phenomenological coefficient related to the spin-orbit coupling constant $\lambda$. The Rashba field shifts the energies of electron states up or down depending on whether their spins are parallel or antiparallel to $\mathbf{B}_{\mathrm{Rashba}}$. Rashba spin-orbit coupling thus lifts the spin-degeneracy of the 2DEG [**Fig. 13(a)].** The spin directions of the two subbands are perpendicular to the electron momentum and wrap around a circle in spin-space when the Fermi surface is traced in momentum space. If an in-plane current is applied, the Fermi surfaces in **Fig. 13(a)** will shift to have more forward-moving states and fewer backward-moving states. Because of the link between spin-direction and electron momentum, the greater number of forward-moving states will cause a non-equililbrium accumulation of down spins close to the larger Fermi surface in **Fig. 13(a)**. When this non-equilibrium accumulation is exchange-coupled to an adjacent magnetic layer, it can apply a torque [Manchon and Zhang, 2009].

In bilayers of ferromagnets with heavy metals, the torques described by two-dimensional models [Manchon and Zhang, 2009, Kim *et al*, 2012; Wang and Manchon, 2012; Pesin and MacDonald, 2012B] survive in three-dimensional calculations. The resulting torques have exactly the same form as those due to the spin Hall effect, Eq. 3.3, with the efficiency factors $\theta_{SH}\eta$ replaced by equivalent factors. A difference is that for torques due to interfacial spin-orbit coupling, the field-like contribution is expected to be larger than the damping-like contribution. In semiclassical calculations [Haney et al., 2013a; Amin and Stiles, 2016a, 2016b], the interfacial spin-orbit interaction acting on the spins passing through, rather than existing in, the two-dimensional interface creates a spin polarization that couples to the magnetization through exchange interaction at the interface. Such calculations allow the torques due to the spin Hall effect and those related to the Edelstein effect to be treated simultaneously. First principles calculations of the torque [Haney *et al*., 2013b, Freimuth *et al*., 2014a, Freimuth *et al*., 2014b] show similar results: both damping-like and field-like torques with contributions both from the bulk and the interface. Experiments in which different types of spacer layers are placed between heavy metal and ferromagnet [Fan *et al*., 2013; Pai *et al*., 2014] suggest that the 3D spin Hall effect mechanism is the most likely explanation for the measured torque, but need to be modeled in more detail.

The conversion between charge currents and spin currents also applies to non-magnetic interfaces with strong spin-orbit coupling. The inverse Edelstein effect, in which an incident spin current is converted to a charge current, was demonstrated experimentally for non-topological two-dimensional states at the Ag/Bi interface [Rojas-Sánchez *et al*., 2013; Zhang *et al*., 2015b]. For Ag/Bi interfaces it has also been demonstrated that spin transfer torques on a neighboring ferromagnetic layer are consistent with interfacial conversion [Jungfleisch *et al*., 2016a].

***D. Interfaces between ferromagnets and topological insulators***

The interfacial (Edelstein) torque discussed in the previous section can be even larger when the same idea is applied to the surface states of a topological insulator [Burkov *et al*., 2010; Culcer *et al*., 2010; Pesin and MacDonald, 2012]. For an ordinary (non-topological) 2DEG, there is partial cancellation between the two subbands of opposite spin helicity [Soumyanarayanan *et al*, 2016]. In topological insulators, there is only a single two-dimensional Fermi surface with the spin direction locked perpendicular to the electron momentum direction [**Fig. 13(b)**]. Current flow within a *topological* 2DEG generates a non-equilibrium spin accumulation with no partial cancellation between subbands. Strong current-induced torques were demonstrated for topological insulator/ferromagnet bilayers in two early experiments, one at room temperature [Mellnik *et al*., 2014] and one at cryogenic temperatures [Fan *et al*., 2014b]. As of yet, however, experiments have not clearly demonstrated that the torque is a two-dimensional surface-state effect as opposed to a bulk effect. In this section, we describe the properties of topological insulators that are relevant to their interfaces with ferromagnets, and then describe measurements on such interfaces.

**1. Spin-momentum locking in topological insulators**

The bismuth and antimony chalcogenides ($Bi_2Se_3$, $Bi_2Te_3$ and $Sb_2Te_3$) and their derivative solid solutions [*e.g.*, $(Bi,Sb)_2Te_3$] are narrow band gap semiconductors with a layered, crystalline rhombohedral structure in the space group $R\bar{3}m$. Strong spin-orbit coupling with band inversion at the bulk Γ-point locks spin and momentum and leads to topologically-protected surface electronic states that cross the bulk band gap [Fu and Kane, 2007; Zhang *et al*., 2009]. These characteristics result in the realization of a three-dimensional topological insulator [Hasan and Kane, 2010; Hasan and Moore, 2011; Qi and Zhang, 2011], whose two-dimensional surface states (neglecting additional corrections such as warping) are described by a massless Dirac Hamiltonian $\mathcal{H} = A(\sigma_x k_y - \sigma_y k_x)$ where $A$ is the interaction energy and the σ's are Pauli spin matrices. The Dirac Hamiltonian implies surface states that are not spin-degenerate and exhibit momentum-space spin-textures similar to those described previously for 2D Rashba systems: the direction of the spin is well defined and perpendicular to the electron momentum direction, as shown in **Fig. 13b**, *i.e*., the two are locked to each other. Spin- and angle-resolved photoemission spectroscopy (spin-ARPES) gives direct experimental evidence for momentum-space spin-textures and massless helical Dirac electrons at the surfaces of the Bi-chalcogenides [Hsieh *et al*., 2009]. Studies have also shown that helical spin textures can be engineered by controlling the quantum coupling between the opposite surfaces of topological insulator films in the ultra thin limit [Landolt *et al*., 2014; Neupane *et al*., 2014]. Remarkably, signatures of the Dirac states and their spin texture persist to room temperature, suggesting "topological spintronics" applications, such as the generation of an inherently spin polarized charge current and its subsequent use for exerting a spin transfer torque on a neighboring ferromagnet [Mellnik *et al*., 2014; Fan *et al*., 2014b]. Recent experiments with the topological insulator α-Sn show large spin transfer torques even at room temperature [Rojas-Sánchez *et al*., 2016].

While spin-ARPES enables direct detection of the spin texture of surface states in 3D topological insulators, measurement of spin-momentum locking using electrical techniques has proved more challenging. It has been theoretically proposed to detect the spin polarization of the surface state current using the voltage generated at a ferromagnetic electrode [Hong *et al*., 2012]; alternatively, a ferromagnetic contact can be used for spin injection into the surface state, followed by detection of the corresponding directional voltages with non-magnetic electrodes [Burkov and Hawthorn, 2010]. The former concept was demonstrated recently by using ferromagnetic tunnel contacts on patterned $Bi_2Se_3$ thin films [Dankert *et al*., 2015; Li *et al*., 2014a]. In these experiments, the ferromagnetic tunnel contact serves as a spin sensitive potentiometer that detects a different spin-dependent electrochemical potential when the magnetization of the contact is parallel or antiparallel to the spin of the electrons in the transport channel. In spite of the extrinsic nature of the samples studied (the chemical potential lies within the bulk conduction band), clear spin-dependent voltage signatures were observed with behavior consistent with spin-momentum locking. Similar results have now been reported in samples with lower carrier density [Ando, 2014; Tang *et al*., 2014; Tian *et al*., 2015] or electrically gated samples [Lee *et al*., 2015], where bulk conduction is significantly reduced.

A qualitatively different approach to the electrical detection of spin-momentum locking has also been developed using spin Hall effect tunneling spectroscopy [Liu *et al*., 2014] to probe spin-textured surface states. These experiments use a tunnel junction geometry that incorporates a ferromagnetic metal (CoFeB) separated from a $Bi_2Se_3$ transport channel by a high resistance MgO tunnel barrier [Liu *et al*., 2015a]. The tunnel barrier serves two purposes: it ensures high efficiency spin polarized tunneling and it physically separates the topological insulator states from the time-reversal breaking effects of the ferromagnet. The geometry allows two complementary measurement schemes: spin-polarized electrons injected from the ferromagnetic electrode into the topological surface states create charge accumulation due to spin-momentum locking; conversely, flowing a charge current in the topological insulator channel generates a spin-dependent voltage across the tunnel junction. The self-consistency between these two measurement schemes provides rigorous evidence for spin-momentum locking and also allows a quantitative measurement of the charge-to-spin conversion efficiency (parametrized as a spin Hall ratio that can be as large as 0.20).

### 2. Diluted moment topological insulators: quantum anomalous Hall effect

The Hall effect, including the spin Hall effect, has an intrinsic contribution due to electrons away from the Fermi level and survives even in the insulating limit at zero temperature. This quantum Hall effect occurs when bulk states at the Fermi level are localized, which may be due to a band gap or a disorder-induced mobility gap. The *anomalous* Hall effect, a consequence of spin-orbit interactions, is present in any ferromagnetic system. Insulators with a non-zero anomalous Hall effect, often referred to as Chern insulators, support spin polarized chiral edge states that enable dissipationless transport accompanied by quantized Hall resistivities. This combination of properties is referred to as the quantum anomalous Hall effect (QAHE). If a 3D topological insulator thin film is made ferromagnetic with a magnetization **M** perpendicular to its surface, the surface states are gapped: $E_k = \pm\sqrt{(Ak_x)^2 + (Ak_y)^2 + (m_z)^2}$, as shown in **Fig. 14a**, and QAHE appears. Magnetization is presently realized by doping of topological insulators with transition metals [Checkelsky *et al.*, 2012; Liu *et al.*, 2012d; Zhang *et al.*, 2012; Chang *et al.*, 2013b; Kou *et al.*, 2013]. **Figure 14b-d** show QAHE observations in such films ($\varrho_{xx}$ near 0 and $\varrho_{xy} = h/e^2$). Ferromagnetic transition temperatures (to date) in quantum anomalous Hall insulators are below 20 K, unsurprising given the low doping and type of chemical bonding which leads to low exchange interactions. The magnetic gap observed in angular-resolved photo-emission spectroscopy is also, unsurprisingly, small (for instance, tens of meV in $(Bi_{1-x}Mn_x)_2Se_3$ thin films with $x$ up to 0.1) and to date not cleanly detected because of (poorly-understood) disorder broadening [Chen *et al.*, 2010; Xu *et al.*, 2012]. Nonetheless, a dip in the spectral weight is seen in angular-resolved photoemission spectra. A recent study however suggests that the observed gap is not magnetic in origin, but may be the result of resonant scattering processes [Sanchez-Barriga *et al.*, 2016].

Spin- and angular-resolved photo-emission spectroscopy measurements have revealed changes in spin texture that accompany the opening of a magnetic gap: [Xu *et al.*, 2012] close to the Γ-point, the surface electron spins cant from in-plane to out-of-plane, creating a "hedgehog spin texture" in momentum space. Further, when the chemical potential of a ferromagnetic topological insulator thin film is tuned within the magnetic gap *via* electrical gating, helical 2D Dirac fermions transition to 1D chiral edge states of a quantum anomalous Hall insulator [Yu *et al.*, 2010a; Chang *et al.*, 2013a; Checkelsky *et al.*, 2014; Kou *et al.*, 2014; Chang *et al.*, 2015a; Chang *et al.*, 2015b; Liu *et al.*, 2016]. The edge state can also be detected through an unconventional angular dependence of the anisotropic magnetoresistance [Kandala *et al.*, 2015].

### 3. Interfaces between topological insulators and ferromagnetic insulators

Heterostructures of topological insulators with ferromagnetic insulators, such as $Bi_2Se_3$/EuS, $(Bi,Sb)_2Te_3$/GdN, $Bi_2Se_3$/YIG or $Bi_2Se_3/Cr_2Ge_2Te_4$, break time-reversal symmetry while still preserving topological symmetry [Wei *et al.*, 2013; Yang *et al.*, 2013; Kandala *et al.*, 2013; Richardella *et al.*, 2015; Lang *et al.*, 2014; Alegria *et al.*, 2014]. Ideally, transport in such bilayers would be entirely carried by the Dirac surface, modified by exchange coupling with the ferromagnetic layer. Although the spin texture and magnetic gap at the buried interface cannot readily be probed using angle-resolved photo-emission spectroscopy, indirect information may be obtained through measurements of quantum corrections to diffusive transport. Diagrammatic calculations show that an unperturbed Dirac surface state should exhibit weak antilocalization, while opening of a gap would yield weak localization [Lu *et al*., 2011]. Indeed, $Bi_2Se_3$ and $(Bi,Sb)_2Te_3$ layers capped by a ferromagnetic layer of GdN show suppression of weak antilocalization [Kandala *et al*., 2013; Richardella *et al*., 2015], while $Bi_2Se_3$/EuS bilayers show a transition to weak localization [Yang *et al*., 2013]. Anomalous Hall effect and polarized neutron reflectometry in $Bi_2Se_3$/EuS [Wei *et al*., 2013; Katmis *et al*., 2016], and magneto-optical Kerr effect and magnetoresistance of $Bi_2Se_3$/YIG [Lang *et al*., 2014] suggest a proximity-induced ferromagnetism in the topological insulator, implying broken time-reversal in the topological insulator surface-derived states resulting from the proximate ferromagnetic insulator.

### 4. Interfaces between topological insulators and ferromagnetic metals: efficient generation of spin transfer torque

As discussed briefly above, spin-textured surface electron states of 3D topological insulators provide a means of generating spin transfer torque in a neighboring ferromagnet [Culcer *et al*., 2010; Mahfouzi *et al.*, 2012; Pesin and MacDonald, 2012]. This phenomenon was measured using bilayers of permalloy (Py)/$Bi_2Se_3$ [Mellnik *et al*., 2014] and of Cr-doped $(Bi,Sb)_2Te_3/(Bi,Sb)_2Te_3$ [Fan *et al*., 2014b]. A microwave frequency current was driven through the Py/$Bi_2Se_3$ heterostructure in the presence of an in-plane magnetic field ($\mathbf{H}=H_x\hat{\mathbf{x}}+H_y\hat{\mathbf{y}}$) misoriented from the direction of the current density ($j\,\hat{\mathbf{y}}$). The current density in the $Bi_2Se_3$ generates a non-equilibrium spin accumulation $\langle S_x\hat{\mathbf{x}}\rangle$ at the interface due to both topological surface states and conventional Rashba-split bulk states. The ensuing diffusion of spins into the ferromagnet creates a spin torque that causes **M** to precess around **H**. This torque is measured as a DC mixing voltage related to the anisotropic magnetoresistance of Py.

When the magnetic field is swept through the condition for ferromagnetic resonance at fixed microwave frequency, the DC mixing voltage shows a line shape that has a field-symmetric and field-antisymmetric component around the resonance field. The former is related to in-plane spin transfer torque, while the latter is related to an out-of-plane torque typically attributed to an Oersted field. In these Py/$Bi_2Se_3$ heterostructures, however, the out-of-plane torque is an order of magnitude larger than can be attributed to an Oersted field and is similar in magnitude to the in-plane torque. These observations, together with the signs of the in-plane and out-of-plane torques, are consistent with a model for spin torque generation by topological surface states [Mellnik *et al*., 2014]. These measurements result in a very large figure of merit for charge to spin conversion: a spin-torque ratio of order unity at room temperature. We caution, however, that present measurements cannot definitively rule out spin-torque generation by bulk states of $Bi_2Se_3$. Subsequent temperature dependent measurements of spin torque ferromagnetic resonance in $Bi_2Se_3/Co_{40}Fe_{40}B_{20}$ bilayers [Wang *et al*., 2015] argued that surface state contributions dominate the spin transfer torque at low temperature. Separation of the contributions of bulk and surface states to spin-torque generation will require experiments in which the chemical potential is tuned from the bulk states into the bulk gap where surface states dominate. A Kubo formula calculation [Sahin and Flatté, 2015] of the spin Hall conductivity in $Bi_xSb_{1-x}$ using Berry curvatures from a tight binding Hamiltonian, predicts a smooth variation of spin Hall ratios as the band structure is varied from trivial to topological.

Measurements of magnetization switching generated by charge currents using a magnetically-doped 3D topological insulator [$(Bi,Sb)_2Te_3$/Cr-$(Bi,Sb)_2Te_3$ bilayers and single Cr-$(Bi,Sb)_2Te_3$ layers with asymmetric interfaces] have been carried out at liquid helium temperatures [Fan et al., 2014b; Fan *et al.,* 2016]. DC measurements of the anomalous Hall effect yield the magnetization of the ferromagnet Cr-$(Bi,Sb)_2Te_3$; second harmonic analysis shows a record-high spin Hall (charge to spin conversion) ratio near 100 (see [Yasuda et al., 2017], however, for an alternative explanation of second harmonic generation in similar heterostructures). Other experiments have probed the charge-to-spin conversion efficiency using spin pumping schemes [Deorani *et al*., 2014; Jamali *et al*., 2015; Shiomi *et al*., 2014; Rojas-Sánchez *et al*., 2016]. Although it is difficult to make quantitative interpretations of these experiments, and in particular to assess the separate roles of the bulk and surface effects, the data are consistent with large figures of merit for charge-to-spin conversion.

### *E. Open questions and new directions*

Although the past several years have seen an explosion in identification and understanding of transport phenomena related to magnetic interfaces, there are many open questions that invite further inquiry. We identify four topics that seem particularly interesting.

*i. Microscopic Theory of Spin-Orbit Torque:* The microscopic theory of transport-current

induced torques exerted on ferromagnetic magnetizations in heterostructures containing a strongly spin-orbit coupled element, either a heavy metal or a topological insulator, is still incomplete. The factors affecting the transmission efficiency for spin currents to cross the interface between a spin-generating material and the magnetic layer are just beginning to be investigated [Zhang *et al.*, 2015c; Pai *et al.*, 2015]. It remains an open question whether alternative approaches to generating torques (*e.g.*, thermal magnon torque, voltage controlled magnetic anisotropy) might be effective. Thus far, demonstrations of spin torque from thermally-generated spin currents have achieved only small-angle magnetic dynamics (precession angles of about 1°) [Choi *et al.,* 2015], but this is an area with many open questions and potential for improvement. For example, measurements of large heat-driven spin currents in materials with weak or no net magnetization are intriguing [Wu *et al.,* 2015b; Wu *et al.,* 2016b]. Progress is necessary to permit materials- (and structures-)by-design strategies. The main issues are related not to the ferromagnetic element or to the strongly spin-orbit coupled element, but instead to their interface. Spin-torques can be calculated unambiguously when spin-orbit interactions play a negligible role on the chemical bonds that link the macrospin to potential sources of spin-current, but this is not generally the case in the systems of maximum interest. Given that the role of spin-orbit interactions is *enhanced* by the reduced symmetry at an interface, it is unlikely that this condition is satisfied at the interface between a ferromagnet and a strongly spin-orbit coupled element. Comparison between theory and experiment suggests that the current-induced torques in heterostructures containing heavy metal layers are related to their intrinsic anomalous Hall effects, and therefore to electronic states away from the Fermi surface, but other experiments suggest strong contributions from interfacial effects such as the Edelstein effect. A complicating factor is that the structure of both heavy metal and ferromagnetic layer, as well as their interface, is clearly important but often not well controlled or analyzed. *Can a rigorous relationship be established between the current-induced spin-torque and fundamental material properties?*

*ii. Energy-scale hierarchy*: We see potential for improved understanding of spin-orbit and exchange driven interfacial torques in understanding the relationship between three regimes (and the associated subfields): (1) conventional spintronics, based on low-frequency spin transport and microwave magnetic dynamics [Tserkovnyak *et al*., 2005], (2) spin caloritronics, where thermally-excited magnetic degrees of freedom at the energy scale set by the ambient temperature play an important role [Bauer *et al*., 2012], and (3) ultrafast dynamics, where transient hot-electron energy scales can exceed those of magnetic excitations (see Sec. VI). Establishing how much physics (e.g., electron-magnon interactions and spin transport carried by electrons and magnons, particularly at interfaces) is shared by these three classes of phenomena may allow for qualitative and quantitative knowledge, such as transport coefficients and interaction parameters, to be transferred between different subfields.

*iii. Room Temperature Quantum Anomalous Hall Effect:* To date, the quantum Hall effect has been achieved only by using very large magnetic fields to break time-reversal symmetry or by magnetically doping topological insulators. Diluted moment topological insulators magnetically order only at low temperature, so the quantum anomalous Hall effect is to date found only below 4 K [Mogi *et al*., 2015]. *Could interfaces between topological insulators, which intrinsically have quantum Hall effects when magnetic, and robust ferromagnetic insulators* result in the quantum anomalous Hall effect in low magnetic field at room temperature, making it potentially important for low power consumption electronic devices? This novel interfacially-induced state would likely present unexpected phenomena, due to interfacially-induced electronic modifications discussed in Sec. II, particularly if correlation effects such as those discussed in the interfacial oxide section II.C can be introduced *via* judicious choice of materials.

*iv. Spintronics with Antiferromagnets:* Antiferromagnets have long played an important role due to exchange bias and spin valve effects, but have novel future potential [Fukami *et al*., 2016a; Baltz *et al*., 2016]. Antiferromagnets have faster collective dynamics (their spin-mode eigenfrequencies are in the terahertz, due to strong exchange forces between spin sub-lattices) and are insensitive to stray magnetic fields. If the spin-modes can be tuned or excited *via* spin-orbit torques, magneto-electric effects, and/or magneto-acoustic effects, this could provide the basis for a deeper understanding of the underlying interactions and for the development of generators and resonant detectors at terahertz frequencies. There is also interest in using insulating antiferromagnets and ferrimagnets for spin transmission and potential low-power memory and logic applications, such as electric field switching of multiferroic antiferromagnets. Progress is being made in heterostructures of ferromagnets, heavy metals, and antiferromagnet layers [Jungwirth *et al*., 2016].

## IV. COMPLEX SPIN TEXTURES INDUCED BY INTERFACES

### *A. Overview*

While it is well known that interfaces influence spin configurations through phenomena such as interface anisotropy, enhanced orbital moments, strain, exchange bias*,* it has only recently been appreciated that interfaces, particularly those where spin-orbit coupling is strong, can fundamentally change the magnetic ground state of a ferromagnet. Most magnetic materials exhibit ferro- or antiferromagnetic *collinear* order due to the nature of exchange coupling, as discussed in Sect. II. However, the interfacial DM interaction (Sec. II.B) can induce *helical* magnetic order with defined chirality and complex spin textures such as spin spirals, skyrmions, and chiral domain walls, with extraordinary properties derived from their chirality and topological nature [Mühlbauer, 2009; Yu *et al.*, 2010b; Zang *et al.,* 2011; Everschor, 2012; Schulz *et al.,* 2012; Yu *et al*., 2012; Emori *et al*., 2013; Ryu *et al*., 2013; Sampaio *et al*., 2013; Nagaosa and Tokura, 2013; Wiesendanger, 2016]. A particularly notable consequence of these complex spin textures is the interplay between charge and spin transport, where spin-orbit coupling leads to emergent (effective) electric and magnetic fields that modify transport and dynamics of the spin textures. This section first discusses chiral magnetic order in bulk materials, then focuses on the statics and dynamics of interface-driven chiral spin textures.

#### 1. Chiral magnetic order and topologically-driven phenomena

Non-collinear spin structures arise in some materials, particularly rare earth-based, due to magnetic anisotropy and long range indirect exchange coupling [Jensen and MacKintosh, 1991]. Helical magnetic order was first observed by neutron diffraction in the non-centrosymmetric (chiral) B20-phase compound MnSi [Ishikawa *et al.*, 1976], and shown theoretically to originate from the Dzyaloshinskii-Moriya (DM) interaction that derives from its chiral crystal structure plus spin-orbit coupling [Bak and Jensen, 1980]. Below the critical magnetic ordering temperature $T_c$, at low magnetic field, the magnetization spontaneously forms a spin spiral with a large period compared to the atomic spacing and handedness set by the sign of the spin-orbit coupling and the chirality of the structure. This helical phase has been extensively studied in metallic B20 silicides and germanides such as $(Fe_xCo_{1-x})Si$, MnSi, and FeGe [Mühlbauer, 2009; Neubauer *et al.*, 2009; Pappas *et al.*, 2009; Yu *et al.*, 2010b; Yu, 2011; Huang and Chien, 2012], as well as the insulating B20 $Cu_2OSeO_3$ oxide [Seki *et al.,* 2012], including thin films and bulk materials.

It was predicted by [Bogdanov and Yablonskii, 1989; Rößler *et al.*, 2006] and subsequently shown experimentally, [Mühlbauer, 2009; Neubauer *et al.*, 2009; Pappas *et al.*, 2009; Yu *et al.*, 2010b; Yu *et al.*, 2011; Huang and Chien, 2012] that the DM interaction could generate not only a helical state, but also spin textures known as skyrmions [see **Figure 15**]. Skyrmions are topological in nature, as discussed in more detail in IV.C, with an integer topological skyrmion number. The novelty brought by the DM interaction is its preference for specific, chiral domain wall structures. Depending on the nature and sign of the DM interaction, it favors "hedgehog" (spins pointing out or in) or "vortex" (winding left or right) type skyrmions. These are often referred to as Néel and Bloch skyrmions respectively, due to similarities to 180° domain wall structures with these names, see **Figs. 16 and 17**, to be discussed in more detail below. Because of the nature of the symmetry breaking of an interface, interfacially-induced skyrmions are hedgehog (chiral Néel) type, while those in the B20 phase are vortex (chiral Bloch) type. Compositional control of B20 phase materials allows tuning of the size and sign of chirality of the Bloch vortex-type skyrmions [Shibata *et al.,* 2013; Siegfried *et al.,* 2015]; tuning of heterostructure parameters similarly control the size and chirality of the interface-induced Néel hedgehog-type [Chen *et al.,* 2013b]. Recent theoretical work looked at the competition between Rashba and Dresselhaus type spin-orbit coupling, and predicted complex phase diagrams of different types of non-collinear spin structures [Rowland *et al.*, 2016].

Since their discovery, skyrmions have generated significant interest, due to both their topological spin structure and their unique interactions with spin and charge currents. In these materials, electrons moving in an electric field experience an effective magnetic field that couples to their magnetic moment due to spin-orbit coupling, leading to emergent electrodynamic phenomena such as the topological Hall effect and the non-linear (the Hall voltage depends nonlinearly on current) topological Hall effect when the skyrmions themselves move under the influence of a current [Neubauer *et al.*, 2009; Schulz *et al.*, 2012]. They exhibit current-driven displacement at current densities far below that needed to drive magnetic domain walls, a result of potential technological significance [Al'Khawaja and Stoof, 2001; Jonietz *et al.*, 2010; Zang *et al.*, 2011; Everschor, 2012; Yu *et al.*, 2012; Iwasaki *et al.*, 2013a; Iwasaki *et al.*, 2013b; Sampaio *et al.*, 2013; Jiang *et al.*, 2015]. They are also predicted, although not yet experimentally shown, to travel from cold to hot in a thermal gradient, similar to domain walls.

A rotational motion of a skyrmion lattice has been observed in a thermal gradient (Jonietz *et al.*, 2010).

Theoretical work and reciprocal and real-space imaging described below have provided significant insights into the nature of spin spirals and skyrmion phases in these materials, and their evolution with temperature and applied field (see **Fig. 15** and Sect. IV.D). The helical phase has zero net magnetic moment in zero applied field. With increasing magnetic field, a canting occurs, followed by a first order metamagnetic transition from the helical antiferromagnetic state to the ferromagnetic state. The skyrmion phase has a net moment and is stabilized by entropy and magnetic energy such that it occurs (in equilibrium) only at non-zero temperature and magnetic field as an intermediate state between the helical and ferromagnetic phases. It is however easily trapped (particularly in materials that are of comparable thickness to the skyrmion size) as a metastable state at lower temperature and field [Huang and Chien, 2012*;* Oike *et al.,* 2015].

### 2. Chiral magnetic order due to interfaces

As discussed in Sec. II.B, the symmetry breaking of interfaces together with spin-orbit coupling can generate DM interactions even for cubic materials (such as Fe) layered with high spin-orbit coupling materials (such as Ir), which result in non-collinear structures like the non-centrosymmetric B20 phase materials discussed just above (helical states, skyrmions, and chiral domain walls). As in the B20 phase materials, the sign of spin-orbit coupling together with the structural breaking of inversion symmetry (up is different than down in a bilayer) leads to a specific chirality (see **Fig. 3**, discussed in Sec. II.B). Spin-polarized scanning tunneling microscopy has allowed direct imaging of spin cycloids in epitaxial ultrathin Mn/W(001), Mn/W(011) and Fe/W(110) films [Bode *et al.*, 2007; Ferriani *et al.*, 2008; Meckler *et al.*, 2009], and of atomic-scale skyrmions in Fe or a Pd/Fe bilayer on Ir(111), including the thermodynamically-expected coexistence of the skyrmion and helical phase at finite applied field [Heinze *et al.*, 2011; Romming *et al.*, 2013]. Inclusion of a higher order exchange interaction (the four-spin exchange interaction) is theoretically believed to be crucial to the ultrathin Fe/Ir skyrmions [Heinze *et al.*, 2011], but it is not clear if this is generally true.

Notably, this same interfacial symmetry breaking plus spin orbit coupling can also produce perpendicular magnetic anisotropy, as in *e.g.,* Co/Pt heterostructures and Co-Pt alloys [Charilaou *et al.*, 2016]. Perpendicular anisotropy, whether interfacially-induced or intrinsic to the structure, in conjunction with long-range dipolar interactions, induces stripe or bubble domains. Magnetic bubbles, in the form of *e.g.,* up-magnetized domains in a down-magnetized background of a perpendicular magnetic anisotropy material even without interfacial DM interaction, are stabilized by magnetic dipolar coupling. These bubbles are bordered by 180° domain walls that are Bloch-type (*i.e.,* vortex-type, in which the direction of spins rotate perpendicular to direction of domain wall), with two possible rotation directions; these have skyrmion number of ±1 and left-right chirality (see **Fig. 17g, h**). In other words, an ideal bubble (*i.e.,* one in which the boundary domain wall lacks Bloch lines) *is* a skyrmion; the diameter of the central ferromagnetically-aligned domain and the width of the domain wall affect the properties of the bubble/skyrmion such as how easily it is moved, but not its topological nature. Dipolar-coupling-induced "skyrmions" are achiral (equal left-right preference), and may form bi-skyrmions (pairs of skyrmions, bound together by as-yet not understood interactions) [Lee *et al.,* 2016].

The *addition* of interfacial DM interactions modifies the nature of a skyrmion/bubble significantly, lowering the energy of Néel-type walls with definite chirality relative to Bloch walls. Interfacial DM interactions arise even in polycrystalline asymmetric stacks of materials with perpendicular magnetic anisotropy such as Pt/CoFe/MgO, Pt/Co/Ni/Co/TaN, and Pt/Co/$AlO_x$ [Emori *et al.*, 2013; Ryu *et al.*, 2013; Pizzini *et al.*, 2014], despite a high degree of interfacial disorder. These materials are of technological interest, since they are magnetic under ambient conditions, readily integrated into room-temperature spintronic devices, and their properties can be tuned by varying layer thicknesses, compositions, and interface materials.

Simulations suggest that skyrmions can be nucleated, stabilized, and manipulated by low currents [Sampaio *et al.*, 2013; Li, 2015; Heinonen *et al.*, 2016]. Recent experiments demonstrating stabilization and current-driven motion of skyrmionic bubbles in thin films [Jiang *et al.*, 2015; Chen *et al.*, 2015b; Moreau-Luchaire *et al.,* 2016; Woo *et al.*, 2016; Boulle *et al.,* 2016; Jiang *et al.,* 2016] suggest that the topic is ripe for further discovery and potential low-power spintronic applications such as skyrmion-based memory (Bobeck, Bonyhard, and Geusic, 1975; Parkin, Hayashi, and Thomas, 2008; Tomasello *et al.*, 2014; X. Zhang *et al.*, 2015; Kang *et al.,* 2016) and logic devices

### *B. Statics of interfacially-induced chiral spin structures*

At the interface of an isotropic magnetic material, the DM interaction vector is $\mathbf{D_{ij}} = D\,\hat{\mathbf{z}} \times \mathbf{u}_{ij}$ with $\hat{\mathbf{z}}$ the unit surface normal, $\mathbf{u}_{ij}$ the unit vector from spins *i* to *j*, and *D* an energy per bond (usually of the order of meV) [Heide *et al.*, 2008]. In continuous form, the DM interaction energy density is [Bogdanov and Yablonskii, 1989; Bogdanov and Rößler, 2001]

$$E_{DMI} = D_m\left[\left(m_x\frac{\partial m_z}{\partial x} - m_z\frac{\partial m_x}{\partial x}\right) + \left(m_y\frac{\partial m_z}{\partial y} - m_z\frac{\partial m_y}{\partial y}\right)\right] = D_m\left[\left(\mathbf{m}\times\partial_y\mathbf{m}\right)_x - \left(\mathbf{m}\times\partial_x\mathbf{m}\right)_y\right] \quad \textbf{[4.1]}$$

with $D_m$ the effective DM interaction micromagnetic constant, with dimensions of surface energy density. The convention adopted here is that the film is in the (*x*,*y*) plane, with $\hat{\mathbf{z}}$ pointing from the substrate to the film. If interfacial DM interactions exist only for the atomic layer at the interface, with an energy per bond *D* (corresponding to the in-plane component of the $\mathbf{D}_{ij}$ vector that is normal to the bond *ij*, see Sec. III), then $D_m = Df/(at)$ where *t* is the film thickness, *a* is the in-plane atomic distance and *f* is a number dependent on lattice type and crystallographic orientation [*e.g., f*=1 for simple cubic (100) plane, $f = \sqrt{3}$ for face centered cubic (111) plane]. For example, for *D* = 1 meV, *a* = 0.2 nm, *f*=1, and *t* = 1 nm, one finds $D_m$ = 0.8 mJ/m$^2$. The sign convention adopted presently [Heide *et al.*, 2008] is that *D* > 0 corresponds to right-handed structures (see **Fig. 17**).

Associated with the energy density is an effective field vector

$$\mathbf{H}_{eff,DM\ interaction} = \frac{2D_m}{\mu_0 M_s}\left(-\frac{\partial m_z}{\partial x}, -\frac{\partial m_z}{\partial y},\ \frac{\partial m_x}{\partial x} + \frac{\partial m_y}{\partial y}\right) \quad \textbf{[4.2]}$$

with an in-plane component along the gradient of $\mathbf{m}_z$. Thus, an interfacial DM interaction favors non-uniform magnetization structures such as cycloids of definite chirality, in contrast to the exchange, anisotropy, and applied field energies, which prefer uniform magnetization. Phase diagrams have been constructed accordingly by [Bogdanov and Hubert, 1994; Kiselev *et al.*, 2011]. Here we focus on domain walls (1D structures) and skyrmions (2D structures).

### 1. 1D domain walls with interfacial DM interactions

For an in-plane magnetized material, Eq. 4.2 shows that the DM interaction only gives rise to an out-of-plane magnetization tilt at a domain wall, which is strongly suppressed by magnetic dipolar coupling energy. For perpendicularly magnetized materials, by contrast, an interfacial DM interaction favors Néel walls (rotation of spins normal to the domain wall) over the magnetic dipole-preferred achiral Bloch walls (see **Fig. 16**). As the magnetic dipolar energy of a Néel wall goes linearly to zero as thickness *t* decreases [Tarasenko *et al.*, 1998] whereas the impact of the interfacial DM interaction $D_m$ grows as *1/t*, DM interactions dominate the magnetic dipole energy for very thin films, *e.g.,* for $D_m$ > 0.1 mJ/m$^2$ for *t*=0.6 nm [Thiaville *et al.*, 2012], as confirmed by imaging [Chen *et al.*, 2013b]. DM interaction-stabilized Néel walls are chiral [Heide *et al.*, 2008]. For $D > 0$, when travelling along a given direction, an up-to-down domain wall has its Néel moment along this direction, and a down-to-up domain wall has its moment oriented oppositely, while for $D < 0$ the converse is true. As their stability and structure is governed by the DM interaction, it has been proposed to call them Dzyaloshinskii domain walls [Thiaville *et al.*, 2012] to distinguish them from classical (achiral) Néel domain walls which are found in thin films with *in-plane* anisotropy and result from dipolar energy alone. Defects known as Bloch lines can form in any of these types of domain walls, where two distinct chiralities meet each other; examples are shown in **Fig. 17i, j** in the domain walls of magnetic bubbles. In the presence of DM interactions, Bloch lines tend to condense into pairs [Yoshimura *et al*, 2015]. In a 2D material, or a 3D material such as a film where the magnetization is uniform through the film thickness, Bloch lines look like points, and are therefore sometimes mistakenly referred to as Bloch points (a Bloch point is a distinct 3D topological entity which will be discussed below). Bloch lines substantially modify the nature and pinning of conventional Bloch and Néel domain walls [Malozemof and Slonczewski, 1979], as well as Dzyaloshinskii domain walls [Yoshimura *et al.*, 2015].

The surface energy for Dzyaloshinskii domain walls is $\sigma = \sigma_0 \pm \pi D_m$ [Dzyaloshinskii, 1965; Heide *et al.*, 2008]. Here, $\sigma_0 = 4\sqrt{AK_{eff}} + \sigma_{ms}$ is the domain wall energy in the absence of DM interaction, with $A$ the micromagnetic exchange constant, $K_{\mathrm{eff}}$ the effective perpendicular anisotropy and $\sigma_{ms}$ the magnetic dipole energy cost of a Néel wall compared to Bloch wall. The $\pm$ sign reflects the dependence on domain wall chirality. Thus, for large enough DM interaction, the total energy of domain walls with the DM interaction-preferred chirality becomes negative, signaling that the uniform magnetic state is no longer stable, as predicted by [Dzyaloshinskii, 1964, 1965]. Note that the same physics takes place for Bloch walls in materials with bulk inversion asymmetry, resulting in the 1D helical phase (see **Fig. 15a**).

The DM interaction modifies not only the micromagnetic energy but also the boundary conditions, leading to spin canting at film edges that can be described as quasi-walls induced by the DM interaction [Rohart and Thiaville, 2013; Meynell *et al.*, 2014]. This causes interactions between nonuniform spin textures and boundaries that can result, *e.g.*, in confinement effects in laterally-constrained geometries [Sampaio *et al.*, 2013; Leonov *et al.*, 2016].

Interfacial DM interactions also strongly affect domain wall dynamics. For magnetic field-driven domain wall motion, the moving domain wall necessarily distorts its shape and as the wall moves faster, the distortion is greater. There is a threshold known as the Walker field $H_W$ (Schryer and Walker, 1974), at which the distortion becomes unstable and steady-state motion impossible. Below $H_W$, the wall motion achieves steady state and above it, the domain wall magnetic structure, seen in the moving frame, precesses as the domain wall moves (Beach, Tsoi, and Erskine, 2008). In the simple model considered by Schryer and Walker (1974), the distortion is a tilt of the plane in which the domain-wall magnetization rotates. When the tilt angle reaches 45° the restoring torque is maximum and the wall magnetic moment becomes unstable. For a one-dimensional domain wall, $H_W = \alpha H_K/2$ where $H_K$ is the effective anisotropy field that stabilizes the orientation of the domain wall magnetization, and $\alpha$ is the Gilbert damping. An interfacial DM interaction stabilizes the domain wall in a Néel orientation, causing $H_W$ to increase with $D_m$ as shown by [Thiaville *et al.*, 2012]. This has been invoked to explain the large domain wall velocities, up to high fields, observed in Pt/Co/$AlO_x$ [Miron *et al.*, 2011b].

Spin-orbit induced spin currents described in Sec. III.C act on chiral Néel walls like a field. This effective field, however, decreases as the moving domain wall magnetization reorients from Néel towards Bloch [Thiaville *et al.*, 2012], so that in a one-dimensional model, Walker breakdown is suppressed and the domain wall velocity saturates at a velocity proportional to $D_m$.

The dynamics of current-driven domain wall motion can be solved by integrating the LLG equation, but generally this has to be numerically solved. Insight into the motion can be gained by assuming a parameterized form of the domain wall structure as done by [Schryer and Walker, 1974] in terms of the displacement of the domain wall $q$, and the in-plane tilt $\Phi$ of the magnetic moment at the center of the domain wall. Extending the original model to include the DM interaction and spin Hall effects [Thiaville *et al.*, 2012; Emori *et al.*, 2013; Khvalkovskiy *et al.*, 2013a] gives the equation of motion for the domain wall velocity $dq/dt$ and the precessional velocity $d\Phi/dt$:

$$\frac{d\Phi}{dt} + \frac{\alpha}{\Delta}\frac{dq}{dt} = \gamma_0 \left( H_z + \frac{\pi}{2} \chi M_s cos\Phi \right) \quad \textbf{[4.3]}$$

$$\frac{1}{\Delta}\frac{dq}{dt} - \alpha \frac{d\Phi}{dt} = \gamma_0 \sin\Phi \left[ \frac{\pi}{2} (H_D + H_x) - H_K \cos\Phi \right] \quad \textbf{[4.4]}$$

where $\Delta = \sqrt{A/K_{eff}}$ is the domain wall width, and $\gamma_0$ and $\alpha$ are the gyromagnetic factor and damping constant in the LLG equation (Eq. 3.2). The fields on the right-hand side are the components of the applied field ($H_x$ along the Néel wall moment and $H_z$ along the easy axis), $H_K$ $=2K_{eff}/(\mu_0 M_s)$ the effective anisotropy field associated with perpendicular anisotropy, $H_D = D_m/(\mu_0 M_s \Delta)$ the DM interaction-induced effective field within the domain wall, and $\chi = g\mu_B \theta_{SH} \eta j/(eM_s t)$ a factor that expresses the spin Hall effect torque for current flowing along the wall normal [Khvalkovskiy *et al.*, 2013b]. For small $H_z$ and $\chi$, it is possible to find a solution with $d\Phi/dt = 0$ and $dq/dt$ = constant, but for larger values no steady state solution exists. The values of these transitions depend on $H_D$, $H_x$, and $H_K$, illustrating that it is possible for the DM interaction to stabilize the steady-state motion of the domain wall, allowing higher steady state velocities.

While these collective coordinate equations accurately describe many key features, they cannot capture distortions of walls such as the non-uniform tilting of a domain wall across the width of a nanostrip [Ryu *et al.*, 2012] when amplified by the DM interaction [Boulle *et al.*, 2013; Emori *et al.*, 2014; Jué *et al.*, 2016b], or the modification of chiral domain walls if Bloch points or lines are present.

### 2. 2D chiral structures: magnetic skyrmions

Magnetic dipolar energy causes a non-uniform magnetic state in thin films with perpendicular magnetic anisotropy. These have two basic 2D structures, called the stripe phase and the bubble phase [Malozemoff and Slonczewski, 1979; Choi *et al.*, 2007], which are similar to domain structures in magnetic garnet films, textures in cholesteric liquid crystals [de Gennes and Prost, 1995], and solutions of amphiphilic molecules. A related domain structure occurs in ultrathin films with large interfacial DM interactions: a cycloidal and a skyrmion phase. Note that while dipole-induced bubbles are surrounded by Bloch domain walls with no chiral preference, skyrmions due to interfacial DM interactions are bounded by Néel domain walls with a single chirality fixed by the sign of $D_m$. DM interaction-induced skyrmions are typically stable down to sizes smaller than dipolar-interaction induced bubbles, and the homochirality of their domain walls together with the increased stability of their structure (due to the DM interaction) make them harder to annihilate or merge.

Skyrmions in ultrathin bilayer or trilayer films were first observed and manipulated by spin-polarized scanning tunneling microscopy [Heinze *et al.*, 2011; Romming *et al.*, 2013], in which their size approaches atomic dimensions. Skyrmions have been observed in single layers close to the spin reorientation transition [Jiang *et al.*, 2015; Boulle *et al.*, 2016] and in multilayer structures where structural inversion asymmetry is created by designing the layer sequence; the larger total film thickness increases magnetic dipolar coupling, which increases stability and size [Moreau-Luchaire *et al.*, 2016; Woo *et al.*, 2016].

Bloch points and lines can form in skyrmionic structures, the former at a point of merging of two vertically-extended skyrmions of the same skyrmion number (discussed below and shown in **Fig. 18**), and the latter as vertically-extended defects (which can be dynamically or thermally induced; note that Bloch lines can only be created in pairs) in a single skyrmion (see **Fig. 17i, j**). Both modify the skyrmion number, and affect the stability and motion of the skyrmion.

### *C. Topological Aspects*

#### 1. Geometrical treatment of noncollinear spin textures

As a non-collinear magnetic texture is described by a unit vector function $\mathbf{m}(\mathbf{r},t)$, it can be viewed at any time $t$ as a mapping of the space (position $\boldsymbol{r}$) to the unit sphere. Both spaces are of the same dimension (2) for ultrathin films, where magnetization is uniform through the thickness. They are moreover of the same topology if the magnetization is uniform at infinity. Continuous transformations (*e.g.*, magnetization dynamics, or a quasi-static evolution under an applied field) distort the mapping continuously without tearing it. Thus, if a texture leads to a complete coverage of the sphere, it cannot be continuously transformed to another that does not, as this would imply tearing the magnetization at some point. The number of times the sphere is wrapped is given by the Chern number, or skyrmion charge $N_{sk}$ [Skyrme, 1958; Skyrme, 1962; Kléman, 1973; Belavin and Polyakov, 1975]:

$$N_{sk} = \frac{1}{4\pi}\int \mathbf{m}\cdot\left(\frac{\partial \boldsymbol{m}}{\partial x}\times\frac{\partial \boldsymbol{m}}{\partial y}\right)dxdy \qquad \textbf{[4.5]}$$

**Figure 17a-f** shows some important skyrmion solutions found by [Belavin and Polyakov, 1975] for the Heisenberg ferromagnet in 2D, considering only exchange energy, with $\mathbf{m}=\pm\,\hat{\mathbf{z}}$ at the core, $\mathbf{m}=\mp\,\hat{\mathbf{z}}$ at infinity, and $m_z = 0$ at a selected distance. DM, anisotropy, and dipolar coupling interactions are not included, so the ground state is a uniformly magnetized ferromagnet, but the states shown are all (meta)stable solutions and are topologically protected, *i.e.* they cannot be continuously deformed to the uniformly magnetized state, and are therefore topological solitons. In **Fig. 17a-f**, the magnetization winds on the unit sphere once, hence $N_{sk} = \pm 1$. Structures with skyrmion number $N_{sk} = +1$ (**a-d**) are topologically equivalent in the sense that they can be continuously deformed into one another. Those with $N_{sk} = -1$ (**e** and **f**) belong to a different topological class; they can be continuously deformed into one another (despite

looking quite different) by rotating all spins 180° around an in-plane direction at 45° between $\hat{\mathbf{x}}$ and $\hat{\mathbf{y}}$, but not into structures with $N_{sk}$ = +1. The direction of **m** at the core defines the polarity $p = \pm 1$.

In skyrmions whose magnetization co-rotates with position around the center (as in Figs. 17**a-d** and **f**), the winding number $S$ = 1, and the chiral angle $\phi$ is defined as the angular difference between the azimuthal angles of **m** and of the spatial vector **r** pointing from the skyrmion core. Chirality (sometimes called helicity) can also be defined by the nature of the skyrmon wall: right or left, Bloch or Néel, referenced to the 180° domain walls shown in **Fig. 16**; $p$ is essential in defining left vs right chirality. For Figs. 17**a-d**, with $p$=1, chiralities are Néel-type right ($\phi$ = 0), left ($\phi$ = π); Bloch-type left ($\phi$ = π /2), and right ($\phi$ = - π /2). For Fig. 17**f**, the chirality is Néel-type ($\phi$ = π), but it is Néel-type *right* (because the magnetization of the wall rotates as in **Fig. 16b** right, with **m** down at the center of the skyrmion as at the back of that figure; mathematically, because $p$ = -1). For **e**, the magnetization counter-rotates, hence $S$ = -1; in this case, chirality is not defined, and the walls are neither Bloch- nor Néel-type.

The skyrmions in **Fig. 17a-f** all have the same exchange energy, no intrinsic size, and a simple relationship between skyrmion and winding number: $S = N_{sk}p$. Reversing the sign of **m** everywhere (e.g., Eq. 4.5) reverses the sign of $N_{sk}$ and $p$ (**Fig. 17f** compared to **a**) but (unsurprisingly) does not change the energy, and does not change $S$. Similar terminology ($S$ and $p$) is used for other magnetic structures such as vortices and bubbles.

DM interactions, anisotropy, and dipolar coupling (demagnetization energy) modify the spin structures, cause an intrinsic skyrmion size, and cause different skyrmions to have different energies (e.g. lowering the energy of Néel-type relative to Bloch-type skyrmions) but the topological nature of the structures is unchanged. As previously discussed, bubbles in materials without DM interactions typically have an extended ferromagnetically-aligned core (shown in **Fig. 17g** and **h**) due to dipolar coupling plus magnetic anisotropy, whereas those with strong DM interaction and low magnetization, hence low dipolar coupling have a continuously varying magnetization direction (**Fig**. **17a-f**), but the skyrmion number is independent of this distinction.

A Bloch line (which looks like a point in 2D projections such as **Fig. 17**) possesses a half-integer (+1/2 or –1/2) skyrmion charge. Bubble-type skyrmions (with extended ferromagnetic cores) with Bloch lines are shown in **Fig. 17i** (4 Bloch lines) and **j** (2 Bloch lines); these add energy and lead to $S$ = -1 or 0; **i** is sometimes called an anti-skyrmion. Notably, the skyrmion in **Fig. 17e**, which also has $N_{sk}$ = -1, looks somewhat similar to a skyrmion with 4 Bloch lines (**Fig. 17i**); high-resolution spin imaging is required to differentiate their domain wall magnetization.

For *easy-plane* anisotropy, topological structures are vortices, shown in **Fig. 17k, l**, whose magnetization **m** lies in the easy *x-y* plane away from the vortex core where $\boldsymbol{m} = \pm\hat{\mathbf{z}}$. Such a pattern covers only one hemisphere, has a skyrmion charge $N_{sk} = \pm 1/2$, and is often referred to as a meron [Phatak *et al*., 2012]. Meron topological stability relies on the assumption that **m** at infinity is in-plane, with a non-zero winding number.

Because $N_{sk}$ is conserved under continuous deformations, a soliton with nontrivial topology can be quite stable. Skyrmion number conservation *can* be violated if the continuous fabric of the magnetization field is torn by a topological defect known as a Bloch point (BP) [Mermin, 1979; Thiaville *et al.*, 2003] (**Fig. 18**). As the basic BP is made of planes with $N_{sk}$ jumping from +1/2 to -1/2, BPs are generally involved in skyrmion creation and annihilation. The simplest process for this is that a BP is "injected" at one surface of the sample, crosses it and exits at another surface (inside the sample, BPs can only be created in pairs with opposite topological numbers). BP injection or pair creation entails an energy barrier [Rohart et al, 2016; Lobanov, 2016], overcome by thermal activation, and/or reduced by structural defects.

Skyrmions of equal and opposite skyrmion number can continuously merge and annihilate each other. For example, an $N_{sk}$ = 0 (**Fig. 17j**) magnetic bubble with two Bloch lines can decay spontaneously when its size is reduced below a critical value, at which the inflationary pressure of long-range dipolar forces becomes too weak to balance the domain wall tension [Ezawa, 2010]. A similar process is thought to be involved in skyrmion number reversal of a vortex (**Fig. 17k, l)** [Van Waeyenberge *et al*., 2006]. When a vortex with $N_{sk}$ = 1/2 is accelerated to a critical speed, it nucleates a vortex-antivortex pair with skyrmion numbers –1/2 and +1/2, respectively. The antivortex and the original vortex approach each other to form a skyrmion ($N_{sk}$ = +1), which shrinks, annihilates (BP involved), and creates a spin-wave explosion [Hertel and Schneider, 2006; Tretiakov and Tchernyshyov, 2007], leaving behind a vortex with opposite skyrmion number $N_{sk}$ = -1/2.

## 2. Relation between dynamics and topology

The deep connection between magnetization dynamics and topology was established by [Thiele, 1973] who translated the LLG equation for a rigidly-moving magnetic soliton (bubble, skyrmion, vortex, etc), where **m** is solely a function of [**r**-**R**($t$)], into an equation of motion for the soliton's position $\mathbf{R}(t)$. Due to the precessional motion of magnetization, their motion is somewhat counterintuitive. When a soliton moves, it experiences a gyrotropic force that is perpendicular to its direction of motion. Thiele's equation expresses the balance of forces acting on the soliton, $\mathbf{F}_g + \mathbf{F}_c + \mathbf{F}_v = 0$, where $\mathbf{F}_g = \mathbf{G} \times \mathbf{v}$ is the gyrotropic force, $\mathbf{v} = d\mathbf{R}/dt$ the soliton velocity, $\mathbf{G} = 4\pi N_{sk}(M_s t/\gamma)\hat{\mathbf{z}}$ the gyrovector that characterizes the gyrotropic force, $t$ the film thickness, $N_{sk}$ the skyrmion number defined above in Eq. 4.5, $\mathbf{F}_c$ is a conservative force derived from the potential (static) energy of the soliton as a function of position, and $\mathbf{F}_v = -B\mathbf{v}$ is a viscous force with a dissipation tensor $B_{ij} = (\alpha M_s t/\gamma) \int \partial_i \mathbf{m} \bullet \partial_j \mathbf{m} \; dxdy$, where $\alpha$ is the Gilbert damping constant, $M_s$ the saturation magnetization, and $\gamma$ was previously defined in Eq. 2.3. Several modifications of the Thiele equation have been recently proposed, to take into account degrees of freedom other than skyrmion position, such as chirality, non-zero mass, breathing modes, and the inhomogeneous nature of the DM interaction (incompletely described by $D_m \propto 1/t$ of Eq. 4.1).

Thus, the gyrotropic force has a topological nature. The gyrotropic force induces rotational dynamics of ferromagnetic solitons with a nonzero skyrmion number [Huber, 1982; Ivanov and Stephanovich, 1989; Papanicolaou and Tomaras, 1991; Moutafis *et al.*, 2009]. Rotational motion has been observed for vortices [Choe *et al.*, 2004] and skyrmions [Büttner *et al.*, 2015]. In a skyrmion crystal, the gyrotropic force affects the spectrum of skyrmion vibrations. Instead of two linearly dispersing phonon branches (longitudinal and transverse), theory predicts a chiral magnetophonon branch (related to skyrmion gyrational motion) with a quadratic dispersion [Zang *et al.*, 2011] and a gapped cyclotron mode [Petrova and Tchernyshyov, 2011]. The magnetophonon branch has been observed in the chiral magnet $Cu_2OSeO_3$ [Onose *et al.*, 2012].

Current-induced domain wall motion in ferromagnetic heterostructures involves two types of torques: the spin-orbit torques of Eq. 3.3 and "conventional" spin-transfer torques due to spin currents flowing through spatially varying magnetizations [Berger, 1984; Zhang and Li, 2004; Beach *et al.*, 2008; Tserkovnyak *et al.*, 2008]. "Conventional" torques have two components, both proportional to charge current density $\mathbf{j}$. One is referred to as adiabatic spin-transfer torque, is directed along $(\mathbf{j} \cdot \nabla)\mathbf{m}$, and can be understood as being due to angular momentum conservation as spins adiabatically follow the magnetization direction. The other is refered to as non-adiabatic spin transfer torque, is directed along $\mathbf{m} \times (\mathbf{j} \cdot \nabla)\mathbf{m}$, and has several underlying physical mecahnisms. The conventional spin-transfer torque generates a force $\mathbf{F}_{STT} = -\mathbf{G} \times \mathbf{u}$, with $\mathbf{u} = -\frac{g\mu_B}{eM_S}P\mathbf{j}$ being the drift velocity of the spin-polarized electrons [Thiaville *et al.*, 2005], where $P$ is the spin polarization of the current. The spin current generated by the spin Hall effect in an adjacent heavy metal exerts a spin-orbit torque, also called a Slonczewski torque, introduced in Eq. 3.3, which can be described as an effective field $\mathbf{H}_{eff,\,SHE} = (\mathbf{m}\times\hat{\mathbf{p}}) / (\mu_0\gamma_0\tau)$ where (as before) $\hat{\mathbf{p}}$ is the direction of the spins induced by the spin Hall effect, and $\tau$ a time that expresses the magnitude of the torque. This torque gives rise to a force $F_{i,SHE} = \frac{M_S\int(\mathbf{m}\times\partial_i\mathbf{m}).\hat{\mathbf{p}}dxdy}{\gamma\tau}$. Its component along the current direction has the same form as the interfacial DM interaction (Eq. 4.1), establishing the direct link between the DM interaction and spin-orbit torque.

### *D. Characterization of complex spin textures*

#### 1. Scattering

Neutron and X-ray scattering provide reciprocal space information about spin textures with nanoscale spatial resolution. Neutron scattering provided the first direct evidence for a helical ground state in bulk MnSi [Ishikawa *et al.*, 1976] and the skyrmion lattice phase [Mühlbauer, 2009; Pappas *et al.*, 2009; Münzer *et al.*, 2010]. Spin dynamics can also be probed; *e.g.,* [Jonietz *et al.*, 2010] observed rotation of a skyrmion lattice in bulk MnSi resulting from spin torque induced by a small applied current. For a recent review of neutron scattering applied to magnetic materials see [Michels, 2014].

Resonant soft X-ray scattering provides complementary and element-specific information. For example, an unexpected existence of two distinct skyrmion sublattices in $Cu_2OSeO_3$ arising from two inequivalent Cu sites with chemically identical coordination numbers but different magnetically-active orbitals was observed with resonant soft X-ray diffraction at the Cu $L_3$ absorption edge [Langner *et al.*, 2014].

#### 2. Imaging

Microscopic insight into the structure and behavior of spin textures can be obtained by direct imaging methods [Hubert and Schäfer, 1998]. Optical microscopies utilize magneto-optical effects [Qiu and Bader, 1999], such as Kerr and Faraday effects. X-ray microscopies [Fischer, 2015] use x-ray dichroism effects which are sensitive to ferro- and antiferromagnetic spin textures, inherently element-specific and quantifiable with respect to magnetic spin and orbital moments [Thole *et al.*, 1992; Carra *et al.*, 1993]. They include real space scanning and full-field X-ray microscopies [Fischer *et al.*, 1997] and X-ray photoelectron emission microscopes [Stöhr *et al.*, 1993] as well as reciprocal space techniques such as X-ray holography [Eisebitt *et al.*, 2004] and various coherent X-ray diffraction imaging techniques [Tripathi *et al.*, 2011; Shi *et al.*, 2016]. The diffraction limit of X-ray microscopies sets the ultimate spatial resolution (nanometers). The inherent time structure of X-ray sources at synchrotrons or X-ray free electron lasers enables time-resolved studies (nanoseconds to femtoseconds). In-plane and out-of-plane components of magnetic textures can be imaged and full 3D structures of magnetic domains have been reported [Da Col *et al.*, 2014; Streubel *et al.*, 2014]. Time-resolved X-ray microscopy studies have revealed mechanisms in magnetic vortices to switch both polarity [Van Waeyenberge *et al.*, 2006] and chirality [Uhlir *et al.*, 2013]. Static X-ray microscopy experiments identified a symmetry-breaking effect in the nucleation of magnetic vortex structures [Im *et al.*, 2012] and observed stochastic behavior in spin-torque-induced domain wall motion [Meier *et al.*, 2007] and depinning processes from notches [Im *et al.*, 2009]. A review article by [Boulle *et al.*, 2011] highlights X-ray photo-electron emission microscopy imaging of current-induced domain wall motion in nanoscale ferromagnetic elements, including submicrometer skyrmions in an ultrathin film, also seen in multilayer films by transmission x-ray microscopy [Moreau-Luchaire *et al.*, 2016; Woo *et al.*, 2016]. For larger skyrmions, magneto-optical microscopy enabled imaging of generation of skyrmions by laterally-inhomogeneous spin-orbit torques [Jiang *et al.*, 2015]. The gigahertz gyrotropic eigenmode dynamics of a single magnetic bubble was imaged with X-ray holography by [Büttner *et al.*, 2015]; the observed trajectory confirmed the skyrmion topology and indicated a skyrmion mass much larger than predicted by existing theories.

Electron microscopy-based techniques include Lorentz transmission electron microscopy (LTEM) [Petford-Long and De Graef, 2012], spin-polarized low-energy electron microscopy (SPLEEM) [Rougemaille and Schmid, 2010], and scanning electron microscopy with polarization analysis (SEMPA) [Chung, Pierce, and Unguris, 2010]; the former is sensitive to the in-plane spin component, and the latter techniques are extremely surface sensitive. A SPLEEM study by [Chen *et al.*, 2013a] characterized the chirality of domain walls in a Co/Ni multilayer with perpendicular anisotropy in contact with Pt or Ir, which induces a DM interaction that stabilizes left-handed or right-handed Néel domain walls respectively [see **Fig**. **19**]. It is notable that the DM interaction, from a single Pt or Ir interface, is strong enough to induce Néel walls with defined chirality in a multilayer stack that would otherwise have achiral Bloch walls.

The magnetic imaging resolution in LTEM is typically on the order 2 nm to 20 nm, with the highest resolution obtained with aberration-corrected instruments [Petford-Long and De Graef, 2012]. Qualitative magnetic information can be obtained using the Fresnel LTEM imaging mode. A quantitative map of chiral spin structures can be achieved by reconstructing the phase shift of the transmitted electrons. This is done *via* a through-focal series of Fresnel images by either the transport-of-intensity equation approach [Petford-Long and De Graef, 2012] or an iterative approach [Koch and Lubk, 2010], or alternatively by off-axis electron holography [Koch and Lubk, 2010]. Phase reconstruction of LTEM images of skyrmion lattices has been reported in thin films of $Fe_{0.5}Co_{0.5}Si$ [Yu *et al.*, 2010b] and in nanowires of MnSi [Yu *et al.*, 2013]. This approach was used by [Phatak *et al.*, 2012] to reveal the structure of novel spin states such as merons in coupled magnetic disks. Similar approaches were used by [Tanigaki *et al.*, 2015] to image vortex cores in stacked ferromagnetic disks, and by [Phatak *et al.*, 2010] to reconstruct the magnetic vector potential around a permalloy square with a Landau domain structure. *In situ* magnetic fields [Petford-Long and De Graef, 2012; Rodríguez *et al.*, 2013] or electric currents can be applied that enable the *local* quasi-static magnetization reversal of a sample to be followed in real-time down to about 40 ms. [Yu *et al.*, 2012] observed near room-temperature motion of skyrmions in FeGe using *in situ* current application and LTEM. [Pollard *et al.*, 2012] combined LTEM with gigahertz applied fields to excite resonant gyrotropic behavior of magnetic vortex cores in permalloy squares and thus obtain high spatial resolution information correlating the radius of core motion with frequency of applied magnetic field.

Scanning probe microscopies including magnetic force microscopy (MFM) (see *e.g.*, [Ferri *et al.*, 2012]) and spin-polarized scanning tunneling microscopy (SP-STM) [Wiesendanger, 2009] have been used to explore the behavior of chiral magnetic spin textures. Spin helices [Bode *et al.*, 2007] and skyrmions [Heinze *et al.*, 2011] induced at surfaces were first observed by SP-STM in ultrathin epitaxial transition metal films; [von Bergmann *et al.*, 2014] presents SP-STM studies of interface-induced chiral spin structures. [Milde *et al.*, 2013] used MFM to explore the destruction of a skyrmion lattice on the surface of a bulk $Fe_{0.5}Co_{0.5}Si$ crystal, and showed that this occurred *via* motion of the skyrmions leading to coalescence and the formation of elongated structures. Writing and deleting of individual skyrmions in an ultrathin film was demonstrated using injected current from an SP-STM probe by [Romming *et al.*, 2013]. MFM combined with additional techniques was used to explore the behavior of artificial skyrmion lattices that are stable at room temperature, fabricated by lithographic patterning of micron-sized Co disks onto either Co/Pt or Co/Pd multilayer films with perpendicular anisotropy [Miao *et al.*, 2014; Gilbert *et al.*, 2015], and to infer the strength of DM interactions from the small skyrmion size in asymmetric Co multilayers [Moreau-Luchaire et al., 2016]. Scanning probe microscopies utilizing nitrogen vacancy centers quantitatively map stray field profiles with a few tens of nanometers spatial resolution [Rondin *et al.*, 2013], allowing identification of individual domain walls, including their chirality, in ultrathin perpendicularly-magnetized films [Tetienne *et al.*, 2014, 2015].

### 3. Magnetotransport

Electrical transport probes of spin textures complement direct imaging techniques and are relevant to potential device applications. Anisotropic magnetoresistance has been used to indirectly probe the presence and structure of domain walls in permalloy, both statically [Hayashi *et al.*, 2007] and dynamically [Hayashi *et al.*, 2008], and more recently, to distinguish Bloch from Néel domain walls in ultrathin Pt/Co/Pt and Pt/Co/$AlO_x$ nanostrips [Franken *et al.*, 2014]. The accumulated Berry phase as conduction electron spins move through a locally-varying magnetic texture gives rise to emergent electrodynamic fields; for example, electrons traversing a skyrmion experience an effective magnetic field that leads to a topological contribution to the Hall effect [Neubauer *et al.*, 2009], similar to an anomalous Hall effect but not proportional to $M(H)$. Dynamic (*i.e.*, time-varying) spin textures generate an effective electric field that has been observed for domain walls [Yang *et al.*, 2009] and skyrmions [Schulz *et al.*, 2012]. The topological Hall effect has been treated mainly in the limit of adiabatic spin tracking and relatively little work has addressed the role of spin-orbit coupling and nonadiabatic spin transport at interfaces. We note that observation of the topological Hall effect is not sufficient to conclude that a skyrmion phase is present, as there are other spin and electron band structure effects that lead to a topological Hall effect [Meynell *et al.,* 2014].

## *E. Dynamics of complex spin textures*

### 1. Domain wall dynamics and current-induced torques in the presence of interfacial spin-orbit coupling

Interfacial spin-orbit coupling in ultrathin films leads to two key effects that change domain wall dynamics qualitatively. For films with perpendicular magnetic anisotropy, with out of plane domains, whether intrinsic to the ferromagnet's structure (*e.g.,* amorphous Tb-Fe or Gd-Co, or hcp Co-Cr) or due to an interfacial effect (*e.g.,* Pt/Co, Co/Ni, Co/MgO), the domain walls become very narrow (≈1nm to 20nm). Moreover, the magnetic dipole energy difference between Néel and Bloch walls is small in ultrathin films [Tarasenko *et al.*, 1998]. In the case of symmetric interfaces, adiabatic spin-transfer torques can drive precessional motion with a very low critical current density independent of the defect-induced pinning potential [Koyama *et al.*, 2011], but the efficiency (velocity per unit current density) is no greater than in thicker in-plane magnetized films.

Large current-induced effective fields [Miron *et al.*, 2009] and high current-driven domain wall velocities have been observed in ultrathin Pt/Co/$AlO_x$ films [Moore *et al.*, 2008; Miron *et al.*, 2011b]. Interestingly, the domain walls move in the opposite direction to the current, a result that runs counter to expectations based on conventional spin-transfer torques. This result arises from two spin-orbit effects that manifest when spatial inversion symmetry is broken. [Haazen *et al.*, 2013] showed that vertical spin currents due to the spin Hall effect in the adjacent heavy metal generate the dominant current-induced torque, and as discussed in Sec. III.C, they can be much stronger than conventional spin-transfer torques based on spin-polarized currents flowing in the bulk. [Thiaville *et al.*, 2012] showed that the spin Hall effect could account for the observed domain wall dynamics in asymmetric structures if the interfacial DM interaction stabilizes chiral Néel domain walls. In this case, the spin Hall effect effective field in the domain wall orients along the easy axis in a direction that alternates from one domain wall to the next so that current drives them in the same direction (see **Fig. 20**). This theory was verified experimentally by measuring the dependence of current-induced velocities [Emori *et al.*, 2013; Ryu *et al.,* 2013] and effective fields [Emori *et al.*, 2014] on in-plane field. [Martinez *et al.*, 2014] contains a detailed micromagnetic and analytical study of current-driven domain wall motion in the presence of in-plane fields.

Although the static domain wall structure in the presence of DM interaction is similar to the classical Néel domain wall, the stabilizing energies are different, which leads to qualitatively different dynamics for Dzyaloshinskii domain walls [Thiaville *et al.*, 2012]. For example, when driven by magnetic field, since the energy difference between Bloch and Néel configurations is determined by the DM interaction rather than the relatively small magnetostatic energy, the Walker breakdown transition occurs at a substantially higher velocity. Moreover, since the DM interaction couples the domain wall magnetization to the domain wall normal, a torque applied to the former tends to reorient the latter. In narrow magnetic tracks, this results in a tilting of the domain wall normal with respect to the nanotrack axis under applied current or magnetic fields [Boulle *et al.*, 2013; Emori *et al.*, 2014], observed experimentally by [Ryu *et al.*, 2012]. This domain wall tilt has complex dynamics [Jué *et al.*, 2016b].

**2. Dynamics of magnetic skyrmions in thin films**

Current-induced rotation and translation of skyrmion lattices by conventional spin-transfer torque has been detected experimentally at current densities as low as $\approx 10^6$ A/m$^2$ in the bulk chiral magnets MnSi and FeGe [Jonietz *et al.*, 2010; Yu *et al.*, 2012]. This threshold current density is several orders of magnitude lower than for current-induced domain wall motion, which has been explained in terms of the weak interaction of the relatively large (tens of nanometers) skyrmions with defects [Iwasaki *et al.*, 2013a; Iwasaki et al., 2013b]. However, at larger current density, above threshold, the current-induced torque arises from the adiabatic spin-transfer torque, and so the velocity per unit current density in chiral magnets is expected to be similar to that of conventional domain walls.

Skyrmion dynamics in thin films have been explored through analytical and micromagnetics treatments, as well as recent experiments. [Sampaio *et al.*, 2013] considered charge current-driven motion of individual skyrmions in a laterally-confined track, and showed that the spin Hall effect (spin-orbit transfer torques described in Sec. III) could lead to substantially more efficient motion compared to conventional spin-transfer torques. That study, which included full degrees of freedom for the skyrmions, showed that skyrmions can pass around a geometrical notch suggesting that pinning should be much less than for domain walls [Fert *et al.,* 2013], similar to modeling results for bulk chiral magnets [Nagaosa and Tokura, 2013]. Experiments in thin-film multilayers, however, find critical current densities similar to those of domain walls [Woo *et al.*, 2016], which can be understood in terms of atomic-scale interface disorder that leads to local variations in DM stength. Nonetheless, skyrmion velocities exceeding 100 m/s have been achieved. It has also been shown experimentally that inhomogeneous spin-orbit torques can generate skyrmions by driving instabilities of the domain wall, similar to surface-tension-driven fluid flows. This was demonstrated by patterned thin film structures, where a constriction resulted in a laterally inhomogeneous current flow and thus laterally-varying magnitudes and directions of spin-orbit torques, as shown in **Fig. 21(a)-(f)** [Jiang *et al.*, 2015]. This behavior was reproduced by micromagnetic simulations [Li, 2015; Heinonen *et al.*, 2016]. Nucleation and annihilation of skyrmions by injected current pulses in an STM is shown in **Fig. 21(e)** [Romming *et al.*, 2013].

Spintronic devices for data storage and logic [Allwood *et al.*, 2005] have been proposed based on manipulating chiral spin structures in magnetic films or nanostructures, including proposals for low-power, high-density information storage and processing devices based on manipulation of individual skyrmions [Kiselev *et al.*, 2011] driven along a magnetic nanotrack by a current-induced spin torque [Fert *et al.,* 2013]. Small skyrmion size enables high bit density and the threshold currents to drive skyrmions are, under some circumstances, several orders of magnitude lower than to propagate domain walls, although we note that current density required to produce rapid steady motion of skyrmions seems likely comparable to conventional domain wall motion. Micromagnetic analysis of ways to ensure smooth motion of skyrmions is discussed by [Zhang *et al.*, 2015e].

The forces described above causes skyrmions to acquire a velocity component orthogonal to the current flow direction, as well as parallel. In a track geometry, this causes skyrmions to approach the edge and then propagate parallel to it [see **Fig. 21(f)**] at a distance dictated by balancing the gyrotropic force and topological repulsion due to tilted magnetization at the edges [Rohart and Thiaville, 2013]. A study by [Tomasello *et al.*, 2014] considered both Néel and Bloch skyrmions driven either by spin Hall effects (spin-orbit torques) or spin-transfer torques, and showed that for Néel skyrmions driven by spin Hall effects, current densities only slightly higher than examined in [Sampaio *et al.*, 2013] expelled the skyrmions at the track edge. Edge roughness and thermal fluctuations decreased the critical current for skyrmion expulsion, which could pose a challenge for track-based spintronics proposals. Simulations by [Diaz and Troncoso, 2015] showed that the chirality of skyrmions affects their dynamics, including the gyrotropic motion; inhomogeneity in DM strength introduces additional effects.

### *F. Open questions and new directions*

Many open questions remain regarding interfacially-stabilized spin textures and in particular magnetic skyrmions; we here highlight a few:

*i. What is the stability of skyrmions, and how is this affected by underlying interactions and*

*material properties?* A quantitative understanding of the stability of skyrmions in ultrathin films, both statically and dynamically, is lacking. Strong exchange interactions are needed to get a high magnetic ordering temperature; that energy competes with DM interaction, limiting the stability range of the skyrmion phase. Interfacially-induced DM interactions favor Néel (hedgehog-type) skyrmions, while intrinsically chiral thin films (*e.g.,* B20 phase) show chiral Bloch type. Magnetic dipolar coupling in films with perpendicular magnetic anisotropy favors Bloch skyrmions, with no preferred chirality. These energies thus compete with each other, and tuning is possible through structure design. Work to date has focused on two classes of materials: (1) heterostructures of ultrathin simple ferromagnets with a high spin-orbit coupled metal (not coincidentally this is also relevant to structures used for spin-orbit torque measurements discussed in Sec. III and V) to yield a high DM interaction to exchange energy ratio, and (2) Co/Pt-type multilayers or amorphous rare earth-transition metal alloys, both with strong perpendicular magnetic anisotropy, with dipole coupling increasing the stability of the skyrmion phase. There is a large range of other materials with different values of the relevant energies that would lead to better understanding of the stability, dynamics, and thermodynamics of the skyrmion lattice. For example, with appropriate choice of materials and/or thickness and/or controlled inhomogeneity in a heterostructure, can skyrmion stability be enhanced, while not adversely affecting its dynamic properties? As discussed in Sec. II and III in the context of new approaches to spin torque, there are many other classes of materials (*e.g.,* surface states of topological insulators and/or antiferromagnets) that could yield stable skyrmion phases and controllably chiral domain walls. Rare earth alloys, known to have strong spin-orbit coupling and helical ground states, are largely unexplored.

Furthermore, while skyrmion stability is commonly discussed in terms of topological protection, the continuous vector field description breaks down at atomic length scales so that topological protection is only approximate even in a perfect material. The creation or annihilation of a skyrmion must occur through injection of a Bloch point (or creation of a pair of Bloch points), for which an energy barrier exists [Rohart *et al.*, 2016; Lobanov *et al.*, 2016]; imaging of Bloch points could enable the study of how this occurs (via thermal activation or perhaps tunneling). The effects of thermal fluctuations and variations in interfacial energy terms on static skyrmion stability, or on the dynamical processes of nucleation and annihilation, are not yet known; to date, thermally activated (Brownian) motion only of large skyrmions with low pinning has been experimentally observed [Jiang *et al.*, 2015].

*ii. What limits the speed and dynamical stability of skyrmions?* The high-speed dynamics of magnetic skyrmions, and their dynamical stability, are not yet well-understood. Some micromagnetic simulations have examined current-driven skyrmion dynamics in ultrathin films, including the effect of skyrmion type (Bloch vs. Néel) and current-induced torque (spin transfer versus spin-orbit torques), as well as the influence of edge roughness. Simulations show instabilities develop at high speeds leading to bubble collapse, such as when the gyrotropic force drives a skyrmion close to the edge of a magnetic track. Bubble domains in conventional bubble materials undergo a transition similar to Walker breakdown at high velocity, in which Bloch lines nucleate and propagate along the bubble domain wall. Analogous processes are likely to occur in skyrmions, although the torques required are likely higher due to additional stabilization by DM interactions. Recent experiments demonstrated current-induced generation [Jiang *et al.*, 2015] and propagation [Woo *et al.*, 2016; Jiang *et al.,* 2016] of individual skyrmions in ultrathin transition metal ferromagnet/heavy metal heterostructures with speeds > 100 m/s. Disorder plays an important role in high-speed dynamics and dynamic stability, and more work is required to understand the potential performance and limitations of proposed skyrmion memory structures.

*iii. How does disorder affect static, dynamic and pinning processes for chiral spin textures in thin films?* The treatment of particularly interfacial but also other types of structural disorder, together with finite temperature effects, is essential to understanding static, dynamic, and pinning processes for chiral spin textures in thin films. Realistic treatments of the nature of the disorder (structural, chemical), its correlation length, and its effects on the various energy terms are required, which will require advanced microstructural characterization, first-principles calculation, micromagnetic simulation, and static and dynamic magnetic measurements.

*iv. How is spin transport affected by non-uniform spin textures?* While the interaction between charge currents and chiral spin textures is of great interest both for manipulating skyrmions and detecting them, an understanding of spin transport in the presence of nonuniform spin textures, accounting for spin-orbit effects, is so far lacking.

*v. How will chiral spin textures respond to ultrafast pulses?* Reversal of magnetization on a femtosecond time scale by a laser pulse has been demonstrated (to be discussed in Sec. VI). There are indications that the symmetric exchange constant $J_{ij}$ and the DM interaction constant $D_{ij}$ are modified on ultrafast time scales through this process. As skyrmion textures originate from a balancing of these energies, chiral spin textures are likely to be affected differently than parallel spin textures, particularly using ultrafast pulses with defined helicity, which will in turn have an impact on femtosecond magnetization processes.

*vi. Is it possible to modulate chiral spin textures by designed structure*? In perovskite oxides, the DM interaction is influenced by the displacement of the oxygen atom between adjacent *B*-site cations. Recent demonstrations of modified oxygen octahedral rotations across interfaces, as described in Sec. II.D, suggest the possibility of designing spatially-varying or modulated DM interaction as a function of depth in oxide heterostructures [Moon *et al.*, 2014b], which could give rise to novel spin texture. Similar design could be done in metallic heterostructures, and in *spatially-modulated* heterostructured B20 phase chiral materials, such as $Mn_{1-x}Fe_xGe$.

## V. LARGE-ANGLE MAGNETIZATION DYNAMICS DRIVEN BY INTERFACIAL TORQUES

The Landau-Lifshitz-Gilbert (LLG) equation, Eq. 2.3, is intrinsically nonlinear, as it contains a cross product of magnetization and the effective field, most terms of which are functions of the magnetization [Gurevich and Melkov, 1996]. Additionally, for large-angle precession, the phenomenological Gilbert damping can itself be a nonlinear function(al) of the dynamic magnetization [Tiberkevich and Slavin, 2007]. These non-linearities play an essential role in all large amplitude magnetization dynamics induced by interfacial spin torques, including magnetic switching, and enable a set of remarkable phenomena predicted by classical nonlinear dynamics and manifested by nanometer-scale magnetic systems at the nanosecond time scale. This section discusses several regimes of non-linear magnetization dynamics that are engendered and/or detected by the interfacial magnetic effects discussed in this article: magnetic switching, spin-torque nano-oscillators, and spin-torque resonant microwave detectors.

### *A. Anti-damping and effective-field torques*

As discussed previously, spin-transfer (Eq. 3.2) and spin-orbit (Eq. 3.3) torques each have two terms that have different effects on magnetization dynamics: conservative "effective-field" torques and dissipative torques (known as damping or anti-damping torques). Effective-field torques modify the energy landscape seen by the magnetization [Slonczewski, 1989], but in steady state do not cause a time-dependent change to the magnetic energy. On the other hand, as illustrated in **Fig. 11**, the dissipative torques change the magnetic energy, either enhancing or counterbalancing the intrinsic magnetization damping [Slonczewski, 1996; Berger, 1996]. Anti-damping torques are the equivalent of an anti-friction in mechanics, or negative resistance in electrical engineering. When the anti-damping torque opposes the intrinsic damping and is larger, the overall damping becomes negative, which means that the torque adds continuously to the magnetic energy, generating large-angle magnetization dynamics. As discussed below, the ultimate dynamics generated by the anti-damping torque depends on sample geometry, applied magnetic field, and various nonlinear effects.

While either anti-damping or effective-field torques can in principle generate interesting and useful types of magnetic dynamics, in general the anti-damping component can be used to excite magnetic dynamics more efficiently, at much lower torque strengths. To reorient a magnetic layer using an effective field torque requires a strength of torque comparable to that from magnetic anisotropy, precisely because an effective-field torque acts like an applied magnetic field. However, anti-damping torques can excite large-angle magnetization reorientation if they compensate the intrinsic magnetic damping. This requires a torque strength that is smaller than that from magnetic anisotropy by a factor of the intrinsic Gilbert damping parameter $\alpha$. This scale factor makes the necessary torque a factor of 10 to $10^4$ times lower than for an effective field because $\alpha$ ranges from $10^{-4}$ to 0.1.

### *B. Magnetization reversal*

When magnetic objects have two equilibrium configurations, the anti-damping torque can switch the magnetization back and forth between them by changing the sign of the current. In magnetoresistive systems, this current-induced magnetization reversal is associated with a change of resistance that can easily be detected, as shown in **Fig. 11b**. This effect was observed in trilayer magnetic structures with the current flowing perpendicularly, initially in metallic pillars such as Co/Cu/Co [Katine *et al*., 2000; Grollier *et al*., 2001] and then in magnetic tunnel junctions with low-resistance tunnel barriers [Huai *et al*., 2004; Fuchs *et al*., 2004; Sato *et al*., 2014]. It was shown subsequently that magnetization switching can also be obtained from an anti-damping torque originating from the spin Hall effect [Liu *et al*., 2012b; Lee *et al*., 2013c], possibly assisted by the effective-field torque arising from the Rashba effect [Miron *et al*., 2010; Miron *et al*., 2011a].

To destabilize magnetization and initiate magnetic switching, the anti-damping torque has to induce magnetization precessions. For in-plane magnetized systems, a magnetization rotates elliptically, alternately tilting in and out-of-plane. The out-of-plane tilts generate a large demagnetizing field that increases the magnetic energy loss which the anti-damping torque has to compensate. For perpendicularly magnetized systems subject to a sufficient out-of-plane-oriented anti-damping torque, the magnetization precesses circularly and only develops in-plane tilts, with smaller associated demagnetizing fields. Therefore, the currents needed to switch perpendicularly magnetized systems *via* anti-damping are smaller than those needed to switch in-plane magnetized materials, one of the reasons that most magnetic materials developed for spin-torque applications have perpendicular magnetic anisotropy [Khvalkovskiy *et al*., 2013b].

An alternative approach for magnetization reversal is to employ a short pulse to apply an impulsive torque, with the switching accomplished after the pulse by a ballistic precessional process with relaxation [Kent *et al.*, 2004]. In this regime, the distinction between anti-damping and effective-field torques becomes less important; precessional switching has been demonstrated using both conventional spin-transfer torque [Lee, Pribiag, *et al*., 2009; Papusoi *et al*., 2009; Liu *et al*., 2010] and the effective field from voltage-controlled magnetic anisotropy [Shiota *et al*., 2012]. This type of precessional switching requires less energy to achieve switching (because the applied pulses can be very short), although the applied currents or voltages are generally much larger than for anti-damping switching. Achieving completely reliable switching by the pulse technique can be challenging, however, because it is necessary to avoid both under-precession and over-precession, even in the presence of thermal fluctuations.

Spin-torque-induced magnetization switching is the writing mechanism envisaged for a new generation of Magnetic Random Access Memory (MRAM). Contrary to most resistive switching memory systems, spin torque is a purely electronic mechanism, giving it high endurance and cyclability compared to other proposed technologies for next-generation memory. Other possibilities for this technology include Boolean and non-Boolean logic [Prenat *et al*., 2009; Ohno *et al*., 2010; Niemier *et al*., 2011; Lakys *et al*., 2012] and advanced computing schemes [Roy *et al*., 2015]. The challenge for spin torque applications is to reduce the current needed for switching without degrading thermal stability. Several solutions are currently being investigated. First, spin torques can be assisted by additional effects such as thermal [Bandiera *et al*., 2011] or electric-field-induced [Shiota *et al*., 2012] torques. Second, the amplitude of spin torques can be increased, by using spin-orbit torques from the spin Hall effect in heavy metals [Liu *et al*., 2012b] or topological insulators [Mellnik *et al*., 2014], as discussed in Sec. III.

There is a complication in using giant spin-orbit torques to switch heterostructures with perpendicular magnetic anisotropy, as needed for high-density magnetic devices, associated with the fact that the anti-damping component of spin-orbit torque is ordinarily in the sample plane (Fig. 6). This in-plane torque does not by itself favor either magnetic configuration (up or down). Deterministic switching requires breaking this up-down equivalence, initially achieved by applying a magnetic field parallel or antiparallel to the current [Miron *et al*., 2011a; Liu *et al*., 2012a; Fan *et al*., 2014b]. This in-plane field leads to a zero torque state that is tilted out of the plane in a direction that depends on the sign of the field and the sign of the current. With removal of the current, magnetization relaxes in the direction of its tilt. In this way, the ultimate direction of magnetization can be controlled by the direction of the current. This process and several for switching in-plane magnetization are discussed in [Fukami et al, 2016b].

There are other approaches to switching without an applied field. Spin-orbit switching has used exchange bias from a neighboring metallic antiferromagnet [Lau *et al*., 2015; Fukami *et al*., 2016a; van den Brink *et al*., 2016], engineered anisotropy to tilt the magnetization slightly away from perpendicular [You *et al*., 2015; Torrejon et al., 2015], or the magnetic dipole interaction with a nearby magnetic layer [Smith *et al*., 2016]. In-plane torque can drive *non-uniform* switching [Miron *et al*., 2011a; Liu *et al*., 2012a], through nucleation of a reversed domain and spin-torque-driven domain wall propagation [Lee *et al*., 2014]. This mechanism however becomes ineffective for magnetic memory bits smaller than a few tens of nanometers, where domain wall nucleation becomes difficult [Zhang *et al*., 2015f]. Efficient anti-damping switching would be enabled by a *perpendicular component* of the anti-damping spin-orbit torque, without need for an additional field. There are currently two proposals for this: (1) use a ferromagnetic layer with tilted magnetization and strong spin-orbit interactions as the source of the spin-orbit torque (replacing the heavy metal layer) [Taniguchi *et al.*, 2015], or (2) use a non-magnetic single-crystal layer with low crystal symmetry (*e.g*., lacking two-fold rotational symmetry about the out-of-plane axis) so that an out-of-plane anti-damping torque becomes symmetry-allowed [MacNeill *et al*., 2016].

### *C. Spin-torque nano-oscillators*

When a magnetic system has only one equilibrium position (*e.g.*, in the presence of a large applied magnetic field), magnetization switching is no longer possible. In this case, if an anti-damping torque drives the effective magnetic damping to negative values, the magnetization will oscillate [Kiselev *et al*., 2003; Rippard *et al*., 2004]. At small DC currents the oscillations occur at small amplitudes around the original equilibrium position with frequency ranging between 100 MHz to tens of gigahertz depending on the magnetic configuration and the applied magnetic field [Bonetti *et al*., 2009]. With increasing DC current, anti-damping and the amplitude of oscillations increase. The possible magnetization trajectories are defined by the magnetic potential set by the conservative forces (external and demagnetizing fields, effective-field spin torque).

Spin-torque nano-oscillators convert these DC-current-induced magnetization precessions into voltage oscillations by utilizing magneto-resistive effects, resulting in microwave generation at the oscillation frequency. The amplitude of the generated microwave power is set by the magneto-resistance ratio; several microwatts have been produced using state-of-the-art MgO-based magnetic tunnel junctions [Tsunegi *et al*., 2014].

Many different types of spin-torque nano-oscillators have been studied, with different combinations for the magnetic configurations of the free and fixed layers (uniform in-plane or out-of-plane, vortex *etc.*) [Dussaux *et al.*, 2010] and a variety of geometries (*e.g.,* laterally confined oscillators [Kiselev *et al*., 2003], or point contact on an extended free layer [Rippard *et al*., 2004]) (see **Fig. 22**). Spin-torque nano-oscillators have mostly used the conventional spin torque from spin-polarized charge currents in trilayer structures, but more recent work used spin-orbit anti-damping torque to generate sustained oscillations in a ferromagnetic metal [Liu *et al*., 2012c; Demidov *et al*., 2012; Demidov *et al*., 2014] or a ferromagnetic insulator [Collet *et al*., 2016].

Spin-torque nano-oscillators distinguish themselves from other types of auto-oscillators by their strong nonlinearities. A particularly important type of nonlinearity for thin-film magnetic samples is that the precession frequency generally depends strongly on the magnitude of the precession angle [Gurevich and Melkov, 1996; Slavin and Tiberkevich, 2008], so that magnetic nano-oscillators based on layered magnetic nano-structures and driven by interfacial torques [Kiselev *et al*., 2003; Bertotti *et al.*, 2009; Slavin and Tiberkevich, 2009] are strongly *non-isochronous*. This nonlinear frequency shift can be detrimental for some applications, because spin-torque nano-oscillators are subject to thermal fluctuations that can cause variations in the precession amplitude as a function of time. When the frequency depends on the precession amplitude, any thermally-induced perturbation of precession amplitude results in a frequency fluctuation in the spin-torque nano-oscillator output, thereby increasing the emission linewidth (or equivalently, the phase noise) [Slavin and Tiberkevich, 2009]. Decreasing the associated emission linewidth can be done by constraining the phase fluctuations *via* phase-locked loops [Keller *et al*., 2009], by choosing a direction of the bias magnetic field corresponding to the minimum nonlinearity [Thadani *et al*., 2008], or through dynamic coupling of several magnetic elements [Kaka *et al*., 2005; Mancoff *et al*., 2005]. It has recently been shown that spin transfer-driven coupled vortex dynamics can give rise to emission linewidths below 50 kHz at room temperature [Locatelli *et al*., 2011].

The same magnetic nonlinearities that lead to disadvantageous linewidth broadening can also provide unique benefits. They confer spin-torque nano-oscillators with an ability to change frequency when current is varied, and also to self-synchronize by electrical [Grollier *et al.*, 2006] or magnetic couplings [Sani *et al*., 2013] when organized in interacting assemblies. The nonlinear frequeny shift can also help generate an interesting type of oscillator mode, a "magnetic bullet," that otherwise would not be stable. This mode is formed when a spin-polarized charge current is applied through a nano-contact to a magnetic thin film magnetized in-plane. In this case, the frequency decreases with increasing precession amplitude and, since the spin wave spectrum in the film has a gap determined by the magnitude of an applied in-plane magnetic field $H$, the nonlinearity can shift the frequency of the excited spin wave *below the spectrum of propagating spin waves*, making the excited spin wave mode *self-localized* and *standing*. Thus the standing mode does not lose energy by propagation of spin waves away from the contact, allowing it to have a low threshold for excitation by the interfacial spin-transfer torque. The existence of bullet modes was first predicted by [Slavin and Tiberkevich, 2005] and demonstrated experimentally by [Bonetti *et al*., 2010]. Another type of localized dynamical excitation, the "magnetic droplet," occurs when a spin-polarized charge current is applied through a magnetic nano-contact to a magnetic film with perpendicular anisotropy [Ivanov and Kosevich, 1976; Ivanov and Kosevich, 1977; Hoefer *et al.*, 2010; Mohseni *et al*., 2013]. Magnetic droplets possess intriguing non-trivial internal dynamics [Hoefer *et al.*, 2010; Mohseni *et al*., 2013].

Compared to other types of voltage-controlled oscillators, spin-torque nano-oscillators have a number of potential practical advantages. The materials currently used (CoFeB and MgO) are compatible with complementary metal-oxide semiconductor (CMOS) technology, tunable over hundreds of MHz by variation of bias current or applied magnetic field, very agile (they can change frequency in nanoseconds) and can operate at room temperature. Some applications, such as telecommunication technologies, demand a very high spectral purity; the challenge will be to decrease the emission linewidth below a kilohertz, which is likely to go together with decreased frequency tunability. However, spin torque nano-oscillators are intrinsically more suitable for applications where noise can be tolerated, and large nonlinearities and tiny oscillator size are required. This is relevant to computing schemes inspired by neural synchronization in the brain, such as associative memories [Macia *et al.*, 2011; Chen *et al.*, 2015a]. This requires networks with tens of synchronized oscillators interfaced with CMOS to perform data processing. The challenge is both technological (enhanced magneto-resistance, stronger coupling between oscillators, achieving tunable coupling, *etc.*) and scientific (to understand and manipulate the complex network dynamics of coupled spin-torque nano-oscillators).

### *D. Spin-torque resonators as detectors of microwave radiation (spin-torque diodes)*

Application of an alternating current $I_{AC}$ to a nanoscale magnetic tunnel junction or a multilayer spin valve results in the generation of a direct voltage $V_{DC}$ by the junction [Tulapurkar *et al.*, 2005; Sankey *et al.*, 2006]. This current-rectifying property, known as the spin-torque diode effect, may find use in microwave signal detection because of its high rectification efficiency in the gigahertz frequency range [Cheng *et al.*, 2013; Shiota *et al.*, 2014]. Two contributions to $V_{DC}$ can be identified. The first contribution, $V_{DC} = \frac{I_{AC}\delta R_{AC} cos\phi}{2}$, arises from mixing of the magnetic tunnel junction (MTJ) resistance oscillations $\delta R_{AC}$ and $I_{AC}$, where $\phi$ is the phase difference between these two oscillations [Sankey *et al.*, 2006]. The resistance oscillations originate from magnetization precession driven by the current-induced torques [Nozaki *et al.*, 2012; Zhu *et al.*, 2012]. The maximum in rectification efficiency has a resonant character and is achieved at a frequency near the ferromagnetic resonance frequency of the magnetic tunnel junction free layer. The second contribution, $V_{DC} = I_{DC}\delta R_{DC}$, is observed when a direct current $I_{DC}$ is applied to the junction in addition to $I_{AC}$. This additional voltage is due to a current-induced shift $\delta R_{DC}$ of time-average resistance of the magnetic tunnel junction from its equilibrium value [Sankey *et al.*, 2008; Miwa *et al.*, 2014]. Such a resistance shift can arise from either a nonlinear shift of the center of magnetization precession trajectory [Miwa *et al.*, 2014] or from excitation of non-adiabatic stochastic resonance of magnetization [Cheng *et al.*, 2010].

Magnetic tunnel junctions have demonstrated microwave detection efficiencies as high as 25,000 V/W, greatly exceeding the limit of 3,800 V/W that can be achieved with conventional semiconductor diodes [Cheng *et al*., 2013].

Apart from the potential for sensitive microwave signal detection, the spin torque diode effect is also useful in studies of interfacial torques and the properties of spin waves in nanoscale ferromagnets. The magnitude and symmetry of the resonance peak in $V_{DC}(f)$ can be used to quantify the magnitude and direction of both anti-damping and field-like components of spin torques [Sankey *et al*., 2008; Kubota *et al.*, 2008], as well as the torque due to voltage-controlled magnetic anisotropy [Zhu *et al*., 2012]. Typically, multiple peaks are observed in $V_{DC}$ as a function of the frequency $f$ of the drive current. These peaks arise from excitation of different spin wave eigenmodes [Sankey *et al*., 2006; Gonçalves *et al*., 2013]. Quantitative analysis of $V_{DC}(f)$ spectra can provide measurements of magnetic anisotropy, exchange stiffness, and magnetic damping of the magnetic layers within magnetic tunnel junctions [Fuchs *et al*., 2007].

### *E. Additional consequences of nonlinear magnetic dynamics*

Nonlinearities allow nanometer-scale magnetic systems to manifest a rich array of dynamical regimes, for example, frequency doubling and the appearance of sum and difference frequencies in the spectrum of magnetization oscillations when the external driving signal contains more than one harmonic component [Gurevich and Melkov, 1996], stochastic resonance [Grigorenko *et al*., 1994; Locatelli *et al*., 2014], magnetic solitons [Kosevich *et al.*, 1990; Kalinikos *et al.*, 1983], chaos [Wigen, 1994; Petit-Watelot *et al*., 2012], delayed feedback [Tiberkevich *et al*., 2014], and synchronization [Kaka *et al*., 2005; Mancoff *et al*., 2005].

Here we discuss two properties of magnetic systems that are particularly important to nonlinear dynamics: the multi-mode character of dynamic magnetic excitations and nonlinear damping. Apart from the main excited mode $\omega_0(\mathbf{k}_0)$ of magnetization precession or ferromagnetic resonance (which could be a spatially uniform mode, $\mathbf{k}_0 = 0$), other spatially non-uniform spin wave modes (or magnons) with dispersion laws $\omega(\mathbf{k})$ can also be excited in nanoscale magnetic samples, and different modes can interact with each other [Gurevich and Melkov, 1996; L'vov, 1994]. [Suhl, 1957] was first to explain theoretically that multi-magnon interaction processes with conservation laws for frequency $\omega$ (energy) and wave vector $\mathbf{k}$

$$n\omega_0 = \omega(\mathbf{k}_1) + \omega(\mathbf{k}_2) \qquad \textbf{[5.1]}$$

$$n\mathbf{k}_0 = \mathbf{k}_1 + \mathbf{k}_2$$

lead to an instability of the uniform magnetization precession mode, causing it to lose energy by exciting lower-frequency spatially non-uniform spin waves when the precession amplitude reaches a threshold value. This is an example of parametric resonance, in which the order of the parametric instability was $n$ = 1 and 2 for processes considered by [Suhl, 1957]. Parametric resonance can also excite precession in magnetic nano-oscillators; *e.g.*, a pumping signal at frequency $2\omega$ can generate precession at $\omega$ when the applied power is beyond a threshold determined primarily by the intrinsic magnetic damping [Gurevich and Melkov, 1996]. In the absence of interfacial torques, thresholds for parametric processes in metal films are ordinarily prohibitively high due to high magnetic dissipation in metals. However, interface-driven anti-damping torques can substantially reduce the effective magnetic damping, greatly reducing this power threshold [Urazhdin *et al.*, 2010; Edwards *et al*., 2012].

Finally, we point out that the form of magnetic damping used in the LLG Equation 2.3 (*i.e.*, a constant value for the phenomenological Gilbert damping parameter α) implicitly assumes a particular form of nonlinear contribution that is not appropriate in all regimes of large-angle precession [Tiberkevich and Slavin, 2007]. In the Gilbert model, magnetic damping is assumed proportional to precession frequency and when, with increasing precession angle, the precession experiences a nonlinear frequency shift, so does the magnetic damping. In particular, for an in-plane-magnetized magnetic film, magnetic damping should decrease with increasing precession angle. Such behavior of the nonlinear dissipation is *inconsistent* with typical magnon-electron and magnon-phonon relaxation processes, which are nonlinear (either three - or four-particle processes) because their intensity *increases* with increasing number of excited magnons. This inconsistency is unimportant in macroscopic systems where multi-magnon interaction processes (Eq. 5.1) are allowed, since these nonlinear magnon-magnon processes make the effective dissipation of a particular magnon mode nonlinear long before the nonlinearity of the dissipative Gilbert term, describing the net energy loss from the overall magnonic system, becomes important. However, in nano-magnetic samples, small sizes make all relevant frequencies discrete, creating difficulties for the conservation laws of multi-magnon interacting processes (Eq. 5.1) and necessitating an understanding of the correct nonlinear structure of the dissipative term in the LLG equation (Eq. 2.3). To address this, [Tiberkevich and Slavin, 2007] developed a nonlinear phenomenological generalization of the Gilbert model.

### *F. Open questions and new directions*

Challenges remain in understanding the nonlinear magnetic dynamics discussed in this section, and in using these for practical applications. The Landau-Lifshitz-Gilbert equation (Eq. 2.3 and its extensions 3.2 and 3.3) is almost universally used as the starting point for analyzing nonlinear magnetization dynamics. However, this is an approximate model, with largely empirical constants. It assumes that the magnetic motion is locally coherent, describable by a single local magnetization vector of fixed length; it is not easily capable of accounting for incoherent excitations (*e.g.*, short wavelength magnons).

*i. Exploration of other nonlinear phenomena or other dynamical regimes, such as relaxation oscillations or high dimensional chaos*: Such features may be observed in single spin-torque nano-oscillators with strong feedback [Khalsa *et al.*, 2015], or in assemblies of interacting oscillators [Flovik et al, 2016]. Investigations of complex dynamical regimes in magnetic systems is particularly interesting thanks to the underlying rich phase diagram and to the possibility of inducing transitions between different regimes through application of local spin-torques and magnetic fields.

*ii. Long-range collective spin transport:* "Spin superconductivity" has been suggested in low damping, easy-plane insulating magnets [Halperin and Hohenberg, 1969; Sonin, 1978; König *et al.*, 2001; Sonin, 2010; Takei *et al.*, 2014; Bender *et al.*, 2014; Chen *et al.*, 2014a], where interfacial spin Seebeck and Hall phenomena are likely critical to creating and detecting this collective spin-transport phenomena. In contrast to charge superconductivity, spin superconductivity is not based on an exact gauge symmetry but on spin-rotational symmetry that can be violated by damping or crystalline anisotropies. Proposals for achieving this state include use of spin-Seebeck pumping to counteract effects of damping and generate a magnon bosonic condensate [Bender *et al.*, 2012 and 2014] and achievement of sufficient excitation so that planar dynamics are not quenched by anisotropies [Sonin, 2010].

*iii. Beyond Landau-Lifshitz-Gilbert: the low-dissipation regime:* While we believe we understand the microscopic mechanisms responsible for magnetic damping in the low amplitude limit, it is likely that additional processes need to be included to describe magnetic dynamics for low-damping materials, in which strongly-nonlinear dynamics are most easily excited. These processes may depend on the degree of nonlinear coupling between different spin wave modes. Differences in nonlinear damping as a function of sample size and dimension govern even whether or not an anti-damping torque is capable of generating large-angle magnetic precession [Tiberkevich and Slavin, 2007; Demidov *et al*., 2011; Duan *et al*., 2014]. Nonlinear modes excited by spin-transfer torques have been observed for much larger sample sizes in low-damping materials like yttrium iron garnet (YIG) than in higher-damping metallic systems [Jungfleisch *et al.*, 2016b].

*iv. Beyond Landau-Lifshitz-Gilbert: strong driving, large systems, long coherence times, and thermal fluctuations:* Modeling of large systems with long coherence times, subject to thermal fluctuations, overwhelms current hardware and software, even with approaches like graphical processing units. Agreement between micromagnetic simulations and spin-torque oscillator experiments is often good at drive levels close to the thresholds for exciting magnetic dynamics, but agreement is generally poor for strong driving, where magnetization dynamics are most nonlinear. Whether this is a shortcoming of our understanding of nonlinear damping, nonlinear torques, or physics beyond the LLG equation is currently unknown, and alternative approaches are needed. Basic linear properties of spin wave excitation modes (*e.g.*, eigenfrequency, magnetization distribution) in complex magnetic nano-systems traditionally treated analytically (see *e.g.,* [Kalinikos and Slavin, 1986]) should be evaluated numerically, while the *nonlinear* interaction of modes, which require prohibitively long simulation times, need to be analyzed analytically [L'vov, 1994], while accounting for restrictions imposed by the quasi-Hamiltonian equations of magnetization dynamics. Progress may come from a hybrid formalism combining the quantitative accuracy of numerical methods with the qualitative clarity and predictive power of analytical techniques developed in the classical theory of nonlinear magnetization dynamics (*e.g.*, [Bertotti *et al.*, 2009]).

*v. Novel circuits and architectures utilizing spin-torque nano-oscillators*: Both traditional Boolean and non-Boolean logic architectures seem worthy of further research, as well as various forms of neuromorphic computation and associative memories [Grollier, Querlioz and Stiles, 2016].

## VI. INTERFACIAL EFFECTS IN ULTRAFAST MAGNETIZATION DYNAMICS

### *A. Introduction*

Using light to probe, modify, and control magnetic properties has long been important. Demand for ever-faster data storage, memory and processing has fueled efforts to find ever-faster ways to control magnetic states. The development of sub 100 fs laser pulses created the possibility of probing *and* controlling magnetism in entirely new ways, at sub-picosecond ("ultrafast") time scales, where the dynamics expressed in LLG Eq. 2.3 should not apply. The pioneering observation of sub-picosecond demagnetization in nickel after excitation by a 60 fs laser pulse [Beaurepaire *et al*., 1996] led to intriguing and controversial observations on an increasingly broad range of materials. The strongest perturbation in light-matter interactions is the AC electric field acting on electrons, such that a femtosecond laser pulse leads to an effective heating of the electron gas far above the temperature of the lattice due to the slow electron-phonon coupling time constant. This electron-light interaction, however, conserves electron spin, so it is unclear how such electron heating could cause collapse of magnetic order on femtosecond time scales.

A wide range of laser-induced phenomena in other metallic systems were subsequently found, including launching of precessional modes [Ju *et al*., 1999; Ju *et al*., 2000; van Kampen *et al*., 2002], and induced magnetic phase transitions [Thiele *et al.*, 2004; Ju *et al.,* 2004]. More recent discoveries include deterministic switching by single femtosecond pulses of circularly polarized light [Stanciu *et al*., 2007] and remarkable (and unexpected) *helicity-independent* toggle-switching of magnetization in ferrimagnetic rare earth-transition metal alloys [Ostler *et al.,* 2012]. These observations raised questions about magneto-optical interaction mechanisms, the role of interfacial spin-orbit coupling and symmetry breaking in these ultrafast processes, and the potential to engineer materials or optical processes to achieve new functionalities.

As discussed in Sec. II, spins in magnetically-ordered materials are dominated by three interactions: the comparatively weak dipolar interaction of spins with each other and with external magnetic fields, the generally-stronger spin-orbit interaction (which causes effective interactions between spin and electric fields), and the exchange spin-spin interaction which is typically the strongest force in magnetism. **Figure 23** shows the relative magnetic field, time and energy scales for magnetic systems. Magnetic fields lower than ≈0.3 T do not significantly affect spin dynamics in the ultrafast sub 100 ps time-domain. In most laser-induced experiments, ultrafast magnetization dynamics are dominated by spin-orbit and exchange interactions. Only at longer times, when precessional dynamics emerge, do dipolar coupling and anisotropy play a major role.

Both spin-orbit and exchange interactions are significantly modified at the interfaces of magnetic heterostructures, as discussed in Sec. II. Light-induced switching of magnetization has been observed only in multi-sublattice magnets, metallic multilayers, and chemically inhomogeneous alloys suggesting that intersublattice and interlayer exchange interactions, and possibly interfacial spin-orbit-interactions, are crucial. An entirely different effect of interfaces and finite size effects on ultrafast magnetization dynamics arises due to the mobility of optically excited carriers, which can drive spin-currents across interfaces [Battiato *et al.,* 2010], providing a non-local mechanism for fast changes of magnetization.

This section describes the important but still poorly understood role of interfaces in ultrafast laser-induced magnetization dynamics, focusing primarily on ferromagnetic and ferrimagnetic metallic films and heterostructures. Although interesting results on laser-induced magnetization dynamics in antiferromagnets, ferromagnetic semiconductors, and insulators have been reported [Kirilyuk *et al.,* 2010], studies to date focus on bulk materials. In Sec. VI.B the basics of femtosecond demagnetization are introduced. Sections VI.C–VI.E address non-local phenomena due to laser-induced spin currents, laser-induced precessional dynamics, and all-optical switching. Section VI.F concludes with an outlook on future research.

### *B. Ultrafast demagnetization*

**Figure 24a** shows the time evolution of the magnetization of Ni following an ultrafast pulse, measured using time-resolved MOKE of time-delayed probe pulses. The rapid (femtosecond) quenching of magnetization is visible, followed by its partial recovery upon cooling down from the transient excited state. This process can be described phenomenologically by a three-temperature model [Beaurepaire *et al*., 1996], which describes the energy flow between three separate sub-systems: the charge of the electrons, their spins, and the lattice [see **Figs. 24(b)** and (c)]. The temperature of each sub-system is a measure of the excess thermal energy in the respective reservoir. After absorption of the femtosecond-laser pulse, internal thermalization of the excited electron system proceeds within 100 fs. The excess heat is then transferred to the lattice degree of freedom *via* electron-phonon coupling, causing electron-lattice equilibration within ≈ 0.5 ps to 2 ps. In parallel, part of the excess energy in the electron and lattice sub-systems flows into the spin sub-system, causing spin excitations such as magnons, lowering the magnetization, and increasing the associated spin temperature. Finally the three temperatures converge once the system has achieved a new thermal equilibrium. On still longer time scales, heat flows out of the metal film to the substrate.

The surprising aspect is not the reduction of magnetic moment itself, but rather the time scale at which this occurs. For a localized system this process is subject to both energy and total angular momentum conservation. Thus, the quest is to find channels that change magnetic moment $\mathbf{M} \approx g_L\mathbf{L} + g_s\mathbf{S} \approx \mathbf{L}+2\mathbf{S}$, while conserving total angular momentum. The latter includes electronic orbital **L** and spin **S** momenta, angular momentum carried by phonons and the laser light field, *i.e.* $\mathbf{L}+\mathbf{S}+ \mathbf{L}_{\mathrm{phonon}}+ \mathbf{L}_{\mathrm{light}}$ (although $\mathbf{L}_{\mathrm{light}}$ is expected to be small). Note that total angular momentum is only conserved for the crystal as a whole, and care has to be taken for a *local* interpretation because of crystal field effects [Töws and Pastor, 2015]. An obvious channel for the "missing" angular momentum is $\mathbf{L}_{\mathrm{phonon}}$ but this requires a spin-lattice relaxation time of picoseconds, which is inconsistent with femtosecond observations. This raised doubts as to whether demagnetization was occurring at all, or whether the observation was an artifact of MOKE becoming inapplicable in this strongly non-equilibrium situation. Indeed "optical artifacts" due to dichroic state-filling effects were observed in specific anomalous cases [Koopmans *et al.*, 2000], and accounted for theoretically [Oppeneer and Liebsch, 2004].

However, techniques developed since 2000 provide evidence that time-resolved MOKE *does* provide a proper measurement of $M$(t) even within the first hundreds of femtoseconds after laser excitation under many standard experimental conditions. Time-resolved photoelectron spectroscopy showed the exchange splitting drops initially, followed by a recovery [Rhie *et al.*, 2003]. Terahertz radiation, emitted due to the rapid collapse of magnetization, was detected [Beaurepaire *et al*., 2004]. Most importantly, since 2007 X-ray magnetic circular dichroism (XMCD) with femtosecond time resolution [Stamm *et al.,* 2007] allows separate probes of orbital and spin moments (**Fig. 25a**); both decreased at approximately the same rate, indicative of an overall increase of thermal disorder in an equilibrium-like fashion [Boeglin *et al*., 2010]. Femtosecond-XMCD also enabled element specific studies (**Fig. 25b**), crucial to understanding switching phenomena in rare earth-transition metal ferrimagnets [Radu *et al*., 2011], and to resolving subtle non-equilibrium anomalies in ferromagnetic alloys like permalloy [Mathias *et al*., 2012].

The underlying mechanism(s) of the ultrafast loss of magnetization remain unclear. Early work hinted at an important role played by highly-excited electrons, *e.g.*, opening a channel for Stoner excitations [Scholl *et al*., 1997]. However, because the lifetime of electron Volt-excited electrons is at most ≈10 fs, this idea seems incompatible with the continuous demagnetization during the first hundreds of femtoseconds [Roth *et al*., 2012]. Other mechanisms based on electronic processes involving spin-orbit scattering and magnon excitations [Carpene *et al*., 2008] have been questioned based on conservation of total angular momentum in the electronic system. Several theories suggested transfer of angular momentum directly between photons and electrons [Zhang and Hübner, 2000; Bigot *et al.,* 2009], but the angular momentum of the laser light field is too small [Koopmans *et al.,* 2000]; also, these theories do not explain the continued demagnetization after 100 fs. Recently, it was shown [Töws and Pastor, 2015] that the crystal field potential together with spin-orbit coupling can provide very efficient relaxation of spin angular momentum.

Another class of theories looks at semi-phenomenogical approaches. Three models, despite differences in microscopic interpretation, display similar results for the demagnetization dynamics [Atxitia and Chubykalo-Fesenko, 2011]. In an atomistic LLG approach, the phenomenological concept of Gilbert damping is transferred from mesoscopic to atomic scale, providing a channel for transferring angular momentum from atomic spin precession to the heat bath of the lattice [Kazantseva *et al*., 2008]. Interfaces are known to impact damping on mesocopic scales by altering the coupling between spin waves and conduction electrons [Berger, 2001]. Because interfaces can dramatically alter the spin wave lifetimes of all wavelengths [Qin *et. al.,* 2013], the LLG description implies interfaces can play a central role in ultrafast demagnetization by impacting damping at atomistic scales. A similar continuum description solves the Landau-Lifshitz-Bloch (LLB) equations including statistically averaged magnetic fluctuations within a two-temperature model describing rapid energy transfer between electrons and lattice [Atxitia *et al*., 2007; Atxitia and Chubykalo-Fesenko, 2011]. The third approach, a microscopic three-temperature model, derives magnetization dynamics by defining a microscopic model Hamiltonian and solving Boltzmann rate equations [Koopmans *et al.,* 2005; Koopmans *et al.,* 2010]. A crucial process in this three-temperature model is Elliott-Yafet spin-flip scattering in which angular momentum is transferred in an electron-phonon scattering event accompanied by emission or absorption of a phonon. This microscopic three-temperature model has successfully reproduced experimental data [Roth *et al.,* 2012], with spin-flip probability agreeing with *ab initio* calculations [Carva *et al.,* 2011], despite ongoing debate about their interpretation [Schellekens and Koopmans, 2013a].

[Müller *et al*., 2009] compared the dynamics of a transition metal ferromagnet with oxides and Heusler high-spin-polarization ferromagnets. The dynamics were reported to be slower for materials with high spin-polarization due to blocking of Elliott-Yafet spin-flip scattering, compatible with the microscopic three-temperature model [Koopmans et al., 2010]. Bovensiepen and co-workers found a two-step demagnetization in the rare earth ferromagnets Gd and Tb [Melnikov *et al*., 2008; Wietstruk *et al*., 2011], where partial demagnetization in the first 2 ps is followed by much slower demagnetization over tens of picoseconds. This two-step demagnetization has been suggested to be generic for ferromagnets with a demagnetization time that is longer than the electron-phonon equilibration time, due to large magnetic moment, low Curie temperature or low spin-flip probability [Koopmans *et al*., 2010]. [Wietstruk *et al.*, 2011] suggest a different mechanism, while [Kimling *et al.,* 2014] emphasize the importance of the magnetic heat capacity.

In the above, magnetization dynamics was analyzed as a local phenomenon. Strong modification of dynamics, however, or even novel phenomena can be anticipated due to interfaces between layers or in laterally heterogeneous granular, chemically segregated or patterned systems, which may introduce additional spin-flip scattering. Recent work comparing demagnetization of pure Co films to Co/Pt multilayers shows significant enhancement of spin-flip scattering driven by large orbital moments at the interfaces [Kuiper *et al.,* 2014].

### *C. Laser-induced nonlocal ultrafast phenomena*

The femtosecond heating and subsequent thermalization of electrons in a metal are shown in **Fig. 26a**. Initially, there are hot electrons with energies far above the Fermi level [Aeschlimann *et al*., 1997] that decay by exciting other electrons [Knorren *et al*., 2000] leading to thermalization of the laser heated electronic system on ≈100 fs timescale, which then relaxes to the lattice temperature on picosecond timescale [Rhie *et al.,* 2003]. The hot electron lifetime depends on spin, with majority spin lifetimes exceeding those of minority spin electrons by up to a factor of two [Knorren *et al*., 2000; Aeschlimann *et al*., 1997]. Theoretical modeling shows a "superdiffusive" spin transport of mainly majority spins away from the excitation region [Battiato *et al.*, 2010], shown in **Fig. 26(b)**, embedded between ballistic spin motion at early time and a purely diffusive later regime when the electronic distribution is close to thermal equilibrium with the lattice (see **Fig. 24**). The latter process can also result in a large (diffusive) spin-dependent Seebeck effect (Sec. III.B) over large distances [Choi *et al.*, 2015].

**Figure 27a** shows superdiffusive spin currents detected in Au films *via* non-linear second harmonic generation [Melnikov *et al*., 2011]. Ballistic Fe spins injected into a Au layer travelling close to the Au Fermi velocity arrive at the Au back interface within hundreds of femtoseconds while a diffusive component was detected at times up to 1 ps, in qualitative agreement with wave diffusion calculations [Kaltenborn *et al.*, 2012]. Variation of Fe layer thickness shows that the active injection region is an ≈1 nm thick Fe layer at the Fe/Au interface [Melnikov *et al*., 2011; Alekhin *et al*., 2015; Melnikov *et al*., 2015], implying that superdiffusive transport is of limited importance for ultrafast demagnetization of significantly thicker ferromagnetic films, an observation supported by recent demagnetization experiments in Ni films [Schellekens *et al.*, 2013]. [Ando *et al*., 2011; Hoffmann, 2013] used the inverse spin Hall effect to detect superdiffusive spin currents in non-magnetic layers. Transient superdiffusive spin currents cause a terahertz electromagnetic pulse with polarization given by the transverse charge current in the spin Hall layer [Kampfrath *et al*., 2013; Seifert *et al.,* 2016].

In a conventional spin-transfer torque magnetic structure, a spin polarized current exerts a torque on a magnetic layer ultimately switching its direction. The use of strong, ultrashort non-equilibrium spin currents could create new ways for spin-transfer torque switching. [Schellekens *et al*., 2014] and [Choi *et al*., 2014] demonstrated spin-torque-induced precession dynamics driven by spin currents. These experiments utilize two ferromagnetic layers, one with in-plane and the other with out-of-plane magnetization, separated by Cu [Choi *et al*., 2014] or Cu and Pt [Schellekens *et al*., 2014] spacer layers (**Fig. 27b**). Although the induced precession angles are small due to limited spin angular momentum transfer (several percent), such experiments present a unique tool to understand and optimize angular momentum transfer through interfaces.

Although deemed important, the role of transient spin accumulation at interfaces is currently not clearly established. The longer ballistic mean free paths for majority electrons should lead to minority spin accumulation at a ferromagnetic interface layer upon injection of an unpolarized current from an adjacent non-magnetic metal layer. However, reports for Au/Ni layers that this could lead to ultrafast demagnetization of 15 nm thick Ni films [Eschenlohr *et al.*, 2013] remain controversial [Eschenlohr *et al*., 2014; Khorsand *et al*., 2014]. [He *et al*., 2013] observed that spin tunneling through MgO spacers influenced ultrafast demagnetization of adjacent CoFeB layers. Control of femtosecond demagnetization in a CoFeB-based magnetic tunnel junction was demonstrated by tuning the voltage applied to the junction [Savoini *et al*., 2014a].

Magnetic switching by superdiffusive current has been reported by [Graves et al., 2013] in an amorphous Gd-Fe-Co alloy. The process depends on ultrafast demagnetization followed by magnetization reversal, as described below for all-optical switching, and is believed to depend on chemical segregation into Gd-rich and Fe-rich nanoregions on a ≈10 nm length scale which enable spin current to flow from one region to the other. This result highlights the importance of understanding the structure of nominally similar amorphous materials. Future advances in X-ray nano-spectroscopy offer the opportunity to also determine effects of non-local transport currents on local valence level populations [Kukreja *et al.,* 2015].

### *D. Dynamics in coupled magnetic systems*

We turn now to how optical pulses affect inter-layer exchange coupling and how spins in multilayers respond to femtosecond excitations. [Ju *et al.*, 1998; Ju *et al.*, 1999; Ju *et al.*, 2000] observed modulation of exchange coupling on a picosecond timescale in ferromagnet/ antiferromagnet exchange-coupled NiFe/NiO bilayers excited with 120 fs laser pulses by comparing the time-resolved magneto-optical response of the NiFe layer to ‘‘bare’’ epitaxial NiFe thin films without NiO. The authors observed modulation of the exchange coupling on a picosecond timescale. They also found that "unpinning" of this exchange bias led to coherent magnetization rotation in the NiFe film, with large modulation ($\Delta M_z/M_S \approx 0.5$) on a time scale of 100 ps. This was the first observation of ferromagnetic resonance induced in a magnetic system with a short laser pulse and is similar to the long-time behavior shown in **Fig. 24(a)**.

Excitation of an exchange-coupled NiFe/FeMn structure with a 9 ps laser pulse showed reduction of the exchange-bias field to ≈ 50 % of its initial value within 20 ps [Weber *et al.*, 2005; Weber *et al.*, 2005a, Weber *et al.*, 2005b]. The fast quenching was followed by a slower recovery of the bias field, with relaxation time ≈170 ps. [Dalla Longa *et al.*, 2008; Dalla Longa *et al.*, 2010] used magneto-optical Kerr effect as a function of pump-probe delay time to estimate the time scale of laser-induced exchange-bias quenching in a polycrystalline Co/IrMn bilayer to be 0.7±0.5 ps. The fast decrease in exchange coupling upon laser heating is attributed to spin disorder at the interface. Permanent changes of exchange bias as a result of femtosecond laser excitation were reported in [Seu and Reilly, 2008] and [Porat *et al.*, 2009].

The ability to locally excite ferromagnetic resonance in magnetic heterostructures [van Kampen *et al.*, 2002] is an effective means for characterization of individual layers and the interfaces between them. [Hicken *et al.*, 2003] showed that analysis of laser-induced precession in a spin valve heterostructure as a function of the orientation and magnitude of the applied magnetic field, yields information about the *g* factor, demagnetizing field, exchange bias field and magnetic anisotropy within an individual ferromagnetic layer, as well as the coupling between layers, that is similar to conventional ferromagnetic resonance. A great advantage of time-resolved techniques over frequency domain techniques is the ability to measure spin precession even for large damping. Spatial resolution requires combining the optical excitation with optical imaging or by using a small optical spot, enabling investigation of how damping of laser-induced precession is affected by interlayer exchange interactions and modified spin-orbit interactions at the interfaces [Engebretson *et al*., 2005; Djordjevic *et al.*, 2006; Weber *et al.*, 2006; Zhang *et al*., 2010; Fan *et al*., 2014a]. Analysis of spin precession induced by femtosecond laser excitation was similarly used to reveal a modified spin-orbit interaction and interfacial effects in ferromagnetic exchange-coupled multilayers (*e.g.,* [Barman *et al.*, 2007; Michalski *et al*., 2007, Rzhevsky *et al.*, 2007; Ren *et al*., 2008]).

In the above studies, interfacial effects in the exchange coupled magnetic multilayers were due to ultrafast laser-induced heating. A way to study ultrafast interfacial dynamics *without* heating was suggested by [Scherbakov *et al*., 2010; Jager *et al*., 2013]. Ultrashort (≈10 ps) strain pulses injected into heterostructures by a femtosecond laser pulse propagate coherently over a distance of ≈100 μm, causing strain-induced spin precession [Scherbakov *et al.*, 2010; Jager *et al*., 2013]. This emerging ability to control and probe ultrafast strain in magnetic heterostrutures may reveal the physics of a broad range of phenomena such as magnetostriction, magneto-structural phase transitions, and interfacial magnetism.

Ultrafast laser excitations have been shown to trigger reorientation of spins over 90° in antiferromagnetic orthoferrites and a scenario for using this phenomenon for all-optical magnetic switching was proposed by [Kimel *et al.*, 2004]. Spin reorientation of the antiferromagnet in an exchange bias ferromagnet/antiferromagnet heterostructure would create a magnetic torque on the ferromagnet, promoting reversal of its magnetization [Le Guyader *et al*., 2013].

### *E. All-optical switching of magnetic films and nanostructures*

The demonstration that light can switch magnetization without applied magnetic field is an important outcome of ultrafast optical research. In pioneering work, [Stanciu *et al.*, 2007] showed fully deterministic magnetization switching in a ferrimagnetic amorphous Gd-Fe-Co alloy film using 40 fs optical pulses. In the initial work, the final direction of magnetization was determined by the helicity of the optical pulse, as shown in **Fig. 28a**, where scanning a laser beam across the sample while modulating the polarization of the beam between left- and right-circular pulses yields a magnetic bit pattern. This remarkable phenomenon has been termed all-optical switching or, more precisely, all-optical *helicity-dependent* switching, in which the final magnetization depends on the optical helicity. In addition to providing insight into the ultrafast response of magnetic systems, all-optical switching may lead to technological breakthroughs in applications such as heat-assisted magnetic recording [Shiroishi *et al*., 2009; Stipe *et al.*, 2010; Wu *et al.*, 2013]. This process appears energy efficient; an energy lower than 10 fJ is expected to be sufficient to switch a $20 \times 20$ nm$^2$ area of magnetic material.

Most experimental and theoretical studies of all-optical switching focus on amorphous (*a*-) rare earth-transition metal alloys of varying composition, where the net magnetization results from the rare earth and transition metal subnetworks which are antiferromagnetically exchange coupled, forming a *ferrimagnet* (shown in **Fig. 25b**). Although these materials have a well-defined $T_c$, the rare earth and transition metal moments have different temperature dependencies, resulting in compensation $M_s$=0 for a rare earth concentration ≈ 0.25 mol/mol (25 at.%) at a compensation temperature ($T_{comp}$). Under suitable growth conditions, these alloys have significant perpendicular magnetic anisotropy. All-optical switching is generally observed in a narrow composition range where $T_{comp}$ is close to the starting temperature [Kirilyuk *et al.*, 2013; Mangin *et al.*, 2014; Medapalli *et al.*, 2014]. There is also a separate temperature where the angular momentum compensates which is likely relevant to ultrafast processes [Kirilyuk *et al.*, 2013; Krivoruchko, 2014].

Recent focus has been on other alloys and heterostructures where optical and magnetic properties can be tuned, including *a*-Tb-Co, *a*-Tb-Fe and *a*-Tb-Fe-Co alloys [Alebrand, Gottwald *et al*., 2012; Hassdenteufel *et al*., 2013; Alebrand *et al.*, 2014; Hassdenteufel *et al.*, 2014a; Hassdenteufel *et al.*, 2014b; Mangin *et al.*, 2014; Savoini *et al.*, 2014b], rare earth/transition metal multilayers (which are generally amorphous; *e.g.*, *a*-Tb/Co, *a*-Gd/Fe, *a*-Ho/$Co_xFe_{1-x}$) and heterostructures [Mangin *et al.*, 2014; Schubert *et al.*, 2014], as well as *a*-Dy- and *a*-Ho-transition metal alloys [Mangin *et al.*, 2014], where all-optical *helicity-dependent* switching appears to be qualitatively similar to *a*-Gd-Fe-Co, despite the significant orbital moments of Tb, Dy and Ho (Gd has S=7/2, L=0), which results in increased spin-orbit interactions and increased magnetic anisotropy. Notably, all-optical switching appears to depend more on volume-average magnetic properties than specific properties of individual layers as demonstrated by [Schubert *et* *al*., 2014]; an *a*-$Tb_{36}Fe_{64}$/*a*-$Tb_{19}Fe_{81}$ heterostructure with zero net magnetization due to antiparallel interfacial exchange coupling exhibits all-optical switching despite not being observed for single *a*-$Tb_{36}Fe_{64}$ or *a*-$Tb_{19}Fe_{81}$ films.

Recent computational [Evans *et al*., 2014] and experimental [Mangin *et al*., 2014] work has shown that all-optical switching occurs in artificial ferrimagnetic systems that have compensation temperatures but are rare earth-free. This is achieved by using the interfacial properties of thin transition metal layers to create perpendicular surface anisotropy and interlayer exchange coupling (as described in Sec. II). By tuning layer properties, it is possible to achieve synthetic ferrimagnets with a tunable compensation temperature in heterostructures such as Pd/$[Co/Ir/Co/Ni/Pt/Co/Ir]_N$/Pd [Mangin *et al*., 2014].

All-optical helicity-*dependent* switching was also demonstrated in *ferromagnetic* thin films such as [Co/Pt] multilayers and granular $L1_0$ FePt in a C matrix [Lambert *et al*., 2014]. Shown in **Fig. 28(b)** are the results for a $[Co(0.4\ nm)/Pt(0.7\ nm)]_3$ multilayer where the final state of the magnet is dependent on the helicity of the light. Deterministic all-optical switching is observed only in the thin film limit (less than a few nanometers) and this appears to be related to suppressing demagnetization by creating stripe domains during the optical heating and cooling process [Lambert *et al*., 2014]. Whether these experiments can be understood without taking into account the known strong interfacial spin-orbit interaction of Co/Pt and consequent likely non-collinear spins, *i.e.,* in the spirit of the model suggested by [Vahaplar *et al.*, 2012] is unclear. In principle, differential heating could arise from magnetic circular dichroism, thermal gradients, superdiffusive spin currents or direct interaction with the polarization of the light. However, the simulations performed in [Vahaplar *et al.*, 2012] clearly showed that heat-assisted switching of a ferromagnet can occur only if the effective magnetic field generated by 60 fs laser pulse reaches 20 T and lasts at least 250 fs. The origin of such a strong optically-induced effective magnetic field in Co/Pt is unclear. The strong spin-orbit interaction at the interface can increase the inverse Faraday effect, change the Curie temperature, magnetic anisotropy and even spin texture. Elucidating the role played by the interfacial spin-orbit interaction in all-optical switching is a challenge for future studies.

The detailed processes of all-optical switching are still unclear, but it has notably *not* been observed in single-element films, but only in alloys or heterostructures that mix *3d* and *4f* elements (*e.g.*, Fe and Gd); *3d* and *4d* elements (*e.g.*, Co and Pd); or *3d* and *5d* (*e.g.*, Fe or Co and Pt or Ir). Ultrafast laser-induced processes are driven by two main forces: spin-orbit and exchange interactions, and reasonable to conclude that interfacial and inter-sublattice exchange interactions and interfacial spin-orbit interactions play a critical role in all-optical switching.

*Helicity-dependent* switching shown in **Fig. 28** results from helicity-dependent absorption, which favors one magnetic configuration over the other [Khorsand *et al*., 2012]. *Helicity-independent* toggle switching is in some ways a more remarkable result, and has been observed in few systems to date. It is observed in *a*-Gd-Fe-Co alloys where the two sublattices (the rare earth dominated by 4*f* electrons, the transition metals by 3*d* electrons) demagnetize at different time-scales [Radu *et al*., 2011]. As shown in **Fig. 25(b)**, ultrafast optical excitation brings *a*-Gd-Fe-Co alloys into a strongly non-equilibrium state with nearly demagnetized Fe and hardly demagnetized Gd. The strong ferromagnetic Fe-Fe exchange interaction drives a fast relaxation from this state. Due to angular momentum transfer between Fe and Gd sublattices, the relaxation first results in a transient ferromagnetic state where Gd and Fe are parallel. Afterwards the Gd moment reemerges in the antiparallel orientation due to the weaker Fe-Gd exchange interaction, but both Fe and Gd moments are now in the opposite direction from the initial state. This toggle mode of reversal is observed *even for linear polarization*, and appears specific to ferrimagnetic materials with two non-equivalent and antiferromagnetically coupled sublattices [Ostler *et al*., 2012]. This switching was explained within a general theoretical framework [Mentink *et al*., 2012], and reproduced by both atomistic LLG [Ostler *et al*., 2012] and the microscopic three-temperature model with two coupled spin baths [Schellekens and Koopmans, 2013b]. Further measurements on *a*-Gd-Fe-Co alloys support this interpretation [Mekonnen *et al*., 2013; Hashimoto *et al*., 2014; Medapalli *et al*., 2014; Le Guyader *et al*., 2015].

All-optical switching is limited to the spot size of the laser heating, too large to compete with bit densities of conventional storage. In heat-assisted magnetic data storage applications aiming for terabit/inch$^2$ densities, Au plasmonic waveguides reduce the laser beam diameter to magnetic bit dimensions [Stipe *et al*., 2010; Wu *et al*., 2013]. This process was mimicked with Au nanoantennas patterned onto high anisotropy *a*-Tb-Fe-Co films, where all-optical switching of 50 nm bits was toggled by individual femtosecond laser pulses [Liu *et al*., 2015b]. The X-ray magnetic holography there employed can be extended to X-ray free electron lasers offering the possibility to image all-optical switching with a single X-ray pulse [Wang *et al*., 2012a].

### F. Open questions and new directions

The field of ultrafast, non-equilibrium magnetism has evolved dramatically over the last two decades. Elucidating the channels that allow ultrafast angular momentum transfer remains central. The discovery of all-optical magnetization reversal in a broad class of metallic alloys and multilayers, but not all materials, showed that interfaces are crucial. The role of interfaces, of exchange coupling across those interfaces, of spin-orbit coupling and the DM interactions induced by interfaces, and how the chiral structures induced by these interactions are affected by ultrafast pumps are all topics in their infancy. It is clear that novel experimental techniques probing new materials with well-defined interfacial structures are needed. We focus here on a few questions:

*i. Can we observe and understand ultrafast angular momentum transfer between spin and lattice?* Quantitative theoretical description of femtosecond magnetism is still embryonic, creating great opportunities for development of ultrafast magnetism theory. Are transient magnetic states fundamentally different in local moment systems such as rare earth compared to transition metal ferromagnets? How is angular momentum transfer affected by interfaces, and specifically does the modification of interactions, including crystal fields, at interfaces play an important role? This latter question requires ultrafast measurement of structural and magnetic response, at nanometer spatial scales and with interface sensitivity.

*ii. How do ultrafast pumps and probes interact with interfacially-induced magnetic states?* Since the spin-orbit interaction is lower energy, and hence slower, than the exchange interaction, in systems where the underlying magnetic state is driven by a competition between these two energies, as in interfacially-induced chiral spin structures, how will these states interact with ultrafast pumps, particularly those of defined helicity?

*iii. Can we probe and control the ultrafast dynamics associated with exchange interactions? How do interfaces affect exchange-driven dynamics?* Harnessing the exchange interaction will enable the fastest magnetic switching. Experiments in high magnetic fields will enable disentanglement of spin dynamics from exchange interaction dynamics, as high magnetic fields can substantially suppress spin dynamics, leaving exchange dynamics unaffected.

*iv. What are the elementary spin-dependent scattering processes that impact superdiffusive spin transport? How are non-equilibrium spin currents affected by interfaces in the ultrafast regime, and can interfaces provide a new means for control?* The effects of interfaces on

equilibrium and near-equilibrium spin currents has been studied for decades (*e.g.*, Bass and Pratt, 2007]. Ultrafast measurements suggest a separation of charge and spin degrees of freedom that may affect our understanding of spin diffusion lengths and spin flip scattering at interfaces. As a first step towards the goal of understanding the interaction of interfaces and non-equilibrium spin currents, time resolved femtosecond X-ray spectroscopy has identified electronic states responsible for transport, scattering and spin accumulation at interfaces [Kukreija *et al*., 2015].

*v. Observation of phonons excited in ultrafast magnetization processes:* While electronic and spin degrees of freedom have been accessible since the first observation of ultrafast demagnetization, we have been blind to the phonons excited in this process. Femtosecond hard X-ray scattering with X-ray free electron lasers [Trigo *et al.*, 2013] and novel femtosecond electron diffraction and imaging [Zhu and Dürr, 2015] promise the ability to examine the phonons excited in ultrafast magnetization processes.

## VII. CONCLUDING REMARKS

Interfacial effects have long played a crucial role in magnetism, both in the development of underlying scientific principles and in the many technologies in which magnetism plays a critical role. Many important effects in magnetism are either intrinsically interfacial phenomena or strongly enhanced at interfaces, for example exchange bias, giant magnetoresistance, chiral domain walls, and spin currents with no accompanying charge current. Interfacial engineering enables potentially novel fundamental behavior by blending competing and/or complementary states, for example topological insulators and ferromagnets. Heterostructures break spatial inversion symmetry, which combined with the time reversal symmetry breaking of ferromagnetism (or antiferromagnetism) yields properties that had previously been seen only in very limited classes of bulk materials, but now can be explored and developed in a straightforward way by judicious choice of materials. The ability to design heterostructure interfaces enables studies of chiral magnetic states, suggests new interactions with the static and dynamic magnetic state of the ferromagnet, and inspires hope of even more dramatic phenomena such as a room temperature quantum anomalous Hall effect and spin superconductivity. The developments discussed in this article push the boundaries of experiment and theory. The long-standing success of the Landau-Lifshitz-Gilbert equation, despite the empirical nature of many of its constants, particularly those associated with damping, has enabled understanding of many dynamic phenomena, but is not relevant to the femtosecond regime. Ultrafast studies of more complex magnetic structures than the simple ferromagnets studied to date are likely to lead to new phenomena. As yet unimagined phenomena will certainly be discovered as we explore heterostructures of materials beyond the relatively well-known materials that make up a large portion of existing research reviewed here.

These scientific advances are already leading to new magnetic-state control strategies that are directly relevant to the magnetic recording and memory industries. These currently involve the use of heavy metals and ferromagnets, and may in the future include antiferromagnets and topological insulators. They have the potential to enable faster, more energy efficient manipulation of magnetic states and potentially to create fundamentally new magnetic states.

Studies of non-linearities associated with large amplitude magnetization dynamics, including switching, could be a verdant area for future work, particularly in the low damping regime. Ultrafast, non-equilibrium manipulation of ferromagnets has yielded remarkable but still incompletely understood phenomena and has the potential to change our understanding of the fundamental quantum mechanical interactions that underpin magnetism, and to provide new mechanisms for magnetization control magnetism. Whether interfaces and spin-orbit coupling at those interfaces, which is somewhat understood in the static and dynamic (down to picosecond) regimes, can be used to modify ultrafast processes is an important open question.

Interfaces are critical to much current magnetism research. Open questions and promising areas for future research have been identified at the end of each section of this article. An earlier shift in semiconductor physics from the study of bulk phenomena to the understanding and manipulation of surfaces, interfaces, and inhomogeneities led to electronic devices that revolutionized our economy and our science, as epitomized in Kroemer's "the interface is the device" Nobel Prize address. In this spirit we hope that our article modestly echoes Kroemer in asserting that *"the interface is the magnet"*.

## ACKNOWLEDGMENTS

This paper benefited greatly from a three-day workshop on this topic. We thank the U.S. Department of Energy (DOE), Office of Science, Office of Basic Energy Sciences (BES), Materials Science and Engineering Division (DMSE), the DMSE Council members, and specifically DMSE Director Linda Horton, Condensed Matter and Materials Physics Area Lead Jim Horwitz, and Experimental Condensed Matter Physics Program Manager Mick Pechan, for sponsoring the workshop and helping with the concept development that led to this review article, Teresa Crockett at DMSE for organizational assistance, and Suzanne Kokosz at Argonne National Lab for her editing assistance. FH, PF thank DOE BES DMSE DE-AC02-05-CH11231 NEMM program, and Rich Wilson for valuable discussions. IKS thanks DOE BES DMSE DE-FG02-87ER-45332. CL thanks DOE BES DMSE DE-FG02-06ER46275, and NSF DMR-1507048 and DMR-1420013. TJ thanks the European Research Council (ERC) Advanced Grant 268066, Czech Republic Ministry of Education Grant LM2011026 and Czech Republic Grant 14-37427. INK thanks DOE BES SC0012670. OT thanks DOE BES DMSE DE-FG02-08ER46544. BLZ thanks NSF DMR-1410247. JPH thanks NSF DMR-1420451. GB thanks DOE BES DMSE DE-SC0012371. AT thanks Agence Nationale de la Recherche ANR-14-CE26-0012. HAD thanks DOE BES DMSE DE-AC02-76SF00515. EEF thanks DOE BES DE-SC0003678. JG thanks ERC Consolidator Grant 682955. DCR thanks NSF DMR-1406333.

## REFERENCES

Abarra, E. N., K. Takano, F. Hellman, and A. E. Berkowitz, 1996, “Thermodynamic Measurements of Magnetic Ordering in Antiferromagnetic Superlattices”, Phys. Rev. Lett. **77**, 3451.

Adamo, C., *et al.*, 2009, “Effect of biaxial strain on the electrical and magnetic properties of (001) $La_{0.7}Sr_{0.3}MnO_3$ thin films”, Appl. Phys. Lett. **95**, 112504.

Aeschlimann, M., M. Bauer, S. Pawlik, W. Weber, R. Burgermeister, D. Oberli, and H.C. Siegmann, 1997, “Ultrafast spin-dependent electron dynamics in fcc Co”, Phys. Rev. Lett. **79**, 5158.

Agrawal, M., V. I. Vasyuchka, A. A. Serga, A. Kirihara, P. Pirro, T. Langner, M. B. Jungfleisch, A. V. Chumak, E. Th. Papaioannou, and B. Hillebrands, 2014, “Role of bulk-magnon transport in the temporal evolution of the longitudinal spin-Seebeck effect”, Phys. Rev. B **89**, 224414.

Ahn, C., *et al.*, 2006, “Electrostatic modification of novel materials”, Rev. Mod. Phys. **78**, 1185.

Alebrand, S., M. Gottwald, M. Hehn, D. Steil, M. Cinchetti, D. Lacour, E.E. Fullerton, M. Aeschlimann, and S. Mangin, 2012, “Light-induced magnetization reversal of high-anisotropy TbCo alloy films”, Appl. Phys. Lett. **101**, 162408.

Alebrand, S., U. Bierbrauer, M. Hehn, M. Gottwald, O. Schmitt, D. Steil, E. E. Fullerton, S. Mangin, M. Cinchetti, and M. Aeschlimann, 2014, “Subpicosecond magnetization dynamics in TbCo alloys”, Phys. Rev. B **89**, 144404.

Alegria, L. D., H. Ji, N. Yao, J. J. Clarke, R. J. Cava, and J. R. Petta, 2014, “Large anomalous Hall effect in ferromagnetic insulator-topological insulator heterostructures”, Appl. Phys. Lett. **105,** 53512.

Alekhin, A., D. Bürstel, A. Melnikov, D. Diesing, and U. Bovensiepen, 2015, "Ultrafast Laser-excited Spin Transport in Au/Fe/MgO (001): Relevance of the Fe Layer Thickness”, Ultrafast Magnetism I Book Series: Springer Proceedings in Physics **159**, 241. J.-Y. Bigot et al. (eds), Ultrafast Magnetism I, Springer Proceedings in Physics Vol. 159 (Springer International Publishing Switzerland 2015).

Al'Khawaja, U., and H. T. C. Stoof, 2001, "Skyrmions in a ferromagnetic Bose-Einstein condensate”, Nature **411**, 918.

Allwood, D. A., G. Xiong, C. C. Faulkner, D. Atkinson, D. Petit, and R. P. Cowburn, 2005, "Magnetic Domain-Wall Logic”, Science **309**, 1688-1692.

Ambrose, T, and C. L. Chien, 1996, “Finite-Size Effects and Uncompensated Magnetization in Thin Antiferromagnetic CoO Layers”, Phys. Rev. Lett. **76**, 1743.

Amin, V, M. D. Stiles, 2016a, “Spin Transport at Interfaces with Spin-Orbit Coupling: Phenomenology,” Phys. Rev. B **94**, 104420.

Amin, V, M. D. Stiles, 2016b, “Spin Transport at Interfaces with Spin-Orbit Coupling: Formalism,” Phys. Rev. B **94**, 104419.

Anderson, P. W., 1950, “Antiferromagnetism. Theory of superexchange interaction”, Phys. Rev. **79**, 350.

Ando, K., 2014, “Dynamical generation of spin currents”, Semicond. Sci. and Tech. **29**, 043002.

Ando, K., S. Fujita, J. Ito, S. Yuasa, Y. Suzuki, Y. Nakatani, T. Miyazaki, H. Yoda, 2014, “Spin-transfer torque magnetoresistive random-access memory technologies for normally off computing (invited)”, J Appl. Phys. **115,** 172607.

Ando K., S. Takahashi, K. Harii, K. Sasage, J. Ieada, S. Maekawa, and E. Saitoh, 2008, “Electric manipulation of spin relaxation using the spin Hall effect”, Phys. Rev. Lett. **101**, 036601.

Ando, K, *et al.*, 2011, “Inverse spin-Hall effect induced by spin pumping in metallic system”, J. Appl. Phys. **109**, 103913.

Ariando, X. *et al.*, 2011, “Electronic phase separation at the $LaAlO_3/SrTiO_3$ interface”, Nat. Commun. **2**, 188.

Aso, R., D. Kan, Y. Shimakawa, and H. Kurata, 2014, “Octahedral tilt propagation controlled by A-site cation size at perovskite oxide heterointerfaces”, Cryst. Growth Des. **14**, 2128.

Atxitia, U., and O. Chubykalo-Fesenko, 2011, “Ultrafast magnetization dynamics rates within the Landau-Lifshitz-Bloch model”, Phys. Rev. B **84**, 144414.

Atxitia, U., O. Chubykalo-Fesenko, N. Kazantseva, D. Hinzke, U. Nowak, and R.W. Chantrell, 2007 “Micromagnetic modeling of laser-induced magnetization dynamics using the Landau-Lifshitz-Bloch equation”, Appl. Phys. Lett. **91**, 232507.

Avery, A. D., Sultan, R., Bassett, D., Wei, D., and B. L. Zink, 2011, “Thermopower and resistivity in ferromagnetic thin films near room temperature”, Phys. Rev. B **83** 100401.

Avery, A. D., M. R. Pufall, and B. L. Zink, 2012, “Observation of the Planar Nernst Effect in Permalloy and Nickel Thin Films with In-Plane Thermal Gradients”, Phys. Rev. Lett. **109**, 196602.

Azevedo, A., L. H. Vilela-Leão, R. I. Rodriguez-Suarez, A. B. Oliveira, and S. M. Rezende, 2005, “DC effect in ferromagnetic resonance: Evidence of the spin pumping effect?”, J. Appl. Phys. **97**, 10C715.

Baibich, M. N., J. M. Broto, A. Fert, F. Nguyen Van Dau, F. Petroff, P. Etienne, G. Creuzet, A. Friederich, and J. Chazelas, 1988, “Giant Magnetoresistance of (001)Fe/(001)Cr Magnetic Superlattices”, Phys. Rev. Lett. **61**, 2472.

Bak, P., and M. H. Jensen, 1980, "Theory of helical magnetic structures and phase transitions in MnSi and FeGe”, J. Phys. C: Sol. State Phys. **13**, L881.

Bakun, A. A., B. P. Zakharchenya, A. A. Rogachev, M. N. Tkachuk, and V. G. Fleisher, 1984, “Observation of a surface photocurrent caused by optical orientation of electrons in a semiconductor”, JETP Lett. **40**, 1293. (English translation of Pis’ma Zh. Eksp. Teor. Fiz. **40**, 464)

Baltz, V., A. Manchon, M. Tsoi, T. Moriyama, T. Ono, and Y. Tserkovnyak, 2016, “Antiferromagnetism:

the next flagship magnetic order for spintronics?", arXiv:1606.04284.

Bandiera, S., R. C. Sousa, M. Marins de Castro, C. Ducruet, C. Portemont, S. Auffret, L. Vila, I. L. Prejbeanu, B. Rodmacq, and B. Dieny, 2011, "Spin transfer torque switching assisted by thermally induced anisotropy reorientation in perpendicular magnetic tunnel junctions", Appl. Phys. Lett. **99**, 202507.

Barbier A., C. Mocuta, W. Neubeck, M. Mulazzi, F. Yakhou, K. Chesnel, A. Sollier, C. Vettier, and F. de Bergevin, 2004, "Surface and Bulk Spin Ordering of Antiferromagnetic Materials: NiO(111)", Phys. Rev. Lett., **93**, 257208.

Barman A., S. Wang, O. Hellwig, A. Berger, E. E. Fullerton, and H. Schmidt, 2007, "Ultrafast magnetization dynamics in high perpendicular anisotropy $[Co/Pt]_n$ multilayers", J. Appl. Phys. **101**, 09D102.

Bass, J. and W. P. Pratt, 1999, "Current-perpendicular (CPP) magnetoresistance in magnetic metallic multilayers," J. Magn. Magn. Mater., **200**, 274.

Bass, J. and W. P. Pratt, 2007, " Spin-diffusion lengths in metals and alloys, and spin-flipping at metal/metal interfaces: an experimentalist's critical review", J. Phys.: Condens. Matter **19**, 183201.

Battiato, M., K. Carva, and P. Oppeneer, 2010, "Superdiffusive Spin Transport as a Mechanism of Ultrafast Demagnetization", Phys. Rev. Lett. **105**, 027203.

Bauer, G. E. W., E. Saitoh, and B. J. van Wees, 2012, "Spin caloritronics", Nat. Mater. **11**, 391.

Bauer, U., L. Yao, A. J. Tan, P. Agrawal, S. Emori, H. L. Tuller, S. van Dijken, and G. S. D. Beach, 2015, "Magneto-ionic control of interfacial magnetism", Nat. Mater. **14**, 174.

Beach, G. S. D., M. Tsoi, and J. L. Erskine, 2008, "Current-induced domain wall motion", J. Magn. Magn. Mater. **320**, 1272.

Beaurepaire, E., J. C. Merle, A. Daunois, and J.-Y. Bigot, 1996, "Ultrafast spin dynamics in ferromagnetic nickel", Phys. Rev. Lett. **76**, 4250.

Beaurepaire, E., G. M. Turner, S. M. Harrel, M. C. Beard, J. Y. Bigot, and C. A. Schmuttenmaer, 2004, "Coherent terahertz emission from ferromagnetic films excited by femtosecond laser pulses", Appl. Phys. Lett. **84**, 3465.

Behnia, K., 2015, *Fundamentals of Thermoelectricity* (Oxford University Press, Oxford).

Belavin, A. A., and A. M. Polyakov, 1975, "Metastable states of two-dimensional isotropic ferromagnets", JETP Lett. **22**, 245. (English translation of Pis'ma Zh. Eksp. Teor. Fiz. **22**, 503.)

Bender, S. A., R. A. Duine, and Y. Tserkovnyak, 2012, " Electronic pumping of quasiequilibrium Bose-Einstein-condensed magnons", Phys. Rev. Lett. **108**, 246601.

Bender, S. A., R. A. Duine, A. Brataas, and Y. Tserkovnyak, 2014, "Dynamic phase diagram of dc-pumped magnon condensates", Phys. Rev. B. **90**, 094409.

Beneud, F. and P. Monod, 1978, “The Elliott relation in pure metals”, Phys. Rev. B. **18**, 2422.

Berger, K., 1970, “Side-Jump Mechanism for the Hall Effect of Ferromagnets”, Phys. Rev B **2**, 4559.

Berger, L., 1984, “Exchange interaction between ferromagnetic domain wall and electric current in very thin metallic films”, J. Appl. Phys., **55**. 1954.

Berger, L., 1996, “Emission of spin waves by a magnetic multilayer traversed by a current”, Phys. Rev. B **54**, 9353.

Berger, L., 2001, "Effect of interfaces on Gilbert damping and ferromagnetic resonance linewidth in magnetic multilayers", J. Appl. Phys. **90,** 4632.

Berkowitz, A. E., and K. Takano, 1999, “Exchange anisotropy – a review”, J. Magn. Magn. Mater. **200**, 552.

Bert, J. A., B. Kalisky, C. Bell, M. Kim, Y. Hikita, H. Y. Hwang, and K. A. Moler, 2011, “Direct imaging of the coexistence of ferromagnetism and superconductivity at the $LaAlO_3/SrTiO_3$ interface”, Nat. Phys. **7**, 767.

Bertotti, G., I. D. Mayergoyz, and C. Serpico, 2009, *Nonlinear Magnetization Dynamics in Nanosystems* (Elsevier Series in Electromagnetism, Elsevier).

Bhattacharya, A., and S. J. May, 2014, “Magnetic oxide heterostructures”, Annu. Rev. Mater. Res. **44**, 65.

Bi, C.; Y. Liu, T. Newhouse-Illige, M. Xu, M. Rosales, J. W. Freeland, O. Mryasov, S. Zheng, S. G. E. te Velthuis, and W. G. Wang, 2014, “Reversible control of Co magnetism by voltage-induced oxidation”, Phys. Rev. Lett. **113**, 267202.

Bigot, J.-Y., M. Vomir, and E. Beaurepaire, 2009, “Coherent ultrafast magnetism induced by femtosecond laser pulses”, Nat. Phys. **5**, 515.

Bihlmayer, G., O. Rader, and R. Winkler, 2015, “Focus on the Rashba effect,” New. J. Phys. 17, 050202.

Binasch, G., P. Grünberg, F. Saurenbach, and W. Zinn, 1989, “Enhanced magnetoresistance in layered magnetic structures with antiferromagnetic interlayer exchange”, Phys. Rev. B **39**, 4828(R).

Binder K and P. C. Hohenberg, 1974, "Surface effects on magnetic phase transitions" Phys. Rev. B **9**, 2194.

Binder K, 1983, *Phase Transitions and Critical Phenomena*, edited by C. Domb and J. L. Lebowitz Vol. 8 (Academic, New York).

Binder K. and D. P. Landau, 1990, "Critical phenomena at surfaces" Physica A **163**,17.

Binder K. and E. Luijten, 2001, “Monte Carlo tests of renormalization-group predictions for critical phenomena in Ising models”, Phys. Rep. **344**,179.

Biskup, N., J. Salafranca, V. Mehta, M. P. Oxley, Y. Suzuki, S. J. Pennycook, S. T. Pantelides, and M. Varela, 2014, “Insulating ferromagnetic $LaCoO_{3-\delta}$ films: A phase induced by ordering of oxygen vacancies”, Phys. Rev. Lett. **112**, 087202.

Blatt, F. J., D. J. Flood, V. Rowe, P. A. Schroeder, and J. E. Cox, 1967, “Magnon-Drag Thermopower in Iron”, Phys. Rev. Lett. **18**, 395.

Bode, M., M. Heide, K. von Bergmann, P. Ferriani, S. Heinze, G. Bihlmayer, A. Kubetzka, O. Pietzsch, S. Blügel, and R. Wiesendanger, 2007, "Chiral magnetic order at surfaces driven by inversion asymmetry”, Nature **447**, 190.

Boeglin, C, E. Beaurepaire, V. Halte, V. Lopez-Flores, C. Stamm, N. Pontius, H. A. Dürr, and J. Y. Bigot, 2010, “Distinguishing the ultrafast dynamics of spin and orbital moments in solids”, Nature **465**, 458.

Bogdanov, A. N., and D. A. Yablonskii, 1989, "Thermodynamically stable ‘vortices’ in magnetically ordered crystals. The mixed state of magnets”, JETP **68**, 101. (English translation of Zh. Eksp. Teor. Fiz. **95**, 178.)

Bogdanov, A. and A. Hubert, 1994, “Thermodynamically stable magnetic vortex states in magnetic crystals”, J. Magn. Magn. Mater. **138,** 255.

Bogdanov, A. N., and U. K. Rößler, 2001, "Chiral Symmetry Breaking in Magnetic Thin Films and Multilayers”, Phys. Rev. Lett. **87**, 037203.

Bonetti, S., P. Muduli, F. Mancoff, and J. Åkerman, 2009, "Spin torque oscillator frequency versus magnetic field angle: The prospect of operation beyond 65 GHz”, Appl. Phys. Lett. **94**, 102507.

Bonetti, S., V. Tiberkevich, G. Consolo, G. Finocchio, P. Muduli, F. Mancoff, A. Slavin, and J. Akerman, 2010, “Experimental Evidence of Self-Localized and Propagating Spin Wave Modes in Obliquely Magnetized Current-Driven Nanocontacts”, Phys. Rev. Lett. **105,** 217204.

Boona, S. R. and J. P. Heremans, 2014, “Magnon thermal mean free path in yttrium iron garnet”, Phys. Rev. B **90**, 064421.

Boona, S. R, R. C. Myers, and J. P. Heremans, 2014, “Spin caloritronics”, Energy Environ. Sci. **7**, 885.

Borisevich A. Y., *et al.*, 2010. “Suppression of Octahedral Tilts and Associated Changes in Electronic Properties at Epitaxial Oxide Heterostructure Interfaces”, Phys. Rev. Lett. **105**, 087204.

Bornemann, S., O. Šipr, S. Mankovsky, S. Polesya, J. B. Staunton, W. Wurth, H. Ebert, and J. Minár, 2012 “Trends in the magnetic properties of Fe, Co, and Ni clusters and monolayers on Ir(111), Pt(111), and Au(111)”, Phys. Rev. B **86**, 104436.

Boross, P., B. Do´ra, A. Kiss, and F. Simon, 2013, "A unified theory of spin-relaxation due to spin-orbit coupling in metals and semiconductors", Sci. Rep. 3, 3233 DOI: 10.1038/srep03233.

Boulle, O., G. Malinowski, and M. Kläui, 2011, "Current-induced domain wall motion in nanoscale ferromagnetic elements”, Mater. Sci. Eng.: R: Rep. **72**, 159.

Boulle, O., S. Rohart, L. D. Buda-Prejbeanu, E. Jué, I. M. Miron, S. Pizzini, J. Vogel, G. Gaudin, and A. Thiaville, 2013, "Domain Wall Tilting in the Presence of the Dzyaloshinskii-Moriya Interaction in Out-of-Plane Magnetized Magnetic Nanotracks", Phys. Rev. Lett. **111**, 217203.

Boulle, O., *et al.*, 2016, "Room-temperature chiral magnetic skyrmions in ultrathin magnetic nanostructures", Nat. Nanotech. **11**, 449.

Bousquet, E., and N. Spaldin, 2011, "Induced magnetoelectric repsonse in *Pnma* perovskites", Phys. Rev. Lett. **107**, 197603.

Brataas, A., A. D. Kent, and H. Ohno, 2012, "Current-induced torques in magnetic materials", Nat. Mater. **11**, 372.

Brataas , A., Y. V. Nazarov, and G. E. W. Bauer, 2000, "Finite-Element Theory of Transport in Ferromagnet–Normal Metal Systems," Phys. Rev. Lett. **84**, 2481.

Brataas A., Y. Nazarov, and G. Bauer, 2001, "Spin-transport in multi-terminal normal metal-ferromagnet systems with non-collinear magnetizations," Eur. Phys. J. B **22,** 99.

Brey, L., J. Fernández-Rossier, and C. Tejedor, 2004, "Spin depolarization in the transport of holes across GaMnAs/GaAlAs/p-GaAs", Phys. Rev. B **70**, 235334.

Brinkman, A., M. Huijben, M. van Zalk, U. Huijben, U. Zeitler, J. C. Maan, W. G. van der Wiel, G. Rijnders, D. H. A. Blank, and H. Hilgenkamp, 2007, "Magnetic effects at the interface between non-magnetic oxides", Nat. Mater. **6**, 493.

Brown, W. F. Jr., 1963, *Micromagnetics*, (Wiley, New York).

Bruno, P., and C. Chappert, 1991, "Oscillatory coupling between ferromagnetic layers separated by a nonmagnetic metal spacer", Phys. Rev. Lett. **67**, 1602.

Burkov, A. A., and D. G. Hawthorn, 2010, "Spin and charge transport on the surface of a topological insulator", Phys. Rev. Lett. **105,** 066802.

Butler, W. H., X.-G. Zhang, T. C. Schulthess, and J. M. MacLaren, 2001, "Spin-dependent tunneling conductance of Fe|MgO|Fe sandwiches", Phys. Rev. B **63**, 054416.

Büttner, F., *et al.*, 2015, "Dynamics and inertia of skyrmionic spin structures", Nat. Phys. **11**, 225.

Bychkov, Y. A., and E. I. Rashba, 1984 "Properties of a 2D electron gas with lifted spectral degeneracy", JETP Lett. **39**, 78 (1984). (English translation of Pis'ma Zh. Eksp. Teor. Fiz. **39,** 66.)

Cable, J. W., M. R. Khan, G. P. Felcher, and I. K. Schuller, 1986, "Macromagnetism and micromagnetism in Ni-Mo metallic superlattices", Phys. Rev. B. **34**, 1643.

Camley R. E., and J. Barnaś, 1989, "Theory of giant magnetoresistance effects in magnetic layered structures with antiferromagnetic coupling", Phys. Rev. Lett. **63**, 664.

Carbone, C., E. Vescovo, O. Rader, W. Gudat, and W. Eberhardt, 1993, "Exchange split quantum well states of a noble metal film on a magnetic substrate", Phys. Rev. Lett. **71**, 2805.

Carcia, P. F., 1988, "Perpendicular magnetic anisotropy in Pd/Co and Pt/Co thin-film layered structures", J. Appl. Phys. **63**, 5066.

Carcia, P. F., A. D. Meinhaldt, and A. Suna, 1985, “Perpendicular magnetic anisotropy in Pd/Co thin film layered structures”, Appl. Phys. Lett. **47**, 178.

Carpene, E., E. Mancini, C. Dallera, M. Brenna, E. Puppin, and S. De Silvestri, 2008, “Dynamics of electron-magnon interaction and ultrafast demagnetization in thin iron films”, Phys. Rev. B **78**, 174422.

Carra, P., B. T. Thole, M. Altarelli, and X. Wang, 1993, "X-ray circular dichroism and local magnetic fields”, Phys. Rev. Lett. **70**, 694.

Carva, K., M. Battiato and P. M. Oppeneer, 2011, “*Ab Initio* Investigation of the Elliott-Yafet Electron-Phonon Mechanism in Laser-Induced Ultrafast Demagnetization”, Phys. Rev. Lett. **107**, 207201.

Chakhalian, J., J. W. Freeland, H.-U. Habermeier, G. Cristiani, G. Khaliullin, M. van Veenendaal, and B. Keimer, 2007, “Orbital reconstruction and covalent bonding at an oxide interface”, Science **318**, 1114.

Chakhalian, J., *et al.*, 2006, “Magnetism at the interface between ferromagnetic and superconducting oxides”, Nat. Phys. **2**, 244.

Chakhalian, J., A. J. Millis, and J. Rondinelli, 2012, “Whither the oxide interface”, Nat. Mater. **11**, 92.

Chang, C.-Z., J. Zhang, X. Feng, *et al.*, 2013a, “Experimental Observation of the Quantum Anomalous Hall Effect in a Magnetic Topological Insulator”, Science **340,** 167.

Chang, C.-Z., J. Zhang, M. Liu, *et al.*, 2013b, “Thin Films of Magnetically Doped Topological Insulator with Carrier-Independent Long-Range Ferromagnetic Order”, Adv. Mater. **25,** 1065.

Chang, C.-Z., W. Zhao, D. Y. Kim, P. Wei, J. K. Jain, C. Liu, M. H. W. Chan, and J. S. Moodera, 2015a, “Zero-Field Dissipationless Chiral Edge Transport and the Nature of Dissipation in the Quantum Anomalous Hall State”, Phys. Rev. Lett. **115**, 057206.

Chang, C.-Z., W. Zhao, D. Y. Kim, H. Zhang, B. A. Asaaf, D. Heiman, S.-C. Zhang, C.-X. Liu, M. H. W. Chan, and J. S. Moodera, 2015b, “High-precision realization of robust quantum anomalous Hall state in a hard ferromagnetic topological insulator”, Nat. Mater. **14**, 473.

Chanthbouala, A., R. Matsumoto, J. Grollier, V. Cros, A. Anane, A. Fert, A. V. Khvalkovskiy, K. A. Zvezdin, K. Nishimura, Y. Nagamine, H. Maehara, K. Tsunekawa, A. Fukushima, and S. Yuasa, 2011, "Vertical-current-induced domain-wall motion in MgO-based magnetic tunnel junctions with low current densities”, Nat. Phys. **7**, 626.

Chapman, J. N., and M. R. Scheinfein, 1999, “Transmission electron microscopies of magnetic nanostructures”, J. Magn. Magn. Mater. **200**, 729.

Chappert, C., 2008, “Metal spintronics: Electronics free of charge”, Nat. Phys. **4**, 837.

Chappert, C., K. L. Dang, P. Beauvillain, H. Hurdequint, and D. Renard, 1986, “Ferromagnetic resonance studies of very thin cobalt films on a gold substrate”, Phys. Rev. B **34**, 3192.

Charilaou, M. and F. Hellman, 2013, “Mean-field simulation of metal oxide antiferromagnetic films and multilayers”, Phys. Rev. B 87, 184433.

Charilaou, M., and F. Hellman, 2014, “Anomalous magnetic thermodynamics in uncompensated collinear antiferromagnets”, Europhys. Lett. **107**, 27002.

Charilaou, M., C. Bordel, and F. Hellman, 2014, “Magnetization switching and inverted hysteresis in perpendicular antiferromagnetic superlattices”, Appl. Phys. Lett. **104**, 212405.

Charilaou, M., C. Bordel, P.-E Berche, B. B. Maranville, P. Fischer, F. Hellman, 2016, “Magnetic properties of ultrathin discontinuous Co/Pt multilayers: Comparison with short-range ordered and isotropic CoPt3 films”, Phys. Rev. B **93**, 224408.

Charilaou, M., and F. Hellman, 2015, “Roughness effects in uncompensated antiferromagnets”, J. Appl. Phys. **117**, 083907.

Charilaou M. and F. Hellman, 2015b, “Surface-induced phenomena in uncompensated collinear antiferromagnets”, J. Phys.: Condens. Matter **27**, 086001.

Chazalviel, J.-N., 1975, “Spin-dependent Hall effect in semiconductors”, Phys. Rev. B **11**, 3918.

Chazalviel, J.-N., and I. Solomon, 1972, “Experimental evidence of the anomalous Hall effect in a nonmagnetic semiconductor”, Phys. Rev. Lett. **29**, 1676.

Checkelsky, J. G., J. Ye, Y. Onose, Y. Iwasa, and Y. Tokura, 2012, “Dirac-fermion-mediated ferromagnetism in a topological insulator”, Nat. Phys. **8,** 729.

Checkelsky, J. G., R. Yoshimi, A. Tsukazaki, K. S. Takahashi, Y. Kozuka, J. Falson, M. Kawasaki, and Y. Tokura, 2014, “Trajectory of the anomalous Hall effect towards the quantized state in a ferromagnetic topological insulator”, Nat. Phys. **10,** 731.

Chen, A., J. Hutchby, V. Zhirnov, and G. Bourianoff, 2015a, *Emerging nanoelectronic devices*, (Wiley, New York).

Chen, G., T. Ma, A. T. N’Diaye, H. Kwon, C. Won, Y. Wu, and A. K. Schmid, 2013a, "Tailoring the chirality of magnetic domain walls by interface engineering”, Nat. Commun. **4**, 2671.

Chen, G., A. Mascaraque, A. T. N’Diaye, and A. K. Schmid, 2015b, “Room temperature skyrmion ground state stabilized through interlayer exchange coupling”, Appl. Phys. Lett. **106**, 242404.

Chen, G., J. Zhu, *et al.*, 2013b, "Novel Chiral Magnetic Domain Wall Structure in Fe/Ni/Cu(001) Films”, Phys. Rev. Lett. **110**, 177204.

Chen, H., A. D. Kent, A. H. MacDonald, and I. Sodemann, 2014a, "Nonlocal transport mediated by spin supercurrents”, Phys. Rev. B **90**, 220401.

Chen, H., Q. Niu, and A. H. MacDonald, 2014b, "Anomalous Hall Effect Arising from Noncollinear Antiferromagnetism", Phys. Rev. Lett. **112**, 017205.

Chen, Y. L., *et al.*, 2010, "Massive Dirac Fermion on the Surface of a Magnetically Doped Topological Insulator", Science **329**, 659.

Cheng, R, J. Xiao, Q. Niu, and A. Brataas, 2014, "Spin Pumping and Spin-Transfer Torques in Antiferromagnets", Phys. Rev. Lett. **113**, 057601.

Cheng, X., J. A. Katine, G. E. Rowlands, and I. N. Krivorotov, 2013, "Nonlinear ferromagnetic resonance induced by spin torque in nanoscale magnetic tunnel junctions", Appl. Phys. Lett. **103**, 082402.

Cheng, X., C. T. Boone, J. Zhu, and I. N. Krivorotov, 2010, "Nonadiabatic Stochastic Resonance of a Nanomagnet Excited by Spin Torque", Phys. Rev. Lett. **105**, 047202.

Cheong, S.-W., and M. Mostovoy, 2007, "Multiferroics: A magnetic twist for ferroelectricity", Nat. Mater. **6**, 13.

Cherifi, R. O., *et al.*, 2015, "Electric-field control of magnetic order above room temperature", Nat. Mater. **13**, 345.

Chernyshov, A., M. Overby, X. Liu, J. K. Furdyna, Y. Lyanda-Geller, and L. P. Rokhinson, 2009, "Evidence for reversible control of magnetization in a ferromagnetic material by means of spin-orbit magnetic field", Nat. Phys. **5**, 656.

Chmaiseem, O., B. Dabrowski, S. Kolesnik, J. Mais, D. E. Brown, R. Kruk, P. Prior, B. Pyles, and J. D. Jorgensen, 2001, "Relationship between structural parameters and the Néel temperature in $Sr_{1-x}Ca_xMnO_3$ ($0 \leq x \leq 1$) and $Sr_{1-y}Ba_yMnO_3$ ($y \leq 0.2$)", Phys. Rev. B **64**, 134412.

Cho, J., *et al.*, 2015, "Thickness dependence of the interfacial Dzyaloshinskii-Moriya interaction in inversion symmetry broken systems", Nat. Commun. **6**, 7635.

Choe, S.-B., Y. Acremann, A. Scholl, A. Bauer, A. Doran, J. Stöhr, and H. A. Padmore, 2004, "Vortex Core-Driven Magnetization Dynamics", Science **304**, 420.

Choi, G.-M., C.-H. Moon, B.-C. Min, K.-J. Lee, and D. G. Cahill, 2015, "Thermal spin-transfer torque driven by the spin-dependent Seebeck effect in metallic spin-valves", Nat. Phys. **11**, 576.

Choi, G. M., B. C. Min, K. J. Lee, and D. G. Cahill, 2014, "Spin current generated by thermally driven ultrafast demagnetization", Nat. Commun. **5**, 4334.

Choi, J., J. Wu, C. Won, Y. Z. Wu, A. Scholl, A. Doran, T. Owens, and Z. Q. Qiu, 2007, "Magnetic Bubble Domain Phase at the Spin Reorientation Transition of Ultrathin Fe/Ni/Cu (001) Film", Phys. Rev. Lett. **98**, 207205.

Chung, S. H., D. T. Pierce, and J. Unguris, 2010, "Simultaneous measurement of magnetic vortex polarity

and chirality using scanning electron microscopy with polarization analysis (SEMPA)”, Ultramicroscopy **110**, 177.

Ciccarelli, C., L. P. Zârbo, A. C. Irvine, R. P. Campion, B. L. Gallagher, J. Wunderlich, T. Jungwirth, and A. J. Ferguson, 2012, “Spin gating electrical current”, Appl. Phys. Lett. **101**, 122411.

Ciccarelli, C., L. Anderson, V. Tshitoyan, A. J. Ferguson, F. Gerhard, C. Gould, L.W. Molenkamp, J. Gayles, J. Železný, L. Šmejkal, Z. Yuan, J. Sinova, F. Freimuth and T. Jungwirth, 2016, "Room-temperature spin–orbit torque in NiMnSb”, Nat. Phys. **12**, 855.

Coey, J. M. D., and S. A. Chambers, 2008, “Oxide dilute magnetic semiconductors – fact or fiction?”, MRS Bull. **33**, 1053.

Coey, J. M. D., M. Viret, and S. von Molnar, 1999, “Mixed-valence manganites”, Adv. Phys. **48**, 167.

Coey, J. M. D., 2010, “Magnetism and Magnetic Materials”, Cambridge University Press.

Coey, J.M.D., M. Venkatesan and P. Stamenov, 2016a, J. Phys.: Cond. Mat. **28** 485001.

Coey, J.M.D., K. Ackland, M. Venkatesan and S. Sen, 2016b, Nat. Phys. advance online publication.

Collet, M., *et al.*, 2016, "Generation of coherent spin-wave modes in Yttrium Iron Garnet microdiscs by spin-orbit torque”, Nat, Commun. **7**, 10377 (2016).

Costa, A. T., R. B. Muniz, S. Lounis, A. B. Klautau, and D. L. Mills, 2010, “Spin-orbit coupling and spin waves in ultrathin ferromagnets: The spin-wave Rashba effect”, Phys. Rev. B **82**, 014428.

Costache, M. V., G. Bridoux, I. Neumann, and S. O. Valenzuela, 2012, “Magnon-drag thermopile”, Nat. Mater. **11**, 199.

Crangle, J. and W. R. Scott, 1965, “Dilute Ferromagnetic Alloys”, J. Appl. Phys. **36**, 921.

Crépieux, A., and C. Lacroix, 1998, “Dzyaloshinsky–Moriya interactions induced by symmetry breaking at a surface”, J. Magn. Magn. Mater. **182**, 341.

Culcer, D., E. Hwang, T. Stanescu, and S. Das Sarma, 2010, “Two-dimensional surface charge transport in topological insulators”, Phys. Rev. B **82**, 155457.

Da Col, S., *et al.*, 2014, "Observation of Bloch-point domain walls in cylindrical magnetic nanowires”, Phys. Rev. B **89**, 180405.

Daalderop, G. H. O., P. J. Kelly, and M. F. H. Schuurmans, 1990, “First-principles calculation of the magnetocrystalline anisotropy energy of iron, cobalt, and nickel”, Phys. Rev. B **41**, 11919.

Daalderop, G. H. O., P. J. Kelly, and F. J. A. den Broeder, 1992, “Prediction and confirmation of perpendicular magnetic anisotropy in Co/Ni multilayers”, Phys. Rev. Lett. **68**, 682.

Daalderop, G. H. O., P. J. Kelly, and M. F. H. Schuurmans, 1994, “Magnetic anisotropy of a free-standing Co monolayer and of multilayers which contain Co monolayers”, Phys. Rev. B **50**, 9989.

Dagotto, E., 2005, “Complexity in Strongly Correlated Electronic Systems”, Science **309**, 257.

Dahlberg, E.D., and R. Proksch, R., 1999, “Magnetic microscopies: The new additions”, J. Magn. Magn.

Mater. **200**, 720.
Dalla Longa, F., J. T. Kohlhepp, W. J. M. de Jonge, and B. Koopmans, 2008, “Laser-induced magnetization dynamics in Co⁄IrMn exchange coupled bilayers”, J. Appl. Phys. **103**, 07B101.
Dalla Longa, F., J. T. Kohlhepp, W. J. M. de Jonge, and B. Koopmans, 2010, “Resolving the genuine laser-induced ultrafast dynamics of exchange interaction in ferromagnet/antiferromagnet bilayers”, Phys. Rev. B **81**, 094435.
Dang, H. T., and A. J. Millis, 2013, “Designing ferromagnetism in vanadium oxide based superlattices”, Phys. Rev. B **87**, 184434.
Dankert, A., J. Geurs, M. V. Kamalakar, and S. P. Dash, 2015, “Room Temperature Electrical Detection of Spin Polarized Currents in Topological Insulators”, Nano. Lett. 15, 7976.
Daughton, J. M., 1992, “Magnetoresistive memory technology”, Thin Solid Films **216**, 162.
Daughton, J. M., 1999, “GMR applications”, J. Magn. Magn. Mater. **192**, 334.
de Gennes, P. G., and J. Prost, 1995, *The physics of liquid crystals* (Oxford, New York).
de Groot, R. A., F. M. Mueller, P. G. v. Engen, and K. H. J. Buschow, 1983, "New Class of Materials: Half-Metallic Ferromagnets." Phys. Rev. Lett. **50**, 2024.
de la Venta, J., S. Wang, T. Saerbeck, J. G. Ramirez, I. Valmianski, and I. K. Schuller, 2014, “Coercivity enhancement in $V_2O_3$/Ni bilayers driven by nanoscale phase coexistence”, Appl. Phys. Lett. **104**, 062410.
Demasius, K.-U., T. Phung, W. Zhang, B. P. Hughes, S.-H. Yang, A. Kellock, W. Han, A. Pushp, and S. S. P. Parkin, 2016, “Enhanced spin–orbit torques by oxygen incorporation in tungsten films”, Nature Communications **7**, 10644.
Demidov, V. E., H. Ulrichs, S. V. Gurevich, S. O. Demokritov, V. Tiberkevich, A. N. Slavin, A. Zholud, and S. Urazhdin, 2014, “Synchronization of spin Hall nano-oscillators to external microwave signals”, Nat. Commun. **5**, 3179.
Demidov, V. E., S. Urazhdin, E. R. J. Edwards, M. D. Stiles, R. D. McMichael, and S. O. Demokritov, 2011, "Control of Magnetic Fluctuations by Spin Current”, Phys. Rev. Lett. **107**, 107204.
Demidov, V. E., S. Urazhdin, H. Ulrichs, V. Tiberkevich, A. Slavin, D. Baither, G. Schmitz, and S. O. Demokritov, 2012, “Magnetic nano-oscillator driven by pure spin current”, Nat. Mater. **11**, 1028.
Demokritov, S., E. Tsymbal, P. Gr¨unberg, W. Zinn, and I. K. Schuller, 1994, "Magnetic-dipole mechanism for biquadratic interlayer coupling”, Phys. Rev B **49**, 720.
Demokritov, S. O., 1998, "Biquadratic interlayer coupling in layered magnetic systems”, J. Phys. D **31**, 925.
den Broeder, F. J. A., W. Hoving, and P. J. H. Bloemen, 1991, “Magnetic anisotropy of multilayers”, J.

Magn. Magn. Mater. **93**, 562.

Deorani, P., J. Son, K. Banerjee, N. Koirala, M. Brahlek, S. Oh, and H. Yang, 2014, “Observation of inverse spin Hall effect in bismuth selenide”, Phys. Rev. B **90,** 094403.

Dhoot, A. S., C. Israel, X. Moya, N. D. Mathur, and R. H. Friend, 2009, “Large electric field effect in electrolyte-gated manganites”, Phys. Rev. Lett. **102**, 136402.

Diaz, S.A. and R. E. Troncoso, 2016, "Controlling skyrmion helicity via engineered Dzyaloshinskii-Moriya interactions,", J. Phys.: Condens. Matter 28, 426005.

Di, K., V. L. Zhang, H. S. Lim, S. C. Ng, M. H. Kuok, X. Qiu, and H. Yang, 2015, “Asymmetric spin-wave dispersion due to Dzyaloshinskii-Moriya interaction in an ultrathinh Pt/CoFeB film”, Appl. Phys. Lett. **106**, 052403.

Diehl H. W., 1986, *Phase Transitions and Critical Phenomena* edited by C. Domb and J. L. Lebowitz Vol. 10 (Academic, New York) p. 75.

Dieny, B., 1994, “Giant magnetoresistance in spin-valve multilayers”, J. Magn. Magn. Mater. **136**, 335.

Dieny, B., *et al.*, 2010, “Spin-transfer effect and its use in spintronics components”, Int. J. Nanotechn. **7**, 591.

Dieny, B., V. S. Speriosu, S. S. P. Parkin, B. A. Gurney, D. R. Wilhoit, and D. Maur, 1991, “Giant magnetoresistive in soft ferromagnetic multilayers”, Phys. Rev. B **43**, 1297.

Dietl, T., 2010, “A ten-year perspective on dilute magnetic semiconductors and oxides”, Nat. Mater. **9**, 965.

Djordjevic M., G. Eilers, A. Parge, M. Münzenberg, and J. S. Moodera, 2006, “Intrinsic and nonlocal Gilbert damping parameter in all optical pump-probe experiments”, J. Appl. Phys. **99**, 08F308.

Donahue, M., and Porter, D., 1999, “OOMMF User’s Guide, Version 1.0”, Interagency Report NISTIR 6376, (NIST, Gaithersburg, MD); http://math.nist.gov/oommf/.

Dong, S., K. Yamauchi, S. Yunoki, R. Yu, S. Liang, A. Moreo, J.-M. Liu, S. Picozzi, and E. Dagotto, 2009, “Exchange bias driven by the Dzyaloshinskii-Moriya interaction and ferroelectric polarization at *G*-type antiferromagnetic perovskite interfaces”, Phys. Rev. Lett. **103**, 127201.

Dosch H., 1992, *Critical Phenomena at Surfaces and Interfaces: Evanescent x-ray and Neutron Scattering* Springer Tracts in Modern Physics Vol. 126, (Springer, Berlin).

Dresselhaus, G., 1955, “Spin-orbit coupling effects in zinc blende structures*”, Phys. Rev.* **100,** 580.

Duan, C.-G., J. P. Velev, R. F. Sabirianov, Z. Zhu, J. Chu, S. S. Jaswal, and E. Y. Tsymbal, 2008, “Surface magnetoelectric effect in ferromagnetic metal films”, Phys. Rev. Lett. **101**, 137201.

Duan, Z., A. Smith, L. Yang, B. Youngblood, J. Lindner, V. E. Demidov, S. O. Demokritov, and I. N. Krivorotov, 2014, "Nanowire spin torque oscillator driven by spin orbit torques”, Nat. Commun. **5**, 5616.

Dupé, B., M. Hoffmann, C. Paillard, and S. Heinze, 2014, “Tailoring magnetic skyrmions in ultra-thin transition metal films”, Nat. Commun. **5**, 4030.

Dussaux, A., *et al.*, 2010, "Large microwave generation from current-driven magnetic vortex oscillators in magnetic tunnel junctions”, Nat. Commun. **1**, 8.

D’yakonov, M. I., and V. I. Perel’, 1971, “Possibility of orienting electron spins with current”, JETP Lett. **13**, 467. (English translation of Pis’ma Zh. Eksp. Teor. Fiz. **13**, 657.)

Dzyaloshinskii, I. A., 1957, “A thermodynamic theory of “weak” ferromagnetism of antiferromagetics”, JETP **5**, 1259 [J. Phys. Chem. Sol. **4**, 241 (1958)].

Dzyaloshinskii, I. E., 1965, "Theory of helicoidal structures in antiferromagnets III”, JETP **20**, 655.

Dzyaloshinskii, I.E., 1964, “Theory of Helicoidal Structures in Antiferromagnets. I. Nonmetals,” JETP **19**, 960. (English translation of Zh. Eksp. Teor. Fiz. **46**, 1420).

Edelstein, V. M., 1990, “Spin polarization of conduction electrons induced by electric current in two-dimensional asymmetric electron system”, Solid State Commun. **73**, 233.

Edwards, D. M., J. Mathon, R. B. Muniz, and M. S. Phan, 1991, “Oscillations of the exchange in magnetic multilayers as an analog of de Haas–van Alphen effect”, Phys. Rev. Lett. **67**, 493.

Edwards, E. R. J., H. Ulrichs, V. E. Demidov, S. O. Demokritov, and S. Urazhdin, 2012, "Parametric excitation of magnetization oscillations controlled by pure spin current”, Phys. Rev. B **86**, 134420.

Eisebitt, S., J. Luning, W. F. Schlotter, M. Lorgen, O. Hellwig, W. Eberhardt, and J. Stohr, 2004, "Lensless imaging of magnetic nanostructures by X-ray spectro-holography”, Nature **432**, 885.

Elliott, R. J., 1954, “Theory of the effect of spin-orbit coupling on magnetic resonance in some semiconductors”, Phys. Rev. **96**, 266.

Emori, S., U. Bauer, S.-M. Ahn, E. Martinez, and G. S. D. Beach, 2013, “Current-driven dynamics of chiral ferromagnetic domain walls”, Nat. Mater. **12**, 611.

Emori, S., E. Martinez, K.-J. Lee, H.-W. Lee, U. Bauer, S.-M. Ahn, P. Agrawal, D. C. Bono, and G. S. D. Beach, 2014, "Spin Hall torque magnetometry of Dzyaloshinskii domain walls”, Phys. Rev. B **90**, 184427.

Engebretson, D. M., W. A. A, Macedo, I. K. Schuller, P. A. Crowell, and C. Leighton, 2005, “Time domain dynamics of the asymmetric magnetization reversal in exchange biased bilayers”, Phys. Rev. B **71** 184412.

Engel, B. N., C. D. England, R. A. Van Leeuwen, M. H. Wiedmann, and C. M. Falco, 1991, “Interface magnetic anisotropy in epitaxial superlattices”, Phys. Rev. Lett. **67**, 1910.

Eschenlohr, A., M. Battiato, P. Maldonado, N. Pontius, T. Kachel, K. Holldack, R. Mitzner, A. Föhlisch,

P. M. Oppeneer and C. Stamm, 2013, “Ultrafast spin transport as key to femtosecond demagnetization”, Nat. Mater. **12**, 332.

Eschenlohr, A., M. Battiato, P. Maldonado, N. Pontius, T. Kachel, K. Holldack, R. Mitzner, A. Föhlisch, P. M. Oppeneer, and C. Stamm, 2014, “Reply to 'Optical excitation of thin magnetic layers in multilayer structures,'” Nat. Mater. **13**, 102.

Escorcia-Aparicio, E. J., J. H. Wolfe, H. J. Choi, W. L. Ling, R. K. Kawakami, and Z. Q. Qui, 1999, “Modification of the magnetic properties of Fe/Cr(001) by controlling the compensation of a vicinal Cr(001) surface”, J. Appl. Phys. **85**, 4961.

Essenberger F., S. Sharma, J. K. Dewhurst, C. Bersier, F. Cricchio, L. Nordström and E. K. U. Gross, 2011, "Magnon spectrum of transition-metal oxides: calculations including long-range magnetic interactions using the LSDA + U method", Phys. Rev. B **84,** 174425.

Evans, R. F. L., T. A. Ostler, R. W. Chantrell, I. Radu, and T. Rasing, 2014, “Ultrafast thermally induced magnetic switching in synthetic ferrimagnets”, Appl. Phys. Lett. **104**, 082410.

Everschor, K., 2012, "Rotating skyrmion lattices by spin torques and field or temperature gradients”, Phys. Rev. B **86**, 054432.

Ezawa, M., 2010, "Giant Skyrmions Stabilized by Dipole-Dipole Interactions in Thin Ferromagnetic Films”, Phys. Rev. Lett. **105**, 197202.

Fan, X., J. Wu, Y. Chen, M. J. Jerry, H. Zhang, and J. Q. Xiao, 2013, “Observation of the nonlocal spin-orbital effective field”, Nat. Commun. **4**, 1799.

Fan Y., X. Ma, F. Fang, J. Zhu, Q. Li, T. P. Ma, Y. Z. Wu, Z. H. Chen, H. B. Zhao, and G. Lüpke, 2014a, “Photoinduced spin angular momentum transfer into an antiferromagnetic insulator”, Phys. Rev. B **89**, 094428.

Fan, Y., P. Upadhyaya, *et al.*, 2014b, “Magnetization switching through giant spin–orbit torque in a magnetically doped topological insulator heterostructure”, Nat. Mater. **13,** 699.

Fan, Y., X. Kou, *et al.*, 2016, “Electric-field control of spin-orbit torque in a magnetically doped topological insulator”, Nat. Nanotech. 11, 352.

Farle, M., K. Baberschke, U. Stetter, A. Aspelmeier, and F. Gerhardter, 1993, “Thickness-dependent Curie temperature of Gd(0001)/W(110) and its dependence on the growth conditions”, Phys. Rev. B **47**, 11571(R).

Ferri, F. A., M. A. Pereira-da-Silva, and E. Marega, Jr., 2012, *Magnetic Force Microscopy: Basic Principles and Applications* DOI: 10.5772/34833. INTECH open access book, see: https://www.intechopen.com/books/atomic-force-microscopy-imaging-measuring-and-manipulating-surfaces-at-the-atomic-scale/magnetic-force-microscopy-basic-principles-and-applications.

Ferriani, P., K. von Bergmann, E. Y. Vedmedenko, S. Heinze, M. Bode, M. Heide, G. Bihlmayer, S.

Blügel, and R. Wiesendanger, 2008, "Atomic-Scale Spin Spiral with a Unique Rotational Sense: Mn Monolayer on W(001)", Phys. Rev. Lett. **101**, 027201.

Fert, A., 1990, "Magnetic and Transport Properties of Metallic Multilayers", Mater. Sci. Forum **59**, 439.

Fert, A., V. Cros, and J. Sampaio, 2013, "Skyrmions on the track", Nat. Nano. **8**, 152.

Fert, A., and P. M. Levy, 1980, "Role of Anisotropic Exchange Interactions in Determining the Properties of Spin-Glasses", Phys. Rev. Lett. **44**, 1538.

Fidler, J., and T. Schrefl, 2000, "Micromagnetic modelling - the current state of the art", J. Phys. D: Appl. Phys. **33**, R135.

Fiederling, R., M. Keim, G. Reuscher, W. Ossau, G. Schmidt, A. Waag, and L. W. Molenkamp, 1999, "Injection and detection of a spin-polarized current in a light-emitting diode", Nature **402**, 787.

Fina, I., *et al.*, 2014, "Anisotropic magnetoresistance in an antiferromagnetic semiconductor", Nat. Commun. **5**, 4671.

Fischer, P., G. Schütz, G. Schmahl, P. Guttmann, and D. Raasch, 1997, "Imaging of magnetic domains with the X-ray microscope at BESSY using X-ray magnetic circular dichroism", Z. Phys. B **101**, 313.

Fischer, P., 2015, "X-ray imaging of magnetic structures", IEEE Trans. Magn. **51**, 1.

Fischbacher, T., M. Franchin, G. Bordignon, and H. Fangohr, "A systematic approach to multiphysics extensions of finite-element-based micromagnetic simulations: Nmag", IEEE Trans. Magn. **43**, 2896.

Fister, T. T., H. Zhou, Z. Luo, S. S. A. Seo, S. O. Hruszkewycz, D. L. Proffit, J. A. Eastman, P. H. Fuoss, P. M. Baldo, H. N. Lee, and D. D. Fong, 2014, "Octahedral rotations in strained $LaAlO_3/SrTiO_3$ (001) heterostructures", APL Mater. **2**, 021102.

Fitting Kourkoutis, L., D. A. Muller, Y. Hotta, and H. Y. Hwang, 2007, "Asymmetric interface profiles in $LaVO_3/SrTiO_3$ heterostructures grown by pulsed laser deposition", Appl. Phys. Lett. **91**, 163101.

Fitzsimmons, M. R., *et al.*, 2004, "Neutron scattering studies of nanomagnetism and artificially structured materials", J. Magn. Magn. Mater. **271**, 103.

Fitzsimmons, M. R., *et al.*, 2011, "Upper limit to magnetism in $LaAlO_3/SrTiO_3$ heterostructures", Phys. Rev. Lett. **107**, 217201.

Flovik,V., F. Macià,, and E. Wahlström, 2016 "Describing synchronization and topological excitations in arrays of magnetic spin torque oscillators through the Kuramoto model", Sci. Rep. **6**, 32528.

Flipse, J., F. L. Bakker, A. Slachter, F. K. Dejene, and B. J. van Wees, 2012, "Direct observation of the spin-dependent Peltier effect", Nat. Nanotechn. **7**, 166 (2012).

Flipse, J., F. K. Dejene, D. Wagenaar, G.E.W. Bauer, J. Ben Youssef, and B. J. van Wees, 2014, "Observation of the Spin Peltier Effect for Magnetic Insulators", Phys. Rev. Lett. **113,** 027601.

Fong, C. Y., J. E. Pask, and L. H. Yang, 2013, *Half-metallic materials and their properties* (Imperial College Press, London).

Frangou, L., S. Oyarzún, S. Auffret, L. Vila, S. Gambarelli and V. Baltz, 2016, " Enhanced Spin Pumping Efficiency in Antiferromagnetic IrMn Thin Films around the Magnetic Phase Transition", Phys. Rev. Lett. **116**, 077203.

Franken, J. H., M. Herps, H. J. M. Swagten, and B. Koopmans, 2014, "Tunable chiral spin texture in magnetic domain-walls", Sci. Rep. **4**, 5248.

Frano, A.; *et al.*, 2013, "Orbital control of noncollinear magnetic order in nickel oxide heterostructures", Phys. Rev. Lett. **111**, 106804.

Freeland, J. W., *et al.*, 2010, "Charge transport and magnetization profile at the interface between the correlated metal $CaRuO_3$ and the antiferromagnetic insulator $CaMnO_3$", Phys. Rev. B **81**, 094414.

Freeman A. J. and R. Wu, 1992, "Magnetism in man made materials", J. Magn. & Magn. Mater. **104**, 1.

Freimuth, F., S. Blügel, and Y. Mokrousov, 2014a, "Berry phase theory of Dzyaloshinskii-Moriya interaction and spin-orbit torques", J. Phys.: Conden. Matter **26**, 104202.

Freimuth, F., S. Bluegel, and Y. Mokrousov, 2014b, "Spin-orbit torques in Co/Pt(111) and Mn/W(001) magnetic bilayers from first principles", Phys. Rev. B **90**, 174423.

Fu, L., and C. L. Kane, 2007, "Topological insulators with inversion symmetry", Phys. Rev. B **76,** 045302.

Fuchs, G. D., N. C. Emley, I. N. Krivorotov, P. M. Braganca, E. M. Ryan, S. I. Kiselev, J. C. Sankey, D. C. Ralph, R. A. Buhrman, and J. A. Katine, 2004, "Spin-transfer effects in nanoscale magnetic tunnel junctions", Appl. Phys. Lett. **85**, 1205.

Fuchs, G. D., *et al.*, 2007, "Spin-torque ferromagnetic resonance measurements of damping in nanomagnets", Appl. Phys. Lett. **91**, 062507.

Fukami, S., C. Zhang, S. DuttaGupta, A. Kurenkov, and H. Ohno, 2016a, "Magnetization switching by spin-orbit torque in an antiferromagnet/ferromagnet bilayer system", Nat. Mater. **15,** 535.

Fukami, S., T. Anekawa, C. Zhang, and H. Ohno, 2016b, "A spin–orbit torque switching scheme with collinear magnetic easy axis and current configuration", Nat. Nanotech. **11**, 621.

Fukuma, Y, L. Wang, H. Idzuchi, S. Takahashi, S. Maekawa and Y. Otani, 2011, "Giant enhancement of spin accumulation and long-distance spin precession in metallic lateral spin valves", Nat. Mater. **10**, 527.

Fulde, P., 1995, *Electron correlations in molecules and solids* (Springer-Verlag, New York).

Gajek, M., J. J. Nowak, J. Z. Sun, P. L. Trouilloud, E. J. O'Sullivan, D. W. Abraham, M. C. Gaidis, G. Hu, S. Brown, Y. Zhu, R. P. Robertazzi, W. J. Gallagher, and D. C. Worledge, 2012, "Spin torque switching of 20 nm magnetic tunnel junctions with perpendicular anisotropy", Appl. Phys. Lett. **100**,

132408.

Galanakis, I., and P. Mavropoulos, 2007, "Spin-polarization and electronic properties of half-metallic Heusler alloys calculated from first principles", J. Phys.: Condens. Matter **19**, 315213.

Gallego, J. M., D. Lederman, S. Kim, and I. K. Schuller, 1995, "Oscillatory Behavior of the Transport Properties in Ni/Co Multilayers: A Superlattice Effect", Phys. Rev. Lett. **74**, 4515.

Gambino, R. J., J. Ziegler, and J. J. Cuomo, 1974, Effects of ion radiation damage on the magnetic domain structure of amorphous Gd–Co alloys, Appl. Phys. Lett. **24**, 99.

Gao, L., X. Jiang, S.-H. Yang, J. Burton, E. Tsymbal, and S. Parkin, 2007, "Bias Voltage Dependence of Tunneling Anisotropic Magnetoresistance in Magnetic Tunnel Junctions with MgO and $Al_2O_3$ Tunnel Barriers", Phys. Rev. Lett. **99,** 226602.

Garcia, M. A., E. Fernandez Pinel, J. de la Venta, A. Quesada, V. Bouzas, J. F. Fernandez, J. J. Romero, M. S. Martin Gonzalez, and J. J. Costa-Kramer, 2009, "Sources of experimental errors in the observation of nanoscale magnetism", J. Appl. Phys. **105**, 013925.

Garello, K., I. M. Miron, C. O. Avci, F. Freimuth, Y. Mokrousov, S. Blügel, S. Auffret, O. Boulle, G. Gaudin, and P. Gambardella, 2013, "Symmetry and magnitude of spin-orbit torques in ferromagnetic heterostructures", Nat. Nanotech. **8**, 587.

Garrison, K., Y. Chang, and P. D. Johnson, 1993, "Spin polarization of quantum well states in copper thin films deposited on a Co(001) substrate", Phys. Rev. Lett. **71**, 2801.

Gay, J. G., and R. Richter, 1986, "Spin Anisotropy of Ferromagnetic Films", Phys. Rev. Lett. **56**, 2728.

Gazquez, J., W. Luo, M. P. Oxley, M. Prange, M. A. Torija, M. Sharma, C. Leighton, S. T. Pantiledes, S. J. Pennycook, and M. Varela, 2011, "Atomic resolution imaging of spin-state superlattices in nanopockets within cobaltite thin films", Nano Lett. **11**, 973.

Gazquez, J., S. Bose, M. Sharma, M. A. Torija, S. J. Pennycook, C. Leighton, and M. Varela, 2013, "Lattice mismatch accommodation *via* oxygen vacancy ordering in epitaxial $La_{0.5}Sr_{0.5}CoO_{3-\delta}$ films", APL Mater. **1**, 012105.

Geprägs, S., S. Meyer, S. Altmannshofer, M. Opel, F. Wilhelm, A. Rogalev, R. Gross, and S. T. B. Goennenwein, 2012, "Investigation of induced Pt magnetic polarization in Pt/$Y_3Fe_5O_{12}$ bilayers", Appl. Phys. Lett. **101**, 262407.

Gilbert, D. A., B. B. Maranville, A. L. Balk, B. J. Kirby, P. Fischer, D. T. Pierze, J. Unguris, J. A. Borchers, and K. Liu, 2015, "Realization of Ground State Artificial Skyrmion Lattices at Room Temperature", Nat. Commun. **6**, 8462.

Gilbert, T. L., 2004, "A phenomenological theory of damping in ferromagnetic materials", IEEE Trans. Mag. **40**, 3443.

Giraud, R., M. Gryglas, L. Thevenard, A. Lemaître, and G. Faini, 2005, "Voltage-controlled tunneling

anisotropic magnetoresistance of a ferromagnetic $p^{++}$-(Ga,Mn)As/$n^{+}$-GaAs Zener-Esaki diode", Appl. Phys. Lett. **87**, 242505.

Gomonay, H. V. and V. M. Loktev, 2010, "Spin transfer and current-induced switching in antiferromagnets", Phys. Rev. B **81**, 144427.

Gonçalves, A. M., I. Barsukov, Y.-J. Chen, L. Yang, J. A. Katine, and I. N. Krivorotov, 2013, "Spin torque ferromagnetic resonance with magnetic field modulation", Appl. Phys. Lett. **103**, 172406.

Gorbenko O. Y., S. V Samoilenkov, I. E. Graboy, and A. R. Kaul, 2002, "Epitaxial stabilization of oxides in thin films", Chem. Mater. **14**, 4026.

Goodenough, J. B., 1955, "Theory of the Role of Covalence in Peroskite-Type Manganites [La,M(II)]$MnO_3$", Phys. Rev. **100**, 564.

Gould, C., C. Rüster, T. Jungwirth, E. Girgis, G. Schott, R. Giraud, K. Brunner, G. Schmidt, and L. Molenkamp, 2004, "Tunneling Anisotropic Magnetoresistance: A Spin-Valve-Like Tunnel Magnetoresistance Using a Single Magnetic Layer", Phys. Rev. Lett. **93,** 11; 117203.

Graves, C. E., *et al.*, 2013, "Nanoscale spin reversal by non-local angular momentum transfer following ultrafast laser excitation in ferrimagnetic GdFeCo", Nat. Mater. **12**, 293.

Grigorenko, A. N., P. I. Nikitin, A. N. Slavin, and P. Y. Zhou, 1994, "Experimental Observation of Magnetostochastic Resonance", J. Appl. Phys. **76**, 6335.

Grimsditch, M., M. R. Khan, A. Kueny and I. K. Schuller, 1983, "Collective Behavior of Magnons in Superlattices", Phys. Rev. Lett. **51**, 498.

Grollier, J., V. Cros, and A. Fert, 2006, "Synchronization of spin-transfer oscillators driven by stimulated microwave currents", Phys. Rev. B **73**, 060409(R).

Grollier, J., V. Cros, A. Hamzic, J. M. George, H. Jaffres, A. Fert, G. Faini, J. Ben Youssef, and H. Legall, 2001, "Spin-polarized current induced switching in Co/Cu/Co pillars", Appl. Phys. Lett. **78**, 3663.

Grollier, J., D. Querlioz, and M. Stiles, 2016, "Spintronic Nanodevices for Bioinspired Computing", IEEE Proc. **104**, 2024. DOI: 10.1109/JPROC.2016.2597152

Grutter, A. J., *et al.*, 2013, "Interfacial Ferromagnetism in $LaNiO_3/CaMnO_3$ Superlattices", Phys. Rev. Lett. **111**, 087202.

Gunnarsson, O., and B. I. Lundqvist, 1976, "Exchange and correlation in atoms, molecules, and solids by the spin-density-functional formalism", Phys. Rev. B **13**, 4274.

Guo, G. Y., S. Murakami, T.-W. Chen, and N. Nagaosa, 2008, "Intrinsic spin Hall effect in platinum: First-principles calculations", Phys. Rev. Lett. **100**, 096401.

Gurevich, A. G., and G. A. Melkov, 1996, *Magnetization Oscillations and Waves*. (CRC, New York).

Haazen, P. P. J., E. Murae, J. H. Franken, R. Lavrijsen, H. J. M. Swagten, and B. Koopmans, 2013, "Domain wall depinning governed by the spin Hall effect", Nat. Mater. **12**, 299.

Hahn, C., G. de Loubens, V. V. Naletov, J. Ben Youssef, O. Klein, and M. Viret, 2014, "Conduction of spin currents through insulating antiferromagnetic oxides", Europhys. Lett. **108**, 57005.

Hall, E. E., 1881, "On the "rotational coefficient" in nickel and cobalt", Philos. Mag. **12**, 157.

Halperin, B. I., and P. C. Hohenberg, 1969, "Hydrodynamic theory of spin waves", Phys. Rev. **188**, 898.

Hals, K. M. D., Y. Tserkovnyak, and A. Brataas, 2011, "Phenomenology of Current-Induced Dynamics in Antiferromagnets", Phys. Rev. Lett. **106**, 107206.

Han, W., R. K. Kawakami, M. Gmitra, J. Fabian, 2014, "Graphene Spintronics," Nature Nanotech. **9**, 794.

Haney, P. M., and A. H. MacDonald, 2008, "Current-Induced Torques Due to Compensated Antiferromagnets", Phys. Rev. Lett. **100**, 196801.

Haney, P. M.; H.-W. Lee, K.-J. Lee, A. Manchon, and M. D. Stiles, 2013a, "Current induced torques and interfacial spin-orbit coupling: Semiclassical modeling", Phys. Rev. B **87**, 174411.

Haney, P. M.; H.-W. Lee, K.-J. Lee, A. Manchon, and M. D. Stiles, 2013b, "Current-induced torques and interfacial spin-orbit coupling", Phys. Rev. B **88**, 214417.

Harrison, W. A., 1989, *Electronic Structure and the Properties of Solids: The Physics of the Chemical Bond* (Dover, New York).

Hasan, M. Z., and C. L. Kane, 2010, "Topological Insulators", Rev. Mod. Phys. **82**, 3045.

Hasan, M. Z., and J. E. Moore, 2011, "Three-Dimensional Topological Insulators", Ann. Rev. Cond. Matt. Phys. **2,** 55.

Hashimoto, Y., *et al.*, 2014, "Ultrafast time-resolved magneto-optical imaging of all-optical switching in GdFeCo with femtosecond time-resolution and a mm spatial-resolution", Rev. Sci. Instrum. **85**, 063702.

Hassdenteufel, A., B. Hebler, C. Schubert, A. Liebig, M. Teich, M. Helm, M. Aeschlimann, M. Albrecht, and R. Bratschitsch, 2013, "Thermally assisted all-optical helicity dependent magnetic switching in amorphous $Fe_{100-x}Tb_x$ alloy films", Adv. Mater. **25**, 3122.

Hassdenteufel, A., C. Schubert, J. Schmidt, P. Richter, D. R. T. Zahn, G. Salvan, M. Helm, R. Bratschitsch, and M. Albrecht, 2014a, "Dependence of all-optical magnetic switching on the sublattice magnetization orientation in Tb-Fe thin films", Appl. Phys. Lett. **105**, 112403.

Hassdenteufel, A., C. Schubert, B. Hebler, H. Schultheiss, J. Fassbender, M. Albrecht, and R. Bratschitsch, 2014b, "All-optical helicity dependent magnetic switching in Tb-Fe thin films with a MHz laser oscillator", Opt. Express **22**, 10017.

Hayashi, M., L. Thomas, C. Rettner, R. Moriya, Y. B. Bazaliy, and S. S. P. Parkin, 2007, "Current Driven

Domain Wall Velocities Exceeding the Spin Angular Momentum Transfer Rate in Permalloy Nanowires”, Phys. Rev. Lett. **98**, 037204.

Hayashi, M., L. Thomas, C. Rettner, R. Moriya, and S. S. P. Parkin, 2008, "Real time observation of the field driven periodic transformation of domain walls in Permalloy nanowires at the Larmor frequency and its first harmonic”, Appl. Phys. Lett. **92**, 112510.

He, J., A. Borisevich, S. V. Kalinin, S. J. Pennycook, and S. T. Pantelides, 2010a, “Control of octahedral tilts and magnetic properties of perovskite oxide heterostructures by substrate symmetry”, Phys. Rev. Lett. **105**, 227203.

He, W., T. Zhu, X. Q. Zhang, H. T. Yang and Z. H. Cheng, 2013, “Ultrafast demagnetization enhancement in CoFeB/MgO/CoFeB magnetic tunneling junction driven by spin tunneling current”, Sci. Rep. **3**, 2883.

He, X., Y. Wang, N. Wu, A. N. Caruso, E. Vescovo, K. D. Belaschenko, P. A. Dowben and Ch. Binek, 2010b, “Robust isothermal electric control of exchange bias at room temperature”, Nat. Mater. **9**, 579.

He, C., *et al.*, 2012, “Interfacial Ferromagnetism and Exchange Bias in $CaRuO_3$/$CaMnO_3$ Superlattices”, Phys. Rev. Lett. **109**, 197202.

Heide, M., G. Bihlmayer, and S. Blügel, 2008, "Dzyaloshinskii-Moriya interaction accounting for the orientation of magnetic domains in ultrathin films: Fe/W(110)”, Phys. Rev. B **78**, 140403.

Heide, M., G. Bihlmayer, and S. Blügel, 2009, “Describing Dzyaloshinskii-Moriya spirals from first principles”, Phys. B: Condens. Matter **404**, 2678.

Heinonen, O., W. Jiang, H. Somaily, S. G. E. te Velthuis, and A. Hoffmann, 2016, “Generation of magnetic skyrmion bubbles by inhomogeneous spin-Hall currents”, Phys. Rev. B **93**, 094407.

Heinze, S., K. von Bergmann, M. Menzel, J. Brede, A. Kubetzka, R. Wiesendanger, G. Bihlmayer, and S. Blügel, 2011, "Spontaneous atomic-scale magnetic skyrmion lattice in two dimensions”, Nat. Phys. **7**, 713.

Hellman F. and E. M. Gyorgy, 1992, "Growth-Induced Magnetic Anisotropy in Amorphous Tb-Fe", Phys. Rev. Lett. **68**, 1391.

Heremans, J. P., and S. R. Boona, 2014, “Viewpoint: Putting a New Spin on Heat Flow”, Physics **7**, 71.

Herman, F., and S. Skillman, 1963, "Atomic Structure Calculations", (Prentice-Hall, New Jersey).

Hermsmeier B., J. Osterwalder, D.L. Friedman and C. S. Fadley, 1989, “Evidence for a High-Temperature Short-Range-Magnetic-Order Transition in MnO(001)”, Phys. Rev. Lett. **62**, 478**.**

Hermsmeier B., J. Osterwalder, D.J. Friedman, B. Sinkovic, T. Tran and C.S. Fadley, 1990, “Spin-polarized photoelectron diffraction and valence-band photoemission from MnO(001)", Phys. Rev. B **42,** 11895**.**

Hertel, R., and C. M. Schneider, 2006, "Exchange Explosions: Magnetization Dynamics during Vortex-Antivortex Annihilation", Phys. Rev. Lett. **97**, 177202.

Hicken R. J., A. Barman, V. V. Kruglyak, and S. Ladak, 2003, "Optical ferromagnetic resonance studies of thin film magnetic structures", J. Phys. D: Appl. Phys. **36**, 2183.

Himpsel, F. J., J. E. Ortega, G. J. Mankey, and R. F. Willis, 1998, "Magnetic nanostructures", Adv. Phys. **47**, 511.

Hirsch, J. E., 1999, "Spin Hall effect", Phys. Rev. Lett. **83**, 1834 (1999).

Hoefer, M. A., T. J. Silva, and M. W. Keller, 2010, "Theory for a dissipative droplet soliton excited by a spin torque nanocontact", Phys. Rev. B **82**, 054432.

Hoffman, J., I. C. Tung, B. B. Nelson-Cheeseman, M. Liu, J.W. Freeland, and A. Bhattacharya, 2013, "Charge transfer and interfacial magnetism in $(LaNiO_3)_n/(LaMnO_3)_2$ superlattices", Phys. Rev. B **88**, 144411.

Hoffmann, A., 2013, "Spin Hall Effect in Metals", IEEE Trans. Magn. **49**, 5172.

Hoffmann, A., and S. D. Bader, 2015 "Opportunities at the Frontiers of Spintronics", Phys. Rev. Appl. **4**, 047001.

Hong, S., V. Diep, S. Datta, and Y. P. Chen, 2012, "Modeling potentiometric measurements in topological insulators including parallel channels", Phys. Rev. B **86**, 085131.

Hopster, H., and H. P. Oepen, 2005, *Magnetic microscopy of nanostructures*, (Springer, Berlin).

Hsieh, D., *et al.*, 2009, "A tunable topological insulator in the spin helical Dirac transport regime", Nature **460**, 1101.

Hu, S., H. Itoh, and T. Kimura, 2014, "Efficient thermal spin injection using CoFeAl nanowire", NPG Asia Mater. **6**, e127.

Huai, Y., F. Albert, P. Nguyen, M. Pakala, and T. Valet, 2004, " Observation of spin-transfer switching in deep submicron-sized and low-resistance magnetic tunnel junctions", Appl. Phys. Lett. **84**, 3118.

Huang, F., M. T. Kief, G. J. Mankey, and R. F. Willis, 1994, "Magnetism in the few-monolayers limit: A surface magneto-optic Kerr-effect study of the magnetic behavior of ultrathin films of Co, Ni, and Co-Ni alloys on Cu(100) and Cu(111)", Phys. Rev. B **49**, 3962.

Huang, S. X., and C. L. Chien, 2012, "Extended Skyrmion Phase in Epitaxial FeGe(111) Thin Films", Phys. Rev. Lett. **108**, 267201.

Huang, S. Y., X. Fan, D. Qu, Y. P. Chen, W. G. Wang, J. Wu, T. Y. Chen, J. Q. Xiao, and C. L. Chien, 2012, "Transport Magnetic Proximity Effects in Platinum", Phys. Rev. Lett. **109**, 107204.

Huang, S. Y., W. G. Wang, S. F. Lee, J. Kwo, and C. L. Chien, 2011, "Intrinsic Spin-Dependent Thermal Transport", Phys. Rev. Lett. **107**, 216604.

Hubbard, J., 1963, "Electron Correlations in Narrow Energy Bands", Proc. Roy. Soc. A **276**, 238–257.

Huber, D. L., 1982, "Dynamics of spin vortices in two-dimensional planar magnets", Phys. Rev. B **26**, 3758.

Hubert, A., and R. Schäfer, 1998, *Magnetic Domains* (Springer, Berlin).

Hwang, H. Y., Y. Iwasa, M. Kawasaki, B. Keimer, N. Nagaosa, and Y. Tokura, 2012, "Emergent phenomena at oxide interfaces", Nat. Mater. **11**, 103.

Ijiri, Y., T. C. Schulthess, J. A. Borchers, P. J. van der Zaag, and R. W. Erwin, 2007, "Link between Perpendicular Coupling and Exchange Biasing in $Fe_3O_4$/CoO Multilayers", Phys. Rev. Lett. **99**, 147201.

Im, M.-Y., L. Bocklage, P. Fischer, and G. Meier, 2009, "Direct Observation of Stochastic Domain-Wall Depinning in Magnetic Nanowires", Phys. Rev. Lett. **102**, 147204.

Im, M.-Y., P. Fischer, K. Yamada, T. Sato, S. Kasai, Y. Nakatani, and T. Ono, 2012, "Symmetry breaking in the formation of magnetic vortex states in a permalloy nanodisk", Nat. Commun. **3**, 983.

Imamura, H., P. Bruno, and Y. Utsumi, 2004, "Twisted exchange interaction betwwen localized spins embedded in a one- or two-dimensional electron gas with Rashba spin-orbit coupling", Phys. Rev. B **69**, 121303(R).

Ishikawa, Y., K. Tajima, D. Bloch, and M. Roth, 1976, "Helical spin structure in manganese silicide MnSi", Sol. Sta. Commun. **19**, 525.

Ivanov, B. A., and A. M. Kosevich, 1976, "Bound states of a large number of magnons in a three-dimensional ferromagnet (magnon droplets)", JETP Lett. **24**, 454.

Ivanov, B. A., and A. M. Kosevich, 1977, "Bound states of a large number of magnons in a ferromagnet with a single-ion anisotropy", Sov. Phys. JETP **45**, 1050.

Ivanov, B. A., and V. A. Stephanovich, 1989, "Two-dimensional soliton dynamics in ferromagnets", Phys. Lett. A **141**, 89.

Iwasaki, J., M. Mochizuki, and N. Nagaosa, 2013a, "Current-induced skyrmion dynamics in constricted geometries", Nat. Nanotech. **8**, 742.

Iwasaki, J., M. Mochizuki, and N. Nagaosa, 2013b, "Universal current-velocity relation of skyrmion motion in chiral magnets", Nat.Commun. **4**, 1463.

Jäger, J. V., *et al.*, 2013, "Picosecond inverse magnetostriction in galfenol thin films", Appl. Phys. Lett. **103**, 032409.

Jamali, M., J. S. Lee, J. S. Jeong, F. Mahfouzi, Y. Lv, Z. Zhao, B. K. Nikolić, K. A. Mkhoyan, N. Samarth, and J.-P. Wang, 2015, "Giant Spin Pumping and Inverse Spin Hall Effect in the Presence of Surface and Bulk Spin-Orbit Coupling of Topological Insulator $Bi_2Se_3$", Nano Lett. **15**, 7126.

Jamali, M., K. Narayanapillai, X. Qiu, L. M. Loong, A. Manchon, and H. Yang, 2013, "Spin-Orbit Torques in Co/Pd Multilayer Nanowires", Phys. Rev. Lett. **111**, 246602.

Jan, G., Y.-J. Wang, T. Moriyama, Y.-J. Lee, M. Lin, T. Zhong, R.-Y. Tong, T. Torng, and P.-K. Wang, 2012, "High Spin Torque Efficiency of Magnetic Tunnel Junctions with MgO/CoFeB/MgO Free Layer", Appl. Phys. Express **5**, 093008.

Jaworski, C. M., J. Yang, S. Mack, D. D. Awschalom, J. P. Heremans, and R. C. Myers, 2010, "Observation of the spin-Seebeck effect in a ferromagnetic semiconductor", Nat. Mater. **9**, 898.

Jaworski, C. M., J. Yang, S. Mack, D. D. Awschalom, R. C. Myers, and J. P. Heremans, 2011, "Spin-Seebeck Effect: A Phonon Drive Spin Distribution", Phys. Rev. Lett. **106**, 186601.

Jedema, F. J.; A. T. Filip, and B. J. van Wees, 2001, "Electrical spin injection and accumulation at room temperature in an all-metal mesoscopic spin valve", Nature **410**, 345.

Jedema, F. J.; H. B. Heersche, A. T. Filip, J. J. A. Baselmans and B. J. van Wees, 2002, "Electrical detection of spin precession in a metallic mesoscopic spin valve", Nature **416**, 713.

Jensen, J., and A.R. MacKintosh, 1991, "Rare Earth Magnetism", Clarendon Press, Oxford.

Jeong, J., N. Aetukuri, T. Graf, T. D. Schladt, M. G. Samant, and S. S. P. Parkin, 2013, "Suppression of metal-insulator transition in $VO_2$ by electric field-induced oxygen vacancy formation", Science **339**, 1402.

Ji, Y., A. Hoffmann, J. S. Jiang, J. E. Pearson, and S. D. Bader, 2007, "Non-local spin injection in lateral spin valves", J. Phys. D Appl. Phys. **40**, 1280.

Ji, Y. ; G. J. Strijkers, F. Y. Yang, C. L. Chien, J. M. Byers, A. Anguelouch, Gang Xiao, and A. Gupta, 2001, "Determination of the Spin Polarization of Half-Metallic $CrO_2$ by Point Contact Andreev Reflection," Phys. Rev. Lett. **86**, 5585.

Jiang, W., *et al.*, 2015, "Blowing Magnetic Skyrmion Bubbles", Science **349**, 283.

Jiang, W, X. Zhang, G. Yu, W. Zhang, M. B. Jungfleisch, J. E. Pearson, O. Heinonen, K. L. Wang, Y. Zhou, A. Hoffmann, and S. G. E. te Velthuis, 2016, "Direct Observation of the Skyrmion Hall Effect", Nat. Phys., doi:10.1038/nphys3883.

Jin, H., Z. Yang, R. C. Myers, and J. P. Hereman, 2014, "Spin Seebeck like signal in ferromagnetic bulk metallic glass without platinum contacts", Solid State Commun. **198**, 40.

Johnson, M. T., P. J. H. Bloemen, F. J. A. den Broeder, J. J. de Vries, 1996, "Magnetic anisotropy in metallic multilayers", Rep. Prog. Phys. **59**, 1409.

Johnson, M., and R. H. Silsbee, 1985, ''Interfacial charge-spin coupling: Injection and detection of spin magnetization in metals,'' Phys. Rev. Lett. **55**, 1790.

Jones, R. O., and O. Gunnarsson, 1989, "The density functional formalism, its applications and prospects", Rev. Mod. Phys. **61**, 689.

Jonietz, F., *et al.*, 2010, "Spin Transfer Torques in MnSi at Ultralow Current Densities", Science **330**, 1648.

Ju, G., A. V. Nurmikko, R. F. C. Farrow, R. F. Marks, M. J. Carey, and B. A. Gurney, 1998, “Ultrafast optical modulation of an exchange biased ferromagnetic/antiferromagnetic bilayer”, Phys. Rev. B **58**, R11857(R).

Ju, G., A. V. Nurmikko, R. F. C. Farrow, R. F. Marks, M. J. Carey, and B. A. Gurney, 1999, “Ultrafast Time Resolved Photoinduced Magnetization Rotation in a Ferromagnetic/Antiferromagnetic Exchange Coupled System”, Phys. Rev. Lett. **82**, 3705.

Ju, G., L. Chen, A. V. Nurmikko, R. F. C. Farrow, R. F. Marks, M. J. Carey, and B. A. Gurney, 2000, “Coherent magnetization rotation induced by optical modulation in ferromagnetic/antiferromagnetic exchange-coupled bilayers”, Phys. Rev. B **62**, 1171.

Ju, G. P., J. Hohlfeld, B. Bergman, R. J. M. van de Veerdonk, O. N. Mryasov, J. Y. Kim, X. W. Wu, D. Weller, and B. Koopmans, 2004, “Ultrafast generation of ferromagnetic order *via* a laser-induced phase transformation in FeRh thin films”, Phys. Rev. Lett. **93**, 197403.

Jué, E., A. Thiaville, S. Pizzini, J. Miltat, J. Sampaio, L.D. Buda-Prejbeanu, S. Rohart, J. Vogel, M. Bonfim, O. Boulle, S. Auffret, I.M. Miron, and G. Gaudin, 2016b, “Domain wall dynamics in ultrathin Pt/Co/AlOx microstrips under large combined magnetic fields”, Phys. Rev. B **93**, 014403.

Julliere, M., 1975, “Tunneling between ferromagnetic films”, Phys. Lett. A **54**, 225.

Jungfleisch, M. B., W. Zhang, J. Sklenar, W. Jiang, J. E. Pearson, J. B. Ketterson, and A. Hoffmann, 2016a, “Interface-driven spin-torque ferromagnetic resonance by Rashba coupling at the interface between non-magnetic materials”, Phys. Rev. B. **93** 224419.

Jungfleisch, M. B., *et al.*, 2016b, “Large spin-wave bullet in a ferromagnetic insulator driven by spin Hall effect”, Phys. Rev. Lett. **116**, 057601.

Jungwirth, T., Q. Niu, and A. H. MacDonald, 2002, “Anomalous Hall effect in ferromagnetic semiconductors”, Phys. Rev. Lett. **88**, 207208.

Jungwirth, T. X. Marti, P. Wadley, and J. Wunderlich, 2016, “Antiferromagnetic spintronics”, Nature Nanotech. **11**, 231.

Kaka, S., M. R. Pufall, W. H. Rippard, T. J. Silva, S. E. Russek, and J. A. Katine, 2005, "Mutual phase-locking of microwave spin torque nano-oscillators”, Nature **437**, 389.

Kalinikos, B. A., N. G. Kovshikov, and A. N. Slavin, 1983, “Observation of spin-wave solitons in ferromagnetic films”, Sov. Phys.-JETP Lett. **38**, 413.

Kalinikos, B. A., and A. N. Slavin, 1986, "Theory of Dipole-Exchange Spin Wave Spectrum for Ferromagnetic Films with Exchange Boundary Conditions”, J. Phys. C: Sol. State Phys. **19**, 7013.

Kaltenborn, S., Y.-H Zhu and H.C. Schneider, 2012, “Wave-diffusion theory of spin transport in metals after ultrashort-pulse excitation”, Phys. Rev. B **85**, 235101.

Kampfrath, T., *et al.*, 2013, “Terahertz spin current pulses controlled by magnetic heterostructures”, Nat. Nanotech. **8**, 256.

Kanamori, J., 1959, “Superexchange interaction and symmetry properties of electron orbitals”, J. Phys. Chem. Sol. **10**, 87.

Kandala, A., A. Richardella, S. Kempinger, C.-X. Liu, and N. Samarth, 2015, “Giant anisotropic magnetoresistance in a quantum anomalous Hall insulator”, Nat. Commun. **6**, 7434.

Kandala, A., A. Richardella, D. W. Rench, D. M. Zhang, T. C. Flanagan, and N. Samarth, 2013, “Growth and characterization of hybrid insulating ferromagnet-topological insulator heterostructure devices”, Appl. Phys. Lett. **103**, 202409.

Kang, W., Y. Huang, X. Zhang, Y. Zhou, W. Zhao, 2016, “Skyrmion-Electronics: An Overview and Outlook,” Proc. IEEE 104, 2040.

Karplus, R., and J. M. Luttinger, 1954, “Hall effect in ferromagnetics”, Phys. Rev. **95**, 1154.

Kasuya, T., 1956, “A Theory of Metallic Ferro- and Antiferromagnetism on Zener's Model,” Prog. Theor. Phys. **16**, 45.

Katine, J. A., F. J. Albert, R. A. Buhrman, E. B. Myers, and D. C. Ralph, 2000, "Current-driven magnetization reversal and spin-wave excitations in Co/Cu/Co pillars”, Phys. Rev. Lett. **84**, 3149.

Katmis, F., V. Lauter, F. S. Nogueira, A. A. Assaf, M. E. Jamer, P. Wei, B. Satpati, J. W. Freeland, I. Eremin, D. Heiman, P. Jarillo-Herrero, J. S. Moodera, 2016, “A high-temperature ferromagnetic topological insulating phase by proximity coupling,” Nature **533**, 513.

Kato, Y. K., R. C. Myers, A. C. Gossard, and D. D. Awschalom, 2004, “Observation of the Spin Hall Effect in Semiconductors”, Science **306**, 1910.

Katsnelson, M. I., V. Yu. Irkhin, L. Chioncel, A. I. Lichtenstein, and R. A. de Groot, 2008, "Half-metallic ferromagnets: From band structure to many-body effects," Rev. Mod. Phys. **80**, 315.

Kazantseva, N., D. Hinzke, U. Nowak, R.W. Chantrell, U. Atxitia, and O. Chubykalo-Fesenko, 2008 “Towards multiscale modeling of magnetic materials: Simulations of FePt”, Phys. Rev. B **77**, 184428.

Kehlberger, A., *et al.*, 2015, “Length Scale of the Spin Seebeck Effect”, Phys. Rev. Lett. **115**, 096602.

Keller, M. W., A. B. Kos, T. J. Silva, W. H. Rippard, and M. R. Pufall, 2009, "Time domain measurement of phase noise in a spin torque oscillator”, Appl. Phys. Lett. **94**, 193105.

Kent, A. D., B. Ozyilmaz, and E. del Barco, 2004, "Spin-transfer-induced precessional magnetization reversal", Appl. Phys. Lett. **84**, 3897.

Khajetoorians, A. A., M. Steinbrecher, M. Ternes, M. Bouhassoune, M. dos Santos Dias, S. Lounis, J. Wiebe and R. Wiesendanger, 2016, "Tailoring the chiral magnetic interaction between two individual atoms", Nat. Commun. **7**, 10620.

Khalsa, G.; M. D. Stiles and J. Grollier, 2015 "Critical current and linewidth reduction in spin-torque nano-oscillators by delayed self-injection", Appl. Phys. Lett. **106**, 242402.

Khomskii, D. I., 2014, *Transition metal compounds* (Cambridge, Cambridge).

Khorsand, A.R., M. Savoini, A. Kirilyuk, A.V. Kimel, A. Tsukamoto, A. Itoh, and T. Rasing, 2012, "Role of Magnetic Circular Dichroism in All-Optical Magnetic Recording", Phys. Rev. Lett. **108**, 127205.

Khorsand, A.R., M. Savoini, A. Kirilyuk, and T. Rasing, 2014, "Optical excitation of thin magnetic layers in multilayer structures", Nat. Mater. **13**, 101.

Khvalkovskiy, A. V., *et al.*, 2013b, "Basic principles of STT-MRAM cell operation in memory arrays", J. Phys. D: Appl. Phys. **46**, 074001.

Khvalkovskiy, A. V., V. Cros, D. Apalkov, V. Nikitin, M. Krounbi, K. A. Zvezdin, A. Anane, J. Grollier, and A. Fert, 2013a, "Matching domain-wall configuration and spin-orbit torques for efficient domain-wall motion", Phys. Rev. B **87**, 020402.

Kikkawa, T., K. Uchida, S. Daimon, Y. Shiomi, H. Adachi, Z. Qiu, D. Hou, X.-F. Jin, S. Maekawa, and E. Saitoh, 2013, "Separation of longitudinal spin Seebeck effect from anomalous Nernst effect: Determination of origin of transverse thermoelectric voltage in metal/insulator junctions", Phys. Rev. B **88**, 214403.

Kim, J., J. Sinha, M. Hayashi, M. Yamanouchi, S. Fukami, T. Suzuki, S. Mitani, and H. Ohno, 2013, "Layer thickness dependence of the current-induced effective field vector in Ta|CoFeB|MgO", Nat. Mater. **12**, 240.

Kim, K.-W.; S.-M. Seo, J. Ryu, K.-J. Lee, and H.-W. Lee, 2012, "Magnetization dynamics induced by in-plane currents in ultrathin magnetic nanostructures with Rashba spin-orbit coupling", Phys. Rev. B **85**, 180404.

Kim, S., D. Lederman, J. M. Gallego, and I. K. Schuller, 1996, "Electron localization in Co/Ni superlattices", Phys. Rev. B **54**, R5291.

Kimel, A.V., A. Kirilyuk, A. Tsvetkov, R. V. Pisarev and T. Rasing, 2004, "Laser-induced ultrafast spin reorientation in the antiferromagnet $TmFeO_3$", Nature **429**, 850.

Kimling, J., J. Kimling, R. B. Wilson, B. Hebler, M. Albrecht, and D. G. Cahill, 2014, "Ultrafast demagnetization of FePt:Cu thin films and the role of magnetic heat capacity", Phys. Rev. B **90,** 224408.

Kirilyuk, A., A. V. Kimel, and T. Rasing, 2010, "Ultrafast optical manipulation of magnetic order", Rev. Mod. Phys. **82**, 2731.

Kirilyuk, A., A.V. Kimel, and T. Rasing, 2013, "Laser-induced magnetization dynamics and reversal in ferrimagnetic alloys", Rep. Prog. Phys. **76**, 026501.

Kiselev, N. S., A. N. Bogdanov, R. Schäfer, and U. K. Rößler, 2011, "Chiral skyrmions in thin magnetic films: new objects for magnetic storage technologies?", J. Phys. D **44**, 392001.

Kiselev, S. I., J. C. Sankey, I. N. Krivorotov, N. C. Emley, R. J. Schoelkopf, R. A. Buhrman, and D. C. Ralph, 2003, "Microwave oscillations of a nanomagnet driven by a spin-polarized current", Nature **425**, 380.

Kittel, C., 1949, "On the Gyromagnetic Ratio and Spectroscopic Splitting Factor of Ferromagnetic Substances," Phys. Rev. **76**, 743.

Kléman, M., 1973, "Defect densities in directional media, mainly liquid crystals", Phil. Mag. **27**, 1057.

Klewe, C., T. Kuschel, J.-M. Schmalhorst, F. Bertram, O. Kuschel, J. Wollschläger, J. Strempfer, M. Meinert, and G. Reiss, 2016, "Static magnetic proximity effect in Pt/$Ni_{1-x}Fe_x$ bilayers investigated by x-ray resonant magnetic reflectivity", Phys. Rev. B **93**, 214440.

Knorren, R., K. H. Bennemann, R. Burgermeister, and M. Aeschlimann, 2000, "Dynamics of excited electrons in copper and ferromagnetic transition metals: Theory and experiment", Phys. Rev. B **61**, 9427.

Kobayashi, Y., H. Sato, Y. Aoki, and A. Kamijo, 1994, "The giant magnetoresistance and the anomalous Hall effect in molecular-beam-epitaxy grown Co/Cu superlattices", J. Phys.: Condens. Matter **6**, 7255.

Koch, C. T., and A. Lubk, 2010, "Off-axis and inline electron holography: A quantitative comparison", Ultramicroscopy **110**, 460.

Kohn, W. and L. J. Sham, 1965, "Self-Consistent Equations Including Exchange and Correlation Effects", Phys. Rev. A **140**, A1133.

König, J., M. C. Bønsager, and A. H. MacDonald, 2001, "Dissipationless spin transport in thin film ferromagnets", Phys. Rev. Lett. **87**, 187202.

Konishi Y., Z. Fang, M. Izumi, T. Manako, M. Kasai, H. Kuwahara, M. Kawasaki, K. Terakua, and Y. Tokura, 1999, "Orbital-state-mediated phase-control of manganites", J. Phys. Soc. Jpn. **68,** 3790.

Koopmans, B., M. van Kampen, J. T. Kohlhepp, and W. J. M. de Jonge, 2000, "Ultrafast magneto-optics in nickel: magnetism or optics?", Phys. Rev. Lett. **85**, 844.

Koopmans, B, J. J. M. Ruigrok, F. Dalla Longa, and W. J. M. de Jonge, 2005, “Unifying ultrafast magnetization dynamics”, Phys. Rev. Lett. **95**, 267207.

Koopmans, B., G. Malinowski, F. Dalla Longa, D. Steiauf, M. Faehnle, T. Roth, M. Cinchetti, and M. Aeschlimann, 2010, “Explaining the paradoxical diversity of ultrafast laser-induced demagnetization”, Nat. Mater. **9**, 259.

Kortright, J. B., D. D. Awschalom, J. Stöhr, S. D. Bader, Y. U. Idzerda, S. S. P. Parkin, I. K. Schuller, and H.-C. Siegmann, 1999, “Research frontiers in magnetic materials at soft X-ray synchrotron radiation facilities”, J. Magn. Magn. Mater. **207**, 7.

Kosevich, A., B. Ivanov, and A. Kovalev, 1990, “Magnetic solitons*”,* Phys. Rep. **194**, 117.

Kou, X., *et al.*, 2014, “Erratum: Scale-Invariant Quantum Anomalous Hall Effect in Magnetic Topological Insulators beyond the Two-Dimensional Limit [Phys. Rev. Lett. **113**, 137201 (2014)]”, Phys. Rev. Lett. **113,** 199901.

Kou, X., *et al.*, 2013, “Interplay between Different Magnetisms in Cr-Doped Topological Insulators”, ACS Nano **7,** 9205.

Koyama, T., *et al.*, 2011, "Observation of the intrinsic pinning of a magnetic domain wall in a ferromagnetic nanowire”, Nat. Mater. **10**, 194.

Krishtopenko, S.S. , V. I. Gavrilenko and M. Goiran, 2011, “Theory of g-factor enhancement in narrow-gap quantum well heterostructures”, J. Phys.: Condens. Matter **23**, 385601.

Krivoruchko, V. N., 2014, “Magnetization dynamics of a two-sublattice ferrimagnet near the angular momentum compensation point”, Low Temp. Phys. **40**, 42.

Kubota, H., *et al.*, 2008, “Quantitative measurement of voltage dependence of spin-transfer torque in MgO-based magnetic tunnel junctions”, Nat. Phys. **4**, 37.

Kuch, W., R. Schäfer, P. Fischer, and F. U. Hillebrecht, 2015, *Magnetic microscopy of layered structures*, (Springer).

Kuiper, K. C., T. Roth, A. J. Schellekens, O. Schmitt, B. Koopmans, M. Cinchetti, and M. Aeschlimann, 2014, “Spin-orbit enhanced demagnetization rate in Co/Pt-multilayers”, Appl. Phys. Lett. **105**, 202402.

Kukreja, R., S. Bonetti, Z. Chen, D. Backes, Y. Acremann, J. A. Katine, A. D. Kent, H. A. Dürr, H. Ohldag, and J. Stöhr, 2015, “X-Ray Detection of Transient Magnetic Moments Induced by a Spin Current in Cu”, Phys. Rev. Lett. **115**, 096601.

Kurebayashi, H., *et al.*, 2014, “An antidamping spin-orbit torque originating from the Berry curvature”, Nat. Nanotechn. **9**, 211.

Kuschel, T., C. Klewe, J.-M. Schmalhosrt, F. Bertram, O. Kuschel, T. Schemme, J. Wollschläger, S. Francoul=al, J. Strempfer, A. Gupta, M. Meinert, G. Götz, D. Meier, and G. Reiss, 2015, “Static

Magnetic Proximity Effect in Pt/$NiFe_2O_4$ and Pt/Fe Bilayers Investigated by X-Ray Resonant Magnetic Reflectivity", Phys. Rev. Lett. **115**, 097401.

Lachman, E. O., A. F. Young, A. Richardella, J. Cuppens, H. R. Naren, Y. Anahory, A. Y. Meltzer, A. Kandala, S. Kempinger, Y. Myasoedov, M. E. Huber, N. Samarth, E. Zeldov, 2015, "Visualization of superparamagnetic dynamics in magnetic topological insulators," Science Adv. **1**, e1500740.

Lakys, Y., W. Zhao, J.-O. Klein and C. Chappert, 2012, "MRAM Crossbar based Configurable Logic Block", in *ISACS, 2012* (IEEE, New York).

Lambert, C.-H., *et al.*, 2014, "All-optical control of ferromagnetic thin films and nanostructures", Science **345**, 1337.

Landau, L.D. and E.M. Lifshitz, 1935, "Theory of the dispersion of magnetic permeability in ferromagnetic bodies", Phys. Z. Sowietunion, **8**, 153.

Landolt, G., *et al.*, 2014, "Spin Texture of $Bi_2Se_3$ Thin Films in the Quantum Tunneling Limit", Phys. Rev. Lett. **112**, 057601.

Lang, M., *et al.*, 2014, "Proximity Induced High-Temperature Magnetic Order in Topological Insulator-Ferrimagnetic Insulator Heterostructure", Nano Lett. **14**, 6; 3459.

Langner, M. C., *et al.*, 2014, "Coupled Skyrmion Sublattices in $Cu_2OSeO_3$", Phys. Rev. Lett. **112**, 167202.

Lau, Y.-C., D. Betto, K. Roade, J. M. D. Coey, and P. Stamenov, 2016, "Spin-orbit torque switching without an external field using interlayer exchange coupling", Nat. Nanotech. **11**, 758.

Le Guyader, L., A. Kleibert, F. Nolting, L. Joly, P. Derlet, R. Pisarev, A. Kirilyuk, T. Rasing, and A. Kimel, 2013, Dynamics of laser-induced spin reorientation in Co/SmFeO3 heterostructure, Phys. Rev. B 87 054437.

Lee, H.-J., E. Helgren, and F. Hellman, 2009, "Gate-controlled magnetic properties of the magnetic semiconductor (Zn,Co)O", Appl. Phys. Lett. **94**, 212106.

Lee, H. -J., C. Bordel, J. Karel, D. W. Cooke, M. Charilaou, and F. Hellman, 2013a, "Electron-mediated Ferromagnetic Behavior in CoO/ZnO Multilayers", Phys. Rev. Lett. **110**, 087206.

Lee, J. C. T., J. J. Chess, S. A. Montoya, X. Shi, N. Tamura, S. K. Mishra, P. Fischer, B. J. McMorran, S. K. Sinha, E. E. Fullerton, S. D. Kevan, and S. Roy, 2016, "Synthesizing skyrmion bound pairs in Fe-Gd thin films", Appl. Phys. Lett. **109**, 022402 (2016).

Lee, J. S. A. Richardella, D. Reifsnyder Hickey, K. A. Mkhoyan, and N. Samarth, 2015, "Mapping the chemical potential dependence of current-induced spin polarization in a topological insulator", Phys. Rev. B **92**, 155312.

Lee, J.-S., Xie, Y.W., Sato, H.K., Bell, C., Hikita, Y., Hwang. H.Y. and Kao, C.-C., 2013b, "Titanium $d_{xy}$ ferromagnetism at the $LaAlO_3$/$SrTiO_3$ interface", Nat. Mater. **12**, 703.

Lee, K. I., S. J. Joo, J. H. Lee, K. Rhie, T.-S. Kim, W. Y. Lee, K. H. Shin, B. C. Lee, P. LeClair, J.-S. Lee, and J. H. Park, 2007, "Kondo effect in magnetic tunnel junctions", Phys. Rev. Lett. **98**, 107202.

Lee. K.-S., S.-W. Lee, B.-C. Min, and K.-J. Lee, "Threshold current for switching of a perpendicular magnetic layer induced by spin Hall effect", 2013c, Appl. Phys. Lett. **102**, 112410.

Lee, M., J. R. Williams, S. Zhang, C. D. Frisbie, and D. Goldhaber-Gordon, 2011, "Electrolyte gate-controlled Kondo effect in $SrTiO_3$", Phys. Rev. Lett. **107**, 256601.

Lee, O. J., L. Q. Liu, C. F. Pai, Y. Li, H. W. Tseng, P. G. Gowtham, J. P. Park, D. C. Ralph, and R. A. Buhrman, 2014, "Central role of domain wall depinning for perpendicular magnetization switching driven by spin torque from the spin Hall effect", Phys. Rev. B **89**, 024418.

Lee, O. J., V. S. Pribiag, P. M. Braganca, P. G. Gowtham, D. C. Ralph, and R. A. Buhrman, 2009, "Ultrafast switching of a nanomagnet by a combined out-of-plane and in-plane polarized spin current pulse", Appl. Phys. Lett. **95**, 012506.

Le Guyader, L., M. Savoini, S. El Moussaoui, M. Buzzi, A. Tsukamoto, A. Itoh, A. Kirilyuk, T. Rasing, A.V. Kimel, and F. Nolting, 2015, "Nanoscale sub-100 picosecond all-optical magnetization switching in GdFeCo microstructures", Nat. Commun. **6**, 5839.

Leonov, A. O., Y. Togawa, T. L. Monchesky, A. N. Bogdanov, J. Kishine, Y. Kousaka, M. Miyagawa, T. Koyama, J. Akimitsu, Ts. Koyama, K. Harada, S. Mori, D. McGrouther, R. Lamb, M. Krajnak, S. McVitie, R. L. Stamps, and K. Inoue, 2016, "Chiral surface twists and skyrmion stability in nanolayers of cubic helimagnets", Phys. Rev. Lett. **117**, 087202.

Li, C. H., O. M. J. van't Erve, J. T. Robinson, Y. Liu, L. Li, and B. T. Jonker, 2014a, "Electrical detection of charge-current-induced spin polarization due to spin-momentum locking in $Bi_2Se_3$", Nat. Nanotech. **9,** 218.

Li, L., C. Richter, J. Mannhart, and R. C. Ashoori, 2011, "Coexistence of magnetic order and two-dimensional superconductivity at $LaAlO_3/SrTiO_3$ interfaces", Nat. Phys. **7**, 762.

Li, P., D. Ellsworth, H. Chang, P. Janantha, D. Richardson, F. Shah, P. Phillips, T. Vijayasarathy, and M. Wu, 2014b, "Generation of pure spin currents *via* spin Seebeck effect in self-biased hexagonal ferrite thin films", Appl. Phys. Lett. **105**, 242412.

Li, S.-Z., 2015, "Generation of skyrmions by chopping magnetic chiral stripe domains with an electric current", arXiv:1510.07353.

Lin, W., K. Chen, S. Zhang, and C. L. Chien, 2016, " Enhancement of Thermally Injected Spin Current

through an Antiferromagnetic Insulator", Phys. Rev. Lett. **116**, 186601.

Lin, C., S. Okamoto, and A. J. Millis, 2006, “Dynamical mean-field study of model double-exchange superlattices”, Phys. Rev. B **73**, 041104 (R).

Lin, S.-Z., 2016, ”Edge instability in a chiral stripe domain under an electric current and skyrmion generation,” Phys. Rev. B **94**, 020402(R).

Liu, H., D. Bedau, D. Backes, J. A. Katine, J. Langer, and A. D. Kent, 2010, "Ultrafast switching in magnetic tunnel junction based orthogonal spin transfer devices”, Appl. Phys. Lett. **97**, 242510.

Liu, L., C.-T. Chen, and J. Z. Sun, 2014, “Spin Hall effect tunneling spectroscopy”, Nature Phys. **10**, 561.

Liu, L., T. Moriyama, D. C. Ralph, and R. A. Buhrman, 2011, “Spin-torque ferromagnetic resonance induced by the spin Hall effect”, Phys. Rev. Lett. **106**, 036601.

Liu, L., A. Richardella, I. Garate, Y. Zhu, N. Samarth, and C.-T. Chen, 2015a, “Spin-polarized tunneling study on spin-momentum locking in topological insulators”, Phys. Rev. B **91**, 235437.

Liu, L. Q., O. J. Lee, T. J. Gudmundsen, D. C. Ralph, and R. A. Buhrman, 2012a, “Current-induced switching of perpendicularly-magnetized magnetic layers using spin torque from the spin Hall effect”, Phys. Rev. Lett. **109**, 096602.

Liu, L. Q., C.-F. Pai, Y. Li, H.W. Tseng, D. C. Ralph, and R. A. Buhrman, 2012b, “Spin torque switching with the giant spin Hall effect of tantalum”, Science **336**, 555.

Liu, L. Q., C.-F. Pai, D. C. Ralph, and R. A. Buhrman, 2012c, “Magnetic Oscillations Driven by the Spin Hall Effect in 3-Terminal Magnetic Tunnel Junction Devices”, Phys. Rev. Lett. **109**, 186602.

Liu, M., J. Zhang, *et al.*, 2012d, “Crossover between Weak Antilocalization and Weak Localization in a Magnetically Doped Topological Insulator”, Phys. Rev. Lett. **108**, 036805.

Liu, T.-M., T. Wang, *et al.*, 2015b, “Nanoscale Confinement of All-Optical Magnetic Switching in TbFeCo – Competition with Nanoscale Heterogeneity”, Nano Lett. **15**, 6862.

Liu, M., W. Wang, A. Richardella, A. Kandala, J. Li, A. Yazdani, N. Samarth, N. P. Ong, 2016, "Large discrete jumps observed in the transition between Chern states in a ferromagnetic topological insulator," Science Adv. **2**, e1600167.

Liu, Y., Z. Yuan, R. J. H. Wesselink, A. A. Starikov, M. van Schilfgaarde, and P. J. Kelly, 2015, “Direct method for calculating temperature-dependent transport properties,” Phys. Rev. B **91**, 220405(R).

Lobanov, I.S., H. Jonsson, V.M. Uzdin, 2016, “Mechanism and activation energy of magnetic skyrmion annihilation obtained from minimum energy path calculations”, Phys. Rev. B **94**, 174418.

Locatelli, N., *et al.*, 2014, "Noise-Enhanced Synchronization of Stochastic Magnetic Oscillators”, Phys. Rev. Appl. **2**, 034009.

Locatelli, N., V. V. Naletov, J. Grollier, G. de Loubens, V. Cros, C. Deranlot, C. Ulysse, G. Faini, O. Klein, and A. Fert, 2011, "Dynamics of two coupled vortices in a spin valve nanopillar excited by spin transfer torque", Appl. Phys. Lett. **98**, 062501.

Lou, X., C. Adelmann, S. A. Crooker, E. S. Garlid, J. Zhang, K. S. Madhukar Reddy, S. D. Flexner, C. J Palmstrom, and P. A. Crowell, 2007, "Electrical detection of spin transport in lateral ferromagnet-semiconductor devices", Nat. Phys. **3**, 197.

Lu, H.-Z., J. Shi, and S.-Q. Shen, 2011, "Competition between Weak Localization and Antilocalization in Topological Surface States", Phys. Rev. Lett. **107**, 076801.

Lucassen, M. E., C. H. Wong, R. A. Duine, and Y. Tserkovnyak, 2011, " Spin-transfer mechanism for magnon-drag thermopower", Appl. Phys. Lett. **99,** 262506.

Lund, M. S., M. R. Fitzsimmons, S. Park, and C. Leighton, 2004, "Temperature dependence magnetic interface location in interdiffused exchange biased bilayers", Appl. Phys. Lett. **85**, 2845.

Lusassen, M. E., C. H. Wong, R. A. Duine, and Y. Tserkovnyak, 2011, "Spin-transfer mechanism for magnon-drag thermopower", Appl. Phys. Lett. **99**, 262506.

L'vov, V. S., 1994, *Wave Turbulence Under Parametric Excitation*. (Springer-Verlag, New York).

Macia, F., A. D. Kent, and F. C. Hoppensteadt, 2011, ″Spin-wave interference patterns created by spin-torque nano-oscillators for memory and computation", Nanotech. **22**, 095301.

MacKay, J. F., C. Teichert, D. E. Savage, and M. G. Lagally, 1996, "Element Specific Magnetization of Buried Interfaces Probed by Diffuse X-Ray Resonant Magnetic Scattering", Phys. Rev. Lett. **77**, 3925.

Macke, S., *et al*., 2014, "Element Specific Monolayer Depth Profiling", Adv. Mater. **26**, 6554.

MacNeill, D., G. M. Stiehl, M. H. D. Guimaraes, R. A. Buhrman, J. Park, and D. C. Ralph, 2016, "Control of spin-orbit torques through crystal symmetry in $WTe_2$/ferromagnet bilayers", Nat. Phys. doi:10.1038/nphys3933.

Mahfouzi, F., N. Nagaosa, and B. K. Nikolić, 2012, "Spin-Orbit Coupling Induced Spin-Transfer Torque and Current Polarization in Topological-Insulator/Ferromagnet Vertical Heterostructures", Phys. Rev. Lett. **109**, 166602.

Majkrzak, C. F., J. W. Cable, J. Kwo, M. Hong, D. B. McWhan, Y. Yafet, J. V. Waszcak, and C. Vettier, 1986 "Observation of a Magnetic Antiphase Domain Structure with Long-Range Order in a Synthetic Gd-Y Superlattice", Phys. Rev. Lett. **56**, 2700.

Malinowski, G., F. Dalla Longa, J. H. H. Rietjens, V. Paluskar, R. Huijink, H.J.M. Swagten, and B. Koopmans, 2008, "Control of speed and efficiency of ultrafast demagnetization by direct transfer of spin angular momentum", Nat. Phys. **4**, 855.

Malashevich, A., D. Vanderbilt, 2008, "First Principles Study of Improper Ferroelectricity in $TbMnO_3$", Phys. Rev. Lett. **101**, 037210.

Malozemoff, A. P., and J. C. Slonczewski, 1979, *Magnetic Domain Walls in Bubble Materials* (Academic, New York).

Manchon, A. and S. Zhang, 2009, "Theory of spin torque due to spin-orbit coupling", Phys. Rev. B, **79**, 094422.

Mancoff, F. B., N. D. Rizzo, B. N. Engel, and S. Tehrani, 2005, "Phase-locking in double-point-contact spin-transfer devices", Nature **437**, 393.

Mangin, S., *et al.*, 2014, "Engineered materials for all-optical helicity-dependent magnetic switching", Nat. Mater. **13**, 286.

Mankovsky, S., D. Ködderitzsch, G. Woltersdorf, and H. Ebert, 2013, "First-principles calculation of the Gilbert damping parameter via the linear response formalism with application to magnetic transition metals and alloys," Phys. Rev. B **87**, 014430.

Marti, X., *et al.*, 2014, "Room-temperature antiferromagnetic memory resistor", Nat. Mater. **13**, 367.

Marfunin, A. S., 1979, "Spectroscopy, Luminescence and Radiation Centers in Minerals", p. 92 (Springer, New York).

Martinez, E., S. Emori, N. Perez, L. Torres, and G. S. D. Beach, 2014, "Current-driven dynamics of Dzyaloshinskii domain walls in the presence of in-plane fields: Full micromagnetic and one-dimensional analysis", J. Appl. Phys. **115**, 213909.

Maruyama, T., *et al.*, 2009, "Large voltage-induced magnetic anisotropy change in a few atomic layers of iron", Nat. Nano. **4**, 158.

Marynowski M., W. Franzen, M. El-Batanouny, and V. Staemmler, 1999, "Observation of an extraordinary antiferromagnetic transition on the NiO(100) surface by metastable helium atom diffraction", Phys. Rev. B **60** 6053.

Mathews, M., F. M. Postma, J. C. Lodder, R. Jansen, G. Rijnders, and D. H. A. Blank, 2005, "Step-induced uniaxial magnetic anisotropy of $La_{0.67}Sr_{0.33}MnO_3$ thin films", Appl. Phys. Lett. **87**, 242507.

Mathias, S., *et al.*, 2012, "Probing the timescale of the exchange interaction in a ferromagnetic alloy", Proc. Natl. Acad. Sci. USA **109**, 4792.

Mathon, J., and A. Umerski, 2001, "Theory of tunneling magnetoresistance of an epitaxial Fe/MgO/Fe(001) junction", Phys. Rev. B **63**, 220403(R).

Matthews, J. A., and A. E. Blakeslee, 1974, "Defects in epitaxial multilayers. I. Misfit dislocations", J. Cryst. Growth **27**, 118.

May, S. J., A. B. Shah, S. G. E. te Velthuis, M. R. Fitzsimmons, J. M. Zuo, X. Zhai, J. N. Eckstein, S. D. Bader, and A. Bhattacharya, 2008, "Magnetically asymmetric interfaces in a $LaMnO_3/SrMnO_3$ superlattice due to structural asymmetries", Phys. Rev. B. **77**, 174409.

McGuire, T., and R. Potter, 1975, "Anisotropic Magnetoresistance in Ferromagnetic *3d* Alloys", IEEE

Trans. Mag. **11**, 1018.

Meckler, S., N. Mikuszeit, A. Preßler, E. Y. Vedmedenko, O. Pietzsch, and R. Wiesendanger, 2009, "Real-Space Observation of a Right-Rotating Inhomogeneous Cycloidal Spin Spiral by Spin-Polarized Scanning Tunneling Microscopy in a Triple Axes Vector Magnet", Phys. Rev. Lett. **103**, 157201.

Medapalli, R., I. Razdolski, M. Savoini, A. R. Khorsand, A. M. Kalashnikova, A. Tsukamoto, A. Itoh, A. Kirilyuk, A. V. Kimel, and T. Rasing, "The role of magnetization compensation point for efficient ultrafast control of magnetization in $Gd_{24}Fe_{66.5}Co_{9.5}$ alloy", Euro. Phys. J. B **86**, 183.

Meier, D., T. Kuschel, L. Shen, A. Gupta, T. Kikkawa, K. Uchida, E. Saitoh, J.-M. Schmalhorst, and G. Reiss, 2013a, "Thermally driven spin and charge currents in thin $NiFe_2O_4$/Pt films", Phys. Rev. B **87**, 054421.

Meier, D., D. Reinhardt, M. Schmid, C. H. Back, J.-M. Schmalhorst, T. Kuschel, and G. Reiss, 2013b, "Influence of heat flow directions on Nernst effects in Py/Pt bilayers", Phys. Rev. B **88**, 184425.

Meier, D., D. Reinhardt, M. van Straaten, C. Klewe, M. Althammer, M. Schreier, S. T. B. Goennenwein, A. Gupta, M. Schmid, C. H. Back, J.-M. Schmalhorst, T. Kuschel, and G. Reiss, 2015, "Longitudinal spin Seebeck effect contribution in transverse spin Seebeck effect experiments in Pt/YIG and Pt/NFO", Nat. Commun. **6**, 8211.

Meier, G., M. Bolte, R. Eiselt, B. Krüger, D.-H. Kim, and P. Fischer, 2007, "Direct Imaging of Stochastic Domain-Wall Motion Driven by Nanosecond Current Pulses", Phys. Rev. Lett. **98**, 187202.

Meiklejohn, W. H. and C. P. Bean, 1956, "New Magnetic Anisotropy", Phys. Rev. **102**, 1413.

Mekonnen, A., A. R. Khorsand, M. Cormier, A. V. Kimel, A. Kirilyuk, A. Hrabec, L. Ranno, A. Tsukamoto, A. Itoh, and T. Rasing, 2013, "Role of the inter-sublattice exchange coupling in short-laser-pulse-induced demagnetization dynamics of GdCo and GdCoFe alloys", Phys. Rev. B **87**, 180406(R).

Mellnik, A. R., *et al.*, 2014, "Spin-transfer torque generated by a topological insulator", Nature **511**, 449.

Melnikov, A., I. Radu, A. Povolotskiy, T. Wehling, A. Lichtenstein, and U. Bovensiepen, 2008, "Ultrafast dynamics at lanthanide surfaces: microscopic interaction of the charge, lattice and spin subsystems", J. Phys. D-Appl. Phys. **41**, 164004.

Melnikov, A., I. Razdolski, T. Wehling, E. Papaioannou, V. Roddatis, P. Fumagalli, O. Aktsipetrov, A. Lichtenstein, and U. Bovensiepen, 2011, "Ultrafast Transport of Laser-Excited Spin-Polarized Carriers in Au/Fe/MgO(001)", Phys. Rev. Lett. **107**, 076601.

Melnikov, A., A. Alekhin, D. Bürstel, D. Diesing, T.O. Wehling, I. Rungger, M. Stamenova, S. Sanvito, and U. Bovensiepen. 2015 "Ultrafast Non-local Spin Dynamics in Metallic Bi-layers by Linear and Non-linear Magneto-optics", ULTRAFAST MAGNETISM I Book Series: Springer Proceedings in Physics **159**, 34. J.-Y. Bigot et al. (eds), Ultrafast Magnetism I, Springer Proceedings in Physics Vol. 159 (Springer International Publishing Switzerland 2015).

Mendes, J. B. S., R. O. Cunha, O. Alves Santos, P. R. T. Ribeiro, F. L. A. Machado, R. L. Rodríguez-Suárez, A. Azevedo, and S. M. Rezende, 2014, “Large inverse spin Hall effect in the antiferromagnetic metal $Ir_{20}Mn_{80}$”, Phys. Rev. B **89**, 140406.

Mentink, J. H., J. Hellsvik, D. V. Afanasiev, B. A. Ivanov, A. Kirilyuk, A. V. Kimel, O. Eriksson, M. I. Katsnelson, and T. Rasing, 2012, “Ultrafast Spin Dynamics in Multisublattice Magnets”, Phys. Rev. Lett. **108**, 057202.

Mermin, N.D., 1979, “The topological theory of defects in ordered media”, Rev. Mod. Phys. **51**, 591.

Meyer, A. J. P., and G. Asch, 1961, “Experimental g' and g Values of Fe, Co, Ni, and Their Alloys”, J. Appl. Phys. **32**, S330.

Meynell, S. A., M. N. Wilson, H. Fritzsche, A. N. Bogdanov, and T. L. Monchesky, 2014, “Surface twist instabilities and skyrmion states in chiral ferromagnets”, Phys. Rev. B **90**, 014406.

Miao, B. F., *et al.*, 2014, "Experimental realization of two-dimensional artificial skyrmion crystals at room temperature”, Phys. Rev. B **90**, 174411.

Michalski, S., J. Zhou, R. Skomski, and R. D. Kirby, 2007, “Coupled precession modes in indirect exchange-coupled [Pt⁄Co]–Co thin films”, J. Appl. Phys. **101** 09D115.

Michels, A., 2014, "Magnetic small-angle neutron scattering of bulk ferromagnets”, J. Phys.: Condens. Matter **26**, 383201.

Milde, P., *et al.*, 2013, "Unwinding of a skyrmion lattice by magnetic monopoles”, Science **340**, 1076-1080.

Miller, T., and T.-C. Chiang, 1992, Phys. Rev. Lett. **68**, 3339.

Millis, A. J., T. Darling, and A. Migliori, 1998, “Quantifying strain dependence in "colossal" magnetoresistance manganites”, J. Appl. Phys. **83**, 1588.

Mills, D. L., 1971, "Surface effects in magnetic crystals near the ordering temperature", Phys. Rev. B **3**, 3887.

Min, B. I. and Y-R Jang, 1991, “The effect of the spin-orbit interaction on the electronic structure of magnetic materials”, J. Phys.: Condens. Matter. **3,** 5131.

Miron, I. M., K. Garello, G. Gaudin, P.-J. Zermatten, M.V. Costache, S. Auffret, S. Bandiera, B. Rodmacq, A. Schuhl, and P. Gambardella, 2011a, “Perpendicular switching of a single ferromagnetic layer induced by in-plane current injection”, Nature **476**, 189.

Miron, I. M., G. Gaudin, S. Auffret, B. Rodmacq, A. Schuhl, S. Pizzini, J. Vogel, and P. Gambardella, 2010, “Current-driven spin torque induced by the Rashba effect in a ferromagnetic metal layer”, Nat. Mater. **9**, 230.

Miron, I. M., T. Moore, *et al.*, 2011b, "Fast current-induced domain-wall motion controlled by the Rashba effect", Nat. Mater. **10**, 419-423.

Miron, I. M., P. J. Zermatten, G. Gaudin, S. Auffret, B. Rodmacq, and A. Schuhl, 2009, "Domain Wall Spin Torquemeter", Phys. Rev. Lett. **102**, 137202.

Miwa, S., *et al.*, 2014, "Highly sensitive nanoscale spin-torque diode", Nat. Mater. **13**, 50.

Miyazaki, T., and N. Tezuka, 1995, "Giant magnetic tunneling effect in Fe/$Al_2O_3$/Fe junction", J. Magn. Magn. Mater. **139,** L231.

Mogi, M., R. Yoshimi, A. Tsukazaki, K. Yasuda, Y. Kozuka, K. S. Takahashi, M. Kawasaki, and Y. Tokura, 2015, "Magnetic modulation doping in topological insulators toward higher-temperature quantum anomalous Hall effect", Appl. Phys. Lett. **107**, 182401.

Mohseni, S. M., S. R. Sani, J. Persson, T. N. Anh Nguyen, S. Chung, Ye. Pogoryelov, P. K. Muduli, E. Iacocca, A. Eklund, R. K. Dumas, S. Bonetti, A. Deac, M. A. Hoefer, and J. Akerman, 2013, "Spin torque-generated magnetic droplet soliton," Science **339**, 1295.

Monso, S., B. Rodmacq, S. Auffret, G. Casali, F. Fettar, B. Gilles, B. Dieny, and P. Boyer, 2002, "Crossover from in-plane to perpendicular anisotropy in Pt/CoFe/$AlO_x$ sandwiches as a function of Al oxidation: A very accurate control of the oxidation of tunnel barriers", Appl. Phys. Lett. **80**, 4157.

Moodera, J. S., L. R. Kinder, T. M. Wong, and R. Meservey, 1995, "Large Magnetoresistance at Room Temperature in Ferromagnetic Thin Film Tunnel Junctions", Phys. Rev. Lett. **74**, 3273.

Moodera, J. S., and G. Mathon, 1999, "Spin polarized tunneling in ferromagnetic junctions", J. Magn. Magn. Mater. **200**, 248.

Moon, E. J., P. V. Balachandran, B. J. Kirby, D. J. Keavney, R. J. Sichel-Tissot, C. M. Schleputz, E. Karapetrova, X. M. Cheng, J. M. Rondinelli, and S. J. May, 2014a, "Effect of interfacial octahedral behavior in ultrathin manganite films", Nano Lett. **14**, 2509.

Moon, E. J., R. Colby, Q. Wang, E. Karapetrova, C. M. Schlepütz, M. R. Fitzsimmons, and S. J. May, 2014b, "Spatial control of functional properties *via* octahedral modulations in complex oxide superlattices", Nat. Commun. **5**, 5710.

Moore, T. A., I. M. Miron, G. Gaudin, G. Serret, S. Auffret, B. Rodmacq, A. Schuhl, S. Pizzini, J. Vogel, and M. Bonfim, 2008, "High domain wall velocities induced by current in ultrathin Pt/Co/AlOx wires with perpendicular magnetic anisotropy", Appl. Phys. Lett. **93**, 262504.

Moreau-Luchaire, C., *et al.*, 2016, "Additive interfacial chiral interaction in multilayers for stabilization of small individual skyrmions at room temperature", Nat. Nanotechn. **11**, 444.

Moriya, T., 1960a, "Anisotropic Superexchange Interaction and Weak Ferromagnetism", Phys. Rev. **120**, 91.

Moriya, T., 1960b, "New Mechanism of Anisotropic Superexchange Interaction", Phys. Rev. Lett. **4**, 228.

Moriyama, T., S. Takei, M. Nagata, Y. Yoshimura, N. Matsuzaki, T. Terashima, Y. Tserkovnyak, and T. Ono, 2015, “Anti-damping spin transfer torque through epitaxial nickel oxide”, Appl. Phys. Lett. **106**,162406.

Morrish, A. H., 2001, *The Physical Principles of Magnetism*, (Wiley-IEEE Press, New York).

Moruzzi, V. L., J. F. Janak, A. R. Williams, 1978, *Calculated Electronic Properties of Metals*, (Pergamon Press, New York).

Mosendz, O., J. E. Pearson, F. Y. Fradin, G. E. W. Bauer, S. D. Bader, and A. Hoffmann, 2010, “Quantifying spin Hall angles from spin pumping: Experiments and theory”, Phys. Rev. Lett. **104**, 046601.

Moser, A., K. Takano, D. T. Margulies, M. Albrecht, Y. Sonobe, Y. Ikeda, S. Sun, S. and E. E. Fullerton, 2002, “Magnetic recording: advancing into the future”, J. Phys. D.: Appl. Phys. **35**, R157.

Moser, J., A. Matos-Abiague, D. Schuh, W. Wegscheider, J. Fabian, and D. Weiss, 2007, “Tunneling Anisotropic Magnetoresistance and Spin-Orbit Coupling in Fe/GaAs/Au Tunnel Junctions”, Phys. Rev. Lett. **99**, 056601.

Mott, N. F., 1936, “The Electrical Conductivity of Transition Metals”, Proc. R. Soc. A **153**, 699.

Moutafis, C., S. Komineas, and J. A. C. Bland, 2009, ““Dynamics and switching processes for magnetic bubbles in nanoelements”, Phys. Rev. B **79**, 224429.

Mross, D. F., and H. Johannesson, 2009, “Two-impurity Kondo model with spin-orbit interactions”, Phys. Rev. B **80**, 155302.

Mühlbauer, S., 2009, "Skyrmion lattice in a chiral magnet”, Science **323**, 915.

Müller, G. M., *et al.*, 2009, “Spin polarization in half-metals probed by femtosecond spin excitation”, Nat. Mater. **8**, 56.

Muller, O., and R. Roy, 1974, *The Major Ternary Structural Families* (New York, Springer Verlag).

Münzer, W., *et al.*, 2010, "Skyrmion lattice in the doped semiconductor $Fe_{1-x}Co_xSi$”, Phys. Rev. B **81**, 041203.

Murakami, S., N. Nagaosa, and S.-C. Zhang, 2003, “Dissipationless Quantum Spin Current at Room Temperature”, Science **301**, 1348

Myers, E. B., D. C. Ralph, J. A. Katine, R. N. Louie, and R. A. Buhrman, 1999, “Current-Induced Switching of Domains in Magnetic Mutlilayer Devices”, Science **285**, 867.

Nagaosa, N., J. Sinova, S. Onoda, A. H. MacDonald, and N. P. Ong, 2010, “Anomalous Hall effect”, Rev. Mod. Phys. **82**, 1539.

Nagaosa, N., and Y. Tokura, 2013, "Topological properties and dynamics of magnetic skyrmions”, Nat. Nanotech. **8**, 899-911.

Nakamura, K., R. Shimabukuro, Y. Fujiwara, T. Akiyama, T. Ito, and A. J. Freeman, 2009, “Giant modification of the magnetocrystalline anisotropy in transition-metal monolayers by an external electric field”, Phys. Rev. Lett. **102**, 187201.

Nanda, B. R. K., S. Satpathy, and M. S. Springborg, 2007, “Electron Leakage and Double-Exchange Ferromagnetism at the Interface between a Metal and an Antiferromagnetic Insulator: $CaRuO_3/CaMnO_3$”, Phys. Rev. Lett. **98**, 216804.

Néel, L., 1954, “Anisotropie magnétique superficielle et surstructures d'orientation”, J. Physique Rad. **15** 225.

Néel, L., 1962a, “A magnetostatic problem concerning ferromagnetic films”, C. R. Hebd. Seances Acad. Sci. **255**, 1545.

Néel, L., 1962b, “A new method of coupling the magnetizations of two thin ferromagnetic films”, C. R. Hebd. Seances Acad. Sci. **255**, 1676.

Nembach, H. T., J. M. Shaw, M. Weiler, E. Jué, and T. J. Silva, 2015, “Linear relation between Heisenberg exchange and interfacial Dzyaloshinkii-Moriya interaction in metal films”, Nat. Phys. **11**, 825.

Neubauer, A., C. Pfleiderer, B. Binz, A. Rosch, R. Ritz, P. G. Niklowitz, and P. Böni, 2009, "Topological Hall Effect in the A Phase of MnSi”, Phys. Rev. Lett. **102**, 186602.

Neupane, M., *et al.*, 2014, “Observation of quantum-tunneling-modulated spin texture in ultrathin topological insulator $Bi_2Se_3$ films”, Nat. Commun. **5**, 3841.

Ney, A., P. Poulopoulos and K. Baberschke, 2001, "Surface and interface magnetic moments of Co/Cu(001)", Europhys. Lett. 54, 820.

Niemier, M. T., *et al.*, 2011, "Nanomagnet logic: progress toward system-level integration”, J. Phys.: Condens. Matter **23**, 493202.

Niizeki, T., T. Kikkawa, K. Uchida, M. Oka, K. Z. Suzuki, H. Yanagihara, E. Kita, and E. Saitoh, 2015, “Observation of longitudinal spin-Seebeck effect in cobalt-ferrite epitaxial thin film”, AIP Adv. **5**, 053603.

Niranjan M. K., C.-G. Duan, S. S. Jaswal, and E. Y. Tsymbal, 2010, “Electric Field Effect on Magnetization at the Fe/MgO(001) Interface”, Appl. Phys. Lett. **96**, 222504.

Nogues, J., and I. K. Schuller, 1999, “Exchange bias”, J. Magn. Magn. Mater. **192**, 203.

Nogues, J., J. Sort, V. Langlais, V. Skumryev, S. Surinach, J. S. Munoz, and M. D. Baro, 2005, “Exchange bias in nanostructures”, Phys. Rep. **422**, 65.

Nozaki, T., *et al.*, 2012, "Electric-field-induced ferromagnetic resonance excitation in an ultrathin ferromagnetic metal layer”, Nat. Phys. **8**, 491.

Nuñez, A. S., R. A. Duine, P. Haney, and A. H. MacDonald, 2006, "Theory of spin torques and giant

magnetoresistance in antiferromagnetic metals”, Phys. Rev. B **73**, 214426.

O’Brien, L., M. J. Erickson, D. Spivak, H. Ambaye, R. J. Goyette, V. Lauter, P. A. Crowell, and C. Leighton, 2014, “Kondo physics in non-local metallic spin transport devices”, Nat. Commun. **5**, 3927.

O’Handley, R. C., 2000, *Modern Magnetic Materials: Principles and Applications* (Wiley, New York).

Ohkawa, F. J., and Y. Uemura, 1974, “Quantized surface states of narrow-gap semiconductors”, J. Phys. Soc. Jpn. **37**, 1325.

Ohno, H., D. Chiba, F. Matsukura, T. Omiya, E. Abe, T. Dietl, Y. Ohno, and K. Ohtani, 2000, “Electric-field control of ferromagnetism”, Nature **408**, 944.

Ohno, H., T. Endoh, T. Hanyu, N. Kasai, and S. Ikeda, 2010, "Magnetic Tunnel Junction for Nonvolatile CMOS Logic”, in IEDM 2010, DOI: 10.1109/IEDM.2010.5703329.

Ohno, Y., D. K. Young, B. Beschoten, F. Matsukara, H. Ohno, and D. D. Awschalom, 1999, “Electrical spin injection in a ferromagnetic semiconductor heterostructure”, Nature **402**, 790.

Oike, H., A. Kikkawa, N. Kanazawa, Y. Taguchi, M. Kawasaki, Y. Tokura, and F. Kagawa, 2015, “Interplay between topological and thermodynamic stability in a metastable magnetic skyrmion lattice”, Nature Phys. **12,** 62.

Onose, Y., Y. Okamura, S. Seki, S. Ishiwata, and Y. Tokura, 2012, "Observation of Magnetic Excitations of Skyrmion Crystal in a Helimagnetic Insulator $Cu_2OSeO_3$”, Phys. Rev. Lett. **109**, 037603.

Opitz, J., P. Zahn, J. Binder, and I. Mertig, 2001, “Ab initio calculation of the interlayer exchange coupling in Fe/Au multilayers: The role of impurities at the interface”, Phys. Rev. B **63**, 094418.

Oppeneer, P.M., and A. Liebsch, 2004, “Ultrafast demagnetization in Ni: theory of magneto-optics for non-equilibrium electron distributions”, J. Phys.: Condens. Matter **16**, 5519.

Ortega, J. E., and F. J. Himpsel, 1992, “Quantum well states as mediators of magnetic coupling in superlattices”, Phys. Rev. Lett. **69**, 844.

Ostler, T. A., *et al.* 2012, “Ultrafast heating as a sufficient stimulus for magnetization reversal in a ferrimagnet”, Nat. Commun. **3**, 666.

Pai, C.-F., L. Liu, Y. Li, H. W. Tseng, D. C. Ralph, and R. A. Buhrman, 2012, “Spin transfer torque devices utilizing the giant spin Hall effect of tungsten,” Appl. Phys. Lett. **101**, 122404.

Pai, C.-F.; M.-H. Nguyen, C. Belvin, L. H. Vilela-Leão, D. C. Ralph, and R. A. Buhrman, 2014, “Enhancement of perpendicular magnetic anisotropy and transmission of spin-Hall-effect-induced spin currents by a Hf spacer layer in W/Hf/CoFeB/MgO structures”, Appl. Phys. Lett. **104**, 082407.

Pai, C.-F., Y. Ou, L. H. Vilela-Leao, D. C. Ralph, and R. A. Buhrman, 2015, “Dependence of the efficiency of spin Hall torque on the transparency of Pt/ferromagnetic layer interfaces,” Phys. Rev. B **92**, 064426.

Papanicolaou, N., and T. N. Tomaras, 1991, "Dynamics of magnetic vortices", Nuc. Phys. B **360**, 425.

Pappas, C., E. Lelièvre-Berna, P. Falus, P. M. Bentley, E. Moskvin, S. Grigoriev, P. Fouquet, and B. Farago, 2009, "Chiral Paramagnetic Skyrmion-like Phase in MnSi", Phys. Rev. Lett. **102**, 197202.

Papusoi, C., B. Delat, B. Rodmacq, D. Houssameddine, J.-P. Michel, U. Ebels, R. C. Sousa, L. Buda-Prejbeanu, and B. Dieny, 2009, "100 ps precessional spin-transfer switching of a planar magnetic random access memory cell with perpendicular spin polarizer", Appl. Phys. Lett. **95**, 072506.

Park, B. G., *et al.*, 2011, “A spin-valve-like magnetoresistance of an antiferromagnet-based tunnel junction”, Nat. Mater. **10**, 347.

Park, B. G., J. Wunderlich, D. A. Williams, S. J. Joo, K. Y. Jung, K. H. Shin, K. Olejník, A. B. Shick, and T. Jungwirth, 2008, “Tunneling Anisotropic Magnetoresistance in Multilayer-(Co/Pt)/$AlO_x$/Pt Structures”, Phys. Rev. Lett. **100**, 087204.

Parkin, S. S. P., M. Hayashi, and L. Thomas, 2008, "Magnetic Domain-Wall Racetrack Memory", Science **320**, 190.

Parkin, S. S. P., X. Jiang, C. Kaiser, A. Panchula, K. Roche, and M. Samant, 2003, “Magnetically engineered spintronic sensors and memory”, Proc. IEEE **91**, 661.

Parkin, S. S. P., C. Kaiser, A. Panchula, P. M. Rice, B. Hughes, M. Samant, and S. H. Yang, 2004, “Giant tunnelling magnetoresistance at room temperature with MgO (100) tunnel barriers”, Nat. Mater. **3**, 862.

Parkin, S. S. P., N. More, and K. P. Roche, 1990, “Oscillations in exchange coupling and magnetoresistance in metallic superlattice structures: Co/Ru, Co/Cr, and Fe/Cr”, Phys. Rev. Lett. **64**, 2304.

Payne, A. P., and B. M. Clemens, 1993, “Influence of roughness distributions and correlations on X-ray diffraction from superlattices”, Phys. Rev. B. **47**, 2289.

Pennycook, S. J., H. Zhou, M. F. Chisholm, A. Y. Borisevich, M. Varela, J. Gazquez, T. J. Pennycook, and J. Narayan, 2013, “Misfit accommodation in oxide thin film heterostructures”, Acta. Mater. **61**, 2725.

Pesin, D., and A. H. MacDonald, 2012, “Spintronics and pseudospintronics in graphene and topological insulators”, Nat. Mater. **11**, 409.

Petford-Long, A.K., and M. De Graef, 2012, “Lorentz microscopy”, in *Characterization of Materials*, edited by Kaufmann, E.N. (Wiley, New York), p. 1787-1801.

Petit-Watelot, S., J.-V. Kim, A. Ruotolo, R. M. Otxoa, K. Bouzehouane, J. Grollier, A. Vansteenkiste, B. Van de Wiele, V. Cros, and T. Devolder, 2012, “Commensurability and chaos in magnetic vortex oscillations”, Nat. Phys. **8**, 682.

Petrova, O., and O. Tchernyshyov, 2011, "Spin waves in a skyrmion crystal", Phys. Rev. B **84**, 214433.

Pfeiffer, A., S. Hu, R. M. Reeve, A. Kronenberg, M. Jourdan, T. Kimura, and M. Klaui, 2015, "Spin currents injected electrically and thermally from highly spin polarized $Co_2MnSi$", Appl. Phys. Lett. **107**, 082401.

Phatak, C., A. K. Petford-Long, and M. De Graef, 2010, "Three-Dimensional Study of the Vector Potential of Magnetic Structures", Phys. Rev. Lett. **104**, 253901.

Phatak, C., A. K. Petford-Long, and O. Heinonen, 2012, "Direct Observation of Unconventional Topological Spin Structure in Coupled Magnetic Discs", Phys. Rev. Lett. **108**, 067205.

Pi, U. H., K. W. Kim, J. Y. Bae, S. C. Lee, Y. J. Cho, K. S. Kim, and S. Seo, 2010, "Tiliting the spin orientation induced by Rashba effect in ferromagnetic metal layer", Appl. Phys. Lett. **97**, 162507.

Pizzini, S., *et al.*, 2014, "Chirality-Induced Asymmetric Magnetic Nucleation in Pt/Co/AlOx Ultrathin Microstructures", Phys. Rev. Lett. **113**, 047203.

Pleimling M. and W. Selke, 1998, "Critical phenomena at perfect and non-perfect surfaces" Eur. Phys. J. B **1** 385.

Pleimling M., 2004, "Critical phenomena at perfect and non-perfect surfaces", J. Phys. A: Math. Gen. **37** R79.

Pollard, S. D., L. Huang, K. S. Buchanan, D. A. Arena, and Y. Zhu, 2012, "Direct dynamic imaging of non-adiabatic spin torque effects", Nat. Commun. **3**, 1028.

Porat, A., S. Bar-Ad, and I. K. Schuller, 2009 "Novel laser-induced dynamics in exchange-biased systems", Europhys. Lett. **87**, 67001.

Pothuizen J. J. M., O. Cohen and G. A. Sawatzky, 1995, "Surface Band Gaps and Superexchange Interaction in Transition Metal Oxides", Mater. Res. Soc. Symp. Proc. **401**, 501

Prakash, A., J. Brangham, F. Yang and J. P. Heremans, 2016, "Spin Seebeck effect through antiferromagnetic NiO", Phys. Rev. B **94**, 014427.

Pratt, W.P., Jr., S.F. Lee, J.M. Slaughter, R. Loloe, P.A. Schroeder, and J. Bass, 1991, "Perpendicular giant magnetoresistances of Ag/Co multilayers", Phys. Rev. Lett. **66**, 3060.

Prejbeanu, I. J., S. Bandiera, J. Alvarez-Hérault, R. C. Sousa, B. Dieny, and J.-P. Nozières, 2013, "Thermally assisted MRAMs: ultimate scalability and logic functionalities", J. Phys. D: Appl. Phys. **46**, 074002.

Prenat, G., B. Dieny, W. Guo, M. El Baraji, V. Javerliac, and J.-P. Nozières, 2009, "Beyond MRAM, CMOS/MTJ Integration for Logic Components", IEEE Trans. Magn. **45**, 3400.

Primanayagam, S. N., 2007, "Perpendicular recording media for hard disk drives", J. Appl. Phys. **102**, 011301.

Prinz, G. A, 1999, "Magnetoelectronics applications", J. Magn. Magn. Mater. **200**, 57.

Qi, X.-L., and S.-C. Zhang, 2011, “Topological insulators and superconductors”, Revs. Mod. Phys. **83**, 1057.

Qin, H. J., Kh. Zakeri, A. Ernst, T.-H. Chuang, Y.-J. Chen, Y. Meng, and J. Kirschner, 2013, "Magnons in ultrathin ferromagnetic films with a large perpendicular magnetic anisotropy", Phys. Rev. B **88** 020404.

Qiu, Z. Q., and S. D. Bader, 1999, “Surface magneto-optic Kerr effect (SMOKE)”, J. Magn. Magn. Mater. **200**, 664.

Qu, D., S. Y. Huang, J. Hu, R. Wu, and C. L. Chien, 2013, “Intrinsic Spin Seebeck Effect in Au/YIG”, Phys. Rev. Lett. **110**, 067206.

Radaelli P. G., G. Iannone, M. Marezio, H. Y. Hwang, S. W. Cheong, J. D. Jorgensen, and D. N. Argyriou, 1997, “Structural effects on the magnetic and transport properties of perovskite $A_{1-x}A_x$'$MnO_3$ ($x$=0.25, 0.30)”, Phys. Rev. B **56**, 8265.

Radu, I., *et al.*, 2011, “Transient ferromagnetic-like state mediating ultrafast reversal of antiferromagnetically coupled spins”, Nature **472**, 205.

Ralph, D. C., and M. D. Stiles, 2008, "Spin transfer torques", J. Magn. Magn. Mater. **320**, 1190.

Ramesh, R., and D. Schlom, 2008, “Whither oxide electronics?”, MRS Bull. **33**, 1006.

Ramos, R, *et al.*, 2013, “Observation of the spin Seebeck effect in epitaxial $Fe_3O_4$thin films”, Appl. Phys. Lett. **102**, 072413.

Rashba, E., 1960, “Properties of semiconductors with an extremum loop. 1. Cyclotron and combinational resonance in a magnetic field perpendicular to the plane of the loop”, Sov. Phys. Solid State **2,** 1109. (English translation of Fiz. Tverd. Tela **2**, 1224 (1960)

Rashba, E., and V. I. Sheka, 1959, “Symmetry of Bands with Spin-Orbit Interaction Included”, Fiz. Tverd. Tela: Collected Papers **2**, 62.

Ren, Y., H. Zhao, Z. Zhang, and Q. Y. Jin, 2008, “Ultrafast optical modulation of exchange coupling in FePt/CoFe composite structure”, Appl. Phys. Lett. **92**, 162513.

Renard, J.-P., P. Bruno, R. Mégy, B. Bartenlian, P. Beauvillain, C. Chappert, C. Dupas, E. Kolb, M. Mulloy, J. Prieur, P. Veillet, and E. Vélu, 1996, “Inverse giant magnetoresistance”, J. Appl. Phys. **79**, 5270.

Revaz, B., M.-C. Cyrille, B. Zink, I. K. Schuller and F. Hellman, 2002, “Enhancement of the electronic contribution to the low temperature specific heat of an Fe/Cr magnetic multilayer”, Phys. Rev. B **65**, 094417.

Rezende, S. M., R. L. Rodríguez-Suárez, R. O. Cunha, A. Rodrigues, F. L. A. Machado, G. A. F. Guerra, J. C. L. Ortiz, and A. Azevedo, 2014, “Magnon spin-current theory for the longitudinal spin-Seebeck effect”, Phys. Rev. B **89**, 014416.

Rhie, H. S., H. A. Dürr, and W. Eberhardt, 2003, “Femtosecond electron and spin dynamics in Ni/W (110) films”, Phys. Rev. Lett. **90**, 247201.

Rice, W. D., P. Ambwani, M. Bombeck, J. D. Thompson, C. Leighton, and S. A. Crooker, 2014, “Persistent optically induced magnetism in oxygen-deficient strontium titanate”, Nat. Mater. **13**, 481.

Richardella, A., A. Kandala, and N. Samarth, 2015, “Topological Insulator Thin Films and Heterostructures: Epitaxial Growth, Transport, and Magnetism” in *Topological insulators: Fundamentals and perspectives* edited by F. Ortmann, S. Roche S. O. Valenzuela (Wiley, Weinheim) p. 295.

Rippard, W. H., M. R. Pufall, S. Kaka, S. E. Russek, and T. J. Silva, 2004, "Direct-Current Induced Dynamics in Co90Fe10/Ni80Fe20 Point Contacts”, Phys. Rev. Lett. **92**, 027201.

Rodmacq, B., S. Auffret, B. Dieny, S. Monso, and P. Boyer, 2003, “Crossovers from in-plane to perpendicular anisotropy in magnetic tunnel junctions as a function of the barrier degree of oxidation”, J. Appl. Phys. **93**, 7513.

Rodmacq, B., A. Manchon, C. Ducruet, S. Auffret, and B. Dieny, 2009, “Influence of thermal annealing on the perpendicular magnetic anisotropy of Pt/Co/$AlO_x$ trilayers”, Phys. Rev. B **79**, 024423.

Rodríguez, L. A., *et al.*., 2013, "Quantitative *in situ* magnetization reversal studies in Lorentz microscopy and electron holography”, Ultramicroscopy **134**, 144.

Rohart, S., and A. Thiaville, 2013, "Skyrmion confinement in ultrathin film nanostructures in the presence of Dzyaloshinskii-Moriya interaction”, Phys. Rev. B **88**, 184422.

Rohart, S., J. Miltat, and A. Thiaville, 2016, “Path to collapse for an isolated skyrmion”, Phys. Rev. B, **93**, 214412.

Rojas-Sánchez, J. C., L. Vila, G. Desfonds, S. Gambarelli, J. P. Attané, J. M. De Teresa, C. Magén and A. Fert, 2013, “Spin-to-charge conversion using Rashba coupling at the interface between non-magnetic materials,” Nat. Comm. **4**, 2944.

Rojas-Sánchez, J. C., S. Oyarzún, Y. Fu, A. Marty, C. Vergnaud, S. Gambarelli, L. Vila, M. Jamet, Y. Ohtsubo, A. Taleb-Ibrahimi, P. Le Fèvre, F. Bertran, N. Reyren, J.-M. George, and A. Fert, 2016, "Spin to Charge Conversion at Room Temperature by Spin Pumping into a New Type of Topological Insulator: α-Sn Films", Phys. Rev. Lett. **116**, 096602.

Romming, N., C. Hanneken, M. Menzel, J. E. Bickel, B. Wolter, K. von Bergmann, A. Kubetzka, and R. Wiesendanger, 2013, "Writing and deleting single magnetic skyrmions”, Science **341**, 636.

Romming, N., A. Kubetzka, C. Hanneken, K. von Bergmann, and R. Wiesendanger, 2015, “Field-dependent size and shape of single magnetic skyrmions”, Phys. Rev. Lett. **114**, 177203.

Rondin, L., J. P. Tetienne, S. Rohart, A. Thiaville, T. Hingant, P. Spinicelli, J. F. Roch, and V. Jacques, 2013, "Stray-field imaging of magnetic vortices with a single diamond spin”, Nat. Commun. **4**,

2279.

Rondinelli J. M., S. J. May, and J. W. Freeland, 2012, “Control of octahedral connectivity in perovskite oxide heterostructures: An emerging route to multifunctional materials discovery”, MRS Bull. **37**, 261.

Rondinelli, J. M., and N. A. Spaldin, 2010, “Substrate coherency driven octahedral rotations in perovskite oxide films”, Phys. Rev. B **82**, 113402.

Rondinelli, J. M., M. Stengel, and N. A. Spaldin, 2008, “Carrier-mediated magnetoelectricity in complex oxide heterostructures”, Nat. Nano. **3**, 46.

Rößler, U. K., A. N. Bogdanov, and C. Pfleiderer, 2006, "Spontaneous skyrmion ground states in magnetic metals”, Nature **442**, 797.

Roth, T, A. J. Schellekens, S. Alebrand, O. Schmitt, D. Steil, B. Koopmans, M. Cinchetti and M. Aeschlimann, 2012, “Temperature Dependence of Laser-Induced Demagnetization in Ni: A Key for Identifying the Underlying Mechanism”, Phys. Rev. X **2**, 021006.

Rougemaille, N., and A. K. Schmid, 2010, "Magnetic imaging with spin-polarized low-energy electron microscopy”, Eur. Phys. J. Appl. Phys. **50**, 20101.

Rowland, J., S. Banerjee, and M. Randeria, "Skyrmions in Chiral Magnets with Rashba and Dresselhaus Spin-Orbit Coupling", 2016, Phys. Rev. B **93**, 020404.

Roy, K., D. Fan, X. Fong, Y. Kim, M. Sharad, S. Paul, S. Chatterjee, S. Bhunia, and S. Mukhopadhyay, 2015, “Exploring Spin Transfer Torque Devices for Unconventional Computing”, IEEE J. Emerg. Sel. Topics Circ. Syst. **5**, 5.

Ruderman, M. A. and C. Kittel, 1954, “Indirect Exchange Coupling of Nuclear Magnetic Moments by Conduction Electrons,” Phys. Rev. **96**, 99.

Ryu, K.-S., L. Thomas, S.-H. Yang, and S. Parkin, 2013, “Chiral spin torque at magnetic domain walls”, Nat. Nanotech. **8**, 527.

Ryu, K.-S., L. Thomas, S.-H. Yang, and S. S. P. Parkin, 2012, "Current Induced Tilting of Domain Walls in High Velocity Motion along Perpendicularly Magnetized Micron-Sized Co/Ni/Co Racetracks”, Appl. Phys. Exp. **5**, 093006.

Rzhevsky A., B. Krichevtsov, D. Bürgler, and C. Schneider, 2007, “Magnetization dynamics induced by ultrashort optical pulses in Fe/Cr thin films”, Phys. Rev. B **75** 224434.

Saglam, H., W. Zhang, M. B. Jungfleisch, J. Sklenar, J. E. Pearson, J. B. Ketterson, and A. Hoffmann, 2016, “Spin transport through the metallic antiferromagnet FeMn”, Phys. Rev. B **94**, 140412(R).

Sahin, C., and M. E. Flatté, 2015, “Tunable Giant Spin Hall Conductivities in a Strong Spin-Orbit Semimetal: $Bi_{1-x}Sb_x$”, Phys. Rev. Lett. **114**, 107201.

Saitoh, E., M. Ueda, H. Miyajima, and G. Tatara, 2006, “Conversion of spin current into charge current at room temperature: Inverse spin-Hall effect", Appl. Phys. Lett. **88**, 182509.

Salamon, M. B., S. Sinha, J. J. Rhyne, J. E. Cunningham, R. W. Erwin, J. Borchers, and C. P. Flynn, 1986, "Long-range incommensurate magnetic order in a Dy-Y multilayer", Phys. Rev. Lett. **56**, 259.

Salis, G., Y. Kato, K. Ensslin, D. C. Driscoll, A. C. Gossard and D. D. Awschalom, 2001, "Electrical control of spin coherence in semiconductor nanostructures", Nature **414,** 619.

Salvador, P. A., A.-M. Haghiri-Gosnet, B. Mercey, M. Hervieu, and B. Raveau, 1999, "Growth and magnetoresistive properties of $(LaMnO_3)_m(SrMnO_3)_n$ superlattices", Appl. Phys. Lett. **75**, 2638.

Sampaio, J., V. Cros, S. Rohart, A. Thiaville, and A. Fert, 2013, "Nucleation, stability and current-induced motion of isolated magnetic skyrmions in nanostructures", Nat. Nanotech. **8**, 839.

Sanchez-Barriga, J. *et al*., 2016. "Nonmagnetic band gap at the Dirac point of the magnetic topological insulator $(Bi1\text{-}xMnx)_2Se_3$" Nat. Commun. **7**, 10559.

Sani, S., J. Persson, S. M. Mohseni, Ye Pogoryelov, P. K. Muduli, A. Eklund, G. Malm, M. Kall, A. DM interactiontriev, and J. Åkerman, 2013, "Mutually synchronized bottom-up multi-nanocontact spin–torque oscillators", Nat. Commun. **4**, 2731.

Sankey, J. C., P. M. Braganca, A. G. F. Garcia, I. N. Krivorotov, R. A. Buhrman, and D. C. Ralph, 2006, "Spin-transfer-driven ferromagnetic resonance of individual nanomagnets", Phys. Rev. Lett. **96**, 227601.

Sankey, J. C., Y.-T. Cui, J. Z. Sun, J. C. Slonczewski, R. A. Buhrman, and D. C. Ralph, 2008, "Measurement of the spin-transfer-torque vector in magnetic tunnel junctions", Nat. Phys. **4**, 67.

Santos, T. S., B. J. Kirby, S. Kumar, S. J. May, J. A. Borchers, B. B. Maranville, J. Zarestky, S. G. E. te Velthuis, J. van den Brink, and A. Bhattacharya, 2011, "Delta doping of ferromagnetism in antiferromagnetic manganite superlattices", Phys. Rev. Lett. **107**, 167202.

Sato, H., E. C. I. Enobio, M. Yamanouchi, S. Ikeda, S. Fukami, S. Kanai, F. Matsukura, and H. Ohno, 2014, "Properties of magnetic tunnel junctions with a MgO/CoFeB/Ta/CoFeB/MgO recording structure down to junction diameter of 11 nm", Appl. Phys. Lett. **105**, 062403.

Savoini, M, C. Piovera, *et al.*, 2014a, "Bias-controlled ultrafast demagnetization in magnetic tunnel junctions", Phys. Rev. B **89**, 140402.

Savoini, M., A.H. Reid, *et al.*, 2014b, "Attempting nanolocalization of all-optical switching through nanoholes in an Al-mask", Proc. SPIE, Spintronics VII **9167,** 91672C.

Schattschneider, P., S. Rubino, C. Hebert, J. Rusz, J. Kunes, P. Novak, E. Carlino, M. Fabrizioli, G. Panaccione, and G. Rossi, 2006, "Detection of magnetic circular dichroism using a transmission electron microscope", Nature **441**, 486.

Schellekens, A. J. and B. Koopmans, 2013a, "Comparing ultrafast demagnitization rates between competing models for finite temperature magnetism", Phys. Rev. Lett. **110**, 217204.

Schellekens, A. J. and B. Koopmans, 2013b, "Microscopic model for ultrafast magnetization dynamics of multisublattice magnets," Phys. Rev. B **87**, 020407(R).

Schellekens, A. J., W. Verhoeven, T. N. Vader and B. Koopmans, 2013 "Investigating the contribution of superdiffusive transport to ultrafast demagnetization of ferromagnetic thin films", Appl. Phys. Lett. **102**, 252408.

Schellekens, A. J., K. C. Kuiper, R. R. de Wit, R. R and B. Koopmans, 2014, "Ultrafast spin-transfer torque driven by femtosecond pulsed-laser excitation", Nat. Commun. **5**, 4333.

Scherbakov, A. V., A. S. Salasyuk, A. V. Akimov, X. Liu, M. Bombeck, C. Bruggemann, D. R. Yakovlev, V. F. Sapega, J. K. Furdyna, and M. Bayer, 2010, "Coherent Magnetization Precession in Ferromagnetic (Ga,Mn) As Induced by Picosecond Acoustic Pulses", Phys. Rev. Lett. **105**, 117204.

Schierle E., E. Weschke, A. Gottberg, W. Söllinger, W. Heiss, G. Springholz, and G. Kaindl, 2008, "Antiferromagnetic Order with Atomic Layer Resolution In EuTe(111) Films", Phys. Rev. Lett**., 101** 267202.

Schmid, M., S. Srichandan, D. Meier, T. Kuschel, J.-M. Schmalhorst, M. Vogel, G. Reiss, C. Strunk, and C. H. Back, 2013, "Transverse Spin Seebeck Effect versus Anomalous and Planar Nernst Effects in Permalloy Thin Films", Phys. Rev. Lett. **111**, 187201.

Schmidt, G., D. Ferrand, L. W. Molenkamp, A. T. Filip, and B. J. van Wees, 2000, "Fundamental obstacle for electrical spin injection from a ferromagnetic metal into a diffusive semiconductor", Phys. Rev. B., **62**, R4790.

Scholl, A; L. Baumgarten, R. Jacquemin, and W. Eberhardt, 1997, "Ultrafast spin dynamics of ferromagnetic thin films observed by fs spin-resolved two-photon photoemission", Phys. Rev. Lett. **79**, 5146.

Schreier, M, et al., 2015, "Sign of inverse spin Hall voltages generated by ferromagnetic resonance and temperature gradients in yttrium iron garnet platinum bilayers", J. Phys. D:Appl. Phys. **48**, 025001.

Schryer, N. L., and L. R. Walker, 1974, "The motion of 180° domain walls in uniform DC magnetic fields", J. Appl. Phys. **45**, 5406.

Schubert, C., A. Hassdenteufel, P. Matthes, J. Schmidt, M. Helm, R. Bratschitsch, and M. Albrecht, 2014, "All-optical helicity dependent magnetic switching in an artificial zero moment magnet", Appl. Phys. Lett. **104**, 082406.

Schuller, I. K., 1980, "New Class of Layered Materials", Phys. Rev. Lett. **44**, 1597.

Schuller, I. K., S. Kim, and C. Leighton, 1999, "Magnetic superlattices and multilayers", J. Magn. Magn. Mater. **200**, 571.

Schulz, T., R. Ritz, A. Bauer, M. Halder, M. Wagner, C. Franz, C. Pfleiderer, K. Everschor, M. Garst, and A Rosch, 2012, "Emergent electrodynamics of skyrmions in a chiral magnet", Nat. Phys. **8**, 301.
Schwarz, K, 1986, "$CrO_2$ predicted as a half-metallic ferromagnet", J. Phys. F Met. Phys. **16**, L211.
Seifert, T, *et al.*, 2016, "Efficient metallic spintronics emitters of ultrabroadband terahertz radiation", Nat. Photon. **10**, 483.
Seki, S., X. Z. Yu, S. Ishiwata, and Y. Tokura, 2012, "Observation of skyrmions in a multiferroic material", Science **336**, 198.
Seki, S., T. Ideue, M. Kubota, Y. Kozuka, R. Takagi, M. Nakamura, Y. Kaneko, M. Kawasaki, and Y. Tokura, 2015, "Themal generation of spin current in an antiferromagnet", Phys. Rev. Lett. **115**, 266601.
Seu, K. A., and A. C. Reilly, 2008, "Ultrafast laser excitation of spin waves and the permanent modification of the exchange bias interaction in IrMn/Co", J. Appl. Phys. **103,** 07C104.
Shanavas, K. V., Z. S. Popović, and S. Satpathy, 2014, "Theoretical model for Rashba spin-orbit interaction in d electrons", Phys. Rev. B **90**, 165108.
Shi, X., P. Fischer, V. Neu, D. Elefant, J. C.T. Lee, D. A. Shapiro, M. Farmand, T. Tyliszczak, H.-W. Shiu, S. Marchesini, S. Roy, S. D. Kevan, 2016, "Soft x-ray ptychography studies of nanoscale magnetic and structural correlations in thin SmCo5 films"*,* Appl. Phys. Lett. 108, 094103.
Shibata, K., X. Z. Yu, T. Hara, D. Morikawa, N. Kanazawa, K. Kimoto, S. Ishiwata, Y. Matsui and Y. Tokura, 2013, "Towards control of the size and helicity of skyrmions in helimagnetic alloys by spin–orbit coupling," Nature Nanotech. **8**, 723.
Shick, A. B., S. Khmelevskyi, O. N. Mryasov, J. Wunderlich, and T. Jungwirth, 2010, "Spin-orbit coupling induced anisotropy effects in bimetallic antiferromagnets: A route towards antiferromagnetic spintronics", Phys. Rev. B **81**, 212409.
Shindou R., and N. Nagaosa, 2001, "Orbital Ferromagnetism and Anomalous Hall Effect in Antiferromagnets on the Distorted fcc Lattice", Phys. Rev. Lett. **87**, 116801.
Shiomi, Y., K. Nomura, Y. Kajiwara, K. Eto, M. Novak, K. Segawa, Y. Ando, and E. Saitoh, 2014, "Spin-Electricity Conversion Induced by Spin Injection into Topological Insulators", Phys. Rev. Lett. **113**, 196601.
Shiota, Y., S. Miwa, S. Tamaru, T. Nozaki, H. Kubota, A. Fukushima, Y. Suzuki, Y, and S. Yuasa, 2014, "High-output microwave detector using voltage-induced ferromagnetic resonance", Appl. Phys. Lett. **105**, 192408.
Shiota, Y., T. Nozaki, F. Bonell, S. Murakami, T. Shinjo, and Y. Suzuki, 2012, "Induction of coherent magnetization switching in a few atomic layers of FeCo using voltage pulses", Nat. Mater. **11**, 39.
Shiroishi, Y., K. Fukuda, I. Tagawa, H. Iwasaki, S. Takenoiri, H. Tanaka, H. Mutoh, and N. Yoshikawa, 2009, "Future Options for HDD Storage", IEEE Trans. Mag. **45**, 3816.

Siegfried, S.A., E. V. Altynbaev, N. M. Chubova, V. Dyadkin, D. Chernyshov, E. V. Moskvin, D. Menzel, A. Heinemann, A. Schreyer, and S. V. Grigoriev, 2015, “Controlling the Dzyaloshinskii-Moriya interaction to alter the chiral link between structure and magnetism for $Fe_{1-x}Co_xSi$,” Phys. Rev. B **91**, 184406.

Silva, T. J. and W. H. Rippard, 2008, "Developments in nano-oscillators based upon spin-transfer point-contact devices”, J. Magn. Magn. Mater. **320**, 1260.

Sinkovic B., B. Hermsmeier and C. S. Fadley, 1985, “Observation of Spin-Polarized Photoelectron Diffraction”, Phys. Rev. Lett. **55**, 1227**.**

Sinova, J., D. Culcer, Q. Niu, N. A. Sinitsyn, T. Jungwirth, and A. H. MacDonald, 2004, “Universal Intrinsic Spin Hall Effect”, Phys. Rev. Lett. **92**, 126603.

Sinova, J., S. O. Valenzuela, J. Wunderlich, C. H. Back, T. Jungwirth, 2015, “Spin Hall effects”, Rev. Mod. Phys. **87**, 1213.

Sklenar, J., W. Zhang, M. B. Jungfleisch, W. Jiang, H. Saglam, J. E. Pearson, J. B. Ketterson, and A. Hoffmann, 2016, “Perspective: Interface generation of spin-orbit torques”, J. Appl. Phys. **120**, 180901.

Skyrme, T.H.R., 1958, “A non-linear theory of strong interactions”, Proc. Roy. Soc. A **247**, 260.

Skyrme, T.H.R., 1962, “A unified field theory of mesons and baryons”, Nucl. Phys. **31**, 556.

Slachter, A., F. L. Bakker, J.-P. Adam, and B. J. van Wees, 2010, “Thermally driven spin injection from a ferromagnet into a non-magnetic metal”, Nat. Phys. **6**, 879.

Slavin, A., and V. Tiberkevich, 2005, “Spin wave mode excited by spin-polarized current in a magnetic nanocontact is a standing self-localized wave bullet”, Phys. Rev. Lett. **95**, 237201.

Slavin, A., and V. Tiberkevich, 2008, “Excitation of spin waves by spin-polarized current in magnetic nano-structures”, IEEE Trans. Magn. **44**, 1916.

Slavin, A., and V. Tiberkevich, 2009 "Nonlinear Auto-Oscillator Theory of Microwave Generation by Spin-Polarized Current”, IEEE Trans. Magn. **45**, 1875.

Slonczewski, J. C., 1989, “Conductance and exchange coupling of two ferromagntes separated by a tunneling barrier”, Phys. Rev. B **39**, 6995.

Slonczewski, J. C., 1991, "Fluctuation mechanism for biquadratic exchange coupling in magnetic multilayers”, Phys. Rev. Lett. **67**, 3172.

Slonczewski, J. C., 1995, “Overview of interlayer exchange theory”, J. Magn. Magn. Mater. **150**, 13.

Slonczewski, J. C., 1996 “Current-driven excitation of magnetic multilayers”, J. Magn. Magn. Mater. **159**, L1.

Slonczewski, J. C., 2002, “Currents and torques in metallic magnetic multilayers”, J. Magn. Magn. Mater. **247**, 324.

Smit, J., 1958, “The spontaneous Hall effect in ferromagnetics II”, Physica **24**, 39.

Smith, A. K., M. Jamali, Z. Zhao, and J.-P. Wang, 2016, “External Field Free Effect Device for Perpendicular Magnetization Reversal Using a Composite Structure with Biasing Layer”, arXiv:1603.09624.

Soldatov, I. V., N. Panarina, C. Hess, L. Schultz, and R. Schäfer, 2014, "Thermoelectric effects and magnetic anisotropy of $Ga_{1-x}Mn_xAs$ thin films", Phys. Rev. B **90**, 104423.

Sonin, E. B., 1978, "Analogs of superfluid currents for spins and electron-hole pairs”, JETP **47**, 1091. (English translation of Zh. Eksp. Teor. Fiz. **74**, 2097.)

Sonin, E. B., 2010, "Spin currents and spin superfluidity”, Adv. Phys. **59**, 181.

Soucaille, R., M. Belmeguenai, J. Torrejon, J-V. Kim, T. Devolder, Y. Roussigné, S-M. Chérif, A. Stashkevich, M. Hayashi, J-P. Adam, 2016, “Probing the Dzyaloshinskii-Moriya interaction in CoFeB ultrathin films using domain wall creep and Brillouin light spectroscopy”, Phys. Rev. B **94**, 104431.

Soumyanarayanan, A, B. Reyren, A. Fert, and C. Panagopoulos, 2016, “Emergent phenomena induced by spin-orbit coupling at surfaces and interfaces”, Nature **539**, 509.

Spaldin, N. A., S.-W. Cheong, and R. Ramesh, 2010, “Multiferroics: Past, present, and future”, Phys. Today **63**, 38.

Spurgeon, S. R., *et al.*, 2014, “Thickness-Dependent Crossover from Charge- to Strain-Mediated Magnetoelectric Coupling in Ferromagnetic/Piezoelectric Oxide Heterostructures”, ACS Nano **8**, 894.

Srajer, G., *et al.*, 2006, “Advances in nanomagnetism *via* X-ray techniques”, J. Magn. Magn. Mater. **307**, 1.

Stamm, C., *et al.*, 2007, “Femtosecond modification of electron localization and transfer of angular momentum in nickel”, Nat. Mater. **6**, 740.

Stanciu, C. D., F. Hansteen, A. V. Kimel, A. Kirilyuk, A. Tsukamoto, A. Itoh, and Th. Rasing, 2007, “All-optical magnetic recording with circularly polarized light”, Phys. Rev. Lett. **99**, 047601.

Starikov, A. A., P. J. Kelly, A. Brataas, Y. Tserkovnyak, and G. E. W. Bauer, 2010, “Unified First-Principles Study of Gilbert Damping, Spin-Flip Diffusion, and Resistivity in Transition Metal Alloys,” Phys. Rev. Lett. **105**, 236601.

Stemmer, S., and J. S. Allen, 2014, “Two-dimensional electron gases at complex oxide interfaces”, Annu. Rev. Mater. Res. **44**, 151.

Stiles, M. D., 1999, “Interlayer exchange coupling”, J. Magn. Magn. Mater. **200,** 322.

Stiles, M. D., 2004, "Interlayer Exchange Coupling”,in *Ultrathin Magnetic Structures III*, ed. by B. Heinrich and J.A.C. Bland, (Springer-Verlag, Berlin).

Stipe, B. C., *et al.*, 2010, “Magnetic recording at 1.5 Pb m$^{-2}$ using an integrated plasmonic antenna”, Nat. Photonics **4**, 484.

Stöhr, J., 1999, “Exploring the microscopic origin of magnetic anisotropies with X-ray magnetic circular dichroism (XMCD) spectroscopy”, J. Magn. Magn. Mater. **200**, 470.

Stöhr, J., Y. Wu, B. D. Hermsmeier, M. G. Samant, G. Harp, S. Koranda, D. Dunham, and B. P. Tonner, 1993, "Element-specific Magnetic Microscopy with Circularly Polarized X-rays”, Science **259**, 658.

Stoner, E. C., 1938, "Collective Electron Ferromagnetism", *Proc. Royal Society A: Mathematical, Physical and Engineering Sciences* **165**, 372.

Streubel, R., J. Lee, D. Makarov, M.-Y. Im, D. Karnaushenko, L. Han, R. Schäfer, P. Fischer, S.-K. Kim, and O. G. Schmidt, 2014, "Magnetic Microstructure of Rolled-Up Single-Layer Ferromagnetic Nanomembranes”, Adv. Mater. **26**, 316.

Subramanian M. A.; A. P. Ramirez, and W. J. Marshall, 1999, “Structural tuning of ferromagnetism in a 3D cuprate perovskite”, Phys. Rev. Lett. **82**, 1558.

Suhl, H., 1957, “The theory of ferromagnetic resonance at high signal powers”, J. Phys. Chem. Sol. **1**, 209.

Sulpizio, J. A., S. Ilani, P. Irvin, and J. Levy, 2014, “Nanoscale phenomena in oxide heterostructures”, Annu. Rev. Mater. Res. **44**, 117.

Sun, J. Z., 1999, “Current-driven magnetic switching in manganite trilayer junctions”, J. Magn. Magn. Mater. **202**, 157.

Takahashi, K. S., M. Kawasaki, and Y. Tokura, 2001, “Interface ferromagnetism in oxide superlattices of $CaMnO_3/CaRuO_3$”, Appl. Phys. Lett. **79**, 1324.

Takahashi, S. and S. Maekawa, 2003, “Spin injection and detection in magnetic nanostructures”, Phys. Rev. B. **67**, 052409.

Takei, S., B. I. Halperin, A. Yacoby, and Y. Tserkovnyak, 2014, "Superfluid spin transport through antiferromagnetic insulators, Phys. Rev. B. **90**, 094408.

Tanaka, T., H. Kontani, M. Naito, D. S. Hirashima, K. Yamada, and J. Inoue, 2008, “Intrinsic spin Hall effect and orbital Hall effect in *4d* and *5d* transition metals”, Phys. Rev. B **77**, 165117.

Tang, J., *et al.*, 2014, “Electrical Detection of Spin-Polarized Surface States Conduction in $(Bi_{0.53}Sb_{0.47})_2Te_3$ Topological Insulator”, Nano Lett. **14**, 5423.

Tanigaki, T., Y. Takahashi, T. Shimakura, T. Akashi, R. Tsuneta, A. Sugawara, and D. Shindo, 2015, "Three-Dimensional Observation of Magnetic Vortex Cores in Stacked Ferromagnetic Discs”, Nano Lett. **15**, 1309.

Taniguchi, T., J. Grollier, and M. D. Stiles, 2015, "Spin-Transfer Torques Generated by the Anomalous Hall Effect and Anisotropic Magnetoresistance”, Phys. Rev. Appl. **3**, 044001.

Tarasenko, S. V., A. Stankiewicz, V. V. Tarasenko, and J. Ferré, 1998, "Bloch wall dynamics in ultrathin ferromagnetic films", J. Magn. Magn. Mater. **189**, 19.

Tehrani, S., J. M. Slaughter, E. Chen, M. Durlam, J. Shi, M. DeHerrera, 1999, "Progress and outlook for MRAM technology", IEEE Trans. Magn. **35,** 2814.

Tetienne, J.-P., *et al.*, 2014, "Nanoscale imaging and control of domain-wall hopping with a nitrogen-vacancy center microscope", Science **344**, 1366.

Tetienne, J.-P., T. Hingant, L.J. Martínez, S. Rohart, A. Thiaville, L. Herrera Diez, K Garcia, J.-P. Adam, J.-V. Kim, J.-F. Roch, I.M. Miron, G. Gaudin, L. Vila, B. Ocker, D. Ravelosona and V. Jacques, 2015, "The nature of domain walls in ultrathin ferromagnets revealed by scanning nanomagnetometry", Nat. Commun. **6**, 6733.

Thadani, K. V., G. Finocchio, Z.-P. Li, O. Ozatay, J. C. Sankey, I. N. Krivorotov, Y.-T. Cui, R. A. Buhrman, and D. C. Ralph, 2008, "Strong linewidth variation for spin-torque nano-oscillators as a function of in-plane magnetic field angle", Phys. Rev. B **78** 024409.

Thiaville, A., J. M. García, R. Dittrich, J. Miltat, and T. Schrefl, 2003, "Micromagnetic study of Bloch-point-mediated vortex core reversal", Phys. Rev. B **67**, 094410.

Thiaville, A., Y. Nakatani, J. Miltat, and Y. Suzuki, 2005, "Micromagnetic understanding of current-driven domain wall motion in patterned nanowires", Europhys. Lett. **69**, 990.

Thiaville, A., S. Rohart, É. Jué, V. Cros, and A. Fert, 2012, "Dynamics of Dzyaloshinskii domain walls in ultrathin magnetic films", Europhys. Lett. **100**, 57002.

Thiele, A., 1973, "Steady state motion of magnetic domains", Phys. Rev. Lett. **30**, 230.

Thiele, J. U., M. Buess, and C. H. Back, 2004, "Spin dynamics of the antiferromagnetic-to-ferromagnetic phase transition in FeRh on a sub-picosecond time scale", Appl. Phys, Lett. **85**, 2857.

Thiele C., K. Dörr, O. Bilani, K. Rodel, and L. Schultz, 2007, "Influence of strain on the magnetization and magnetoelectric effect in $La_{0.7}A_{0.3}MnO_3$/PMN-PT(001) ($A$=Sr,Ca)", Phys. Rev. B **75**, 054408.

Thole, B. T., P. Carra, F. Sette, and G. van der Laan, 1992, "X-ray circular dichroism as a probe of orbital magnetization", Phys. Rev. Lett. **68**, 1943.

Thomson, W., 1856, "On the Electro-Dynamic Qualities of Metals:--Effects of Magnetization on the Electric Conductivity of Nickel and of Iron", Proc. Royal Soc. London **8**, 546.

Tian, J., I. Miotkowski, S. Hong, and Y. P. Chen, "Electrical injection and detection of spin-polarized currents in topological insulator $Bi_2Te_2Se$", Sci. Rep. **5**, 14293.

Tiberkevich, V. S., and A. N. Slavin, 2007, "Nonlinear phenomenological model of magnetic dissipation for large precession angles: Generalization of the Gilbert model", Phys. Rev. B **75**, 014440.

Tiberkevich, V. S., R. S. Khymyn, H. X. Tang, and A. N. Slavin, 2014, "Sensitivity to external signals and synchronization properties of a non-isochronous auto-oscillator with delayed feedback", Sci.

Rep. **4** 3873.

Tokura, Y., and N. Nagaosa, 2000, “Orbital physics in transition-metal oxides”, Science **288**, 462.

Tomasello, R., E. Martinez, R. Zivieri, L. Torres, M. Carpentieri, and G. Finocchio, 2014, "A strategy for the design of skyrmion racetrack memories”, Sci. Rep. **4**, 6874.

Tombros, N., C. Jozsa, M. Popinciuc, H. T. Jonkman & B. J. van Wees, 2007, “Electronic spin transport and spin precession in single graphene layers at room temperature,” Nature **448**, 571.

Torija, M.A., M. Sharma, J. Gazquez, M. Varela, C. He, J. Schmitt, J.A. Borchers, M. Laver, S. El-Khatib, and C. Leighton, 2011, “Chemically-driven nanoscopic magnetic phase separation at the $SrTiO_3$(001)/$La_{1-x}Sr_xCoO_3$ interface”, Adv. Mater. **23**, 2711.

Torrejon, J., F. Garcia-Sanchez, T. Taniguchi, J. Sinha, S. Mitani, J.-V. Kim, and M. Hayashi, 2015, “Current-driven asymmetric magnetization switching in perpendicularly magnetized CoFeB/MgO heterostructures”, Phys. Rev. B **91**, 214434.

Töws, W. and G. M. Pastor, 2015, “Many-Body Theory of Ultrafast Demagnetization and Angular Momentum Transfer in Ferromagnetic Transition Metals,” Phys. Rev. Lett. **115**, 217204.

Tretiakov, O., and O. Tchernyshyov, 2007, "Vortices in thin ferromagnetic films and the skyrmion number”, Phys. Rev. B **75**, 012408.

Trigo, M., *et al.*, 2013, “Fourier-transform inelastic X-ray scattering from time- and momentum-dependent phonon-phonon correlations”, Nat. Phys. **9**, 790.

Tripathi, A., J. Mohanty, S. H. Dietze, O. G. Shpyrko, E. Shipton, E. E. Fullerton, S. S. Kim, and I. McNulty, 2011, "Dichroic coherent diffractive imaging”, Proc. Nat. Acad. Sci. **108**, 13393.

Tserkovnyak, Y., A. Brataas, and G. E. W. Bauer, 2002, “Enhanced Gilbert damping in thin ferromagnetic films”, Phys. Rev. Lett. **88**, 117601.

Tserkovnyak, Y., A. Brataas, G. E. W. Bauer, and B. I. Halperin, 2005, “Non-local magnetization dynamics in ferromagnetic heterostructures”, Rev. Mod. Phys. **77**, 1375.

Tserkovnyak, Y,, A. Brataas, and G. E.W. Bauer, 2008, “Theory of current-driven magnetization dynamics in inhomogeneous ferromagnets”, J. Magn. Magn. Mater. **320**, 1282.

Tserkovnyak, Y., D. A. Pesin, and D. Loss, 2015, “Spin and orbital magnetic response on the surface of a topological insulator”, Phys. Rev. B **91**, 041121(R).

Tshitoyan, V., C. Ciccarelli, A. P. Mihai, M. Ali, A. C. Irvine, T. A. Moore, T. Jungwirth, and A. J. Ferguson, 2015, “Electrical manipulation of a ferromagnet by an antiferromagnet”, Phys. Rev. B **92**, 214406.

Tsoi, M., A. G. M. Jansen, J. Bass, W.-C. Chiang, M. Seck, V. Tsoi, and P. Wyder, 1998, “Excitation of a magnetic multilayer by an electric current”, Phys. Rev. Lett. **80**, 4281.

Tsoi, M., A. G. M. Jansen, J. Bass, W.-C. Chiang, V. Tsoi, and P. Wyder, 2000, “Generation and detection of phase-coherent current-driven magnons in magnetic multilayers”, Nature **406**, 46.

Tsujikawa, M. and T. Oda, 2009, “Finite electric field effects in the large perpendicular magnetic anisotropy surface Pt/Fe/Pt (001): A first-principles study”, Phys. Rev. Lett. **102**, 247203.

Tsunegi, S., H. *et al.*, 2014, "High emission power and Q factor in spin torque vortex oscillator consisting of FeB free layer”, Appl. Phys. Exp. **7**, 063009.

Tsymbal, E., and I. Zutic, 2011, Eds., *Handbook of spin transport and magnetism*, (CRC, Boca Raton).

Tucker J. W., 2000, "A Monte Carlo study of thin spin-1 Ising films with surface exchange enhancement", J. Magn. Magn. Mater. **210** 383.

Tulapurkar, A. A., Y. Suzuki, A. Fukushima, H. Kubota, H. Maehara, K. Tsunekawa, D. D. Djayaprawira, N. Watanabe, and S. Yuasa, 2005, "Spin-torque diode effect in magnetic tunnel junctions”, Nature **438**, 339.

Uchida, K., H. Adachi, T. Ota, H. Nakayama, S. Maekawa, E. Saitoh, 2010a, “Observation of longitudinal spin-Seebeck effect in magnetic insulators”, Appl. Phys. Lett. 97, 172505.

Uchida, K., T. Nonaka, T. Ota, and E. Saitoh, 2010c, “Longitudinal spin-Seebeck effect in sintered polycrystalline (Mn,Zn)$Fe_2O_4$”, Appl. Phys. Lett. **97**, 262504.

Uchida, K. T. Nonaka, T. Kikkawa, Y. Kajiwara, and E. Saitoh, 2013, “Longitudinal spin Seebeck effect in various garnet ferrites”, Phys. Rev. B **87**, 104412.

Uchida, K., S. Takahashi, K. Harii, J. Ieda, W. Koshibae, K. Ando, S. Maekawa, and E. Saitoh, 2008, “Observation of the spin Seebeck effect”, Nature **455**, 778.

Uchida, K, J. Xiao, *et al.*, 2010b, “Spin Seebeck Insulator”, Nat. Mater. **9**, 894.

Uchida, K., T. Ota, H. Adachi, J. Xiao, T. Nonaka, Y. Kajiwara, G. E. W. Bauer, S. Maekawa, and E. Saitoh, 2012, "Thermal spin pumping and magnon-phonon-mediated spin-Seebeck effect", J. Appl. Phys. **111**, 103903.

Udvardi, L., and L. Szunyogh, 2009, “Chiral Asymmetry of the Spin-Wave Spectra in Ultrathin Magnetic Films”, Phys. Rev. Lett. **102**, 207204.

Ueda, K., H. Tabata, and T. Kawai, 2001, “Control of magnetic properties in LaCrO3-LaFeO3 artificial superlattices”, J. Appl. Phys. **89**, 2847.

Uhlir, V., M. Urbanek, L. Hladik, J. Spousta, M. Y. Im, P. Fischer, N. Eibagi, J. J. Kan, E. E. Fullerton, and T. Sikola, 2013, "Dynamic switching of the spin circulation in tapered magnetic nanodisks”, Nat. Nanotech. **8**, 341.

Unguris, J., R. J. Celotta, and D. T. Pierce, 1997, “Determination of the Exchange Coupling Strengths for Fe/Au/Fe”, Phys. Rev. Lett. **79**, 2734.

Urazhdin, S., V. Tiberkevich, and A. Slavin, 2010, “Parametric excitation of a magnetic nanocontact by a microwave field”, Phys. Rev. Lett. **105**, 237204.

Uribe-Laverde, M. A., *et al.*, 2014, “X-ray absorption spectroscopy study of the electronic and magnetic proximity effects in $YBa_2Cu_3O_7/La_{2/3}Ca_{1/3}MnO_3$ and $La_{2-x}Sr_xCuO_4/La_{2/3}Ca_{1/3}MnO_3$ multilayers”, Phys. Rev. B **90**, 205135.

Vahaplar, K, *et al.*, 2012, “All-optical magnetization reversal by circularly polarized laser pulses: Experiment and multiscale modeling”, Phys. Rev. B **85**, 104402.

Valenzuela, S. O., and M. Tinkham, 2006, “Direct electronic measurement of the spin Hall effect”, Nature **442**, 176.

Valet, T., and A. Fert, 1993, “Theory of the perpendicular magnetoresistance in magnetic multilayers”, Phys. Rev. B **48**, 7099.

van Bree, J., A. Yu. Silov, P. M. Koenraad, and M. E. Flatté, 2014, “Spin-Orbit-Induced Circulating Currents in a Semiconductor Nanostructure”, Phys. Rev. Lett. **112**, 187201.

van den Brink, A., G. Vermijs, A. Solignac, J. Koo, J. T. Kohlhepp, H. J. M. Swagten, and B. Koopmans, 2016, “Field-free magnetization reversal by spin-Hall effect and exchange bias”, Nat. Commun. **7**, 10854.

van der Merwe, J. H., 1991, “Misfit dislocation generation in epitaxial layers”, Crit. Rev. Sol. Stat. Mater. Res. **17**, 187.

van Kampen, M, C. Jozsa, J. T. Kohlhepp, P. LeClair, L. Lagae, W. J. M. de Jonge, and B. Koopmans, 2002, “All-optical probe of coherent spin waves”, Phys. Rev. Lett. **88**, 227201.

Vansteenkiste, A., J. Leliaert, M. Dvornik, M. Helsen, F. Garcia Sanchez, and B. Van Waeyenberge, 2014, “The design and verification of MuMax3”, AIP Adv. **4**, 107133.

Van Vleck, J. H., 1950, “Concerning the Theory of Ferromagnetic Resonance Absorption,” Phys. Rev. **78**, 266.

Van Waeyenberge, *et al.*., 2006, "Magnetic vortex core reversal by excitation with short bursts of an alternating field”, Nature **444**, 461.

Varela, M., J. Gazquez, and S. J. Pennycook, 2012, “STEM/EELS imaging of complex oxides and interfaces”, MRS Bull. **37**, 29.

Vas’ko, F. T., 1979, “Spin splitting in the spectrum of two-dimensional electrons due to the surface potential”, P. Zh. Eksp. Teor. Fiz. **30**, 574. English translation at JETP Lett. **30**, 541 (1980).

Vaz, C. A. F., J. A. C. Bland, and G. Lauhoff, 2008, “Magnetism in ultrathin film structures”, Rep. Prog. Phys. **71,** 056501.

Vaz, C. A. F., J. Hoffman, Y. Segal, J. W. Reiner, R. D. Grober, Z. Zhang, C. H. Ahn, and F. J. Walker, 2010, “Origin of the magnetoelectronic coupling effect in $Pb(Zr_{0.2}Ti_{0.8})O_3/La_{0.8}Sr_{0.2}MnO_3$ multiferroic heterostructures”, Phys. Rev. Lett. **104**, 127202.

Vélez, S., V. N. Golovach, A. Bedoya-Pinto, M. Isasa, E. Sagaste, M. Abadia, C. Rogero, L. E. Hueso, F.

S. Bergeret and F. Casanova, 2016, "Hanle Magnetoresistance in Thin Metal Films woth Strrong Spin-Orboit Coupling", Phys. Rev. Lett. **116**, 016603.

Visani, C., *et al.*, 2011, "Symmetrical interfacial reconstruction and magnetism in $La_{0.7}Ca_{0.3}MnO_3/YBa_2Cu_3O_7/La_{0.7}Ca_{0.3}MnO_3$ heterostructures", Phys. Rev. B **84**, 060405(R).

von Barth, U. and L. Hedin, 1972, "A local exchange-correlation potential for the spin polarized case. I", J. Phys. C: Solid State Phys. **5,** 1629.

von Bergmann, K., A. Kubetzka, O. Pietzsch, and R. Wiesendanger, 2014, "Interface-induced chiral domain walls, spin spirals and skyrmions revealed by spin-polarized scanning tunneling microscopy", J. Phys.: Condens. Matter **26,** 394002.

Wadley, P, *et al.*, 2016, "Electrical switching of an antiferromagnet", Science **351**, 587.

Wahle, J., N. Blümer, J. Schlipf, K. Held, and D. Vollhardt, 1998, "Microscopic conditions favoring itinerant ferromagnetism", Phys. Rev. B **58**, 12749.

Walter, J., H. Wang, B. Luo, C.D. Frisbie, and C. Leighton, "Electrostatic *versus* electrochemical doping and control of ferromagnetism in ion-gel-gated ultrathin La0.5Sr0.5CoO3-d," ACS Nano **10**, 7799.

Wang, H., C. Du, P. C. Hammel, and F. Yang, 2014, "Antiferromagnetic Spin Transport from $Y_3Fe_5O_{12}$ into NiO", Phys. Rev. Lett. **113**, 097202.

Wang, W.-G., M. Li, S. Hageman, and C. L. Chien, 2011, "Electric-field-assisted switching in magnetic tunnel junctions", Nat. Mater. **11**, 64.

Wang, X., and A. Manchon, 2012, "Diffusive Spin Dynamics in Ferromagnetic Thin Films with a Rashba Interaction", Phys. Rev. Lett. **108,** 117201.

Wang, Y., P. Deorani, K. Banerjee, N. Koirala, M. Brahlek, S. Oh, and H. Yang, 2015, "Topological Surface States Originated Spin-Orbit Torques in $Bi_2Se_3$", Phys. Rev. Lett. **114**, 257202.

Wang, Z.-H., G. Cristiani, and H.-U. Habermeier, 2003, "Uniaxial magnetic anisotropy and magnetic switching in $La_{0.67}Sr_{0.33}MnO_3$ thin films grown on vicinal $SrTiO_3$(100)", Appl. Phys. Lett. **82**, 3731.

Watzman, S. J., R.A., Duine, Y. Tserkovnyak, S. R. Boona, H. Jin, A. Prakash, Y. Zheng and J. P. Heremans, 2016, "Magnon-drag thermopower and Nernst coefficient in Fe and Co", Phys. Rev. B **94**, 144407.

Weber, M. C., H. Nembach, S. Blomeier, B. Hillebrands, R. Kaltofen, J. Schumann, M. J. Carey, and J. Fassbender, 2005, "All-optical probe of magnetization dynamics in exchange biased bilayers on the picosecond timescale", Eur. Phys. J. B **45**, 243.

Weber, M. C., H. Nembach, B. Hillebrands, and J. Fassbender, 2005a, "Modified Gilbert damping due to exchange bias in NiFe/FeMn bilayers", J. Appl. Phys. **97**, 10A701.

Weber, M. C., H. Nembach, B. Hillebrands, and J. Fassbender, 2005b, "Real-time temperature dynamics in exchange-biased bilayers upon laser excitation", IEEE Trans. Magn. **41**, 1089.

Weber, M. C., H. Nembach, B. Hillebrands, M. J. Carey, and J. Fassbender, 2006, "Real-time evidence of two-magnon scattering in exchange-coupled bilayers", J. Appl. Phys. **99**, 08J308.

Wei, P., F. Katmis, B. A. Assaf, H. Steinberg, P. Jarillo-Herrero, D. Heiman, and J. S. Moodera, 2013, "Exchange-Coupling-Induced Symmetry Breaking in Topological Insulators", Phys. Rev. Lett. **110,** 186807.

Weisbuch, C., and C. Hermann, 1977, "Optical detection of conduction electron spin resonance in GaAs", Phys. Rev. B **15**, 816.

Weisheit, M., S. Fahler, A. Marty, Y. Souche, C. Poinsignon, and D. Givord, 2007, "Electric field-induced modification of magnetism in thin-film ferromagnets", Science **315**, 349.

Weller, D., and M. F. Doerner, 2000, "Extremely high-density longitudinal magnetic recording media", Annu. Rev. Mater. Science **30**, 611.

Weller, D., J. Stöhr, R. Nakajima, A. Carl, M. Samant, C. Chappert, R. Megy, P. Beauvillain, and P. Veillet, 1995, "Microscopic Origin of Magnetic Anisotropy in Au/Co/Au Probed with X-Ray Magnetic Circular Dichroism", Phys. Rev. Lett. **75**, 3752.

Wende, H., 2004, "Recent advances in x-ray absorption spectroscopy", Rep. Prog. Phys. **67**, 2105.

Wiesendanger, R., 2009, "Spin mapping at the nanoscale and atomic scale", Rev. Mod. Phys. **81**, 1495.

Wiesendanger, R., 2016, "Nanoscale magnetic skyrmions in metallic films and multilayers: a new twist for spintronics", Nat. Rev. Mater. **1,** 16044.

Wietstruk, M, A. Melnikov, C. Stamm, T. Kachel, N. Pontius, M. Sultan, C. Gahl, M. Weinelt, H. A. Dürr, and U. Bovensiepen, 2011, "Hot-Electron-Driven Enhancement of Spin-Lattice Coupling in Gd and Tb 4f Ferromagnets Observed by Femtosecond X-Ray Magnetic Circular Dichroism", Phys. Rev. Lett. **106**, 127401.

Wigen, P. E., 1994, *Nonlinear Phenomena and Chaos in Magnetic Materials* (World Scientific, Singapore).

Wilhelm, F., *et al.*, 2000, "Layer-Resolved Magnetic Moments in Ni/Pt Multilayers", Phys. Rev. Lett. **85**, 413.

Willmott, P. R., *et al.*, 2007, "Structural basis for the conducting interface between $LaAlO_3$ and $SrTiO_3$", Phys. Rev. Lett. **99**, 155502.

Woo, S., *et al.*, 2016, "Observation of room-temperature magnetic skyrmions and their current-driven dynamics in ultrathin metallic ferromagnets", Nat. Mater. **15**, 501.

Wu, A. Q., *et al.*, 2013, "HAMR Areal Density Demonstration of 1+ Tbpsi on Spinstand", IEEE Trans. Magn. **49,** 779.

Wu, J., *et al.*, 2011, "Direct observation of imprinted antiferromagnetic vortex states in CoO/Fe/Ag(001) discs", Nat. Phys. **7**, 303.

Wu, H., X. Zhang, C. H. Wan, B. S. Tao, L. Huang, W. J. Kong, and X. F. Han, 2016a, "Hanle amgnetoresistance: The role of edge spin accumulation and interfacial spin current", Phys. Rev B **94**, 174407.

Wu, S. M., J. Hoffman, J. E. Pearson, and A. Bhattacharya, 2014, "Unambiguous separation of the inverse spin Hall and anomalous Nernst effects within a ferromagnetic metal using the spin Seebeck effect", Appl. Phys. Lett. **105**, 092409.

Wu, S. M., F. Y. Fradin, J. Hoffman, A. Hoffmann, and A. Bhattacharya, 2015a, "Spin Seebeck devices using local on-chip heating", J. Appl. Phys. **117**, 17C509 .

Wu, S., J. E. Pearson, and A. Bhattacharya, 2015b, "Paramagnetic spin Seebeck effect", Phys. Rev. Lett. **114**, 186602.

Wu, S. M., W. Zhang, A. KC, P. Borisov, J. E. Pearson, J. S. Jiang, D. Lederman, A. Hoffmann, and A. Bhattacharya, 2016b, "Antiferromagnetic spin Seebeck Effect", Phys. Rev. Lett. **116**, 097204.

Wunderlich, J., *et al.*, 2006, "Coulomb Blockade Anisotropic Magnetoresistance Effect in a (Ga,Mn)As Single-Electron Transistor", Phys. Rev. Lett. **97,** 077201.

Wunderlich, J., B. Kaestner, J. Sinova, and T. Jungwirth, 2005, "Experimental Observation of the Spin-Hall Effect in a Two-Dimensional Spin-Orbit Coupled Semiconductor System", Phys. Rev. Lett. **94**, 047204.

Xia, K., P. J. Kelly, G. E. W. Bauer, A. Brataas, and I. Turek, 2002, "Spin torques in ferromagnetic/ normal-metal structures," Phys. Rev. B **65**, 220401(R)

Xiao, J.; A. Zangwill; and M. D. Stiles, 2004, "Boltzmann Test of Slonczewski's Theory of Spin Transfer Torque", Phys. Rev. B **70**, 172405.

Xiao, J., G. E. W. Bauer, K. Uchida, E. Saitoh, and S. Maekawa, 2010, "Theory of magnon-driven spin Seebeck effect", Phys. Rev. B **82**, 099904.

Xu, S.-Y., *et al.*, 2012, "Hedgehog spin texture and Berry's phase tuning in a magnetic topological insulator", Nat. Phys. **8**, 616.

Xu, Y., S. Wang, and K. Xia, 2008, "Spin-Transfer Torques in Antiferromagnetic Metals from First Principles", Phys. Rev. Lett. **100**, 226602.

Yafet, Y, 1963, Solid State Physics vol. 14, New York, USA, Academic Press.

Yamada, Y., K. Ueno, T. Fukumura, H. T. Yuan, H. Shimotani, Y. Iwasa, L. Gu, S. Tsukimoto, Y. Ikuhara, and M. Kawasaki, 2011, "Electrically induced ferromagnetism at room temperature in cobalt-doped titanium dioxide", Science **332**, 1065.

Yamasaki, K., S. Oki, S. Yamada, T. Kanashima, and K. Hamaya, 2015, "Spin-related thermoelectric conversion in lateral spin-valve devices with single-crystalline Co 2 FeSi electrodes", Appl. Phys. Express **8**, 043003.

Yanes, R., J. Jackson, L. Udvardi, L. Szunyogh, and U. Nowak, 2013, "Exchange Bias Driven by Dzyaloshinskii-Moriya Interactions", Phys. Rev. Lett. **111**, 217202.

Yang, H.X., M. Chshiev, B. Dieny, J.H. Lee, A. Manchon, and K.H. Shin, 2011, "First principles investigation of the very large perpendicular magnetic anisotropy at the Fe/MgO and Co/MgO interfaces", Phys. Rev. B **84**, 054401.

Yang, H., A. Thiaville, S. Rohart, A. Fert, and M. Chshiev, 2015, "Anatomy of Dzyaloshinkii-Moriya Interaction at Co/Pt Interfaces", Phys. Rev. Lett. **115**, 267210.

Yang, J. H., Z. L. Li, X. Z. Lu, M.-H. Whangbo, S.-H. Wei, X. G. Gong, and H. J. Xiang, 2012, "Strong Dzyaloshinskii-Moriya Interaction and Origin of Ferroelectricity in $Cu_2OSeO_3$", Phys. Rev. Lett. **109**, 107203.

Yang, Q. I., *et al.*, 2013, "Emerging Weak Localization Effects on Topological Insulator-Insulating Ferromagnet ($Bi_2Se_3$-EuS) Interface", Phys. Rev. B **88**, 081407.

Yang, S., G. Beach, C. Knutson, D. Xiao, Q. Niu, M. Tsoi, and J. Erskine, 2009, "Universal Electromotive Force Induced by Domain Wall Motion", Phys. Rev. Lett. **102**, 067201.

Yasuda, K., A. Tsukazaki, R. Yoshimi, K. Kondou, K. S. Takahashi, Y. Otani, M. Kawasaki, Y. Tokura, 2017, arXiv:1612.06862.

Yia, D. J. Liu, S.-L. Hsua,, L. Zhang, Y. Choig, J.-W. Kim, Z. Chen, J. D. Clarkson, C. R. Serrao, E. Arenholz, P. J. Ryang, H. Xu, R. J. Birgeneau, and R. Ramesh, 2016, "Atomic-scale control of magnetic anisotropy via novel spin–orbit coupling effect in $La_{2/3}Sr_{1/3}MnO_3/SrIrO_3$ superlattices", Proc. Nat. Acad. Sci. **113**, 6397.

Yosida, Y., 1957, "Magnetic Properties of Cu-Mn Alloys," Phys. Rev. **106**, 893.

Yoshimura, Y., et al., 2015, "Soliton-like magnetic domain wall motion induced by the interfacial Dzyaloshinskii-Moriya interaction", Nat. Phys. **12**, 157.

You, L., O. Lee, D. Bhowmik, D. Labanowski, J. Hong, J. Bokor, and S. Salahuddin, 2015, "Switching of perpendicularly polarized nanomagnets with spin orbit torque without an external magnetic field by engineering a tilted anisotropy", Proc. Nat. Acad. Sci **112**, 10310.

Yu, R., W. Zhang, H.-J. Zhang, S.-C. Zhang, X. Dai, and Z. Fang, 2010a, "Quantized Anomalous Hall Effect in Magnetic Topological Insulators", Science **329,** 61.

Yu, X. Z., J. P. DeGrave, Y. Hara, T. Hara, S. Jin, and Y. Tokura, 2013, "Observation of the Magnetic Skyrmion Lattice in a MnSi Nanowire by Lorentz TEM", Nano. Lett. **13**, 3755.

Yu, X. Z., N. Kanazawa, Y. Onose, K. Kimoto,W. Z. Zhang, S. Ishiwata, Y. Matsui, and Y. Tokura, 2011, "Near room-temperature formation of a skyrmion crystal in thin-films of the helimagnet FeGe", Nat. Mater. **10**, 106.

Yu, X. Z., N. Kanazawa, W. Z. Zhang, T. Nagai, T. Hara, K. Kimoto, Y. Matsui, Y. Onose, and

Y. Tokura, 2012, "Skyrmion flow near room temperature in an ultralow current density", Nat. Commun. **3**, 988.

Yu, X. Z., Y. Onose, N. Kanazawa, J. H. Park, J. H. Han, Y. Matsui, N. Nagaosa, and Y. Tokura, 2010b, "Real-space observation of a two-dimensional skyrmion crystal", Nature **465**, 901.

Yuasa, S., T. Nagahama, A. Fukushima, Y. Suzuki and K. Ando, 2004, "Giant room-temperature magnetoresistance in single-crystal Fe/MgO/Fe magnetic tunnel junctions", Nat. Mater. **3**, 868.

Zakeri, Kh., Y. Zhang, J. Prokop, T.-H. Chuang, N. Sakr, W. X. Tang, and J. Kirschner, 2010, "Asymmetric Spin-Wave Dispersion on Fe(110): Direct Evidence of the Dzyaloshinskii-Moriya Interaction", Phys. Rev. Lett. **104**, 137203.

Zang, J., M. Mostovoy, J. H. Han, and N. Nagaosa, 2011, "Dynamics of skyrmion crystals in metallic thin films", Phys. Rev. Lett. **107**, 136804.

Železný, J, H. Gao, K. Výborný, J. Zemen, J. Mašek, A. Manchon, J. Wunderlich, J. Sinova, and T. Jungwirth, 2014, "Relativistic Néel-Order Fields Induced by Electrical Current in Antiferromagnets", Phys. Rev. Lett. **113**, 157201.

Zener, C., 1951, "Interaction between the d-shells in the transition metals. II. Ferromagnetic compounds of manganese with perovskite structure", Phys. Rev. **82**, 403.

Zhang, C., S. Fukami, H. Sato, F. Matsukura, and H. Ohno, 2015f, "Spin-orbit torque induced magnetization switching in nano-scale Ta/CoFeB/MgO", Appl. Phys. Lett. **107**, 012401.

Zhang, D., *et al.*, 2012, "Interplay between ferromagnetism, surface states, and quantum corrections in a magnetically doped topological insulator", Phys. Rev. B **86**, 205127.

Zhang, G. P., and W. Hubner, 2000, "Laser-induced ultrafast demagnetization in ferromagnetic metals", Phys. Rev. Lett. **85**, 3025.

Zhang, H., C.-X. Liu, X.-L. Qi, X. Dai, Z. Fang, and S.-C.Zhang, 2009, "Topological insulators in $Bi_2Se_3$, $Bi_2Te_3$ and $Sb_2Te_3$ with a single Dirac cone on the surface", Nat. Phys. **5**, 438.

Zhang, J. Y., J. Hwang, S. Raghavan, and S. Stemmer, 2013, "Symmetry Lowering in Extreme-Electron-Density Perovskite Quantum Wells", Phys. Rev. Lett. **110**, 256401.

Zhang, S., and Z. Li, 2004, "Roles of Nonequilibrium Conduction Electrons on the Magnetization Dynamics of Ferromagnets", Phys. Rev. Lett. **93**, 127204.

Zhang, W., M. B. Jungfleisch, F. Freimuth, W. Jiang, J. Sklenar, J. E. Pearson, J. B. Ketterson, Y. Mokrousov, and A. Hoffmann, 2015d, "All-electrical manipulation of magnetization dynamics in a ferromagnet by antiferromagnets with anisotropic spin Hall effects", Phys. Rev. B **92**, 144405.

Zhang, W., M. B. Jungfleisch, W. Jiang, Y. Liu, J. E. Pearson, S. G. E. te Velthuis, A. Hoffmann, F. Freimuth, and Y. Mokrousov, 2015a, "Reduced spin-Hall effects from magnetic proximity", Phys. Rev. B **91**, 115316.

Zhang, W., M. B. Jungfleisch, W. Jiang, J. E. Pearson, A. Hoffmann, F. Freimuth, and Y. Mokrousov, 2014, “Spin Hall Effects in Metallic Antiferromagnets”, Phys. Rev. Lett. **113**, 196602.

Zhang, W., M. B. Jungfleisch, W. Jiang, J. E. Pearson, and A. Hoffmann, 2015b, “Spin pumping and inverse Rashba-Edelstein effect in NiFe/Ag/Bi and NiFe/Ag/Sb”, J. Appl. Phys. **117**, 17C727.

Zhang, W., W. Han, X. Jiang, S.-H. Yang, and S. S. P. Parkin, 2015c, “Role of transparency of platinum–ferromagnet interfaces in determining the intrinsic magnitude of the spin Hall effect,” Nat. Phys. **11**, 496.

Zhang, X., G. P. Zhao, H. Fangohr, J. P. Liu, W. X. Xia, J. Xia, and F. J. Morvan, 2015e, "Skyrmion-skyrmion and skyrmion-edge repulsions in skyrmion-based racetrack memory”, Sci. Rep. **5**, 11369.

Zhang, Z., B. Cui, G. Wang, B. Ma, Q. Y. Jin, and Y. Liu, 2010, “Ultrafast laser-induced magnetization precession dynamics in FePt/CoFe exchange-coupled films”, Appl. Phys. Lett. **97**, 172508.

Zhu, J., J. A. Katine, G. E. Rowlands, Y.-J. Chen, Z. Duan, J. G. Alzate, P. Upadhyaya, J. Langer, P. K. Amiri, K. L. Wang, and I. N. Krivorotov, 2012, "Voltage-Induced Ferromagnetic Resonance in Magnetic Tunnel Junctions”, Phys. Rev. Lett. **108**, 197203.

Zhu, J.-G., and C. Park, 2006, “Magnetic tunnel junctions”, Mater. Today **9,** 36.

Zhu, Y., and H. Dürr, 2015, “The future of electron microscopy”, Physics Today **68**, 32.

Zink, B. L., M. Manno, L. O’Brien, J. Lotze, M. Weiler, D. Bassett, S. J. Mason, S. T. B. Goennenwein, M. Johnson, and C. Leighton, 2016, “Efficient spin transport through native oxides of nickel and permalloy with platinum and gold overlayers”, Phys. Rev. B **93** 184401.

Zutic, I., J. Fabian, and S. Das Sarma, 2004, “Spintronics: Fundamentals and applications”, Rev. Mod. Phys. **76**, 323.

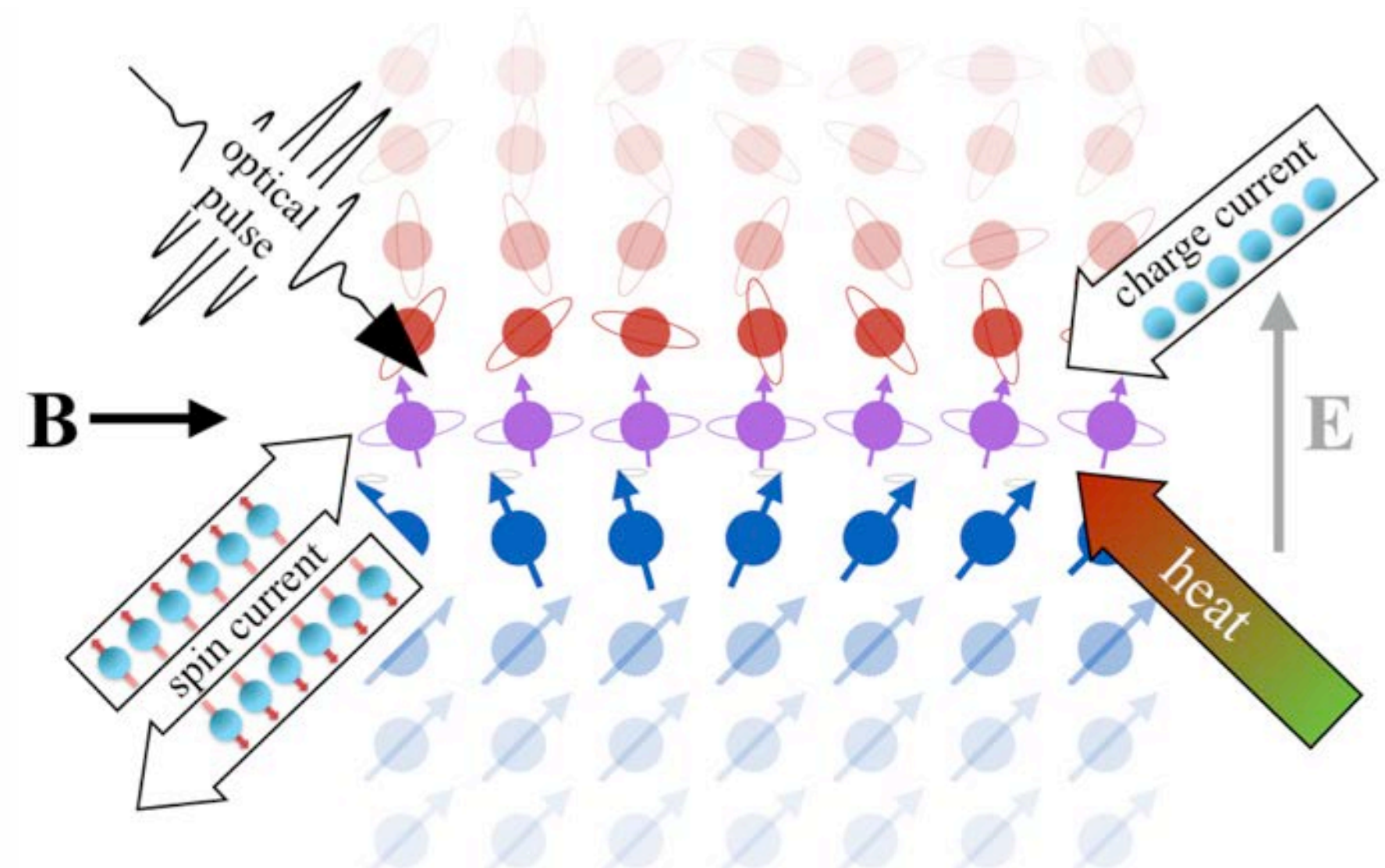

**Figure 1** Interface between a ferromagnet (blue atoms with arrows) and a heavy metal with strong spin-orbit coupling (red atoms with circles). Interface atoms are shown in purple with circles and arrows, schematically indicating interfacial mixing of structure, chemical, magnetic, and electronic states that modify spin and orbital properties on each side, in turn creating new magnetic properties, novel charge and spin transport, and emergent electromagnetic fields. (Time dependent) charge currents, optical pulses, heat, and electric and magnetic fields (directions are illustrative) interact with this heterostructure to produce spin currents, which modify the electronic and magnetic states.

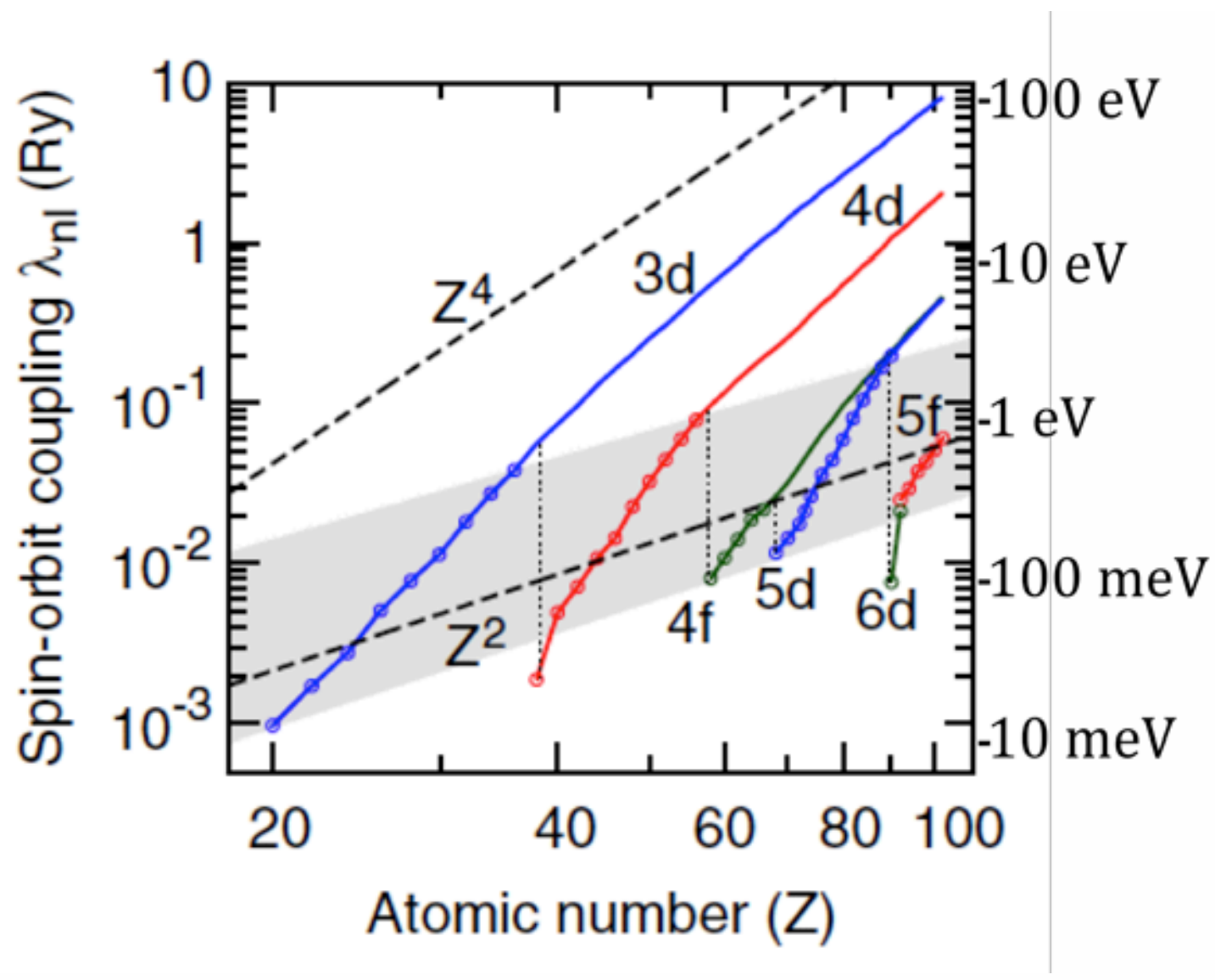

**Figure 2** Dependence of the individual orbital spin-orbit coupling strength $\lambda_{nl}$ for atoms as a function of their atomic number $Z$. Calculated results [Herman and Skillman, 1963] using the Hartree-Fock method (solid colored lines) are compared to the hydrogenic $Z^4$ dependence, which is computed for the 3*d* series (upper dashed line). For the *outermost* electrons (indicated by the circles and the shaded area), which are the relevant electrons in the solid, the quantum numbers $n$, $l$ change with $Z$ and the spin-orbit interaction increases much more slowly, following roughly the Landau-Lifshitz $Z^2$ scaling (lower dashed line), although within each series, the dependence remains closer to $Z^4$. Adapted from [Shanavas *et al.*, 2014].

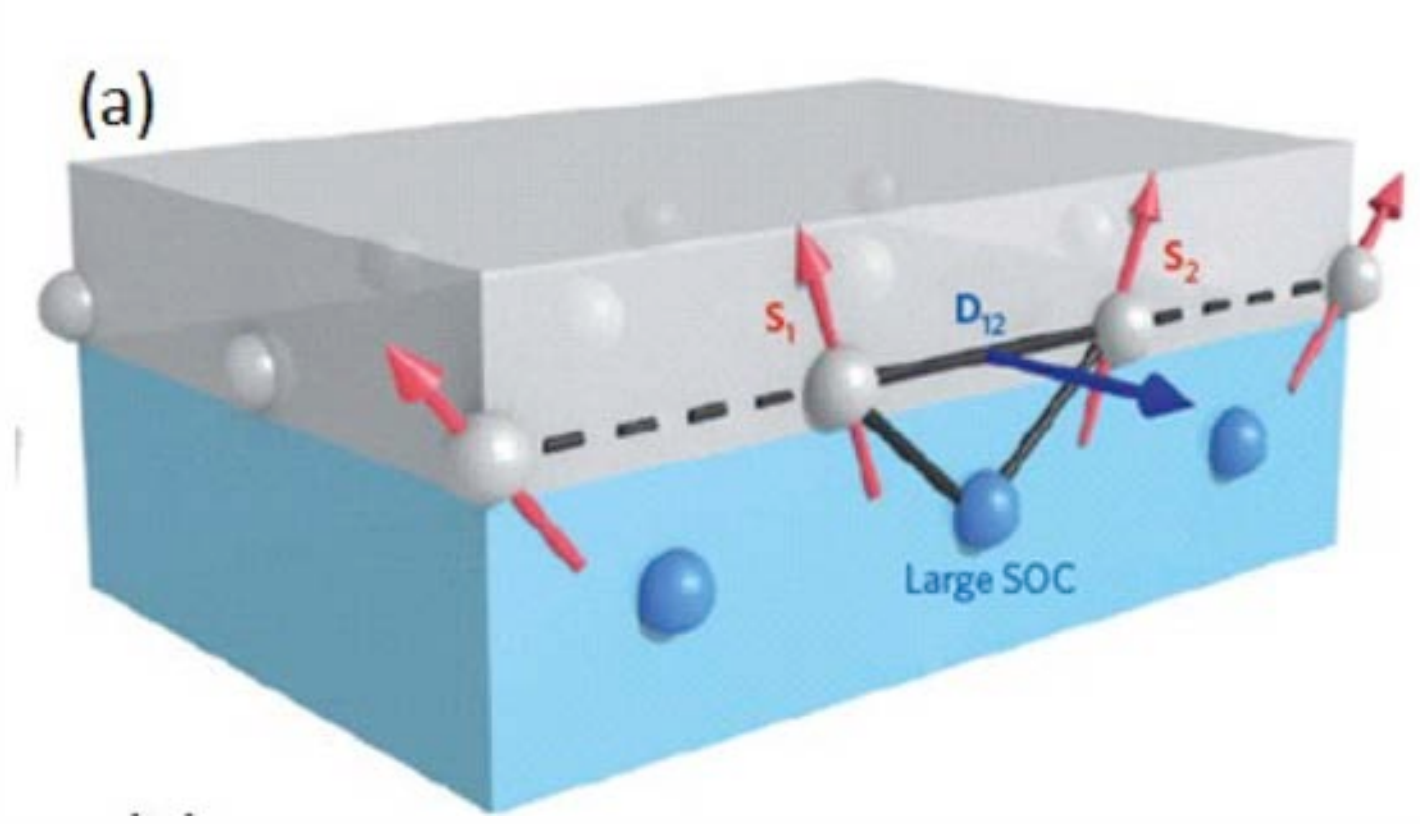

**Figure 3** Schematic of the 3-site mechanism for generating an interfacial Dzyaloshinskii-Moriya interaction [Fert *et al.*, 2013]. Spins $\mathbf{S}_1$ and $\mathbf{S}_2$ in the ferromagnetic (grey, upper) layer couple to each other through overlap of their wave functions with an atom with large spin-orbit coupling (blue, lower layer). This overlap gives rise to a contribution to the energy of the form $\mathbf{D}_{12}\cdot(\mathbf{S}_1\times\mathbf{S}_2)$, where $\mathbf{D}_{12}$ lies in the plane of the interface, in the direction normal to the plane defined by the three atoms. These qualitative properties are dictated by symmetry and identical to those predicted by more detailed non-local band models [e.g. Heinze *et al.*, 2011; Dupé *et al.*, 2014].

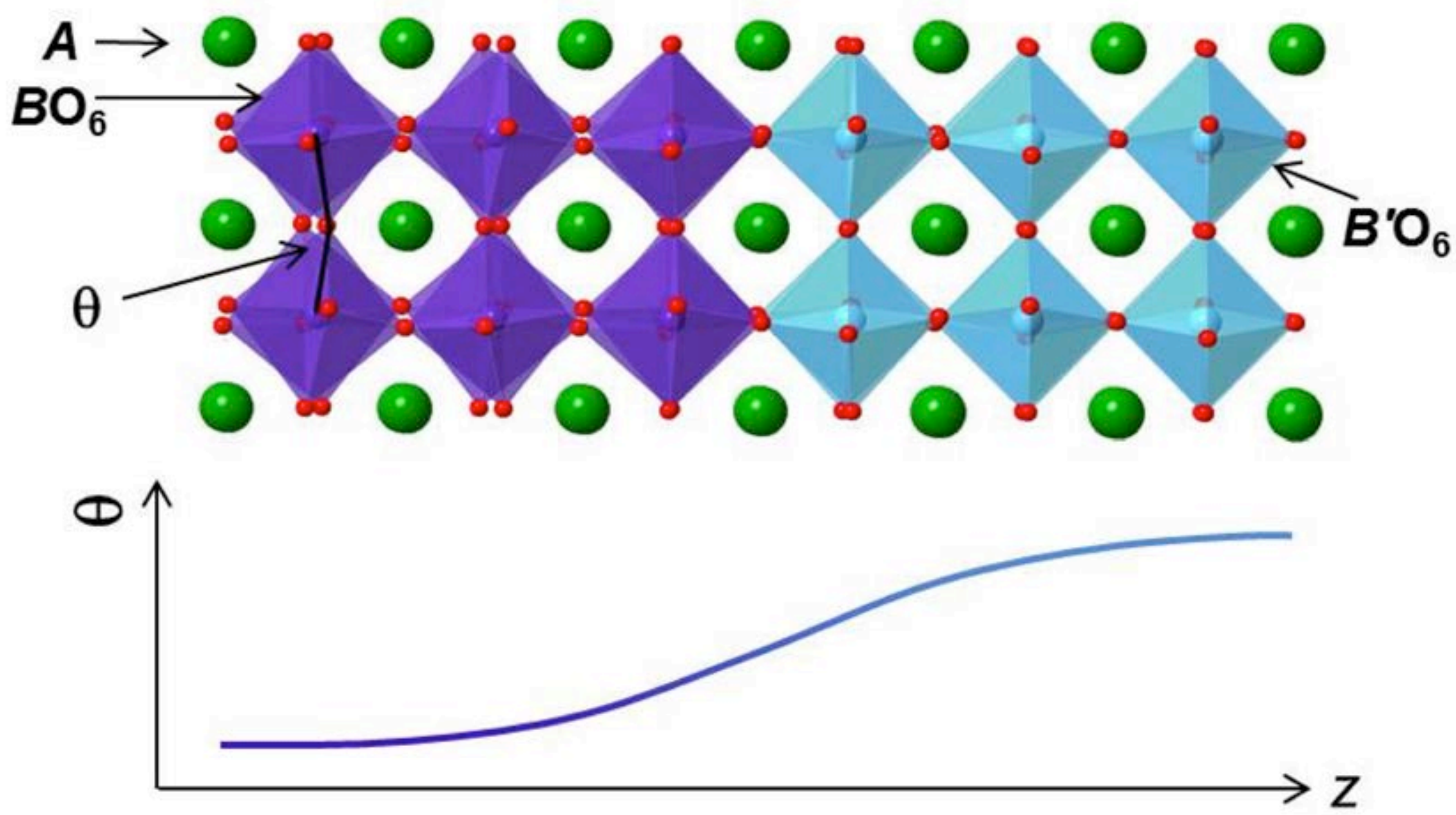

**Figure 4** Schematic of a perovskite $ABO_3$/$AB'O_3$ interface, across which spatial variations in the *B*-O-*B* bond angle *(θ)* can lead to non-bulk-like magnetic behavior. The length scale for coupling of $BO_6$ rotations is typically on the order of 2 to 8 unit cells, depending on the interfaces. The *A*-site cations and oxide anions are depicted as large green and small red spheres, respectively, while the $BO_6$ ($B'O_6$) octahedra are purple (blue). Similar length scales are commonly observed for interfacial charge transfer at oxide interfaces.

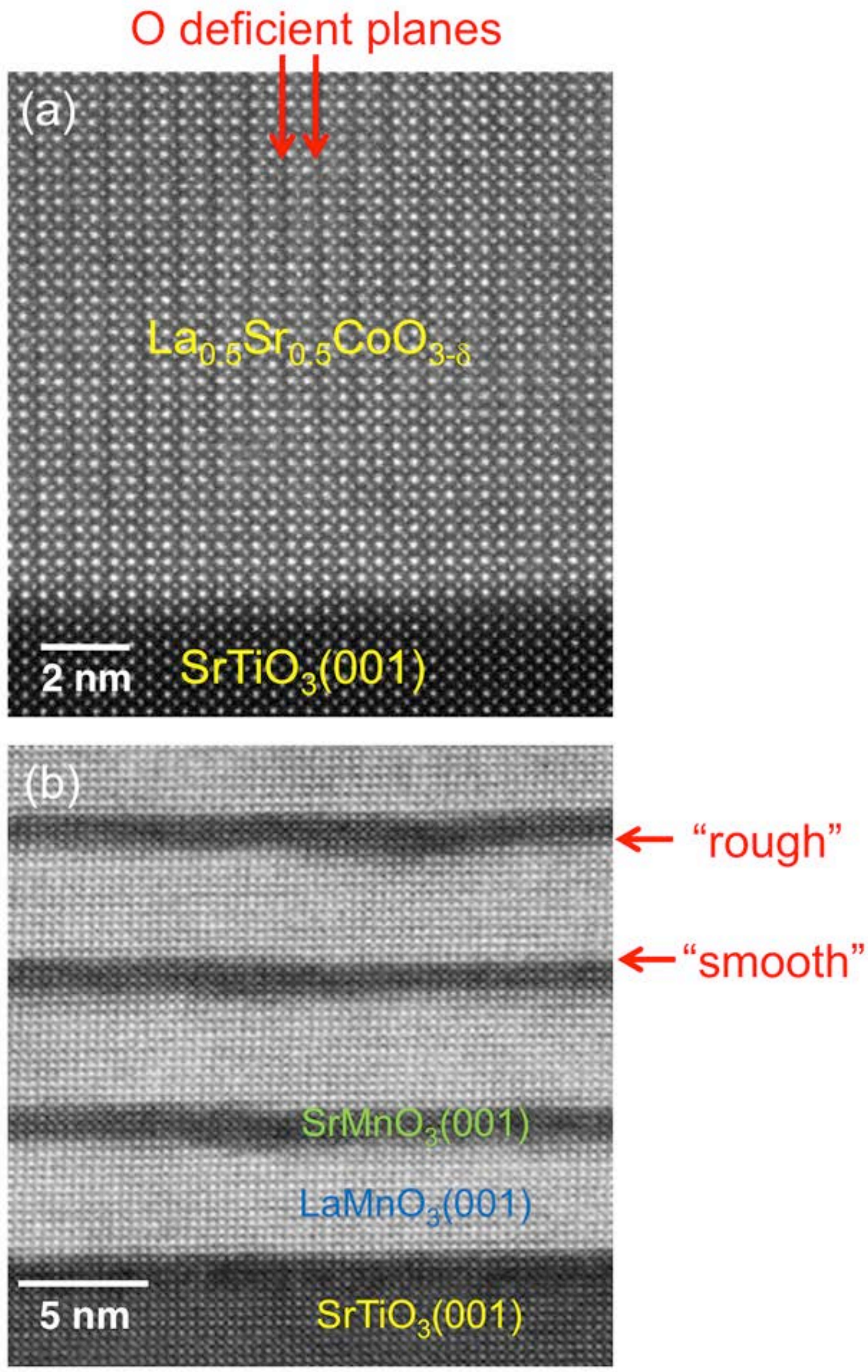

**Figure 5** (a) Scanning transmission electron microscopy image of a $La_{0.5}Sr_{0.5}CoO_{3-\delta}$ film on $SrTiO_3(001)$. Note the in-plane structural modulation due to oxygen vacancy ordering (red arrows at top). Image adapted from [Gazquez *et al*., 2013]. (b) Scanning transmission electron microscopy image of a $SrMnO_3/LaMnO_3$ superlattice on $SrTiO_3$. Note the structural asymmetry between the top and bottom interfaces of the $LaMnO_3$ layers (red arrows on right). Image adapted from [May *et al*., 2008].

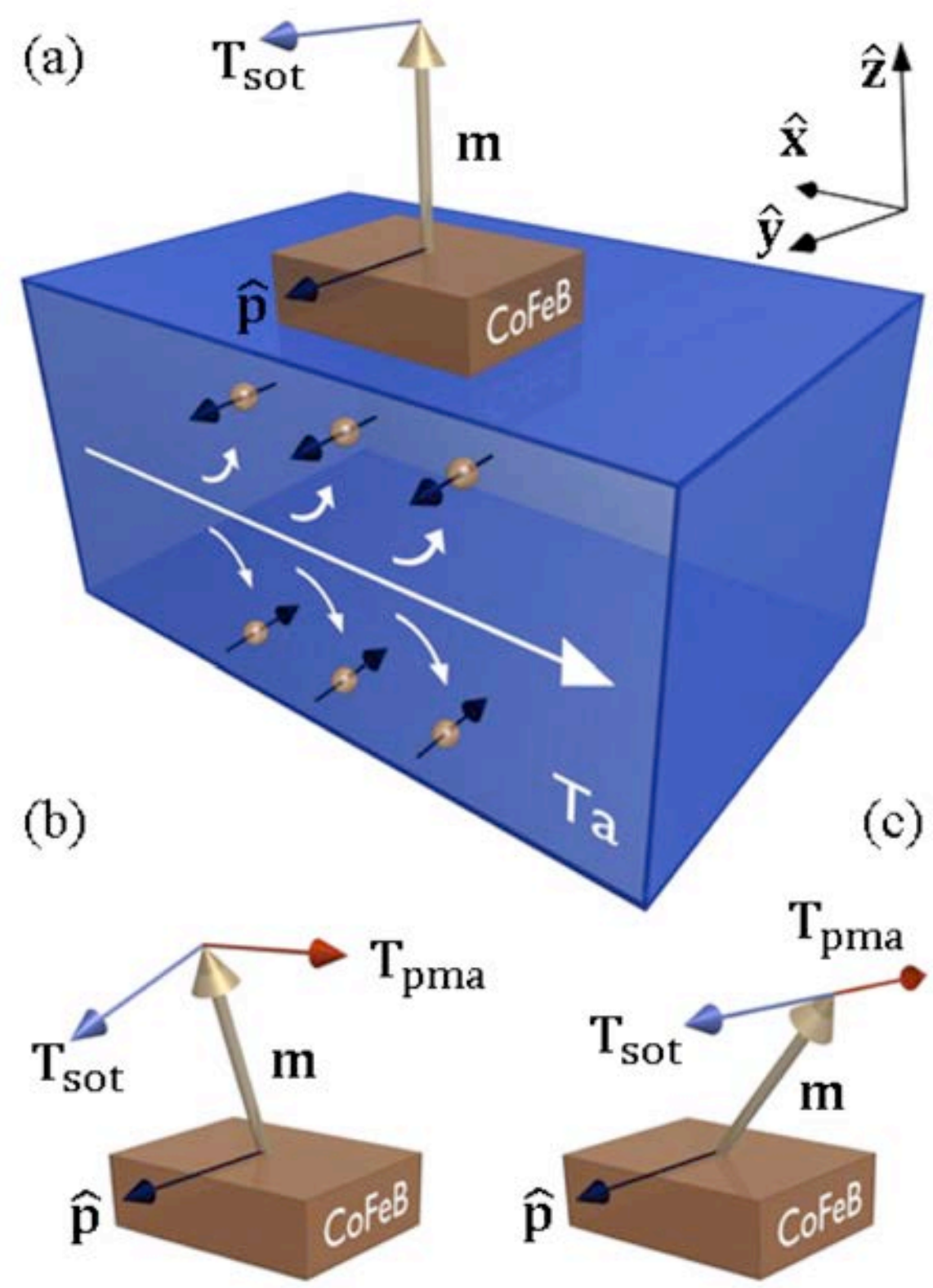

**Figure 6** Schematic of typical spin-orbit torque (dominated by antidamping-like torque of Eq. 3.3). (a) Charge current **j** flows in $\hat{\mathbf{x}}$-direction in a strongly spin-orbit-coupled non-magnetic metal (here, Ta) (large white arrow shows electron flow along $-\hat{\mathbf{x}}$ with electrons represented as gold spheres), creating a bulk spin Hall effect ($\pm\hat{\mathbf{y}}$-polarized spin moments (opposite to the $\mp\hat{\mathbf{y}}$ spin direction), shown with small dark blue arrows, deflected along $\pm\hat{\mathbf{z}}$). This mechanism creates a spin current with spin moments pointing along $\hat{\mathbf{p}}$ **(**$=\hat{\mathbf{y}}$**)** (black arrow) that flows along $\hat{\mathbf{z}}$ into the ferromagnet (here, *a*-Co-Fe-B with perpendicular magnetic anisotropy (PMA) and magnetization **m**) and travels some distance before losing spin-polarization, causing spin-orbit torque $\mathbf{T}_{sot}$ (blue vector) that acts on **m** (gold vector). In response to the *onset* of current and spin flow, **m** tilts in the direction of $\mathbf{T}_{sot}$ *i.e.,* towards $\hat{\mathbf{y}}$, shown in (b). As **m** tilts away from $\hat{\mathbf{z}}$, $\mathbf{T}_{sot}$ also tilts (shown in b) along the component of $\hat{\mathbf{p}}$ perpendicular to **m**, and a new torque develops due to PMA, $\mathbf{T}_{pma}$ (red vector) (one type of effective field, referred to in Eq. 2.3). Below a critical current (which depends on anisotropy strength), **m** precesses with decreasing amplitude (dependent on damping constant $\alpha$) until it reaches a stationary state with **m** tilted in the *x-z* plane where the two torques cancel, as shown in (c). At higher current, the stationary state **m** is along $\hat{\mathbf{p}}=\hat{\mathbf{y}}$. In the absence of an additional symmetry-breaking field (as discussed in the text), **m** never crosses the *x-y* plane, *i.e.*, these torques do not lead to magnetization reversal for uniform **m**.

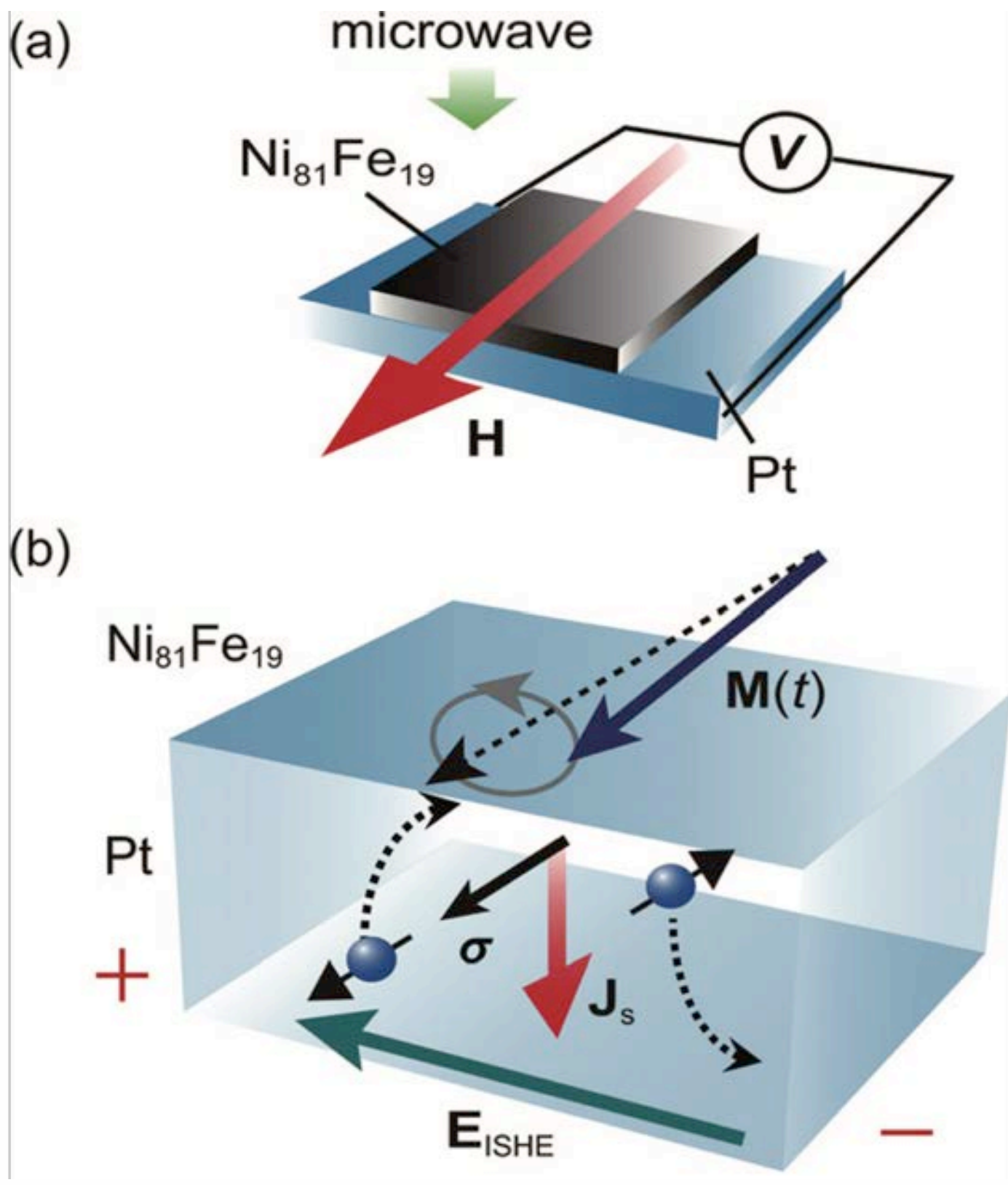

**Figure 7** Schematic of spin pumping from a ferromagnet ($Ni_{81}Fe_{19}$) film into a nonmagnetic metal (Pt). (a) A microwave field causes precession of magnetization **M**(t) around the applied external field **H**, pumping spins into the nonmagnet and generating voltage *V* through the inverse spin Hall effect. (b) The precessing **M**(t) pumps spins into the Pt causing spin current $\mathbf{J}_s$ with spins **σ**, in the directions shown, equivalent to a net flow of moments (black arrows on electrons) oppositely directed to **M**. The moving electrons are deflected (to the right, for Pt which has the opposite sign of SHE to Ta) by the inverse spin Hall effect which creates an emergent (effective) electric field $\mathbf{E}_{ISHE}$ in the direction shown. In an open circuit, shown in (a), the resulting induced transient current causes charge to accumulate at the ends of the sample, indicated by the +/-, giving rise to a real electric field (equal and opposite to $\mathbf{E}_{ISHE}$) and the measured voltage *V*. From [Ando *et al.*, 2011]

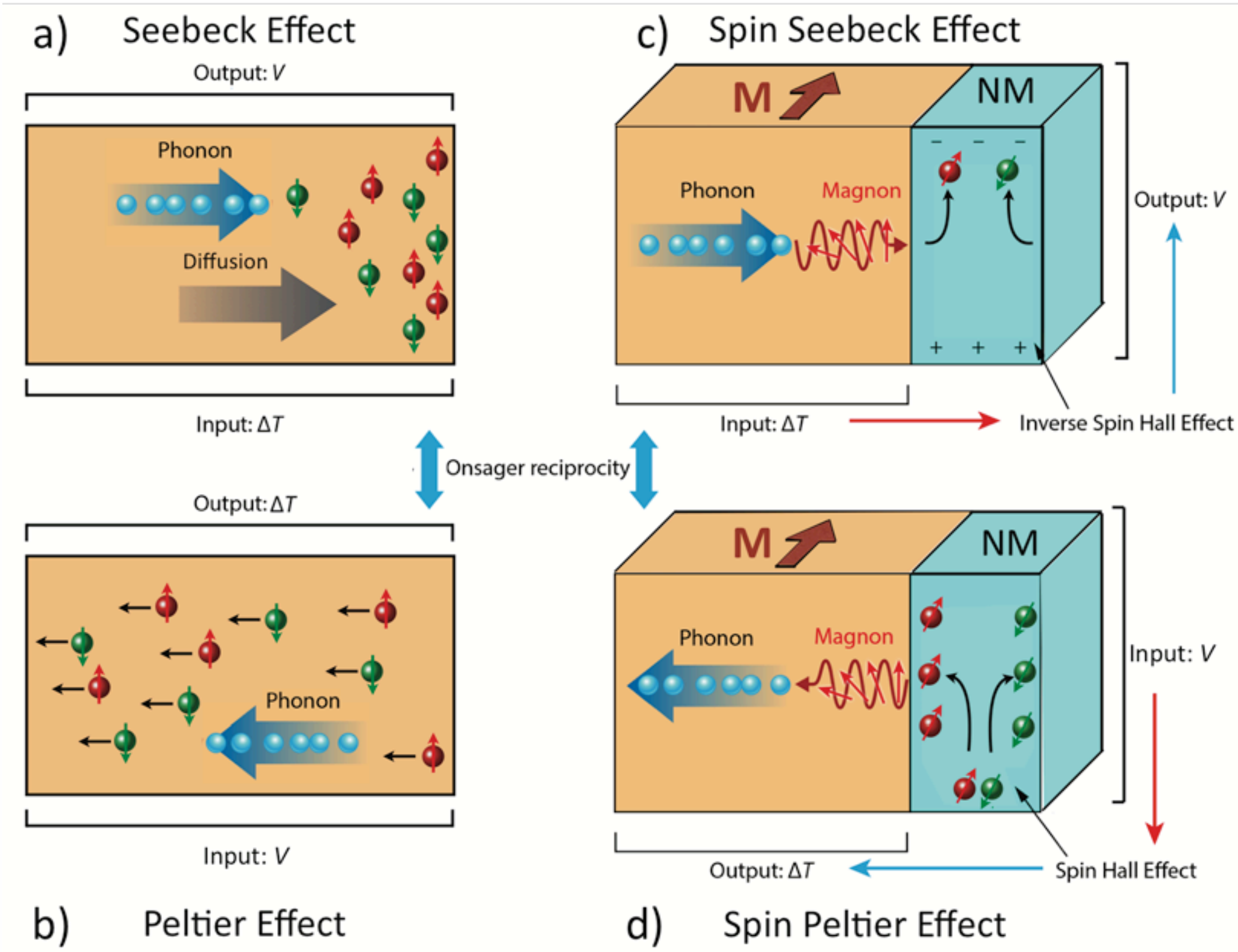

**Figure 8** Schematic illustrations of (a) conventional Seebeck effect with temperature difference $\Delta T$ applied to a metal (hot end on left), electric field $E$ (voltage $V$) generated along $\Delta T$ and (b) conventional Peltier with applied voltage generating a temperature difference. The conventional effects in (a) and (b) are dominated by thermally-induced asymmetric diffusion (indicated by large grey diffusion arrow) of both spin up (red) and spin down (green) electrons in (a), or voltage-induced drift (small black arrows for both spin up and spin down electrons) in (b), though a phonon flux is always present (as indicated schematically) as well as a magnon flux, which can contribute *via* momentum transfer in drag effects. These are compared to (c) spin Seebeck effect with $\Delta T$ applied to a ferromagnet with **M** into the page (either metal or insulator) which produces a spin current and (d) the reciprocal spin Peltier effect. When connected to a non-magnetic metal (NM, blue), $\Delta T$ induces a spin current that is converted into a vertical electric field $\mathbf{E}_{\mathrm{ISHE}}$ and $V_{\mathrm{ISHE}}$ due to the inverse spin Hall effect. In (c) and (d), magnons in the ferromagnet are shown producing the spin current, but different chemical potentials for up and down spins in a metallic ferromagnet would also produce a spin current *if (and only if)* the spin diffusion length were comparable to the sample length. For spin Seebeck and spin Peltier effects, clearly distinguishable only in magnetic insulators, schematics show current understanding of the physical mechanism, which is thermally-driven magnons that cause incoherent spin pumping at the interface. Adapted from [Heremans and Boona, 2014].

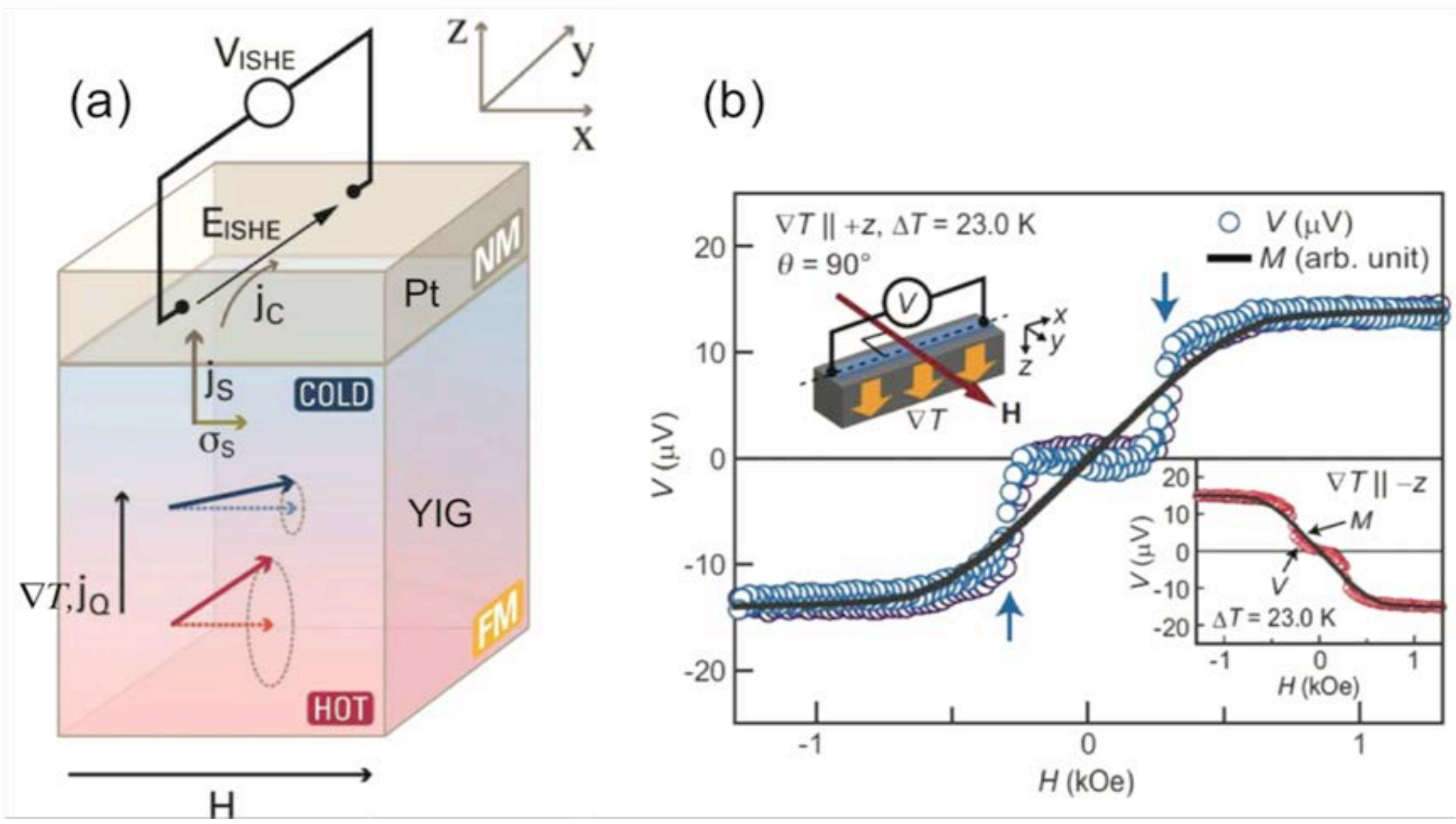

**Figure 9** Longitudinal spin Seebeck effect in the YIG/Pt system. (a) geometry of the experiment [Boona *et al.*, 2014] and (b) data, showing the dependence of the inverse spin Hall effect voltage $V_{ISHE}$ on the applied magnetic field $H$ at room temperature [Uchida *et al.*, 2010a].

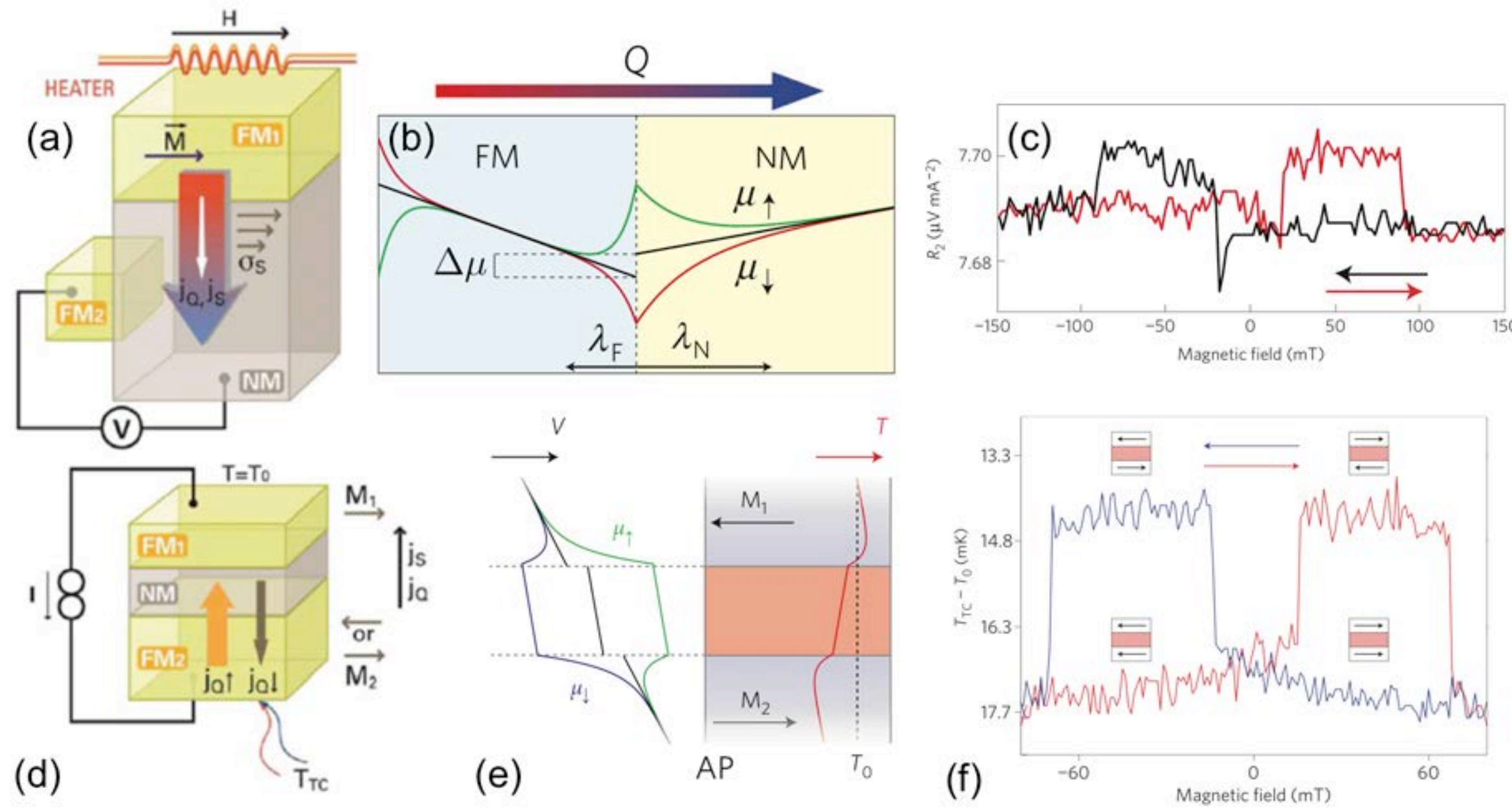

**Figure 10** Spin-dependent Seebeck (panels a-c) and Peltier (panels d-f) effects for conduction electrons: (a) non-local geometry used for observation of spin-dependent Seebeck effect [Boona *et al.*, 2014]; (b) expected spatial dependence of spin-dependent chemical potentials compared to spin diffusion lengths in ferromagnets (FM) and non-magnets (NM) and (c) thermally-driven spin accumulation signal [Slachter *et al.*, 2010]; (d) Structure used for Onsager reciprocal spin-dependent Peltier effect [Boona *et al.*, 2014]; (e) expected spin-dependent chemical potentials for antiparallel alignment of FM elements, clarifying the short length scale probed and (f) current-driven spin-dependent temperature difference [Flipse *et al.,* 2012].

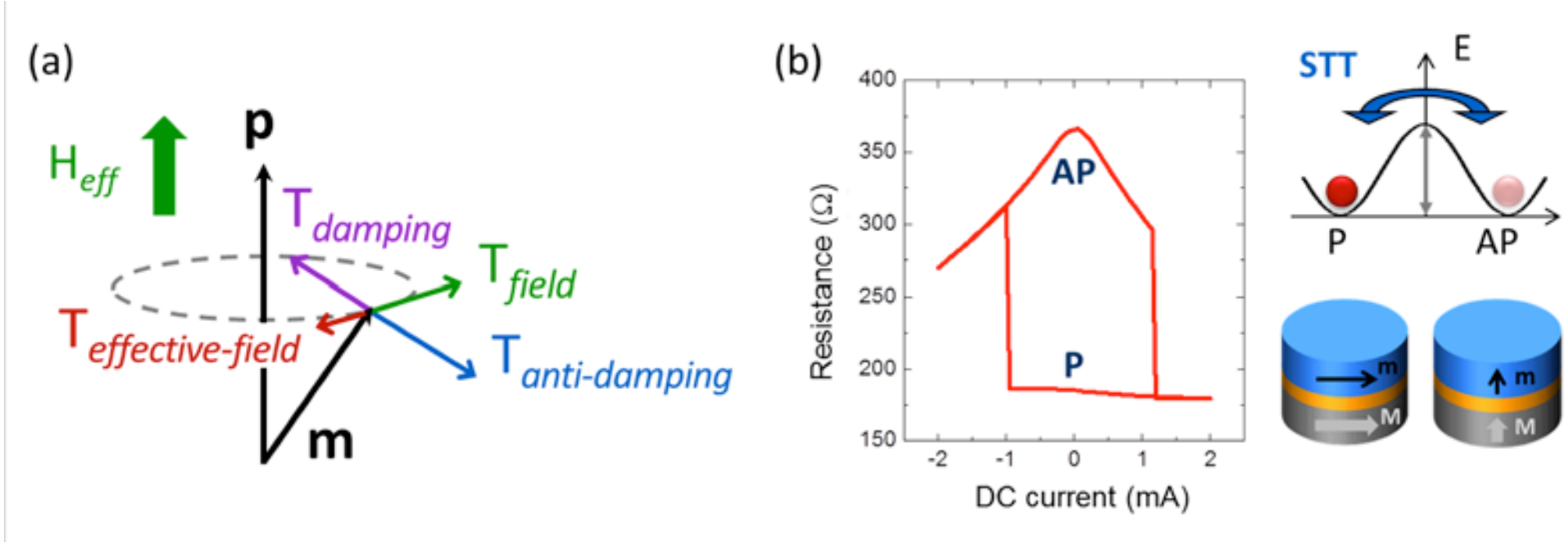

**Figure 11** (a) Torques **T** on a uniform magnetization with direction **m**. The vector **p** indicates the direction of spin polarization of the spin current; in "conventional" spin-transfer, this is along $\mathbf{M}_{\text{fixed}}$ (see Eq. 3.2) and in spin-orbit torque transfer, this is $\hat{\mathbf{p}}$ (see Eq. 3.3) which is in-plane and perpendicular to the charge current. $\mathbf{T}_{\text{field}}$ is the torque associated with $\mathbf{H}_{\text{eff}}$ (which includes applied field, exchange interaction, anisotropy) and $\mathbf{T}_{\text{damping}}$ is the (conventional) Gilbert damping torque, as described in Eq. 2.3. $\mathbf{T}_{\text{effective-field}}$ and $\mathbf{T}_{\text{anti-damping}}$ are the field-like and damping-like (sometimes called Slonczewski) parts of the torque associated with spin-transfer or spin-orbit torques, as discussed in Eqs. 3.2 and 3.3; depending on the sign of the charge current producing the torque, $\mathbf{T}_{\text{anti-damping}}$ may be directed along or opposite to the conventional damping torque $\mathbf{T}_{\text{damping}}$. When they are opposite, as shown, $\mathbf{T}_{\text{anti-damping}}$ offsets $\mathbf{T}_{\text{damping}}$, leading to various dynamic behaviors. (b) Left: Hysteretic switching of the resistance of a magnetic tunnel junction (ferromagnet/non-magnetic insulator/ferromagnet) by *spin-transfer torque* from a direct current. Upper right: energy diagram of spin-transfer torque switching between parallel (P) and antiparallel (AP) states of the two ferromagnets. Lower right: magnetic tunnel junction with in-plane (left cylinder) or perpendicular (right cylinder) magnetic anisotropy can be switched between P and AP configurations using spin-transfer torque; typically one layer has pinned magnetization ($\mathbf{M}_{\text{fixed}}$) while the other can be switched.

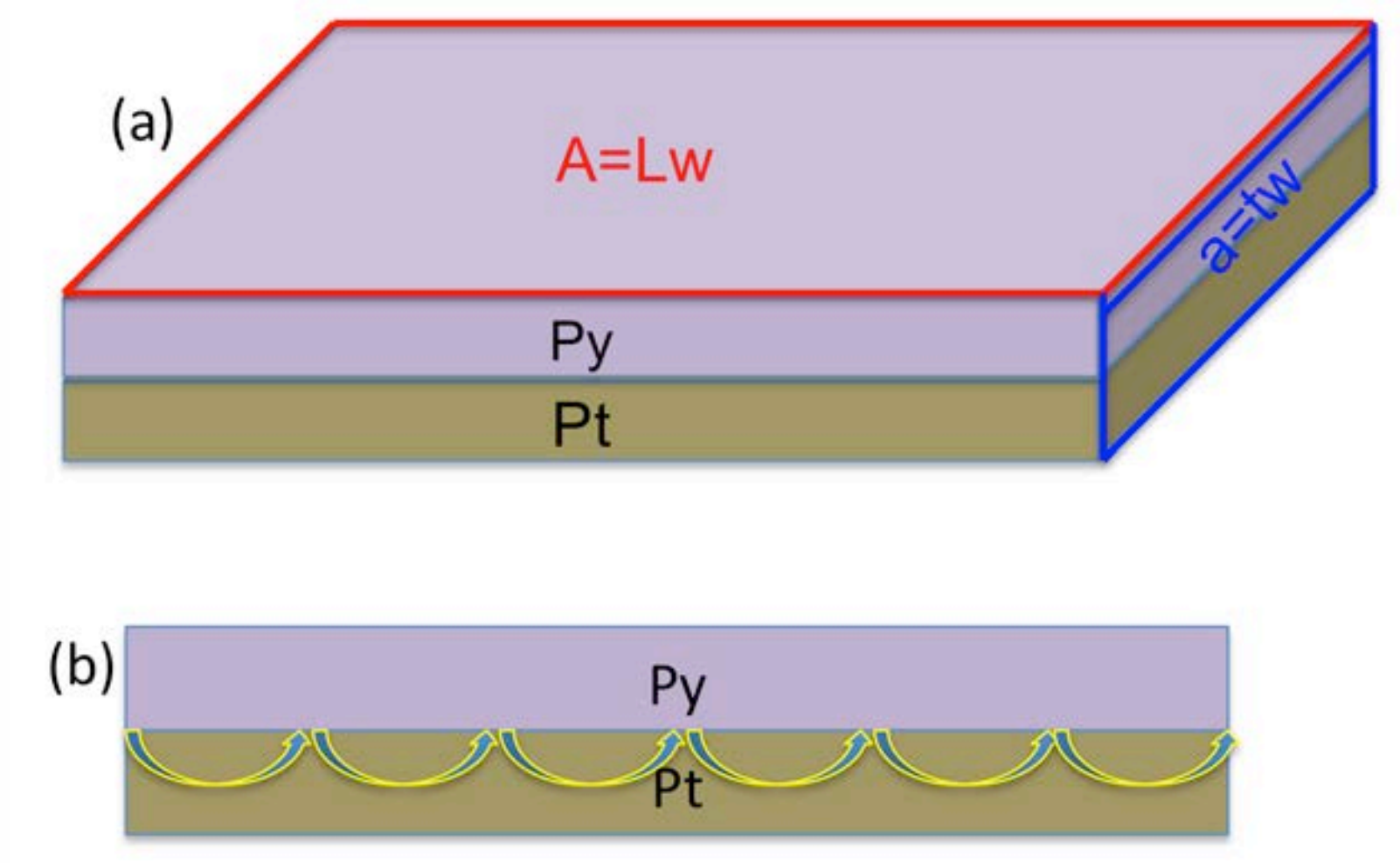

**Figure 12** (a) In a bilayer of a heavy metal and ferromagnet (*e.g.*, Pt/permalloy (Py)) in which spin torque is created by the spin Hall effect, the charge current density $J$ and the spin current density $J_S$ pass through very different areas $a$ and $A$ respectively. (b) High torque efficiency is possible because each electron transfers spin to Py several times, as described in the text.

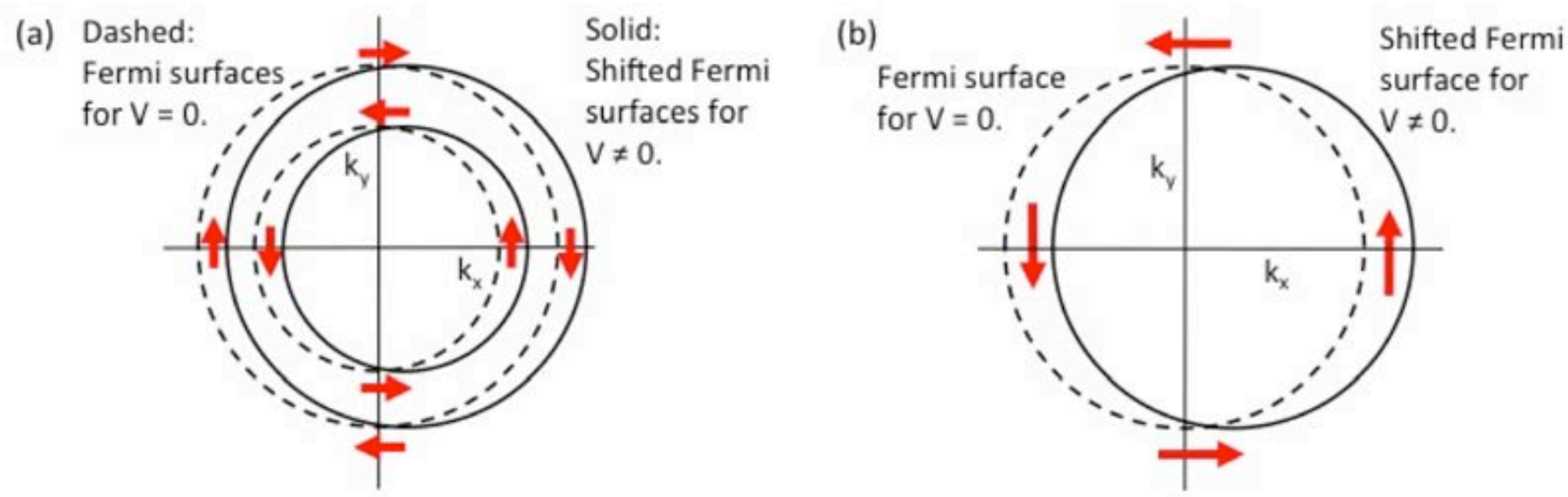

**Figure 13** (a) The Rashba spin-orbit interaction splits the spin-degeneracy of surface or interface state Fermi surfaces. The spin-orientation of a surface band state depends on momentum, and on a given Fermi surface is opposite for opposite momenta because of time-reversal symmetry. An in-plane current increases the occupation probability of states on one side of the Fermi surface and decreases them on the other side, generating a non-equilibrium spin accumulation. If these spins are exchange coupled to an adjacent ferromagnet, they can apply a torque. (b) Topological insulators have an odd number of surface-state Fermi surfaces and in the simplest case, a single Fermi surface. The partial cancellation that occurs between the separate spin-accumulations of weakly spin-split Fermi surfaces in the Rashba interaction case (a) is therefore absent in the topological insulator case (b) and spin accumulations likely larger. In both cases the spin accumulation is required by symmetry to be perpendicular to the current direction, but its sign depends on surface state electronic structure details.

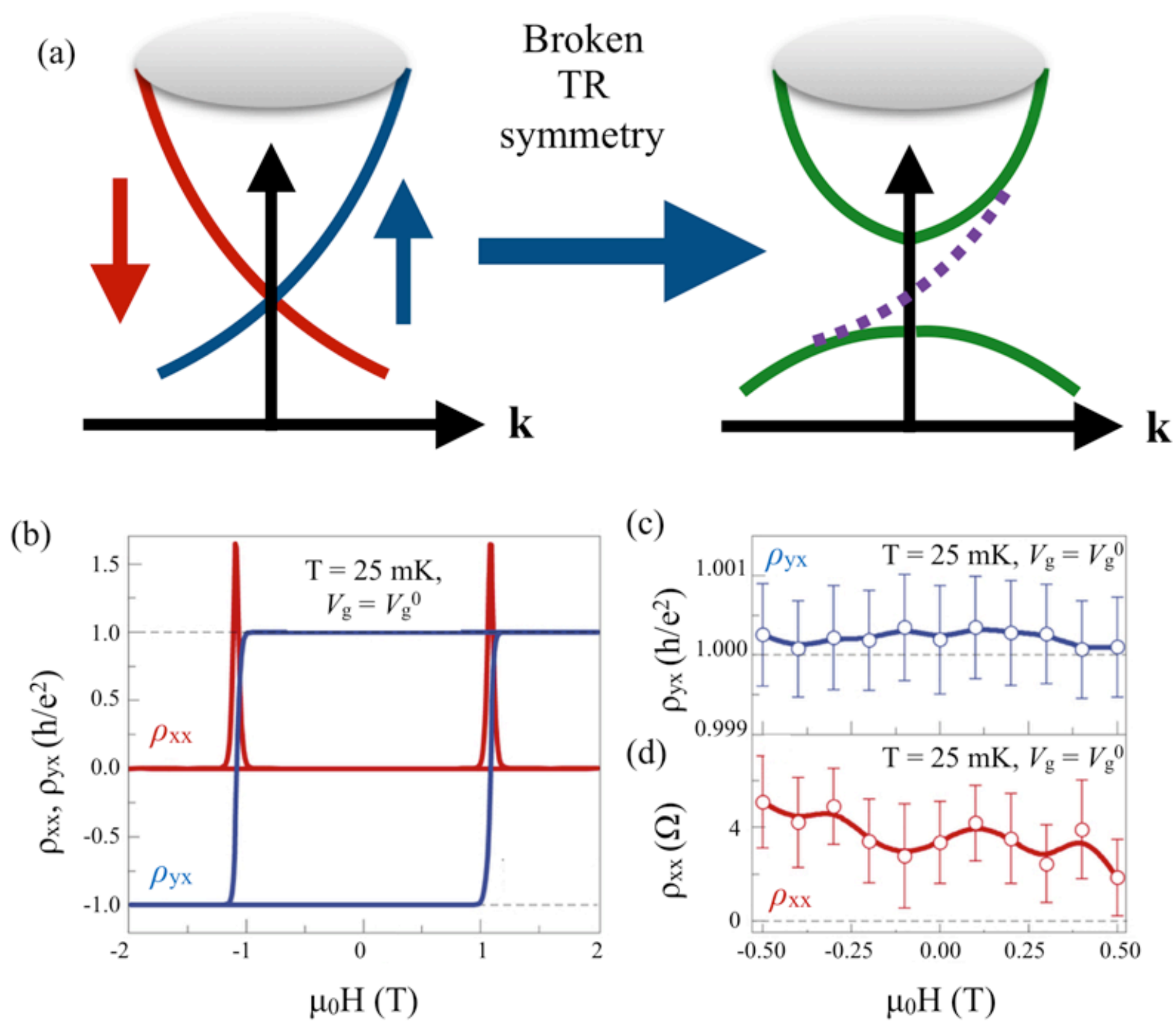

**Figure 14** (a) Illustration of formation of the quantum anomalous Hall effect in a 3D TI film. When time-reversal (TR) symmetry is broken by ferromagnetic ordering in a magnetically-doped 3D TI, the Dirac point in the surface band structure is disrupted by opening of a 'magnetic gap'; the helical spin textured surface states then engender a single chiral edge mode (dashed purple line) characterized by ballistic transport. (b-d) Observation of quantized anomalous Hall effect in thin film V-doped $(Bi,Sb)_2Te_3$: when the electron chemical potential is tuned into the magnetic gap *via* electrostatic gate, the Hall resistance $\rho_{yx} = h/e^2$ (to 1 part in $10^3$) and longitudinal sheet resistance $\rho_{xx}$ is only a few ohms [Chang *et al*., 2015b]. Subsequent experiments demonstrated quantization of Hall resistance to 4 parts in $10^4$ [Liu *et al.*, 2016], despite significant magnetic disorder [Lachman *et al*., 2015].

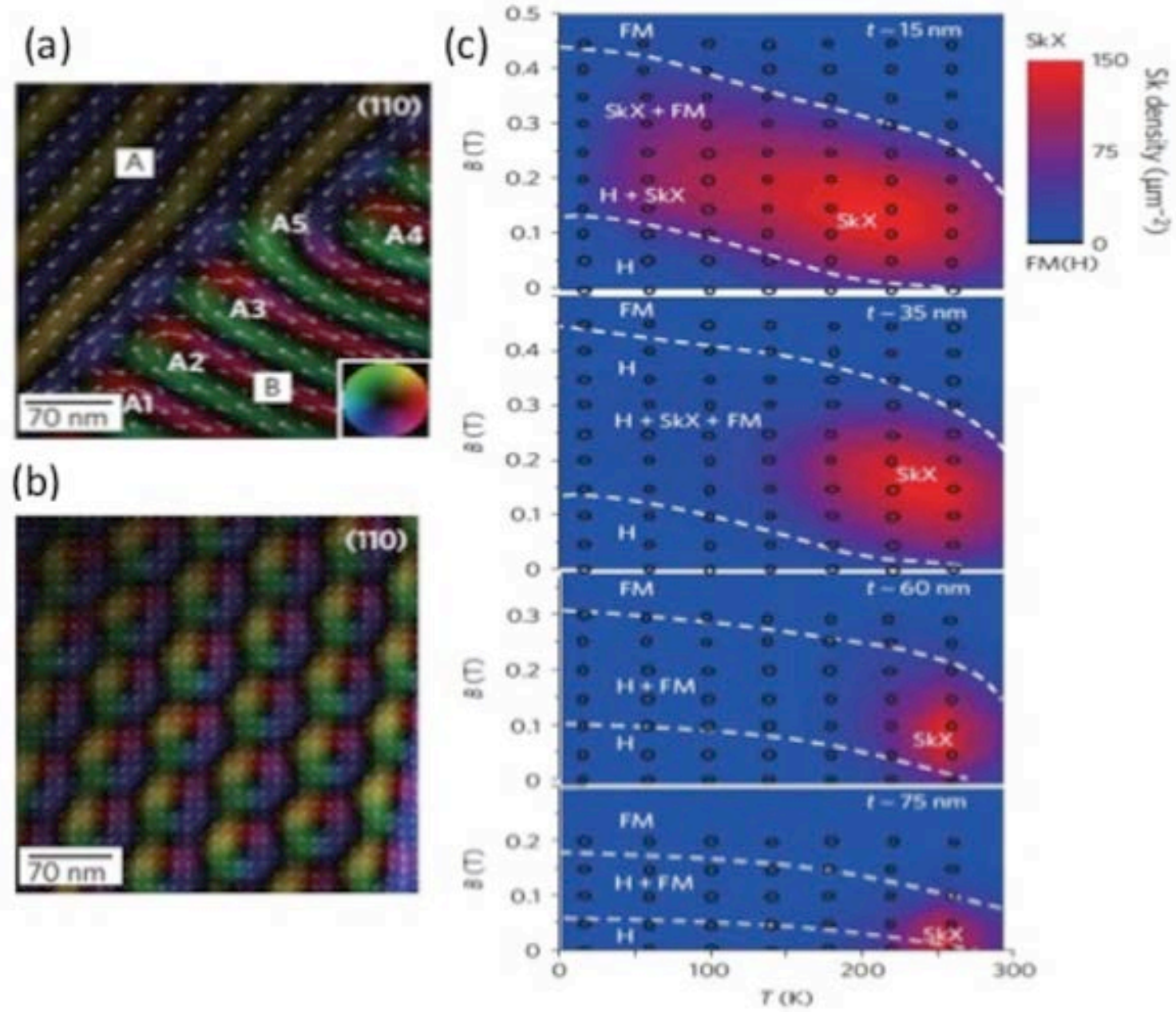

**Figure 15** Lorentz transmission electron microscopy (LTEM) images of B20 FeGe thin films in (a) the helical phase, showing magnetic chirality twinning at a structural twin boundary (A and B regions) and (b) the skyrmion phase induced by 0.1 T magnetic field applied normal to sample plane at 260 K. Color wheel (inset) and white arrows represent the magnetization direction at every point. c) Sample thickness dependence of skyrmion (SkX), helical (H) and ferromagnet (FM) phase diagram in the magnetic field $B$ - temperature $T$ plane. Color bar is the skyrmion density per square micron. [Yu *et al.,* 2011]

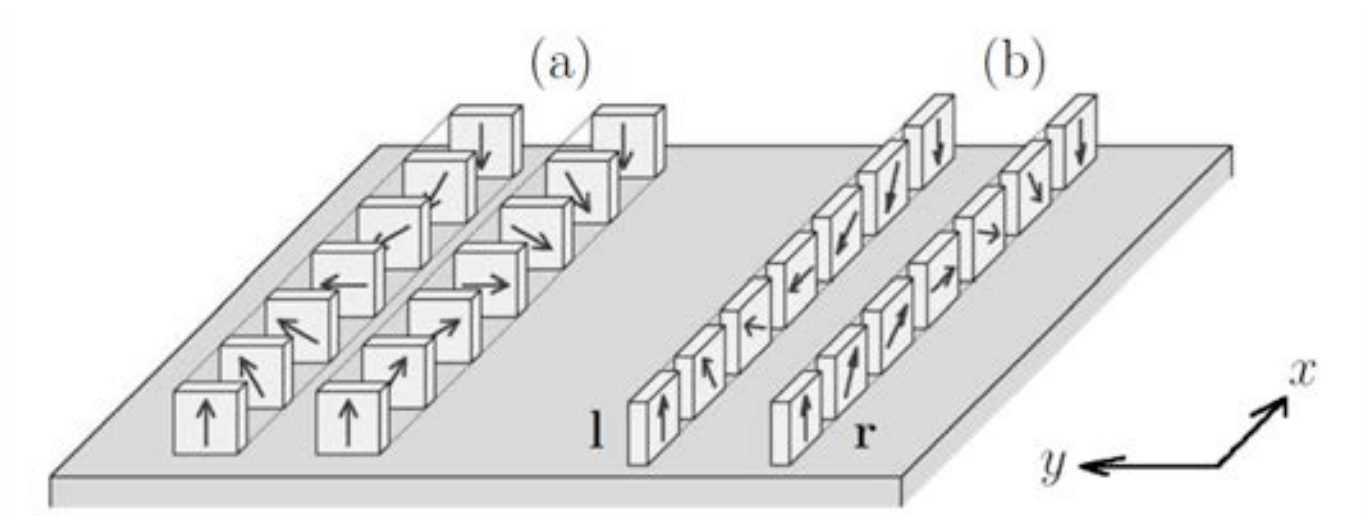

**Figure 16** Four types of 180° domain walls in materials magnetized up at front of figure and down at back. (a) Bloch left- and right-handed walls, and (b) Néel left- and right-handed walls [Heide *et al.*, 2008].

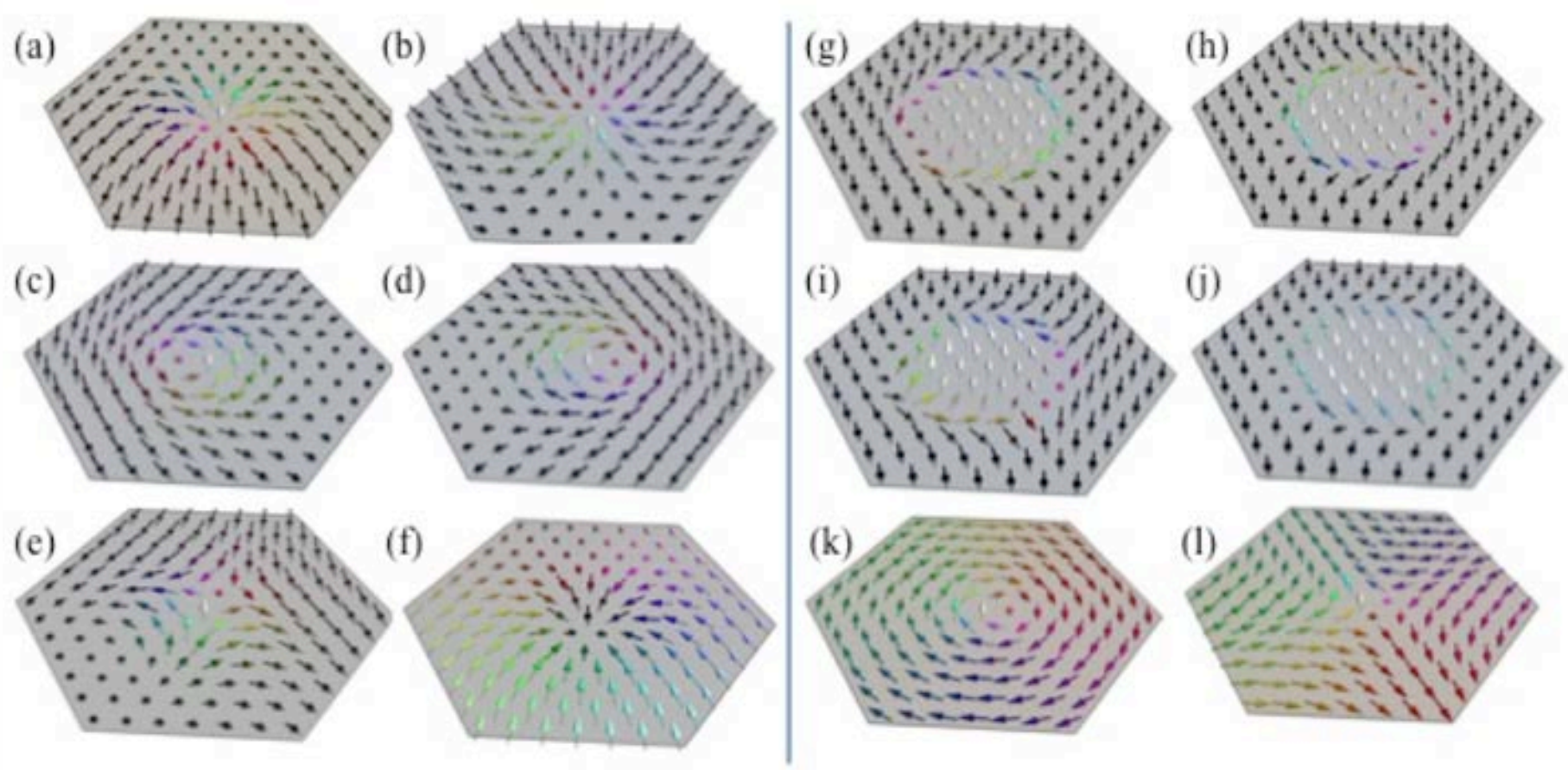

**Figure 17** Topology and chirality of skyrmions and vortices. The heavy metal layer is assumed to be below in order to define chirality. Structures (a)-(f) are skyrmions constructed using the Belavin-Polyakov profile (2D Heisenberg exchange coupling only). (a)-(e) are within a down magnetized background with an up core. (a), (b) are right-handed, left-handed Néel (hedgehog-type) respectively with $N_{sk}$=+1; (c), (d) left-handed, right-handed Bloch (vortex-type) respectively with $N_{sk}$=+1; (e) $N_{sk}$= -1, with no defined chirality (because spin directions and spatial coordinates counter-rotate). (f) is within an up magnetized background with a down core, $N_{sk}$=-1, right-handed Néel skyrmion (note that this structure is the same as (a) with all directions reversed). Structures (g) - (j) are magnetic bubbles within a down magnetized background with an extended up core; (g), (h) have $N_{sk}$=+1, and left-handed/right handed Bloch walls respectively. (i) has 4 Bloch lines at each of which the in-plane moment reverses, reducing $N_{sk}$ by 1/2, leading to $N_{sk}$=-1 (this structure is sometimes called an anti-skyrmion); (j) has 2 Bloch lines and $N_{sk}$=0. Structures (k) and (l) are magnetic vortex and antivortex respectively, both with an up core and **m** in the *x*-*y* plane away from the core and $N_{sk}$= ±1/2; both are sometimes called merons.

.

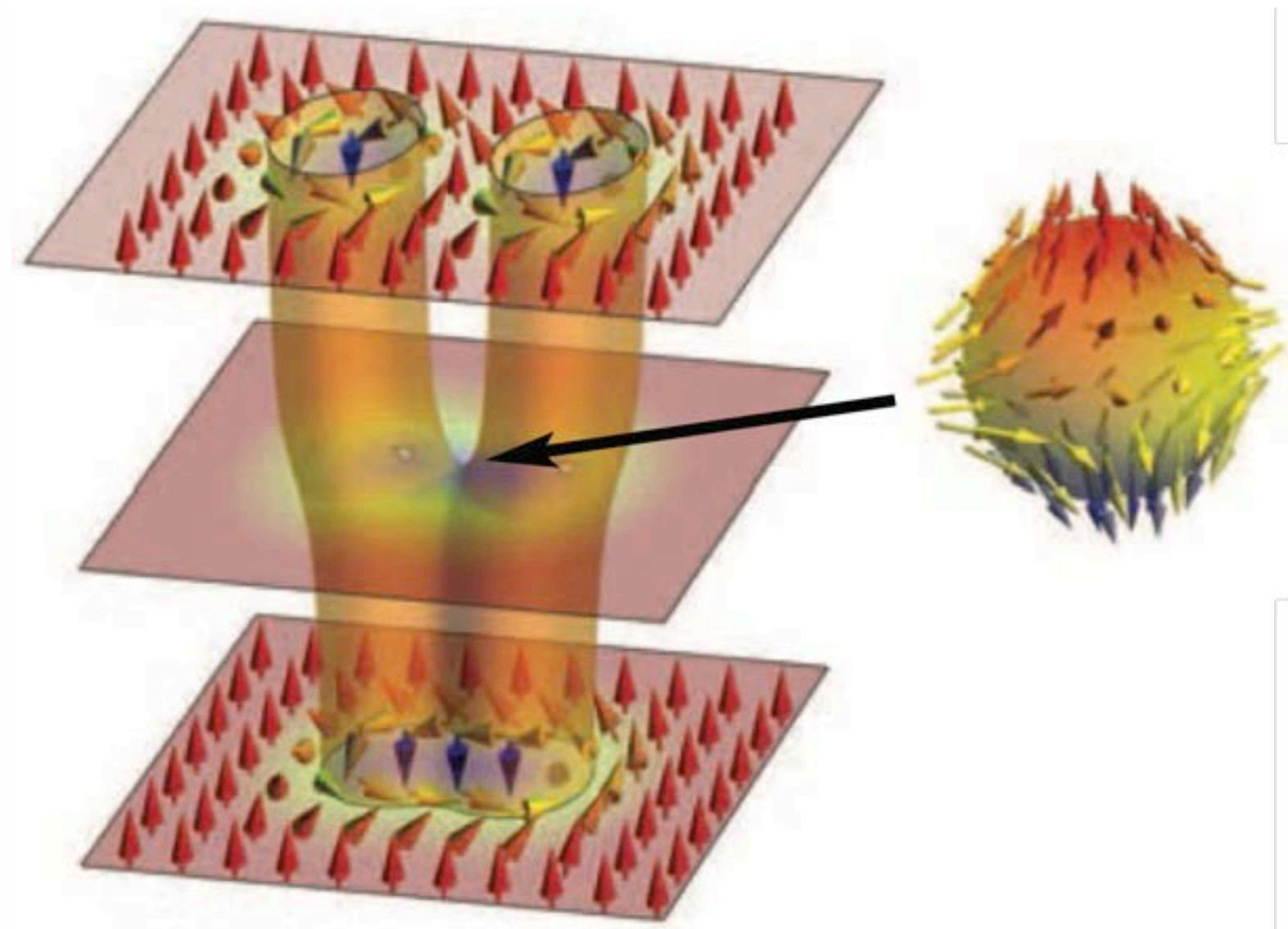

**Figure 18** Schematic merging of two skyrmions. At the merging point the magnetization vanishes at a singular point, the Bloch point (arrow), which acts like the slider of a zipper connecting two vertically-extended skyrmions. [Milde *et al.*, 2013]

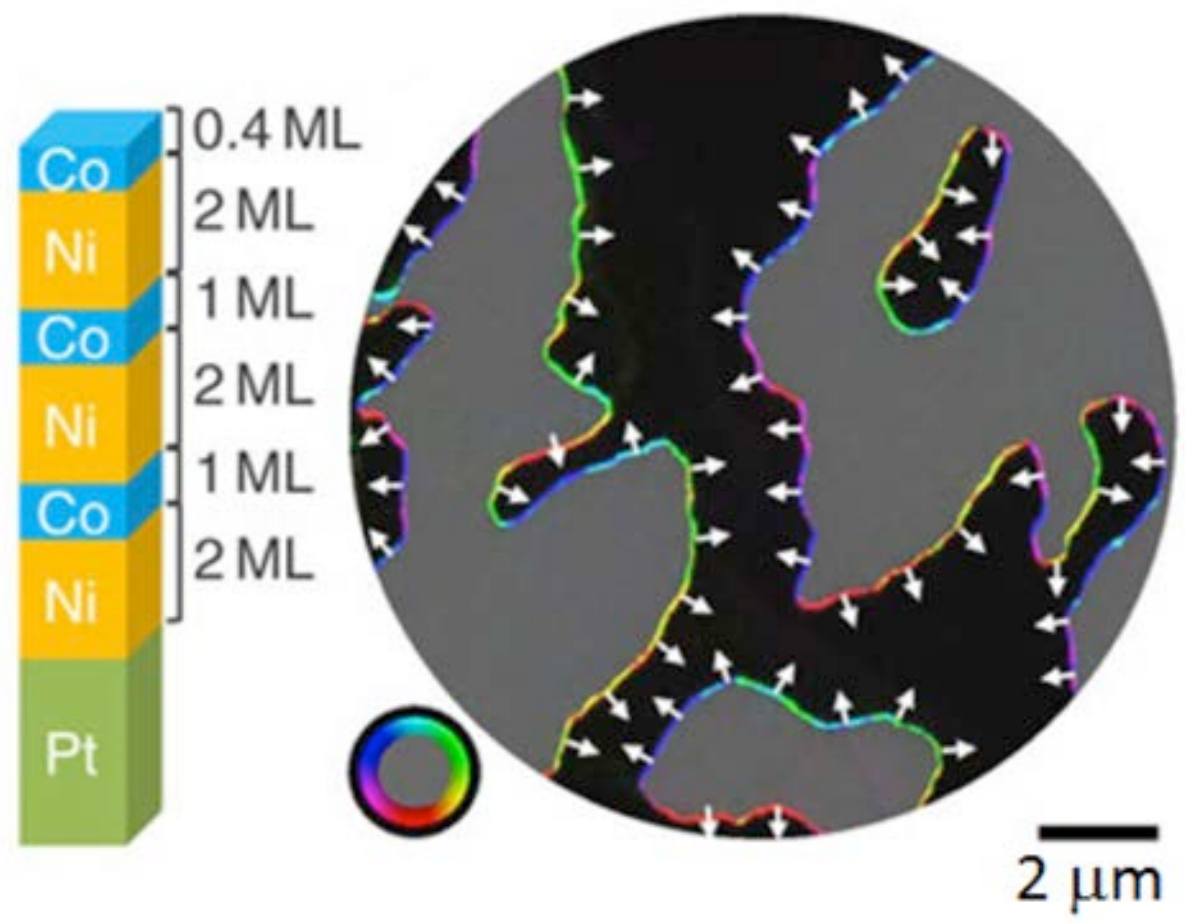

**Figure 19** Chiral right-handed Néel walls in a [Co/Ni] multilayer with perpendicular magnetic anisotropy on a Pt(111) substrate (layer thicknesss as shown). Images taken with spin-polarized low energy electron microscopy (SPLEEM). Up/down magnetic domains shown in grey/black. Color wheel shows in plane direction of spins, with white arrows clarifying direction of chirality based on angle of spin, at the 180° domain walls [Chen *et al.*, 2013a].

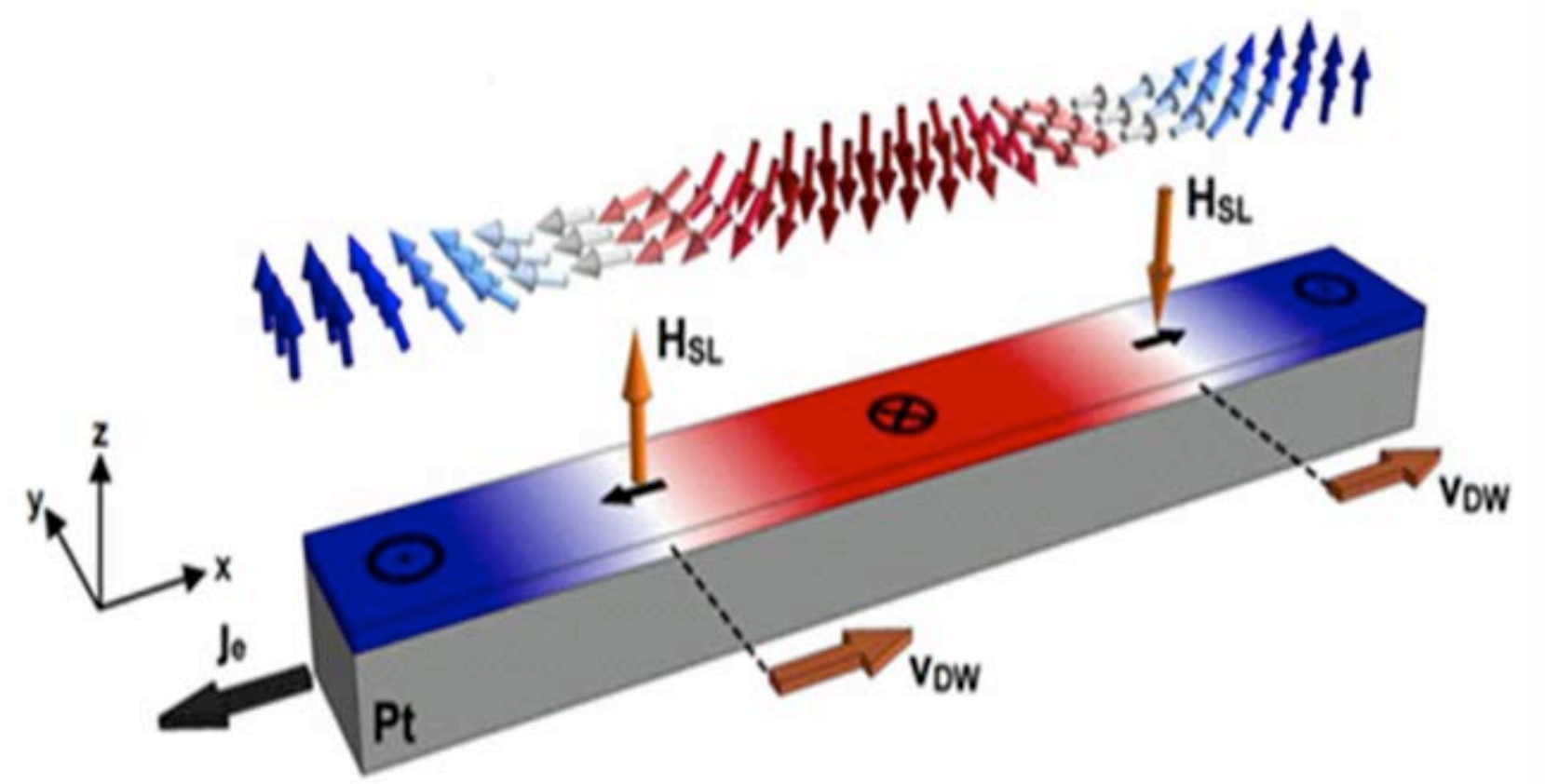

**Figure 20** Schematic showing two left-handed chiral Néel domain walls (also called Dzyaloshinskii domain walls) separating a (red) down-domain from two (blue) up-domains in the top ferromagnetic layer (red, blue, and white arrows show directions of spins in this layer) driven by the Slonczewski-like effective field $\mathbf{H}_{\mathrm{SL}}$ (along ±**z** at each domain wall, as shown) due to charge current $j_c$ (along –**x**, shown with thick black arrow) and resulting spin Hall effect in the underlying Pt layer, which causes spin current along +**z** with +**y**-polarization. This spin current produces oppositely-directed $\mathbf{H}_{\mathrm{SL}}$ due to oppositely-directed spins in the two domain walls, causing *both* domain walls to move with velocity $\mathbf{v}_{\mathrm{DW}}$ along +**x**. [Emori *et al.*, 2013]

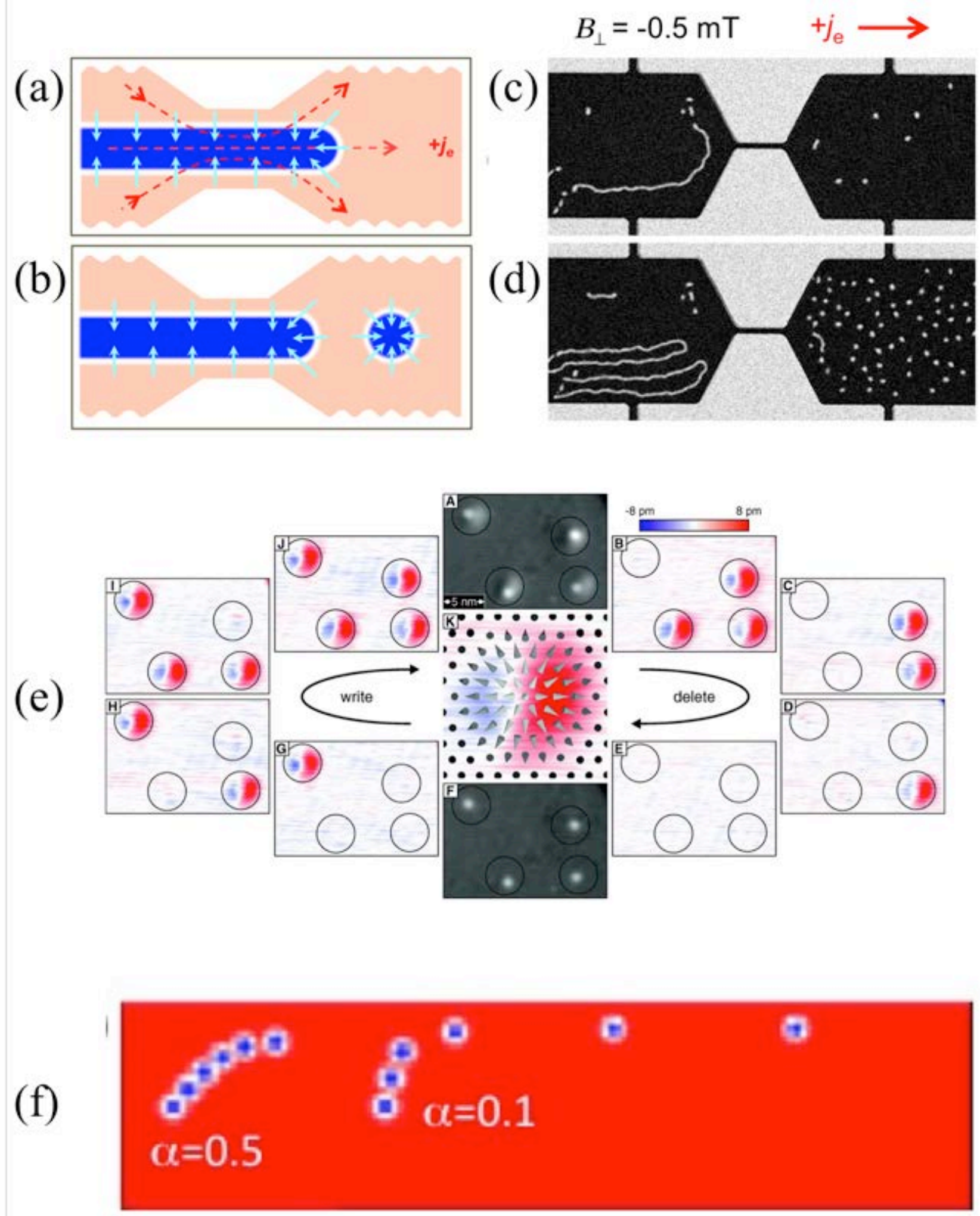

**Figure 21** Current induced nucleation and motion of skyrmions. (a), (b) Schematic of the transformation of stripe domains [dark blue extended areas], with chiral Néel domain walls (small light blue arrows) due to DM interactions, into magnetic skyrmions [circular domain in (b)] in perpendicularly-magnetized Ta/*a*-Co-Fe-B/*a*-$TaO_x$ due to a laterally-inhomogeneous in-plane charge current density [dashed red arrows in (a)] resulting in inhomogeneous spin-orbit torques. (c), (d) Initial and final state of skyrmion generation due to electric charge currents imaged magneto-optically [Jiang *et al.*, 2015]. (e) Scanning tunneling microscopy (STM) image of creation ("writing") and annihilation ("deleting") of individual skyrmions in Pd/Fe bilayer on Ir (111) substrate in an applied magnetic field of 3 T at 4 K, using local spin-polarized tunneling currents from the STM tip. Atomic defects in the film pin the skyrmions [Romming *et al.*, 2013]. (f) Micromagnetic simulations of the trajectory of a spin-orbit-torque-driven skyrmion starting from rest, for two values of damping parameter α, showing gyrotropic motion. The strip is 200 nm wide and six images are shown, every 20 ns, for each *α*.

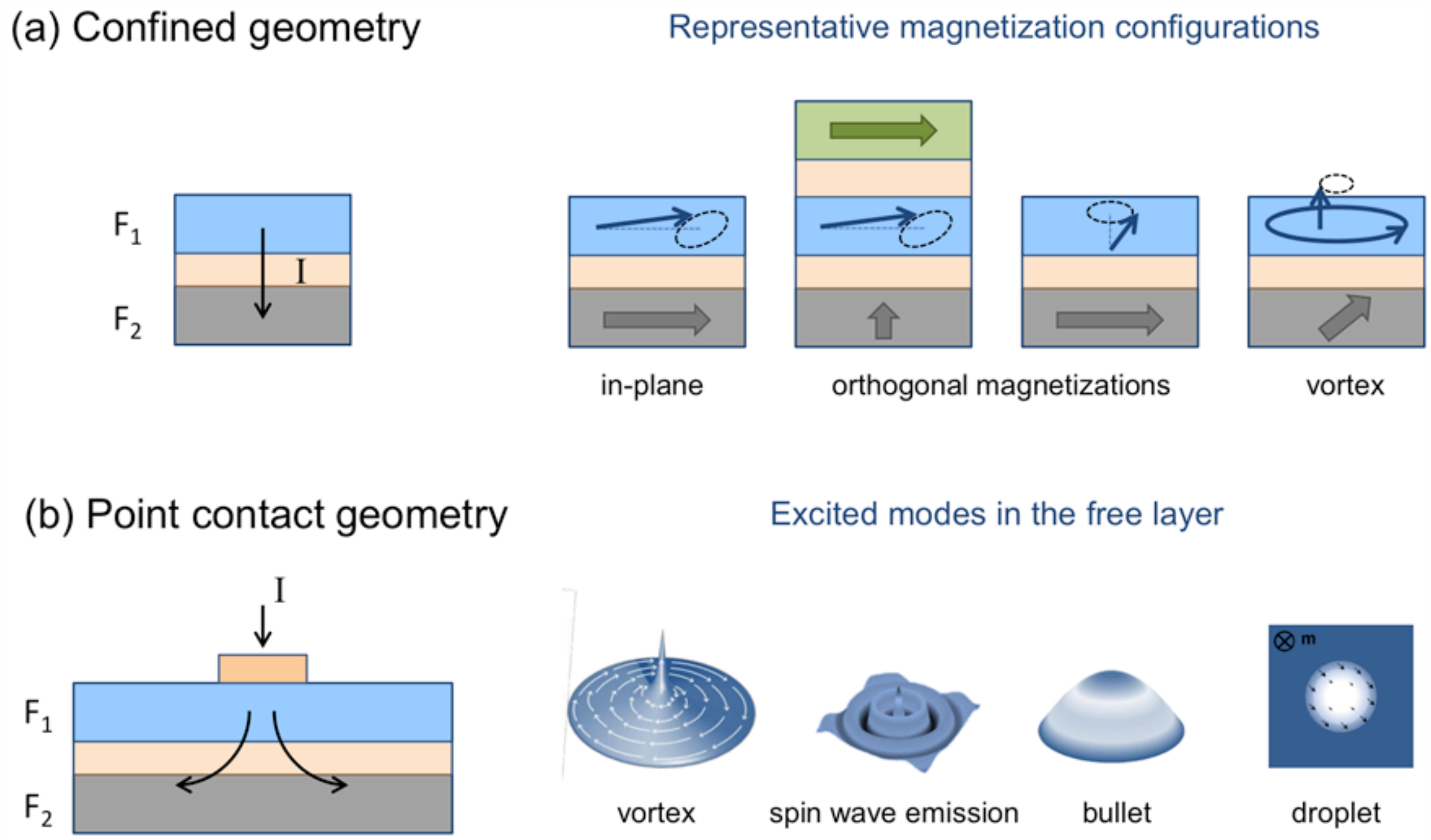

**Figure 22** Different types of spin-torque nano-oscillators. Top: confined geometries with a pillar structure. Bottom: geometries in which current from a point contact excites magnetic dynamics in an unpatterned magnetic free layer.

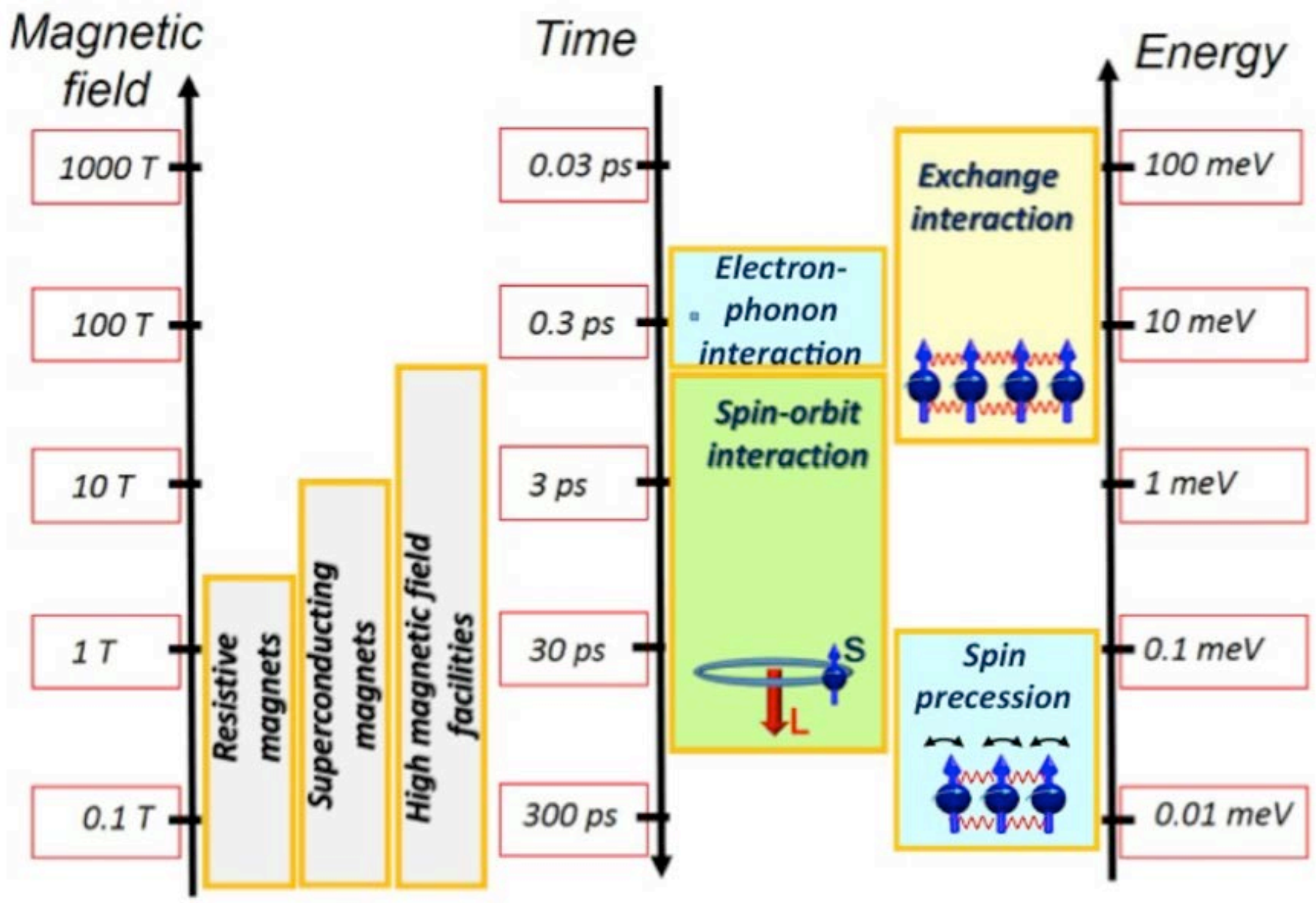

**Figure 23** Fundamental interaction energies and time scales relevant for ultrafast processes. The effective magnetic field associated with the exchange interaction reach 100 T to 1000 T; these fields correspond to the periods of the Larmor precession in the range 30 fs to 300 fs.

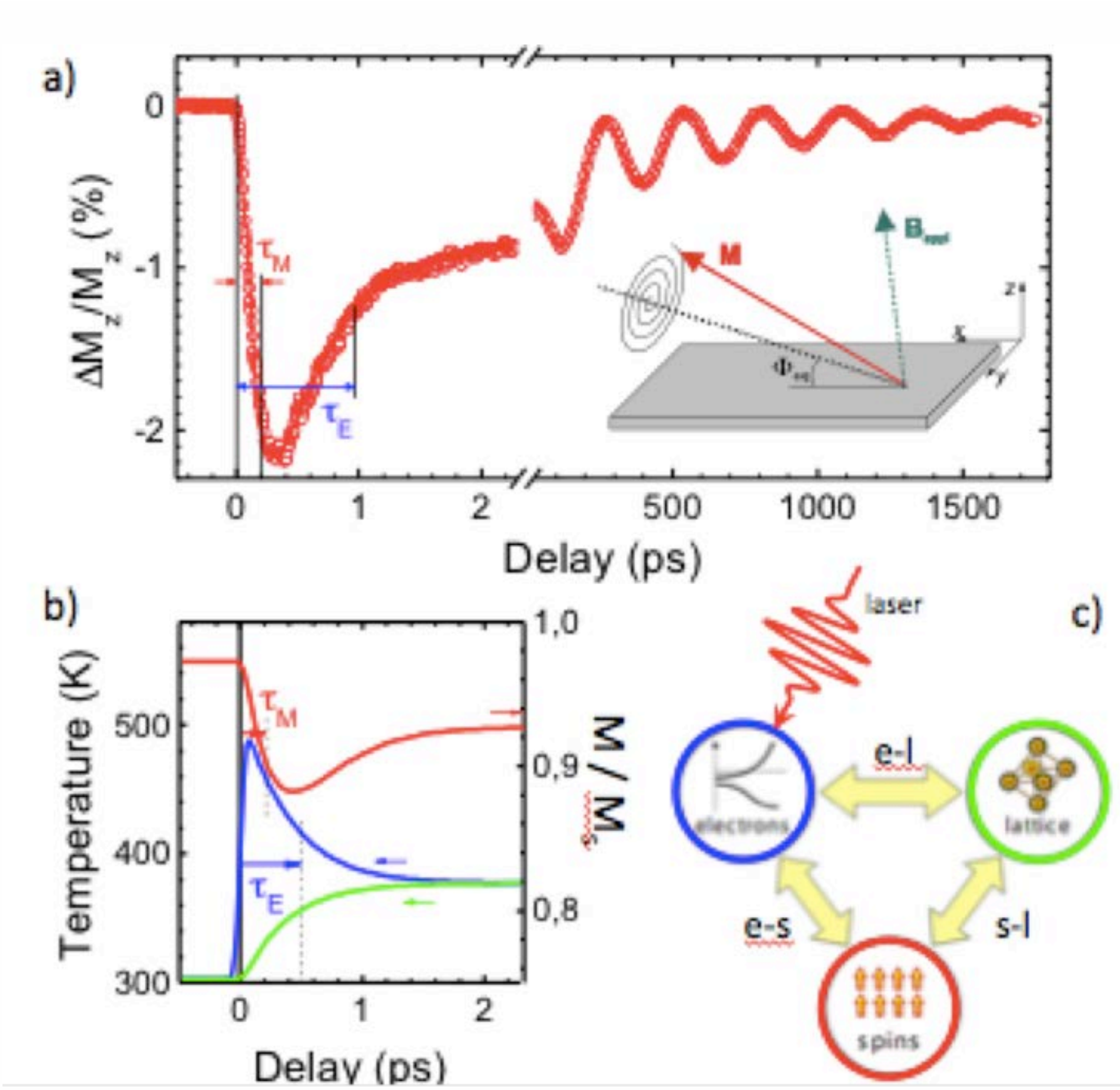

**Figure 24** (a) Time-resolved magneto-optic response of a nickel thin film after excitation by a femtosecond laser pulse [Koopmans *et al.*, 2005], showing partial loss of magnetic order at sub-picosecond timescales ($\tau_M$), followed by recovery due to electron-lattice equilibration ($\tau_E$). A field is applied out-of-plane to cant the magnetization (inset), leading to precessional dynamics at a slower time scale. (b) Example of simulation showing evolution of the magnetization (red, right axis), electron temperature (blue) and lattice temperature (green) [Koopmans *et al.*, 2010]. (c) Schematics of the three-temperature model, showing interactions between spinless electron gas (e), their spins (s) and the lattice (l) after the system is brought out of equilibrium by a laser pulse [Kirilyuk *et al.,* 2010].

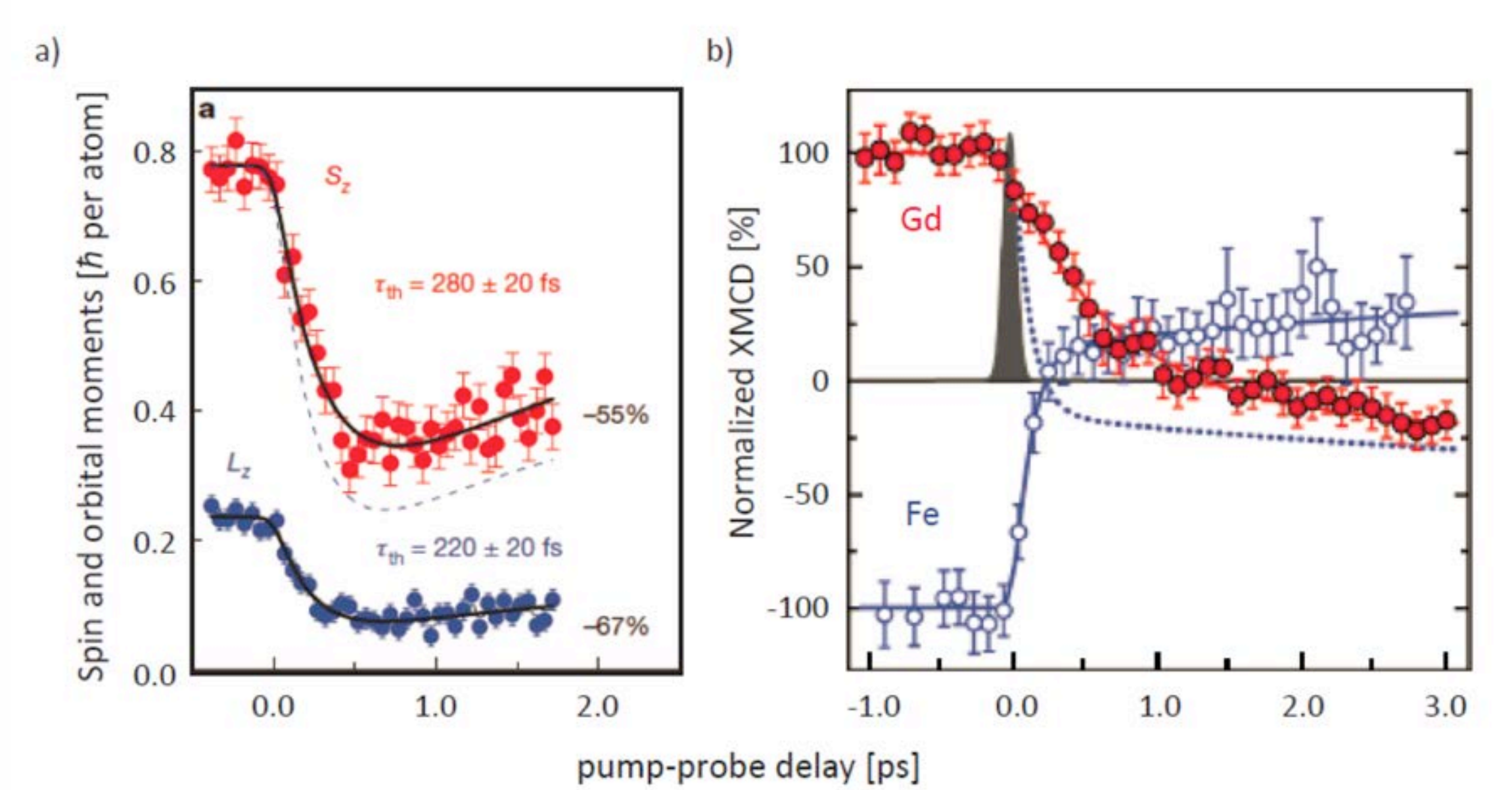

**Figure 25** Femtosecond x-ray pulses showing optically induced magnetization dynamics. (a) Decay of spin, $S_z$, and orbital, $L_z$, moments in CoPd films [Boeglin *et al*., 2010]. (b) Reversal of Gd 4f and Fe 3d magnetic moments in *a*-Gd-Fe-Co alloys [Radu *et al*., 2011].

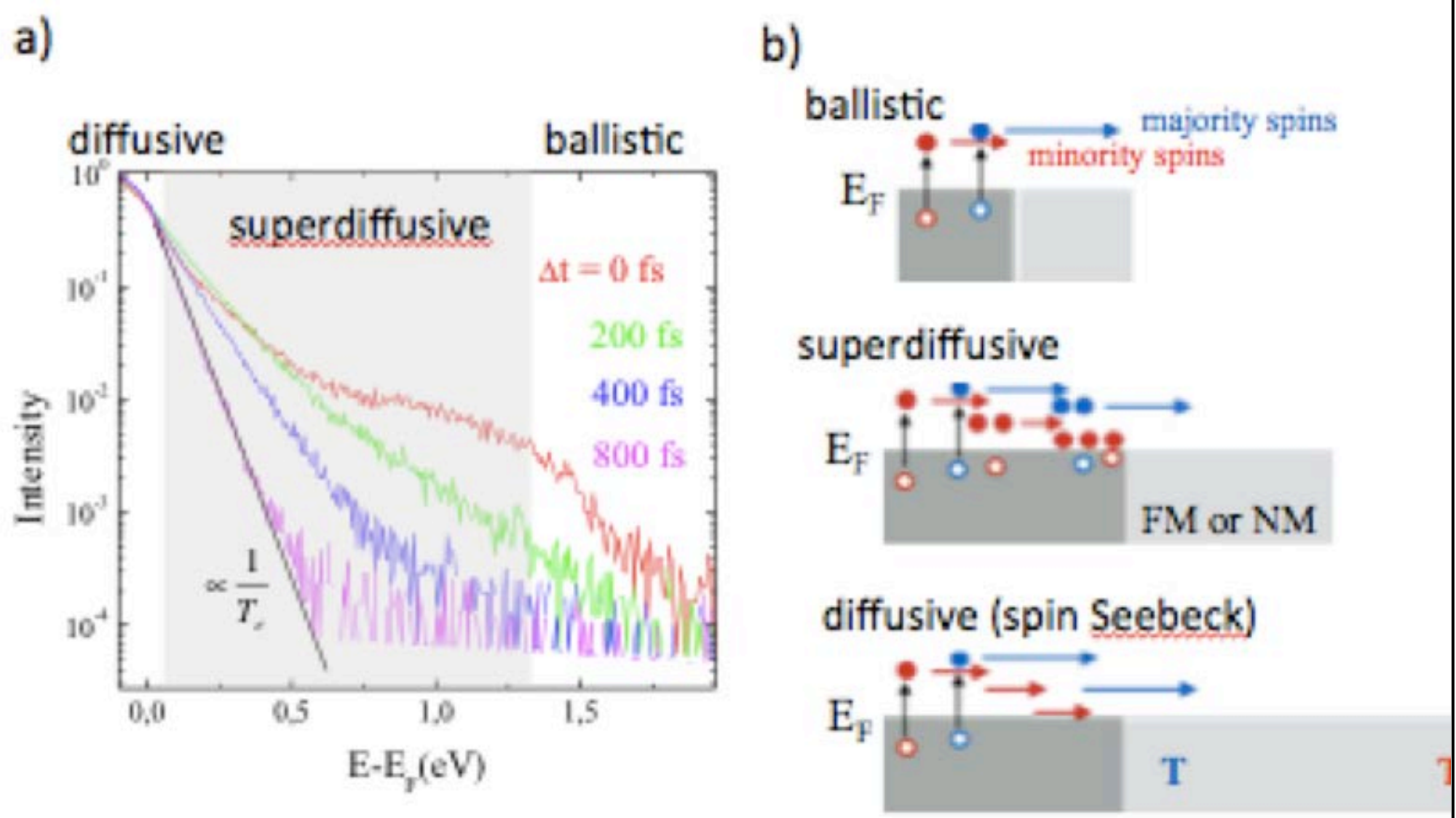

**Figure 26** (a) Time-resolved photoemission can probe the laser-excited hot electron distribution above the Fermi level, $E_F$. Thermalization implies that the electron distribution follows Fermi-Dirac statistics (solid line) and can be described by a temperature, $T_e$. Adapted from [Rhie *et al.*, 2003]. (b) Illustration of electronic scattering processes leading to decay of initially ballistic spin motion. Superdiffusive spin transport (indicated by grey shading in a)) occurs on sub-psec timescales during the crossover from ballistic to diffusive transport. The diffusive regime may also produce spin currents due to resulting temperature gradients (the spin-dependent Seebeck effect, associated with different chemical potentials of up and down spins).

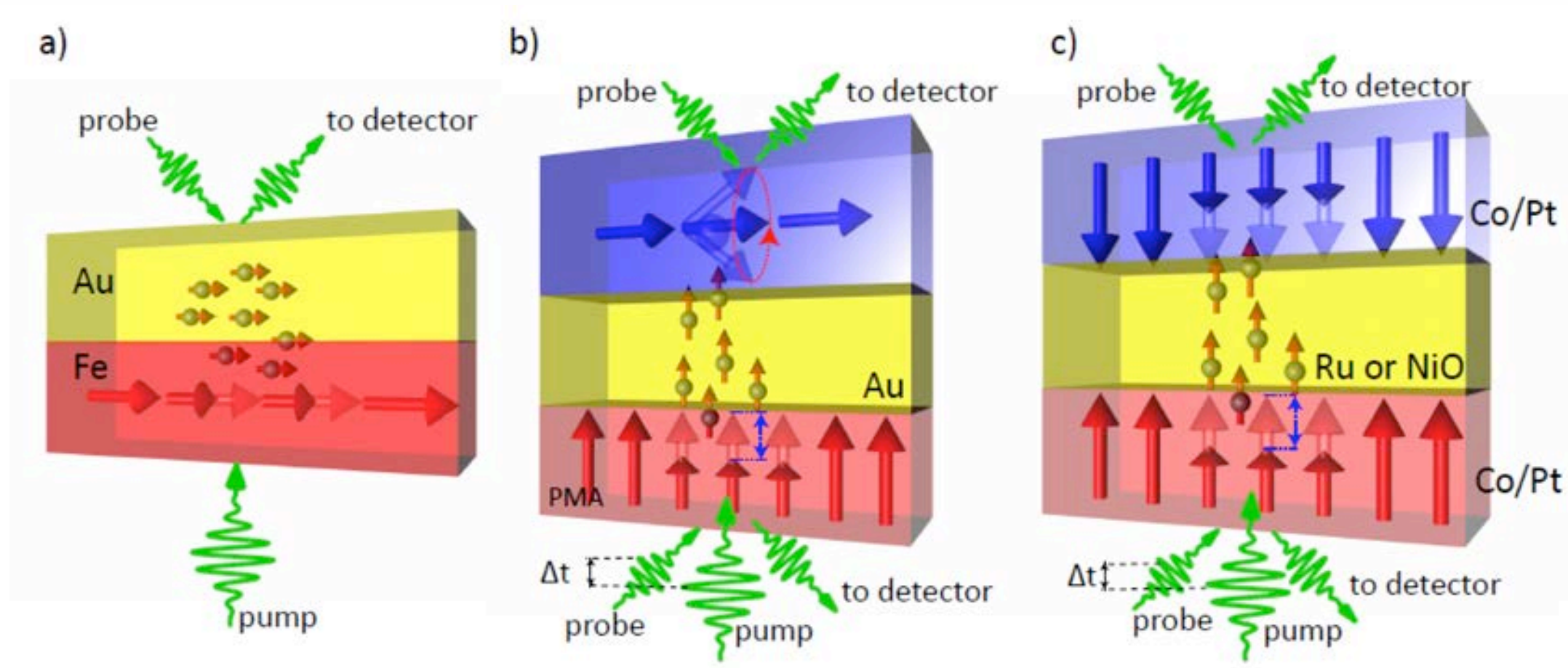

**Figure 27** Schematics of experiments showing superdiffusive spin currents. (a) Demagnetization of Fe film by an ultrashort pump pulse injects spin current into adjacent Au layer with its arrival at the Au backside detected by time-resolved magnetic second harmonic generation, discussed in [Melnikov *et al.*, 2011]. (b) Femtosecond demagnetization of the bottom ferromagnetic layer with perpendicular magnetic anisotropy results in spin-torque-induced precession dynamics in the top ferromagnetic layer driven by superdiffusive spin currents, discussed in [Schellekens *et al.,* 2014; Choi *et al.*, 2014]. (c) Optical pumping heterostructures where femtosecond demagnetization of the bottom Co/Pt multilayer affects demagnetization of the top Co/Pt multilayer for metallic (Ru) but not for insulating (NiO) spacer layers, discussed in [Malinowski *et al.,* 2008]; demagnetization also depends on the relative orientation of the two magnetic layers.

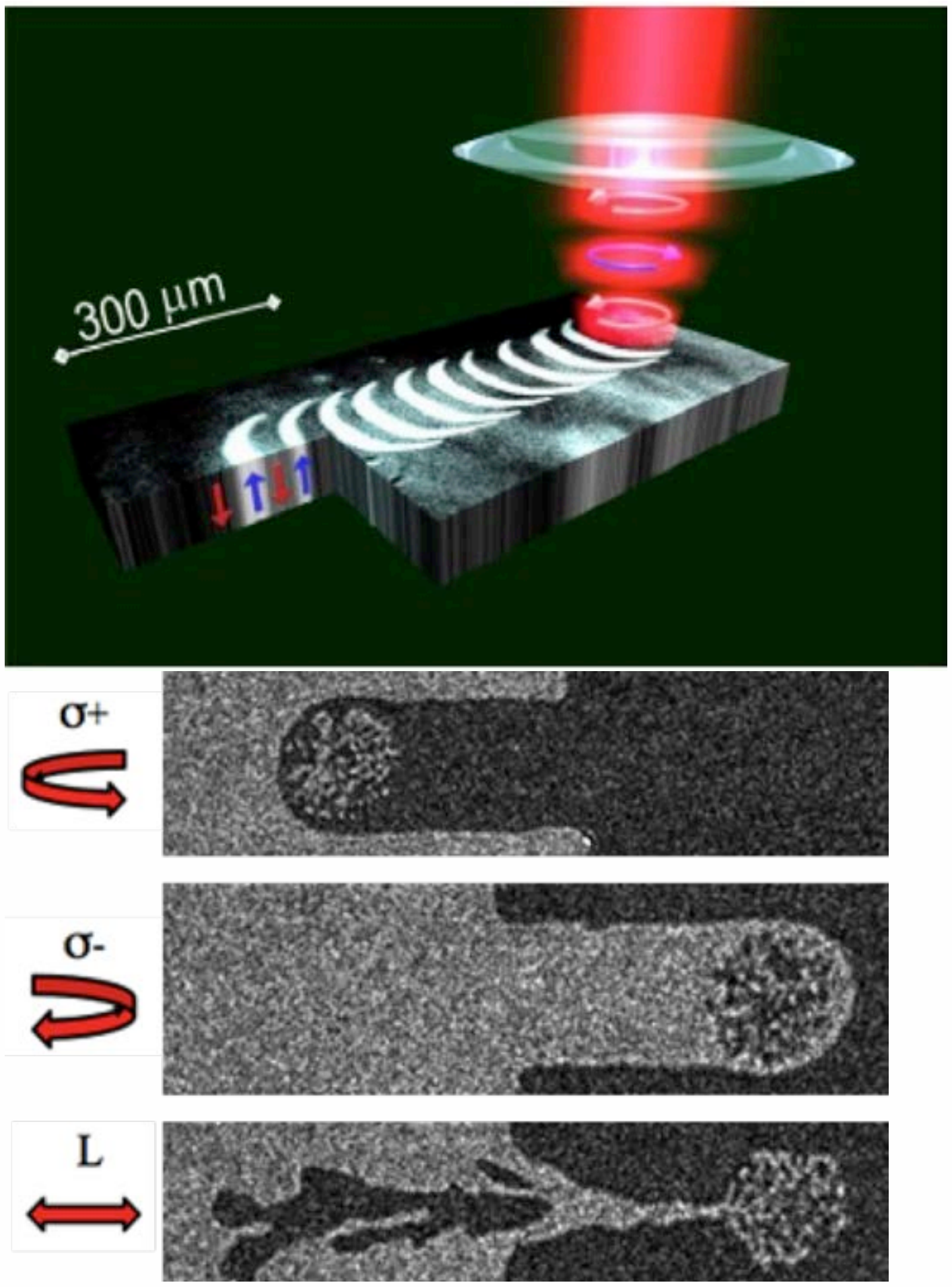

**Figure 28** All-optical switching: (a) scanning a laser beam across the sample and simultaneously modulating its polarization between left- and right-circular pulses yields a magnetic bit pattern in an amorphous Gd-Fe-Co alloy [Stanciu *et al.*, 2007]; (b) magneto-optical images of a $[Co(0.4\ nm)/Pt(0.7\ nm)]_3$ multilayer [Lambert *et al.*, 2014] where sweeping a pulsed laser source with circular polarization determines the final state's magnetic orientation.